\documentclass[iop]{emulateapj}
\usepackage{amsmath}	% Advanced maths commands
\usepackage{amssymb}
\usepackage{xspace}
\usepackage{epstopdf}
\usepackage{url}
\usepackage{array,multirow}

\usepackage{times}
\usepackage[english]{babel}

\usepackage{mathrsfs}
\usepackage{graphicx}
\usepackage{listings}
\usepackage[usenames]{color}

\usepackage{xcolor}
\definecolor{green}{HTML}{33CC33}
\definecolor{red}{HTML}{FF3300}
\definecolor{blue}{HTML}{3333FF}

\usepackage[colorlinks=true,citecolor=blue,urlcolor=green]{hyperref}

\usepackage{natbib,twoopt}
\usepackage[varg]{txfonts}
\usepackage[normalem]{ulem}

\usepackage{blindtext}
\usepackage{lipsum}

\usepackage{subfigure}
\usepackage[export]{adjustbox}
\usepackage[T1]{fontenc}
\usepackage{ae,aecompl}

\usepackage{threeparttable}
\usepackage{booktabs}

\newcommand{\ie}{i.e.\@\xspace} %id est
\newcommand{\eg}{e.g.\@\xspace} %id est
\renewcommand{\arraystretch}{1.}

\renewcommand{\eqref}[1]{Equation~\ref{#1}}
\newcommand{\fref}[1]{Figure~\ref{#1}}
\newcommand{\tref}[1]{Table~\ref{#1}}
\newcommand{\sref}[1]{Section~\ref{#1}}

\DeclareMathOperator\erf{erf}

\newcommand{\numax}{\ensuremath{\nu_{\rm max}}\xspace}
\newcommand{\dnu}{\ensuremath{\Delta\nu}\xspace}
\newcommand{\kp}{\emph{Kepler}\xspace}
\newcommand{\Kp}{\ensuremath{\rm Kp}\xspace}
\newcommand{\teff}{\ensuremath{T_{\rm eff}}\xspace}
\newcommand{\logg}{\ensuremath{\log g}\xspace}
\newcommand{\feh}{\ensuremath{\rm [Fe/H]}\xspace}

\numberwithin{equation}{section}

\makeatletter
\def\maketag@@@#1{\hbox{\m@th\normalfont\normalsize#1}}
\makeatother

\bibpunct{(}{)}{;}{a}{}{,} %% natbib format for A&A and ApJ

\makeatletter
\newcommand*\mysize{%
  \@setfontsize\mysize{5.7}{8.0}%
}
\makeatother

\makeatletter
\newcommand*\tabsize{%
  \@setfontsize\tabsize{7.}{8.0}%
}
\makeatother

\makeatletter
\newcommand\footnoteref[1]{\protected@xdef\@thefnmark{\ref{#1}}\@footnotemark}
\makeatother

\shorttitle{The \emph{Kepler} asteroseismic LEGACY dwarfs sample \textsc{I}}
\shortauthors{M. N. Lund et al.}

\begin{document}

\title{Standing on the shoulders of Dwarfs: the {\it Kepler} asteroseismic LEGACY sample \textsc{I} --- oscillation mode parameters\vspace*{0.3cm}}  

% The list of authors, and the short list which is used in the headers.
% If you need two or more lines of authors, add an extra line using \newauthor
\author{Mikkel~N.~Lund$^{1,2\star}$}
\author{V{\'{\i}}ctor~Silva~Aguirre$^2$}
\author{Guy~R.~Davies$^{1,2}$}
\author{William~J.~Chaplin$^{1,2}$}
\author{J\o rgen~Christensen-Dalsgaard$^2$}
\author{G\"{u}nter~Houdek$^2$}
\author{Timothy~R.~White$^2$}
\author{Timothy~R.~Bedding$^{3,2}$}
\author{Warrick~H.~Ball$^{4,5}$}
\author{Daniel~Huber$^{3,6,2}$}
\author{H.~M.~Antia$^7$}
\author{Yveline~Lebreton$^{8,9}$}
\author{David~W.~Latham$^{10}$}
\author{Rasmus~Handberg$^2$}
\author{Kuldeep Verma$^{7,2}$} 
\author{Sarbani~Basu$^{11}$}
\author{Luca~Casagrande$^{12}$}
\author{Anders~B.~Justesen$^2$}
\author{Hans~Kjeldsen$^2$}
\author{Jakob~R.~Mosumgaard$^2$\vspace*{0.5em}\\}

\affil{\scriptsize \begin{flushleft}$^1$School of Physics and Astronomy, University of Birmingham, Edgbaston, Birmingham, B15 2TT, UK; $^{\star}$\href{mailto:lundm@bison.ph.bham.ac.uk}{lundm@bison.ph.bham.ac.uk}\\
$^2$Stellar Astrophysics Centre, Department of Physics and Astronomy, Aarhus University, Ny Munkegade 120, DK-8000 Aarhus C, Denmark\\
$^3$Sydney Institute for Astronomy (SIfA), School of Physics, University of Sydney, NSW 2006, Australia\\
$^4$Institut f\"{u}r Astrophysik, Georg-August-Universit\"{a}t G\"{o}ttingen, Friedrich-Hund-Platz 1, 37077, G\"{o}ttingen, Germany\\
$^5$Max-Planck-Institut f\"{u}r Sonnensystemforschung, Justus-von-Liebig-Weg 3, 37077, G\"{o}ttingen, Germany\\
$^6$SETI Institute, 189 Bernardo Avenue, Mountain View, CA 94043, USA\\
$^7$Tata Institute of Fundamental Research, Homi Bhabha Road, Mumbai 400005, India\\
$^8$Observatoire de Paris, GEPI, CNRS UMR 8111, F-92195 Meudon, France\\
$^9$Institut de Physique de Rennes, Universit\'{e} de Rennes 1, CNRS UMR 6251, F-35042 Rennes, France\\
$^{10}$Harvard-Smithsonian Center for Astrophysics, 60 Garden Street Cambridge, MA 02138 USA\\
$^{11}$Department of Astronomy, Yale University, PO Box 208101, New Haven, CT 06520-8101, USA\\
$^{12}$Research School of Astronomy and Astrophysics, Mount Stromlo Observatory, The Australian National University, ACT 2611, Australia 
\end{flushleft}}

%%%%%%%%%%%%%%%%%%%%%%%%%%%%%%%%%%%%%%%%%%%%%%%%%%%%%%%%%%%%%%%%%%%%%%%%%%%%%%%%%%%%%%%%%%%%%%%%%%%%%%%%%%%%%%%%%%%%%%%%%%
%%%%%%%%%%%%%%%%%%%%%%%%%%%%%%%%%%%%%%%%%%%%%%%%%%%%%%%%%%%%%%%%%%%%%%%%%%%%%%%%%%%%%%%%%%%%%%%%%%%%%%%%%%%%%%%%%%%%%%%%%%

\begin{abstract}
The advent of space-based missions like \kp has revolutionized the study of solar-type stars, particularly through the measurement and modeling of their resonant modes of oscillation. Here we analyze a sample of 66 \kp main-sequence stars showing solar-like oscillations as part of the \kp seismic LEGACY project. We use \kp short-cadence data, of which each star has at least 12 months, to create frequency power spectra optimized for asteroseismology. 
For each star we identify its modes of oscillation and extract parameters such as frequency, amplitude, and line width using a Bayesian Markov chain Monte Carlo `peak-bagging' approach.   
We report the extracted mode parameters for all 66 stars, as well as derived quantities such as frequency difference ratios, the large and small separations \dnu and $\delta\nu_{02}$; the behavior of line widths with frequency and line widths at \numax with \teff, for which we derive parametrizations; and behavior of mode visibilities. These average properties can be applied in future peak-bagging exercises to better constrain the parameters of the stellar oscillation spectra.
The frequencies and frequency ratios can tightly constrain the fundamental parameters of these solar-type stars, and mode line widths and amplitudes can test models of mode damping and excitation.    
\end{abstract}

%%%%%%%%%%%%%%%%%%%%%%%%%%%%%%%%%%%%%%%%%%%%%%%%%%%%%%%%%%%%%%%%%%%%%%%%%%%%%%%%%%%%%%%%%%%%%%%%%%%%%%%%%%%%%%%%%%%%%%%%%%
%%%%%%%%%%%%%%%%%%%%%%%%%%%%%%%%%%%%%%%%%%%%%%%%%%%%%%%%%%%%%%%%%%%%%%%%%%%%%%%%%%%%%%%%%%%%%%%%%%%%%%%%%%%%%%%%%%%%%%%%%%


%%%%%%%%%%%%%%%%%%%%%%%%%%%%%%%%%%%%%%%%%%%%%%%%%%%%%%%%%%%%%%%%%%%%%%%%%%%%%%%%%%%%%%%%%%%%%%%%%%%%%%%%%%%%%%%%%%%%%%%%%%
%%%%%%%%%%%%%%%%%%%%%%%%%%%%%%%%%%%%%%%%%%%%%%%%%%%%%%%%%%%%%%%%%%%%%%%%%%%%%%%%%%%%%%%%%%%%%%%%%%%%%%%%%%%%%%%%%%%%%%%%%%
%%%%%%%%%%%%%%%%%%%%%%%%%%%%%%%%%%%%%%%%%%%%%%%%%%%%%%%%%%%%%%%%%%%%%%%%%%%%%%%%%%%%%%%%%%%%%%%%%%%%%%%%%%%%%%%%%%%%%%%%%%
%%%%%%%%%%%%%%%%%%%%%%%%%%%%%%%%%%%%%%%%%%%%%%%%%%%%%%%%%%%%%%%%%%%%%%%%%%%%%%%%%%%%%%%%%%%%%%%%%%%%%%%%%%%%%%%%%%%%%%%%%%
\section{Introduction}
\label{sec:intro}

The study of stars and extrasolar planets via the properties of their host stars has experienced a revolution in recent years \citep[][]{2013ARA&A..51..353C,2016arXiv160206838C}. This largely arose from the successful application of asteroseismology using observations from the CoRoT \citep[][]{2009IAUS..253...71B} and \kp missions \citep[][]{2010PASP..122..131G}. This application has been made possible by extracting high-precision parameters from the stellar frequency-power spectra owing to the long time-baseline and photometric quality of these space missions.

Asteroseismology allows the determination of fundamental stellar parameters such as mass, radius, and age through modeling of individual mode frequencies or frequency-difference ratios. The \kp mission has already provided stellar parameters for a number of stars, including planetary hosts, using average seismic parameters \citep[][]{2011Sci...332..213C,2014ApJS..210....1C,2012ApJ...757...99S,2013ApJ...767..127H}, individual frequencies \citep[][]{2010ApJ...713L.164C,2010ApJ...710.1596B,2012ApJ...746..123H,2012ApJ...748L..10M,2014ApJS..214...27M,2015ApJ...811L..37M,2014ApJ...782...14V,2014A&A...570A..54L,2015ApJ...799..170C,2015MNRAS.452.2127S}, and frequency-difference ratios \citep[][]{2013ApJ...769..141S,2015MNRAS.452.2127S,2014A&A...569A..21L}.

The high precision of extracted mode frequencies further allows the study of ionization zones and the convective envelope boundary from acoustic glitches \citep[][]{2007MNRAS.375..861H,2011A&A...529A..63S,2014ApJ...782...18M,2014ApJ...794..114V}, and one may also learn about the physics of the excitation and damping of the oscillation modes from measured mode line widths, amplitudes, and visibilities \citep[][]{1999A&A...351..582H,2006astro.ph.12024H,2005JApA...26..171S,2007A&A...463..297S,2012A&A...540L...7B}.

Frequencies and line widths have been reported for several solar-like and subgiant stars observed by \kp by \citet[][]{2012A&A...537A.134A,2012A&A...543A..54A,2014A&A...566A..20A}, and planet-hosting stars by \citet[][]{2016MNRAS.456.2183D}.
In this paper we analyze a sample of 66 main-sequence (MS) solar-like stars observed for at least 12 months by the \kp mission.
We extracted mode parameters by `peak-bagging'\footnote{First coined by Roger Ulrich circa 1983 (private communication) from the analogy to hill climbing where it refers to reaching the summits of a collection of peaks, the term was later re-introduced by \citet[][]{2003Ap&SS.284..109A}.} the frequency-power spectra of the stars using a Markov chain Monte Carlo (MCMC) method \citep[][]{2011A&A...527A..56H} and used the Bayesian quality control presented by \citet{2016MNRAS.456.2183D}. For each star we report values for the mode frequencies, amplitudes, line widths, and visibilities. Additionally, we provide summary descriptions for each of the above quantities, such as average seismic parameters derived from the frequencies and prescriptions of the mode line widths against frequency. The frequencies reported here will be modeled in the accompanying paper by \citet[][hereafter Paper~II]{2016arXiv161108776S}.
The lessons learned from the presented analysis will be useful for the study of MS solar-like oscillators with the TESS \citep[][]{2014SPIE.9143E..20R} and PLATO \citep[][]{2013arXiv1310.0696R} missions, and the continued analysis of these data from K2 \citep[][]{2015PASP..127.1038C,2016arXiv160807292L,2016MNRAS.tmp.1271L}.

The paper is structured as follows: \sref{sec:sample} describes the target sample, including the preparation of \kp data and spectroscopic properties. \sref{sec:param_est} is devoted to the parameter estimation from the MCMC peak-bagging, including a description of the fitting strategy, the adopted Bayesian quality assurance, and the derivation of frequency difference ratios and their correlations. In \sref{sec:res} we present our results from the peak-bagging for the mode frequencies, focusing specifically on frequency errors and average seismic parameters in \sref{sec:fre}; amplitudes in \sref{sec:ampl}; line widths in \sref{sec:lws}; and visibilities in \sref{sec:visibilities}.
In \sref{sec:exout} we give an example of the output generated for each of the analyzed stars. We conclude in \sref{sec:con}.

%%%%%%%%%%%%%%%%%%%%%%%%%%%%%%%%%%%%%%%%%%%%%%%%%%%%%%%%%%%%%%%%%%%%%%%%%%%%%%%%%%%%%%%%%%%%%%%%%%%%%%%%%%%%%%%%%%%%%%%%%%
%%%%%%%%%%%%%%%%%%%%%%%%%%%%%%%%%%%%%%%%%%%%%%%%%%%%%%%%%%%%%%%%%%%%%%%%%%%%%%%%%%%%%%%%%%%%%%%%%%%%%%%%%%%%%%%%%%%%%%%%%%
%%%%%%%%%%%%%%%%%%%%%%%%%%%%%%%%%%%%%%%%%%%%%%%%%%%%%%%%%%%%%%%%%%%%%%%%%%%%%%%%%%%%%%%%%%%%%%%%%%%%%%%%%%%%%%%%%%%%%%%%%%
%%%%%%%%%%%%%%%%%%%%%%%%%%%%%%%%%%%%%%%%%%%%%%%%%%%%%%%%%%%%%%%%%%%%%%%%%%%%%%%%%%%%%%%%%%%%%%%%%%%%%%%%%%%%%%%%%%%%%%%%%%
\section{Target sample}
\label{sec:sample}

\begin{figure}
\centering
\includegraphics[width=\columnwidth]{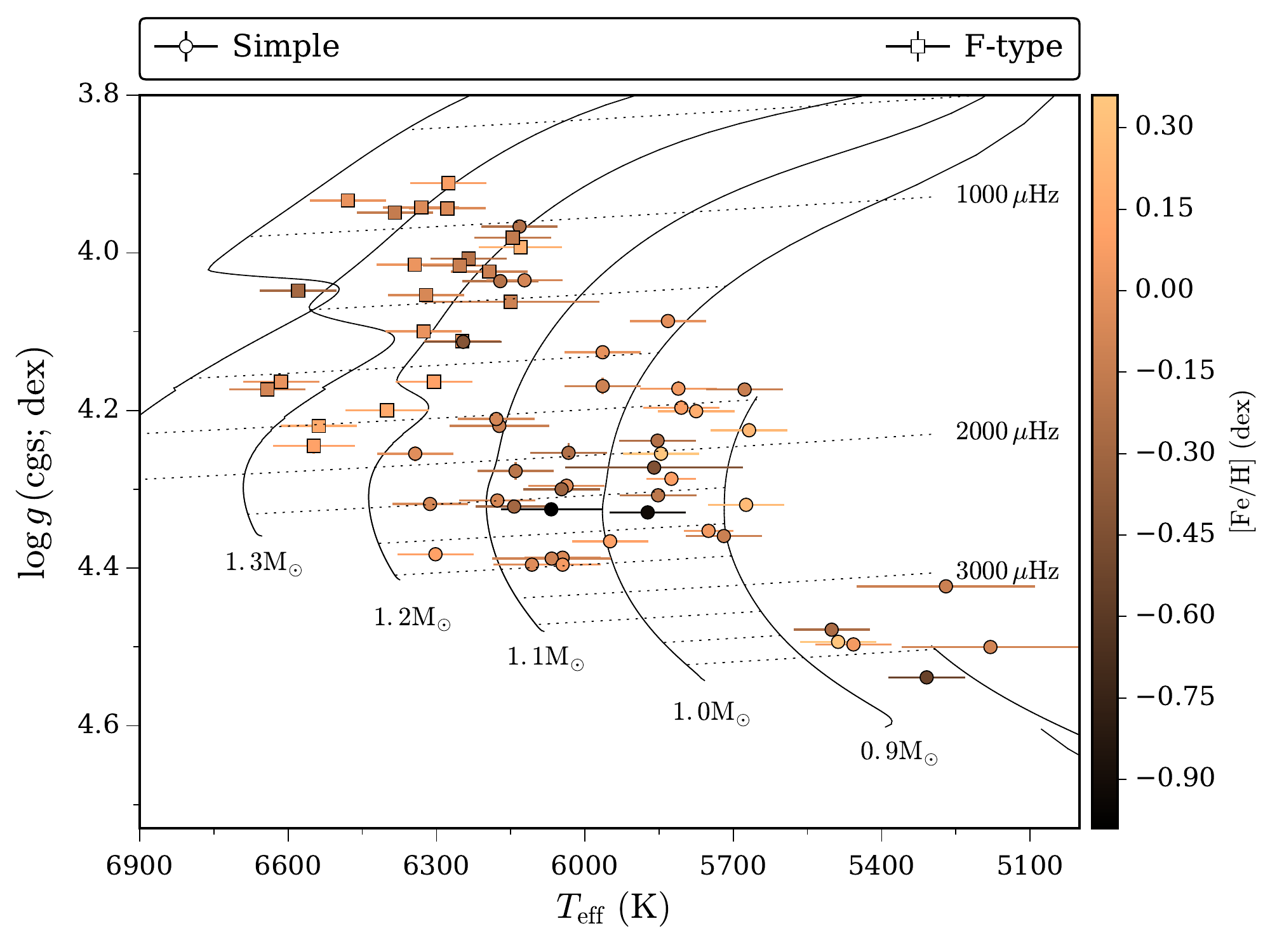}
\caption{Kiel-diagram of sample stars, with \teff and \feh from the spectroscopic input (\tref{tab:target_table}) and $\log g$ from the modeling in Paper~II. Stellar evolutionary tracks have been calculated using the Garching Stellar Evolution Code \citep[GARSTEC;][]{2008Ap&SS.316...99W} adopting $\rm [Fe/H]=0$. The marker type indicates whether the star is considered to be of ``Simple'' or ``F-type'' character according to \citet[][]{2012A&A...543A..54A}. The dotted lines show lines of constant \numax, according to the simple scaling as $\numax \propto g \sqrt{\teff}$, in steps of $250\, \mu \rm Hz$.}
\label{fig:HRdia_grid}
\end{figure}

Our sample consists of 66 solar-type oscillators observed by the \kp satellite, all part of the KASC \citep{2010AN....331..966K} working group 1 (WG1) sample of solar-like p-mode oscillators. All stars have short-cadence (SC; $\Delta t=58.89\, \rm s$) observations with an observing base line of at least 12 months, and represent some of the highest signal-to-noise solar-like oscillators observed by \kp. The sample consists only of main-sequence (MS) and slightly more evolved subgiant stars. These have frequency structures corresponding to the `Simple' or `F-type' categories by \citet[][]{2012A&A...543A..54A}, \ie, none of the stars show obvious bumped dipole modes.
The sample was peak-bagged as part of the \kp dwarf seismic `LEGACY' project, with the asteroseismic modeling of extracted parameters presented in Paper~II. In \fref{fig:HRdia_grid} the sample is shown in a Kiel-diagram (\teff vs. \logg), with parameters adopted from Paper~II; for additional details on the sample see \tref{tab:target_table}.
We note that all targets from the \citet[][]{2014arXiv1403.7046A} study of oscillation mode line widths (which included data up to Quarter 12) are part of our sample, with the exception of KIC 3424541, 3733735, 10355856, and 10909629. These four stars were classified as F-type by \citet[][]{2014arXiv1403.7046A}, but were omitted from our sample because of possible mixed-mode structures.

%%%%%%%%%%%%%%%%%%%%%%%%%%%%%%%%%%%%%%%%%%%%%%%%%%%%%%%%%%%%%%%
%%%%%%%%%%%%%%%%%%%%%%%%%%%%%%%%%%%%%%%%%%%%%%%%%%%%%%%%%%%%%%

%%%%%%%%%%%%%%%%%%%%%%%%%%%%%%%%%%%%%%%%%%%%%%%%%%%%%%%%%%%%%%%%%%%%%%%%%%%%%%%%%%%%%%%%%%%%%%%%%%%%%%%%%%%%%%%%%%%%%%%%%%
%%%%%%%%%%%%%%%%%%%%%%%%%%%%%%%%%%%%%%%%%%%%%%%%%%%%%%%%%%%%%%%%%%%%%%%%%%%%%%%%%%%%%%%%%%%%%%%%%%%%%%%%%%%%%%%%%%%%%%%%%%
%%%%%%%%%%%%%%%%%%%%%%%%%%%%%%%%%%%%%%%%%%%%%%%%%%%%%%%%%%%%%%%%%%%%%%%%%%%%%%%%%%%%%%%%%%%%%%%%%%%%%%%%%%%%%%%%%%%%%%%%%%

% Table made Wed Sep 28 16:52:24 2016
\begin{table*} 
\resizebox*{!}{\textheight}{% 
\begin{threeparttable} 
\centering 
\caption{Parameters for the targets in the studied sample. Metallicities, temperatures, line-of-sight (LOS) velocities, and $v\sin i$ values are adopted from the SPC analysis of the targets unless otherwise indicated (see table notes). 
    Systematic uncertainties of $\pm 59\, \rm K$ ($T_{\rm eff}$) and $\pm 0.062 \, \rm dex$ ([Fe/H]) have been added in quadrature as 
    suggested by \citet[][]{2012ApJ...757..161T}. Values for $\log g$ are adopted from Paper~II. The table lists the KIC value and other popular names given, if any; 
    \kp magnitude (Kp); frequency of maximum amplitude (\numax) and large separation (\dnu) --- for uncertainties in \numax and \dnu see \tref{tab:asym_fit_table}; 
    number of peak-bagged modes; the category according to \citet[][]{2012A&A...543A..54A}; first-last quartes during which the targets were observed in SC, and which quarters were missing in-between.} 
\label{tab:target_table} 
\begin{tabular}{@{}r@{\hskip 2ex}c@{\hskip 2ex}r@{\hskip 7ex}r@{\hskip 7ex}r@{\hskip 2ex}ccc@{\hskip 2ex}c@{\hskip 2ex}l@{\hskip 5ex}r@{\hskip 5ex}r@{\hskip 8ex}r@{\hskip 5ex}r@{}} 
\toprule 
\multicolumn{1}{@{\hskip -2ex}c}{KIC} & Popular & \multicolumn{1}{@{\hskip -4ex}c}{Kp}  & \multicolumn{1}{@{\hskip -5ex}c}{\numax}  & \multicolumn{1}{@{\hskip -2ex}c}{\dnu} & \multicolumn{1}{c}{Number of} & Category & Braketing & Missing & \multicolumn{1}{@{\hskip -5ex}c}{\teff} & \multicolumn{1}{@{\hskip -2ex}c}{[Fe/H]} & \multicolumn{1}{@{\hskip -2ex}c}{\logg} & \multicolumn{1}{@{\hskip -2ex}c}{LOS} & \multicolumn{1}{c}{$v\sin i$}\\  & name & \multicolumn{1}{@{\hskip -4ex}c}{(mag)} & \multicolumn{1}{@{\hskip -5ex}c}{($\rm \mu Hz$)} & \multicolumn{1}{@{\hskip -2ex}c}{($\rm \mu Hz$)} & \multicolumn{1}{c}{modes} &  & Quarters & Quarters & \multicolumn{1}{@{\hskip -5ex}c}{(K)} & \multicolumn{1}{@{\hskip -2ex}c}{(dex)} & \multicolumn{1}{@{\hskip -2ex}c}{(cgs; dex)} & \multicolumn{1}{@{\hskip -2ex}c}{($\rm km\, s^{-1}$)} & \multicolumn{1}{c}{($\rm km\, s^{-1}$)}\\ 
\midrule 
$1435467$ &       & $8.88$ & $1407$ & $70.4$ & $46   $ & F-like & $\rm 5.1-17.2$ & $\rm    $ & $6326 \pm 77$ & $0.01 \pm 0.10$ & $4.100$\rlap{$_{-0.009}^{+0.009}$} &  $-66.52 \pm 0.10$ & $11.90 \pm 0.50$ \\[0.5ex] 
$2837475$ &       & $8.48$ & $1558$ & $75.7$ & $54   $ & F-like & $\rm 5.1-17.2$ & $\rm    $ & $6614 \pm 77$ & $0.01 \pm 0.10$ & $4.163$\rlap{$_{-0.007}^{+0.007}$} &  $-13.28 \pm 0.10$ & $23.30 \pm 0.50$ \\[0.5ex] 
$3427720$ &       & $9.11$ & $2737$ & $120.1$ & $36   $ & Simple & $\rm 5.1-17.2$ & $\rm    $ & $6045 \pm 77$ & $-0.06 \pm 0.10$ & $4.387$\rlap{$_{-0.005}^{+0.004}$} &  $-22.91 \pm 0.10$ & $2.90 \pm 0.50$ \\[0.5ex] 
$3456181$ &       & $9.66$ & $970$ & $52.3$ & $45   $ & F-like & $\rm 5.1-11.3$ & $\rm 6,10$ & $6384 \pm 77$ & $-0.15 \pm 0.10$ & $3.950$\rlap{$_{-0.007}^{+0.005}$} &  $-50.12 \pm 0.10$ & $8.50 \pm 0.50$ \\[0.5ex] 
$3632418$ & Cassie & $8.22$ & $1167$ & $60.7$ & $54   $ & F-like & $\rm 5.1-17.2$ & $\rm    $ & $6193 \pm 77$ & $-0.12 \pm 0.10$ & $4.024$\rlap{$_{-0.007}^{+0.005}$} &  $-19.11 \pm 0.10$ & $8.50 \pm 0.50$ \\[0.5ex] 
$3656476$ & Java  & $9.52$ & $1925$ & $93.2$ & $38   $ & Simple & $\rm 5.1-17.2$ & $\rm 6,10,14$ & $5668 \pm 77$ & $0.25 \pm 0.10$ & $4.225$\rlap{$_{-0.008}^{+0.010}$} &  $-13.29 \pm 0.10$ & $2.30 \pm 0.50$ \\[0.5ex] 
$3735871$ &       & $9.71$ & $2863$ & $123.0$ & $34   $ & Simple & $\rm 5.1-17.2$ & $\rm    $ & $6107 \pm 77$ & $-0.04 \pm 0.10$ & $4.396$\rlap{$_{-0.007}^{+0.007}$} &  $6.90 \pm 0.10$ & $4.80 \pm 0.50$ \\[0.5ex] 
$4914923$ & Vitto & $9.46$ & $1817$ & $88.5$ & $38   $ & Simple & $\rm 5.1-17.2$ & $\rm 6$ & $5805 \pm 77$ & $0.08 \pm 0.10$ & $4.197$\rlap{$_{-0.008}^{+0.010}$} &  $-39.16 \pm 0.10$ & $3.40 \pm 0.50$ \\[0.5ex] 
$5184732$ & Kitty & $8.16$ & $2089$ & $95.5$ & $49   $ & Simple & $\rm 7.1-17.2$ & $\rm    $ & $5846 \pm 77$ & $0.36 \pm 0.10$ & $4.255$\rlap{$_{-0.010}^{+0.008}$} &  $15.41 \pm 0.10$ & $4.00 \pm 0.50$ \\[0.5ex] 
$5773345$ &       & $9.16$ & $1101$ & $57.3$ & $45   $ & F-like & $\rm 6.1-11.3$ & $\rm 9$ & $6130 \pm 84$\textsuperscript{\raisebox{1pt}{\rlap{5}}} & $0.21 \pm 0.09$\textsuperscript{\raisebox{1pt}{\rlap{5}}} & $3.993$\rlap{$_{-0.008}^{+0.007}$} &    &    \\[0.5ex] 
$5950854$ &       & $10.96$ & $1927$ & $96.6$ & $26   $ & Simple & $\rm 5.1-10.3$ & $\rm 6,7.2$ & $5853 \pm 77$ & $-0.23 \pm 0.10$ & $4.238$\rlap{$_{-0.007}^{+0.007}$} &  $-42.49 \pm 0.10$ & $3.00 \pm 0.50$ \\[0.5ex] 
$6106415$ & Perky & $7.18$ & $2249$ & $104.1$ & $49   $ & Simple & $\rm 6.1-16.3$ & $\rm 9,13$ & $6037 \pm 77$ & $-0.04 \pm 0.10$ & $4.295$\rlap{$_{-0.009}^{+0.009}$} &  $-14.80 \pm 0.10$ & $4.90 \pm 0.50$ \\[0.5ex] 
$6116048$ & Nunny & $8.42$ & $2127$ & $100.8$ & $49   $ & Simple & $\rm 5.1-17.2$ & $\rm    $ & $6033 \pm 77$ & $-0.23 \pm 0.10$ & $4.254$\rlap{$_{-0.009}^{+0.012}$} &  $-53.26 \pm 0.10$ & $4.00 \pm 0.50$ \\[0.5ex] 
$6225718$ & Saxo2 & $7.50$ & $2364$ & $105.7$ & $59   $ & Simple & $\rm 6.1-17.2$ & $\rm    $ & $6313 \pm 77$ & $-0.07 \pm 0.10$ & $4.319$\rlap{$_{-0.005}^{+0.007}$} &  $-1.32 \pm 0.10$ & $5.50 \pm 0.50$ \\[0.5ex] 
$6508366$ & Baloo & $8.97$ & $958$ & $51.6$ & $50   $ & F-like & $\rm 5.1-17.2$ & $\rm    $ & $6331 \pm 77$ & $-0.05 \pm 0.10$ & $3.942$\rlap{$_{-0.005}^{+0.007}$} &  $2.62 \pm 0.10$ & $22.50 \pm 0.50$ \\[0.5ex] 
$6603624$ & Saxo  & $9.09$ & $2384$ & $110.1$ & $44   $ & Simple & $\rm 5.1-17.2$ & $\rm    $ & $5674 \pm 77$ & $0.28 \pm 0.10$ & $4.320$\rlap{$_{-0.004}^{+0.005}$} &  $-58.82 \pm 0.10$ & $0.70 \pm 0.50$ \\[0.5ex] 
$6679371$ &       & $8.73$ & $942$ & $50.6$ & $55   $ & F-like & $\rm 5.1-17.2$ & $\rm    $ & $6479 \pm 77$ & $0.01 \pm 0.10$ & $3.934$\rlap{$_{-0.007}^{+0.008}$} &  $-23.58 \pm 0.10$ & $17.30 \pm 0.50$ \\[0.5ex] 
$6933899$ & Fred  & $9.62$ & $1390$ & $72.1$ & $39   $ & Simple & $\rm 5.1-17.2$ & $\rm    $ & $5832 \pm 77$ & $-0.01 \pm 0.10$ & $4.079$\rlap{$_{-0.008}^{+0.009}$} &  $-6.97 \pm 0.10$ & $3.60 \pm 0.50$ \\[0.5ex] 
$7103006$ &       & $8.86$ & $1168$ & $59.7$ & $54   $ & F-like & $\rm 5.1-17.2$ & $\rm    $ & $6344 \pm 77$ & $0.02 \pm 0.10$ & $4.015$\rlap{$_{-0.007}^{+0.007}$} &  $-22.36 \pm 0.10$ & $12.10 \pm 0.50$ \\[0.5ex] 
$7106245$ &       & $10.79$ & $2398$ & $111.4$ & $24   $ & Simple & $\rm 5.1-15.3$ & $\rm    $ & $6068 \pm 102$\textsuperscript{\raisebox{1pt}{\rlap{3}}} & $-0.99 \pm 0.19$\textsuperscript{\raisebox{1pt}{\rlap{3}}} & $4.310$\rlap{$_{-0.010}^{+0.008}$} &    &    \\[0.5ex] 
$7206837$ & Bagheera & $9.77$ & $1653$ & $79.1$ & $45   $ & F-like & $\rm 5.1-17.2$ & $\rm    $ & $6305 \pm 77$ & $0.10 \pm 0.10$ & $4.163$\rlap{$_{-0.008}^{+0.007}$} &  $-18.54 \pm 0.10$ & $9.30 \pm 0.50$ \\[0.5ex] 
$7296438$ &       & $10.09$ & $1848$ & $88.7$ & $32   $ & Simple & $\rm 7.1-11.3$ & $\rm    $ & $5775 \pm 77$ & $0.19 \pm 0.10$ & $4.201$\rlap{$_{-0.009}^{+0.010}$} &  $3.36 \pm 0.10$ & $1.80 \pm 0.50$ \\[0.5ex] 
$7510397$ &       & $7.77$ & $1189$ & $62.2$ & $47   $ & Simple & $\rm 7.1-17.2$ & $\rm 16$ & $6171 \pm 77$ & $-0.21 \pm 0.10$ & $4.036$\rlap{$_{-0.007}^{+0.004}$} &  $-34.10 \pm 0.10$ & $6.40 \pm 0.50$ \\[0.5ex] 
$7680114$ & Simba & $10.07$ & $1709$ & $85.1$ & $41   $ & Simple & $\rm 5.1-17.2$ & $\rm 6,7.2,10$ & $5811 \pm 77$ & $0.05 \pm 0.10$ & $4.172$\rlap{$_{-0.008}^{+0.010}$} &  $-58.93 \pm 0.10$ & $3.00 \pm 0.50$ \\[0.5ex] 
$7771282$ &       & $10.77$ & $1465$ & $72.5$ & $32   $ & F-like & $\rm 5.1-11.3$ & $\rm 6$ & $6248 \pm 77$ & $-0.02 \pm 0.10$ & $4.112$\rlap{$_{-0.007}^{+0.007}$} &  $-0.38 \pm 0.10$ & $8.30 \pm 0.50$ \\[0.5ex] 
$7871531$ &       & $9.25$ & $3456$ & $151.3$ & $35   $ & Simple & $\rm 5.1-17.2$ & $\rm    $ & $5501 \pm 77$ & $-0.26 \pm 0.10$ & $4.478$\rlap{$_{-0.007}^{+0.005}$} &  $-20.65 \pm 0.10$ & $0.90 \pm 0.50$ \\[0.5ex] 
$7940546$ & Akela & $7.40$ & $1117$ & $58.8$ & $58   $ & F-like & $\rm 7.1-17.2$ & $\rm    $ & $6235 \pm 77$ & $-0.20 \pm 0.10$ & $4.000$\rlap{$_{-0.002}^{+0.002}$} &  $-3.03 \pm 0.10$ & $9.10 \pm 0.50$ \\[0.5ex] 
$7970740$ &       & $7.78$ & $4197$ & $173.5$ & $46   $ & Simple & $\rm 6.1-17.2$ & $\rm    $ & $5309 \pm 77$ & $-0.54 \pm 0.10$ & $4.539$\rlap{$_{-0.004}^{+0.005}$} &  $-60.24 \pm 0.10$ & $0.00 \pm 0.50$ \\[0.5ex] 
$8006161$ & Doris & $7.36$ & $3575$ & $149.4$ & $54   $ & Simple & $\rm 5.1-17.2$ & $\rm    $ & $5488 \pm 77$ & $0.34 \pm 0.10$ & $4.494$\rlap{$_{-0.007}^{+0.007}$} &  $-45.56 \pm 0.10$ & $0.70 \pm 0.50$ \\[0.5ex] 
$8150065$ &       & $10.74$ & $1877$ & $89.3$ & $24   $ & Simple & $\rm 5.1-10.3$ & $\rm 6,7.2$ & $6173 \pm 101$\textsuperscript{\raisebox{1pt}{\rlap{3}}} & $-0.13 \pm 0.15$\textsuperscript{\raisebox{1pt}{\rlap{3}}} & $4.220$\rlap{$_{-0.008}^{+0.008}$} &    &    \\[0.5ex] 
$8179536$ &       & $9.46$ & $2075$ & $95.1$ & $39   $ & Simple & $\rm 5.1-11.3$ & $\rm 6$ & $6343 \pm 77$ & $-0.03 \pm 0.10$ & $4.255$\rlap{$_{-0.010}^{+0.010}$} &  $-31.40 \pm 0.10$ & $9.90 \pm 0.50$ \\[0.5ex] 
$8228742$ & Horace & $9.37$ & $1190$ & $62.1$ & $44   $ & Simple & $\rm 5.1-17.2$ & $\rm    $ & $6122 \pm 77$ & $-0.08 \pm 0.10$ & $4.032$\rlap{$_{-0.005}^{+0.004}$} &  $10.71 \pm 0.10$ & $6.10 \pm 0.50$ \\[0.5ex] 
$8379927$ & Arthur & $6.96$ & $2795$ & $120.3$ & $49   $ & Simple & $\rm 2.1-17.2$ & $\rm 2.2,2.3,3,4$ & $6067 \pm 120$\textsuperscript{\raisebox{1pt}{\rlap{1}}} & $-0.10 \pm 0.15$\textsuperscript{\raisebox{1pt}{\rlap{1}}} & $4.388$\rlap{$_{-0.008}^{+0.007}$} &    &    \\[0.5ex] 
$8394589$ &       & $9.52$ & $2397$ & $109.5$ & $44   $ & Simple & $\rm 5.1-17.2$ & $\rm    $ & $6143 \pm 77$ & $-0.29 \pm 0.10$ & $4.322$\rlap{$_{-0.008}^{+0.008}$} &  $22.58 \pm 0.10$ & $6.40 \pm 0.50$ \\[0.5ex] 
$8424992$ &       & $10.32$ & $2534$ & $120.6$ & $22   $ & Simple & $\rm 7.1-10.3$ & $\rm    $ & $5719 \pm 77$ & $-0.12 \pm 0.10$ & $4.359$\rlap{$_{-0.007}^{+0.007}$} &  $-87.63 \pm 0.10$ & $1.30 \pm 0.50$ \\[0.5ex] 
$8694723$ &       & $8.88$ & $1471$ & $75.1$ & $53   $ & Simple & $\rm 5.1-17.2$ & $\rm    $ & $6246 \pm 77$ & $-0.42 \pm 0.10$ & $4.113$\rlap{$_{-0.007}^{+0.009}$} &  $15.88 \pm 0.10$ & $7.10 \pm 0.50$ \\[0.5ex] 
$8760414$ & Pucky & $9.62$ & $2455$ & $117.2$ & $44   $ & Simple & $\rm 5.1-17.2$ & $\rm    $ & $5873 \pm 77$ & $-0.92 \pm 0.10$ & $4.320$\rlap{$_{-0.007}^{+0.003}$} &  $-115.64 \pm 0.10$ & $2.50 \pm 0.50$ \\[0.5ex] 
$8938364$ & Java2 & $10.11$ & $1675$ & $85.7$ & $41   $ & Simple & $\rm 6.1-17.2$ & $\rm    $ & $5677 \pm 77$ & $-0.13 \pm 0.10$ & $4.173$\rlap{$_{-0.007}^{+0.002}$} &  $-68.12 \pm 0.10$ & $2.40 \pm 0.50$ \\[0.5ex] 
$9025370$ &       & $8.85$ & $2989$ & $132.6$ & $28   $ & Simple & $\rm 5.1-17.2$ & $\rm    $ & $5270 \pm 180$\textsuperscript{\raisebox{1pt}{\rlap{2}}} & $-0.12 \pm 0.18$\textsuperscript{\raisebox{1pt}{\rlap{2}}} & $4.423$\rlap{$_{-0.007}^{+0.004}$} &    &    \\[0.5ex] 
$9098294$ &       & $9.76$ & $2315$ & $108.9$ & $34   $ & Simple & $\rm 5.1-17.2$ & $\rm    $ & $5852 \pm 77$ & $-0.18 \pm 0.10$ & $4.308$\rlap{$_{-0.005}^{+0.007}$} &  $-71.72 \pm 0.10$ & $3.00 \pm 0.50$ \\[0.5ex] 
$9139151$ & Carlsberg & $9.18$ & $2690$ & $117.3$ & $35   $ & Simple & $\rm 5.1-17.2$ & $\rm    $ & $6302 \pm 77$ & $0.10 \pm 0.10$ & $4.382$\rlap{$_{-0.008}^{+0.008}$} &  $-29.06 \pm 0.10$ & $5.50 \pm 0.50$ \\[0.5ex] 
$9139163$ & Punto & $8.33$ & $1730$ & $81.2$ & $57   $ & F-like & $\rm 5.1-17.2$ & $\rm    $ & $6400 \pm 84$\textsuperscript{\raisebox{1pt}{\rlap{5}}} & $0.15 \pm 0.09$\textsuperscript{\raisebox{1pt}{\rlap{5}}} & $4.200$\rlap{$_{-0.009}^{+0.008}$} &    &    \\[0.5ex] 
$9206432$ &       & $9.08$ & $1866$ & $84.9$ & $49   $ & F-like & $\rm 5.1-12.3$ & $\rm 7$ & $6538 \pm 77$ & $0.16 \pm 0.10$ & $4.220$\rlap{$_{-0.005}^{+0.007}$} &  $-1.73 \pm 0.10$ & $6.70 \pm 0.50$ \\[0.5ex] 
$9353712$ &       & $10.84$ & $934$ & $51.5$ & $41   $ & F-like & $\rm 5.1-12.3$ & $\rm 6,7.2$ & $6278 \pm 77$ & $-0.05 \pm 0.10$ & $3.943$\rlap{$_{-0.007}^{+0.005}$} &  $-46.67 \pm 0.10$ & $6.80 \pm 0.50$ \\[0.5ex] 
$9410862$ &       & $10.71$ & $2279$ & $107.4$ & $33   $ & Simple & $\rm 5.1-15.3$ & $\rm    $ & $6047 \pm 77$ & $-0.31 \pm 0.10$ & $4.300$\rlap{$_{-0.009}^{+0.008}$} &  $-56.84 \pm 0.10$ & $3.80 \pm 0.50$ \\[0.5ex] 
$9414417$ &       & $9.58$ & $1155$ & $60.1$ & $54   $ & F-like & $\rm 6.1-17.2$ & $\rm 7$ & $6253 \pm 75$\textsuperscript{\raisebox{1pt}{\rlap{6}}} & $-0.13 \pm 0.10$\textsuperscript{\raisebox{1pt}{\rlap{6}}} & $4.016$\rlap{$_{-0.005}^{+0.005}$} &    &    \\[0.5ex] 
$9812850$ &       & $9.47$ & $1255$ & $64.7$ & $49   $ & F-like & $\rm 5.1-17.2$ & $\rm    $ & $6321 \pm 77$ & $-0.07 \pm 0.10$ & $4.053$\rlap{$_{-0.008}^{+0.009}$} &  $31.18 \pm 0.10$ & $12.50 \pm 0.50$ \\[0.5ex] 
$9955598$ &       & $9.44$ & $3617$ & $153.3$ & $31   $ & Simple & $\rm 5.1-17.2$ & $\rm    $ & $5457 \pm 77$ & $0.05 \pm 0.10$ & $4.497$\rlap{$_{-0.007}^{+0.005}$} &  $-28.48 \pm 0.10$ & $1.10 \pm 0.50$ \\[0.5ex] 
$9965715$ &       & $9.34$ & $2079$ & $97.2$ & $40   $ & Simple & $\rm 5.1-13.3$ & $\rm 7,11$ & $5860 \pm 180$\textsuperscript{\raisebox{1pt}{\rlap{2}}} & $-0.44 \pm 0.18$\textsuperscript{\raisebox{1pt}{\rlap{2}}} & $4.272$\rlap{$_{-0.008}^{+0.009}$} &    &    \\[0.5ex] 
$10068307$ &       & $8.18$ & $995$ & $53.9$ & $49   $ & Simple & $\rm 7.1-17.2$ & $\rm    $ & $6132 \pm 77$ & $-0.23 \pm 0.10$ & $3.967$\rlap{$_{-0.004}^{+0.004}$} &  $-14.78 \pm 0.10$ & $6.40 \pm 0.50$ \\[0.5ex] 
$10079226$ &       & $10.07$ & $2653$ & $116.3$ & $31   $ & Simple & $\rm 7.1-10.3$ & $\rm    $ & $5949 \pm 77$ & $0.11 \pm 0.10$ & $4.366$\rlap{$_{-0.005}^{+0.005}$} &  $-37.15 \pm 0.10$ & $4.00 \pm 0.50$ \\[0.5ex] 
$10162436$ &       & $8.61$ & $1052$ & $55.7$ & $51   $ & F-like & $\rm 5.1-17.2$ & $\rm 7,10,11,15$ & $6146 \pm 77$ & $-0.16 \pm 0.10$ & $3.981$\rlap{$_{-0.005}^{+0.005}$} &  $-52.92 \pm 0.10$ & $6.40 \pm 0.50$ \\[0.5ex] 
$10454113$ & Pinocha & $8.62$ & $2357$ & $105.1$ & $54   $ & Simple & $\rm 5.1-17.2$ & $\rm    $ & $6177 \pm 77$ & $-0.07 \pm 0.10$ & $4.314$\rlap{$_{-0.005}^{+0.005}$} &  $-21.22 \pm 0.10$ & $6.10 \pm 0.50$ \\[0.5ex] 
$10516096$ & Manon & $9.46$ & $1690$ & $84.4$ & $40   $ & Simple & $\rm 5.1-17.2$ & $\rm 6,10.1$ & $5964 \pm 77$ & $-0.11 \pm 0.10$ & $4.169$\rlap{$_{-0.010}^{+0.011}$} &  $1.28 \pm 0.10$ & $4.60 \pm 0.50$ \\[0.5ex] 
$10644253$ & Mowgli & $9.16$ & $2900$ & $123.1$ & $34   $ & Simple & $\rm 5.1-17.2$ & $\rm    $ & $6045 \pm 77$ & $0.06 \pm 0.10$ & $4.396$\rlap{$_{-0.007}^{+0.008}$} &  $-18.91 \pm 0.10$ & $3.20 \pm 0.50$ \\[0.5ex] 
$10730618$ &       & $10.45$ & $1282$ & $66.3$ & $39   $ & F-like & $\rm 0-11.3$ & $\rm 6,7.2$ & $6150 \pm 180$\textsuperscript{\raisebox{1pt}{\rlap{2}}} & $-0.11 \pm 0.18$\textsuperscript{\raisebox{1pt}{\rlap{2}}} & $4.062$\rlap{$_{-0.008}^{+0.007}$} &    &    \\[0.5ex] 
$10963065$ & Rudy  & $8.77$ & $2204$ & $103.2$ & $42   $ & Simple & $\rm 2.3-17.2$ & $\rm 3,4,8,9,12,16$ & $6140 \pm 77$ & $-0.19 \pm 0.10$ & $4.277$\rlap{$_{-0.011}^{+0.011}$} &  $-54.95 \pm 0.10$ & $4.50 \pm 0.50$ \\[0.5ex] 
$11081729$ &       & $9.03$ & $1968$ & $90.1$ & $40   $ & F-like & $\rm 5.1-17.2$ & $\rm    $ & $6548 \pm 83$ & $0.11 \pm 0.10$ & $4.245$\rlap{$_{-0.010}^{+0.009}$} &  $0.27 \pm 0.10$ & $24.10 \pm 0.50$ \\[0.5ex] 
$11253226$ & Tinky & $8.44$ & $1591$ & $76.9$ & $58   $ & F-like & $\rm 5.1-17.2$ & $\rm    $ & $6642 \pm 77$ & $-0.08 \pm 0.10$ & $4.173$\rlap{$_{-0.005}^{+0.004}$} &  $10.65 \pm 0.10$ & $14.40 \pm 0.50$ \\[0.5ex] 
$11772920$ &       & $9.66$ & $3675$ & $157.7$ & $27   $ & Simple & $\rm 5.1-17.2$ & $\rm    $ & $5180 \pm 180$\textsuperscript{\raisebox{1pt}{\rlap{2}}} & $-0.09 \pm 0.18$\textsuperscript{\raisebox{1pt}{\rlap{2}}} & $4.500$\rlap{$_{-0.005}^{+0.008}$} &    &    \\[0.5ex] 
$12009504$ & Dushera & $9.32$ & $1866$ & $88.2$ & $43   $ & Simple & $\rm 5.1-17.2$ & $\rm    $ & $6179 \pm 77$ & $-0.08 \pm 0.10$ & $4.211$\rlap{$_{-0.007}^{+0.005}$} &  $12.82 \pm 0.10$ & $7.70 \pm 0.50$ \\[0.5ex] 
$12069127$ &       & $10.70$ & $885$ & $48.4$ & $39   $ & F-like & $\rm 5.1-11.3$ & $\rm 6$ & $6276 \pm 77$ & $0.08 \pm 0.10$ & $3.912$\rlap{$_{-0.005}^{+0.004}$} &  $-25.33 \pm 0.10$ & $6.00 \pm 0.50$ \\[0.5ex] 
$12069424$ & 16 Cyg A & $5.86$ & $2188$ & $103.3$ & $53   $ & Simple & $\rm 6.1-17.2$ & $\rm    $ & $5825 \pm 50$\textsuperscript{\raisebox{1pt}{\rlap{4}}} & $0.10 \pm 0.03$\textsuperscript{\raisebox{1pt}{\rlap{4}}} & $4.287$\rlap{$_{-0.007}^{+0.007}$} &  $-27.35 \pm 0.10$ & $2.80 \pm 0.50$  \\[0.5ex] 
$12069449$ & 16 Cyg B & $6.09$ & $2561$ & $116.9$ & $52   $ & Simple & $\rm 6.1-17.2$ & $\rm    $ & $5750 \pm 50$\textsuperscript{\raisebox{1pt}{\rlap{4}}} & $0.05 \pm 0.02$\textsuperscript{\raisebox{1pt}{\rlap{4}}} & $4.353$\rlap{$_{-0.007}^{+0.005}$} &  $-27.82 \pm 0.10$ & $2.10 \pm 0.50$  \\[0.5ex] 
$12258514$ & Barney & $8.08$ & $1513$ & $74.8$ & $45   $ & Simple & $\rm 5.1-17.2$ & $\rm    $ & $5964 \pm 77$ & $-0.00 \pm 0.10$ & $4.126$\rlap{$_{-0.004}^{+0.003}$} &  $-18.98 \pm 0.10$ & $3.90 \pm 0.50$ \\[0.5ex] 
$12317678$ &       & $8.74$ & $1212$ & $63.5$ & $57   $ & F-like & $\rm 5.1-17.2$ & $\rm    $ & $6580 \pm 77$ & $-0.28 \pm 0.10$ & $4.048$\rlap{$_{-0.008}^{+0.009}$} &  $-58.14 \pm 0.10$ & $8.40 \pm 0.50$ \\[0.5ex] 
\bottomrule
\end{tabular} 
\begin{tablenotes}[normal]  
  \small 
\item NOTES: $T_{\rm eff}$ and $\rm [Fe/H]$ from ($1$) \citet{2012ApJS..199...30P}; ($2$) \citet[][]{2014ApJS..215...19P}; ($3$) the SAGA project \citep[][see \url{http://www.mso.anu.edu.au/saga/saga_home.html}]{2014ApJ...787..110C};     ($4$) \citet{2009A&A...508L..17R}; ($5$) \citet{2014ApJS..210....1C}; or ($6$) \citet{2013ApJ...767..127H}.
 \end{tablenotes} 
 \end{threeparttable}% 
 } 
\end{table*}

%%%%%%%%%%%%%%%%%%%%%%%%%%%%%%%%%%%%%%%%%%%%%%%%%%%%%%%%%%%%%%%%%%%%%%%%%%%%%%%%%%%%%%%%%%%%%%%%%%%%%%%%%%%%%%%%%%%%%%%%%%
%%%%%%%%%%%%%%%%%%%%%%%%%%%%%%%%%%%%%%%%%%%%%%%%%%%%%%%%%%%%%%%%%%%%%%%%%%%%%%%%%%%%%%%%%%%%%%%%%%%%%%%%%%%%%%%%%%%%%%%%%%
%%%%%%%%%%%%%%%%%%%%%%%%%%%%%%%%%%%%%%%%%%%%%%%%%%%%%%%%%%%%%%%%%%%%%%%%%%%%%%%%%%%%%%%%%%%%%%%%%%%%%%%%%%%%%%%%%%%%%%%%%%

%%%%%%%%%%%%%%%%%%%%%%%%%%%%%%%%%%%%%%%%%%%%%%%%%%%%%%%%%%%%%%%%%%%%%%%%%%%%%%%%%%%%%%%%%%%%%%%%%%%%%%%%%%%%%%%%%%%%%%%%%%
%%%%%%%%%%%%%%%%%%%%%%%%%%%%%%%%%%%%%%%%%%%%%%%%%%%%%%%%%%%%%%%%%%%%%%%%%%%%%%%%%%%%%%%%%%%%%%%%%%%%%%%%%%%%%%%%%%%%%%%%%%
\subsection{Data preparation}

For most targets, data were taken continuously from Quarter 5 (Q5) through Q17. To minimize gaps in the time series, data from the initial short quarters (Q0 or Q1) were omitted unless continuous with the subsequent data. \tref{tab:target_table} lists the quarters used for each target.
Light curves were constructed from pixel data downloaded from the KASOC database\footnote{\label{note1}\url{www.kasoc.phys.au.dk}}, using the procedure
developed by S. Bloemen (private comm.) to define pixel masks for aperture photometry. The light curves were then corrected using the KASOC filter \citep[see][]{2014MNRAS.445.2698H}. Briefly, the light curves were first corrected for jumps and concatenated. They were then median filtered using two filters of different widths --- one long, one short --- with the final filter being a weighted sum of the two filters based on the variability in the light curve. For the four \kp objects of interest (KOIs) in the sample (KICs 3632418, 9414417, 9955598, and 10963065) an iterative removal of the planetary transits was performed based on the planetary phase-curve (see \citealt[][]{2014MNRAS.445.2698H} for further details).

The power density spectrum (PDS) returned from the KASOC filter is made from a weighted least-squares sine-wave fitting, single-sided calibrated, normalized to Parseval's theorem, and converted to power density by dividing by the integral of the spectral window \citep[][]{1992PASP..104..413K,1992PhDT.......208K}.

%%%%%%%%%%%%%%%%%%%%%%%%%%%%%%%%%%%%%%%%%%%%%%%%%%%%%%%%%%%%%%%%%%%%%%%%%%%%%%%%%%%%%%%%%%%%%%%%%%%%%%%%%%%%%%%%%%%%%%%%%%
\subsection{Atmospheric and stellar parameters}

\begin{figure}
	\centering
	\includegraphics[width=\columnwidth]{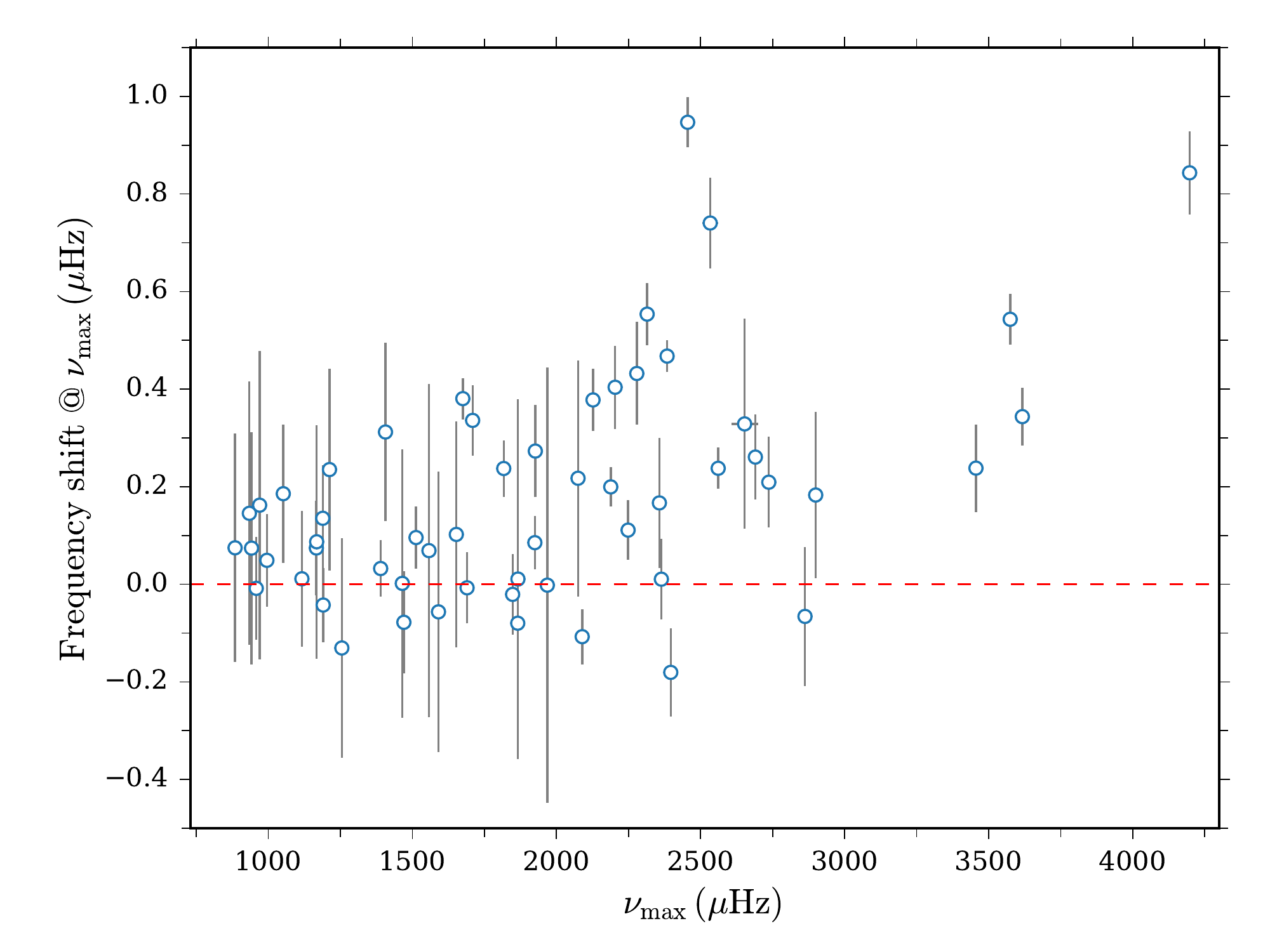}
    \caption{Frequency shift at \numax from line-of-sight velocities of the stars in the sample (see \tref{tab:target_table}). The uncertainty indicates for a given target the minimum frequency
uncertainty of the five radial modes nearest \numax. The frequency uncertainty on individual modes is in several cases lower than the line-of-sight frequency shift. Large or small, the systematic shift can thus cause bias in the modeling if left uncorrected.}
    \label{fig:freshift}
\end{figure}

We have obtained atmospheric parameters from the Stellar Parameters Classification tool \citep[SPC; see][]{2012Natur.486..375B}, with data from the Tillinghast Reflector Echelle Spectrograph \citep[TRES;][]{2007RMxAC..28..129S, furesz_phd} on the 1.5-m Tillinghast telescope at the F.~L.~Whipple Observatory. Information from the SPC analysis is available on the \kp Community Follow-up Observing Program (CFOP) website\footnote{\url{https://cfop.ipac.caltech.edu/home/}}. In the SPC derivation of parameters, \logg values were fixed to the asteroseismic values given in \citet[][]{2014ApJS..210....1C} to decrease the impact on uncertainties from correlations between \teff, \logg, and \feh. We added in quadrature to the derived uncertainties on \teff and \feh systematic uncertainties of $\pm 59\, \rm K$ and $\pm 0.062 \, \rm dex$, as suggested by \citet[][]{2012ApJ...757..161T}.    
For a subset of targets, spectroscopic values were taken from the literature (\tref{tab:target_table}). We also list in \tref{tab:target_table} the line-of-sight (LOS) velocities derived from the SPC analysis, which should be used in any modeling efforts using individual frequencies to account for the Doppler shift of the frequencies \citep[][]{2014MNRAS.445L..94D}. In \fref{fig:freshift} we show the values of these Doppler frequency shifts, which in some cases exceed the uncertainties on the individual frequencies. Even if the frequency shift is small compared to the uncertainties on the mode frequencies, it is systematic and should therefore always be corrected to avoid biases in the stellar modeling. The SPC LOS values have been corrected by $-0.61\, \rm km\, s^{-1}$ to put the velocities onto the IAU system. This correction is primarily accounting for the fact that the CfA library of synthetic spectra does not include the solar gravitational redshift.
Stellar parameters used in this paper, such as masses and radii, are adopted from the modeling effort presented in Paper~II. 

%%%%%%%%%%%%%%%%%%%%%%%%%%%%%%%%%%%%%%%%%%%%%%%%%%%%%%%%%%%%%%%%%%%%%%%%%%%%%%%%%%%%%%%%%%%%%%%%%%%%%%%%%%%%%%%%%%%%%%%%%%
\subsection{Sun-as-a-star data}
\label{sec:sun-as-star}

As part of the project, the Sun was fitted in the same manner as the sample targets (see \sref{sec:param_est}). This was done primarily to test the modeling efforts presented in Paper~II against a known reference, and at the same time to assess the returns from the peak-bagging.
The power spectrum was produced from data from VIRGO\footnote{Variability of Solar Irradiance and Gravity Oscillations} \citep{2009A&A...501L..27F} on-board the \textit{SoHO}\footnote{Solar and Heliospheric Observatory} spacecraft \citep[][]{1995SoPh..162..101F,1997SoPh..170....1F}.
Specifically, a time series was created from a weighted sum of the green and red channels of the VIRGO Sun photometers (SPM) with central wavelengths of 500 nm (green), and 862 nm (red). 
Weights were selected such that the response-function weighted centroid wavelength from the two SPM channels matched that from
the \kp response function (641.7 nm). The two-component light curves were filtered individually using a 30-day median filter and then summed in relative flux units with the appropriate weights (green: 0.785; red: 0.215). The solar time series had a length of 1150 days (corresponding to ${\sim}3.15$ years, or the approximate duration of 13 Quarters). This is the typical time series length for targets in the sample. To find the level to which the spectrum should be degraded, the magnitude distribution was computed for the sample, including also stars that have a mixed-mode character. The median magnitude of $\Kp\approx9.17$ closely matches that of KIC 9139151 and so noise was added to the solar time series to match the level of this star.

The solar data set will primarily be used for estimates relating to frequencies, such as \dnu and \numax, but not for analysis of line widths, amplitudes, or visibilities. This is because one cannot, with the simple weighting of relatively narrow band filters done here, assume that the measurements of amplitudes and visibilities adhere strictly to what would be observed with \kp.

%%%%%%%%%%%%%%%%%%%%%%%%%%%%%%%%%%%%%%%%%%%%%%%%%%%%%%%%%%%%%%%%%%
%%%%%%%%%%%%%%%%%%%%%%%%%%%%%%%%%%%%%%%%%%%%%%%%%%%%%%%%%%%%%%%%%%

%%%%%%%%%%%%%%%%%%%%%%%%%%%%%%%%%%%%%%%%%%%%%%%%%%%%%%%%%%%%%%%%%%%%%%%%%%%%%%%%%%%%%%%%%%%%%%%%%%%%%%%%%%%%%%%%%%%%%%%%%%
%%%%%%%%%%%%%%%%%%%%%%%%%%%%%%%%%%%%%%%%%%%%%%%%%%%%%%%%%%%%%%%%%%%%%%%%%%%%%%%%%%%%%%%%%%%%%%%%%%%%%%%%%%%%%%%%%%%%%%%%%%
%%%%%%%%%%%%%%%%%%%%%%%%%%%%%%%%%%%%%%%%%%%%%%%%%%%%%%%%%%%%%%%%%%%%%%%%%%%%%%%%%%%%%%%%%%%%%%%%%%%%%%%%%%%%%%%%%%%%%%%%%%
%%%%%%%%%%%%%%%%%%%%%%%%%%%%%%%%%%%%%%%%%%%%%%%%%%%%%%%%%%%%%%%%%%%%%%%%%%%%%%%%%%%%%%%%%%%%%%%%%%%%%%%%%%%%%%%%%%%%%%%%%%
\section{Parameter estimation}
\label{sec:param_est}

%%%%%%%%%%%%%%%%%%%%%%%%%%%%%%%%%%%%%%%%%%%%%%%%%%%%%%%%%%%%%%%%%%%%%%%%%%%%%%%%%%%%%%%%%%%%%%%%%%%%%%%%%%%%%%%%%%%%%%%%%%
%%%%%%%%%%%%%%%%%%%%%%%%%%%%%%%%%%%%%%%%%%%%%%%%%%%%%%%%%%%%%%%%%%%%%%%%%%%%%%%%%%%%%%%%%%%%%%%%%%%%%%%%%%%%%%%%%%%%%%%%%%
\subsection{Oscillation spectrum model}
\label{sec:osc_model}

To model the power spectrum, we described each oscillation mode as a Lorentzian function ($L_{nlm}$), which corresponds to the shape of a stochastically-excited and intrinsically-damped mode \citep[][]{1953tht..book.....B,1988ApJ...328..879K}:
\begin{equation}
L_{nlm}(\nu) = \frac{ \mathcal{E}_{l m}(i_{\star}) \tilde{V}_l^2 S_{n 0} }{1 + \frac{4}{\Gamma^2_{n l}} \left(\nu - \nu_{n l} + m\nu_{s} \right)^2  } \, .
\label{eq:lor}
\end{equation}
Each mode is characterized by the frequency $\nu_{n l}$ of the zonal ($m=0$) component, a height ${H_{n l m}=\mathcal{E}_{l m}(i_{\star}) \tilde{V}_l^2 S_{n 0}}$, a FWHM mode width $\Gamma_{n l}$, and a rotational splitting $\nu_{s}$ \citep[assumed constant with frequency; see][for a discussion of the impact of differential rotation on the constancy of $\nu_{s}$]{2014ApJ...790..121L}. In $H_{n l m}$, $\mathcal{E}_{l m}(i_{\star})$ is the geometrical factor that sets the relative visibilities between the $2l+1$ (azimuthal) $m$-components as a function of the stellar inclination $i_{\star}$ \citep[see,][]{1977AcA....27..203D, 2003ApJ...589.1009G};
$\tilde{V}_l^2$ denotes the squared visibility (power units) of a non-radial mode relative to a radial mode at the same frequency, \ie,
\begin{equation}\label{eq:visi}
\tilde{V}^ 2_l = (V_l / V_0)^2\quad {\rm and} \quad \tilde{V}_{\rm tot}^ 2 = \sum_l \tilde{V}^2_l\, ,
\end{equation}
from the spatial filtering resulting from integrating the intensity for a mode of a given degree over the stellar surface; $S_{n 0}$ then denotes the height of the radial mode of order $n$.
The use of $\tilde{V}^ 2_l$ assumes equipartition of energy between modes of different angular degrees, thus only with a dependence on frequency. This is a good assumption for stochastically excited low degree high order acoustic modes, observed for many lifetimes \citep[see, \eg,][]{1982MNRAS.198..141C}.

The full model to fit to the power spectrum is then given by a series of the Lorentzian functions in \eqref{eq:lor} as
\begin{equation}
\mathcal{P}(\nu) = \eta^2(\nu)\left( \sum\limits_{n=n_0}^{n_{\rm max}} \sum\limits_{l=l_0}^{l_{\rm max}} \sum\limits_{m=-l}^{l} L_{nlm}(\nu) + N(\nu) \right) + W\, .
\label{eq:spec}
\end{equation}
Here $N(\nu)$ denotes the adopted background function; $W$ is a constant white shot-noise component; $\eta^2(\nu)$ describes the apodization of the signal power at frequency $\nu$ from the ${\sim} 1$-minute sampling of the temporal signal \citep[see, \eg,][]{2011ApJ...732...54C,2014A&A...570A..41K}, and is given by:
\begin{equation}
\eta^2(\nu) = \frac{\sin^2(x)}{x^2}\,\,\, {\rm with} \,\,\, x=\pi\nu\Delta t \, ,
\end{equation}
where $\Delta t$ gives the integration time for the observations\footnote{In \kp $\Delta t$ equals the sampling time wherefore $x$ sometimes is given as $x= \frac{\pi \nu}{2\nu_{\rm nq}}$, where $\nu_{\rm nq}$ is the Nyquist frequency --- this is, however, an imprecise definition, because $\nu_{\rm nq}$ relates to the sampling time whereas the apodization relates to the integration time.}.
For the background we used the function \citep{1993ASPC...42..111H,1994SoPh..152..247A}:
\begin{equation}\label{eq:harvey2}
N(\nu) = \sum_{i=1}^2\frac{\xi_i\sigma_{i}^2 \tau_i}{1 + (2\pi\nu\tau_i)^{\alpha_i}}\, ,
\end{equation}
which characterizes a temporal signal from granulation having an exponentially decaying autocovariance, with a power of the temporal decay rate as $-2/\alpha_i$; $\tau_i$ gives the characteristic time scale of the $i$th background component; $\sigma_i$ the corresponding root-mean-square (\textsc{rms}) variation of the component in the time domain. The normalization constants $\xi_i$ are such that the integral (for positive frequencies) of the background component equals $\sigma_i^2$, in accordance with the Parseval-Plancherel theorem \citep[see, \eg,][]{2009A&A...495..979M,2013ApJ...767...34K,2014A&A...570A..41K}.

In fitting \eqref{eq:spec} to the power spectrum, we varied the mode amplitude (square-root of integrated mode power) rather than the mode height to decrease the correlation with $\Gamma_{n l}$ \citep[][]{1994A&A...289..649T}. To obtain the height ($S$) in power density units from the varied amplitude ($A$) we used the relation \citep{2006MNRAS.371..935F,2008A&A...485..813C}:
\begin{equation}
S_{nl} \approx 2A_{nl}^2/\pi\Gamma_{nl}\, .
\label{eq:height}
\end{equation}
This is a valid approximation for a single-sided power spectrum when the modes are well resolved, \ie, when the observing duration $T_{\rm obs}$ greatly exceeds the mode life time $2/\pi\Gamma$.
We note that $A_{nl}$ and $\Gamma_{nl}$ were varied for radial modes only ($l=0$), and then linearly interpolated to the frequencies of the non-radial modes.
The fitting of the power spectrum then finally involved estimating the parameters ${{\bf\Theta} = \{\nu_s, i_{\star}, W, \tau_i, \sigma_i, \alpha_i, A_{n0}, \Gamma_{n0}, \nu_{nl}, \tilde{V}_l \}}$.

%%%%%%%%%%%%%%%%%%%%%%%%%%%%%%%%%%%%%%%%%%%%%%%%%%%%%%%%%%%%%%%%%%%%%%%%%%%%%%%%%%%%%%%%%%%%%%%%%%%%%%%%%%%%%%%%%%%%%%%%%%
%%%%%%%%%%%%%%%%%%%%%%%%%%%%%%%%%%%%%%%%%%%%%%%%%%%%%%%%%%%%%%%%%%%%%%%%%%%%%%%%%%%%%%%%%%%%%%%%%%%%%%%%%%%%%%%%%%%%%%%%%%
\subsection{Fitting strategy}
\label{sec:strat}

Parameters were estimated in a Bayesian manner from a global peak-bagging fit to the power spectrum including all parameters ${\bf\Theta}$ \citep[see, \eg,][]{2011A&A...527A..56H}.
This was done by mapping the posterior probability of the parameters ${\bf\Theta}$ given the data $D$ and any prior information $I$, which from Bayes' theorem is given as:
\begin{equation}\label{eq:bayes010}
p({\bf\Theta}| D,I) = \frac{{p({\bf\Theta}|I)p(D|{\bf\Theta},I)}}{p(D|I)}  \, . 
\end{equation}
Here $p({\bf\Theta}|I)$ is the prior probability assigned to the parameters ${\bf\Theta}$ given $I$, and $p(D|{\bf\Theta},I)$ is the likelihood of the observed data $D$ given the parameters ${\bf\Theta}$. $p(D|I)$, known as the \emph{evidence}, is given by the integral of the numerator over the full parameter space, and thus acts as a normalization. The evidence is unnecessary in the mapping of the relative posterior distribution, so we end up mapping:
\begin{equation}
\ln p({\bf\Theta}| D,I) = \ln p({\bf\Theta}|I) + \ln \mathcal{L}({\bf\Theta}) + C\, ,
\end{equation}
where logarithmic units are adopted for numerical stability, and $C$ is a constant.
Assuming a $\chi^2$ 2-d.o.f. statistic for the power spectrum relative to the limit spectrum in \eqref{eq:spec} \citep[][]{1994A&A...287..685G}, the logarithm of the likelihood for a given observed power, $O_j$, relates to the limit spectrum, $\mathcal{P}(\nu_j ; {\bf\Theta})$, as \citep[see][]{1986ssds.proc..105D,1990ApJ...364..699A,1994A&A...289..649T}:
\begin{equation}
\rm ln \mathcal{L}({\bf\Theta}) = -\sum_j{\left\lbrace ln \mathcal{P}(\nu_j ; {\bf\Theta}) + \frac{O_j}{\mathcal{P}(\nu_j ; {\bf\Theta})} \right\rbrace}.
\label{eq:prob}
\end{equation}

Mapping of \eqref{eq:bayes010} was performed using an affine invariant ensemble Markov chain Monte Carlo (MCMC) sampler \citep[see][]{2010GodmanWeare}, specifically via the \texttt{Python} implementation \texttt{emcee} by \citet[][]{2013PASP..125..306F}.
For a given fit we employed 500 so-called \emph{walkers} that were initiated by sampling from the priors of the model parameters (see \sref{sec:prior}). Each of these was run for at least 2000 steps. We further adopted parallel tempering using five temperatures, with tempering parameters determined according to \citet[][]{2009A&A...506...15B}, and a thinning of the MCMC chains by a factor of 10. As part of the post-processing, the appropriate burn-in for a given target and whether sufficient mixing had been achieved was determined by (1) visual inspection of the chain traces, (2) using the Geweke diagnostic \citep[][]{MR1380276}, and (3) by assessing the length of the chain compared to the autocorrelation time (giving the number of independent draws from the target distribution).

Final parameter estimates were obtained from the median (frequencies) or mode (amplitudes, line widths, and visibilities) of the marginalized posterior probability density functions (PDFs) --- with the MCMC sampling the marginalization is obtained naturally and the PDF for a given parameter is simply given by the normalized distribution of the samples of the parameter.
A measure for the parameter uncertainty is given by the credible interval as the interval spanning the $68.27\%$ highest probability density (HPD) of the PDF.

%%%%%%%%%%%%%%%%%%%%%%%%%%%%%%%%%%%%%%%%%%%%%%%%%%%%%%%%%%%%%%%%%%%%%%%%%%%%%%%%%%%%%%%%%%%%%%%%%%%%%%%%%%%%%%%%%%%%%%%%%%
\subsubsection{Mode identification and initial guesses}
\label{sec:modeid}

\begin{figure}
	\centering
	\includegraphics[width=\columnwidth]{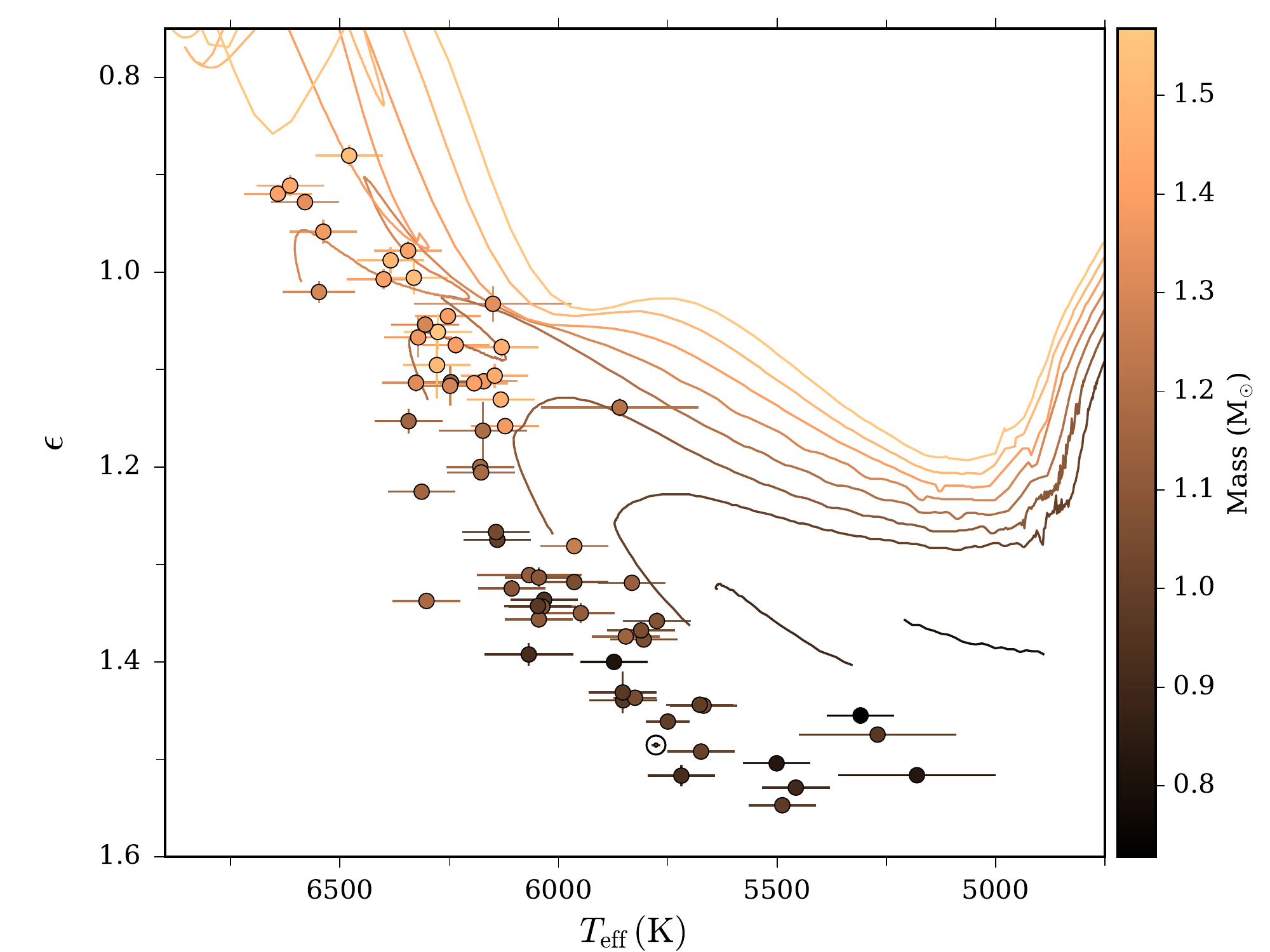}
    \caption{Measured values of $\epsilon$ against \teff. The color indicates the modeled mass of the stars using the results from the BASTA pipeline (Paper~II). Shown are also $\epsilon$-evolutionary tracks from \citet[][]{2011ApJ...743..161W}, calculated from ASTEC evolutionary tracks with $Z_0=0.017$ \citep[][]{2008Ap&SS.316...13C}, for masses going from $0.8\, \rm M_{\odot}$ to $1.6\, \rm M_{\odot}$ in steps of $0.1\, \rm M_{\odot}$.}
    \label{fig:dnu_epsilon}
\end{figure}

\begin{figure*}
    \centering
    \begin{subfigure}
        \centering
        \includegraphics[width=0.32\textwidth]{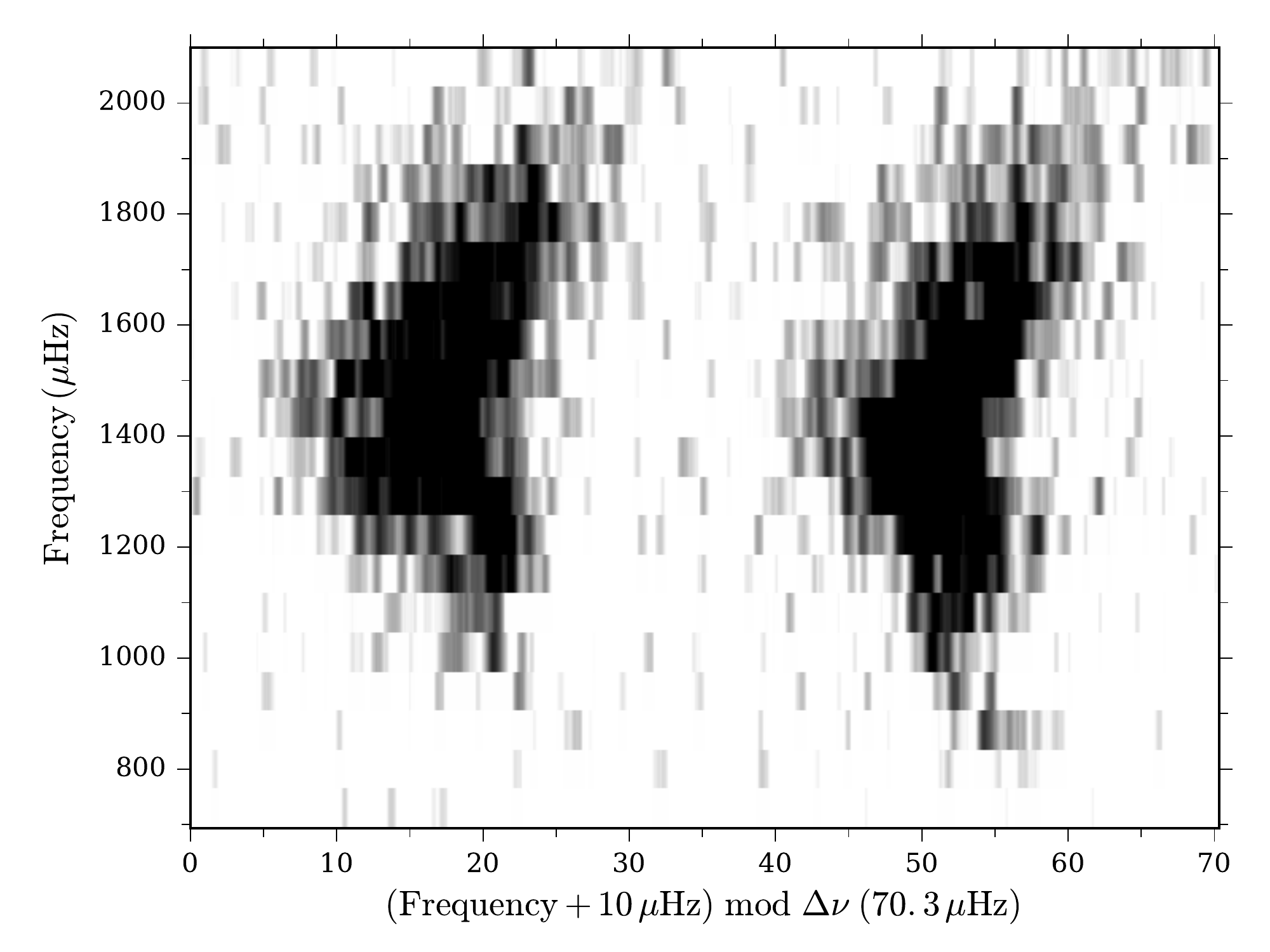}
    \end{subfigure}\hfill
    \begin{subfigure}
        \centering
        \includegraphics[width=0.32\textwidth]{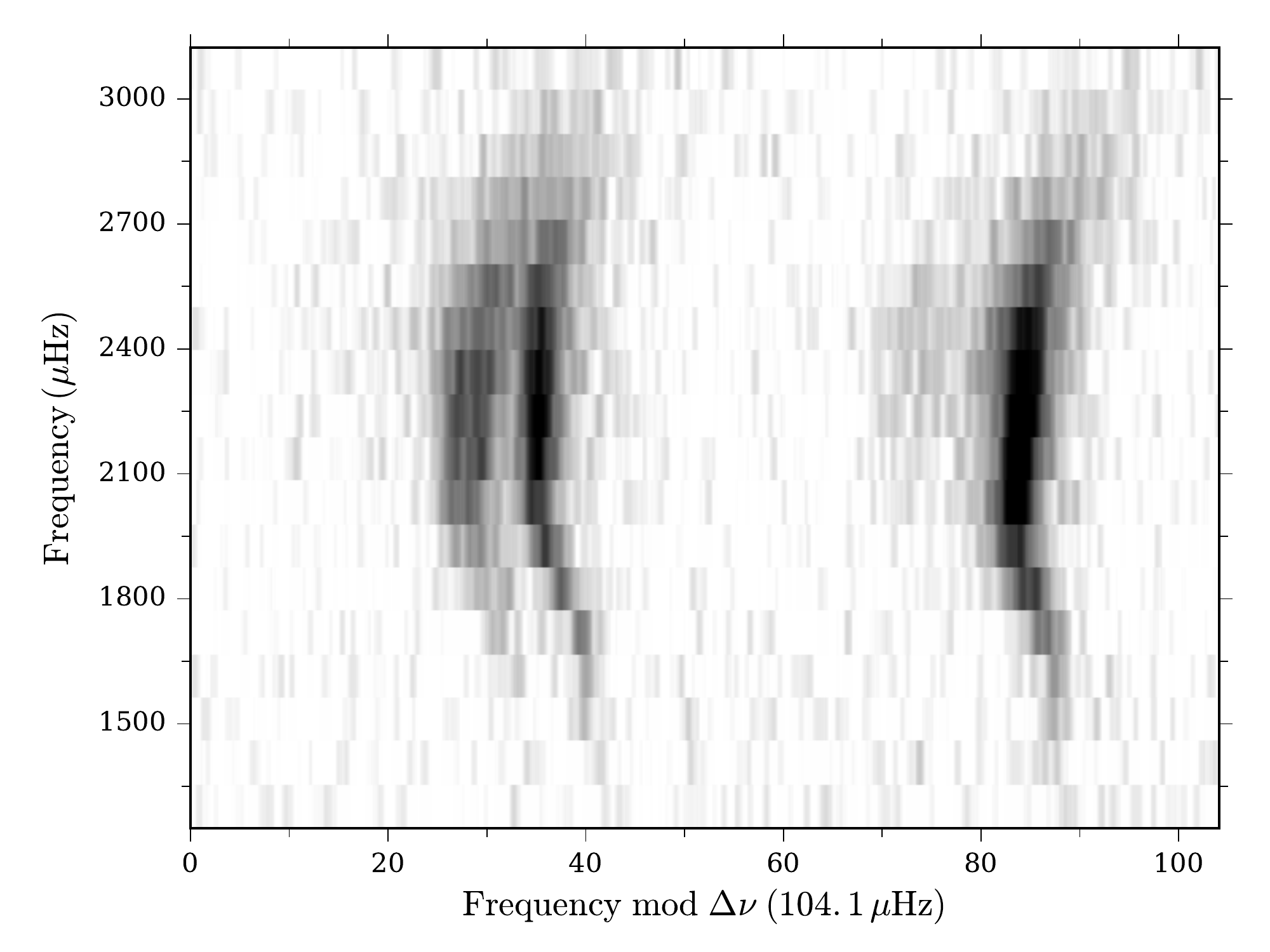}
    \end{subfigure}\hfill
    \begin{subfigure}
        \centering
        \includegraphics[width=0.32\textwidth]{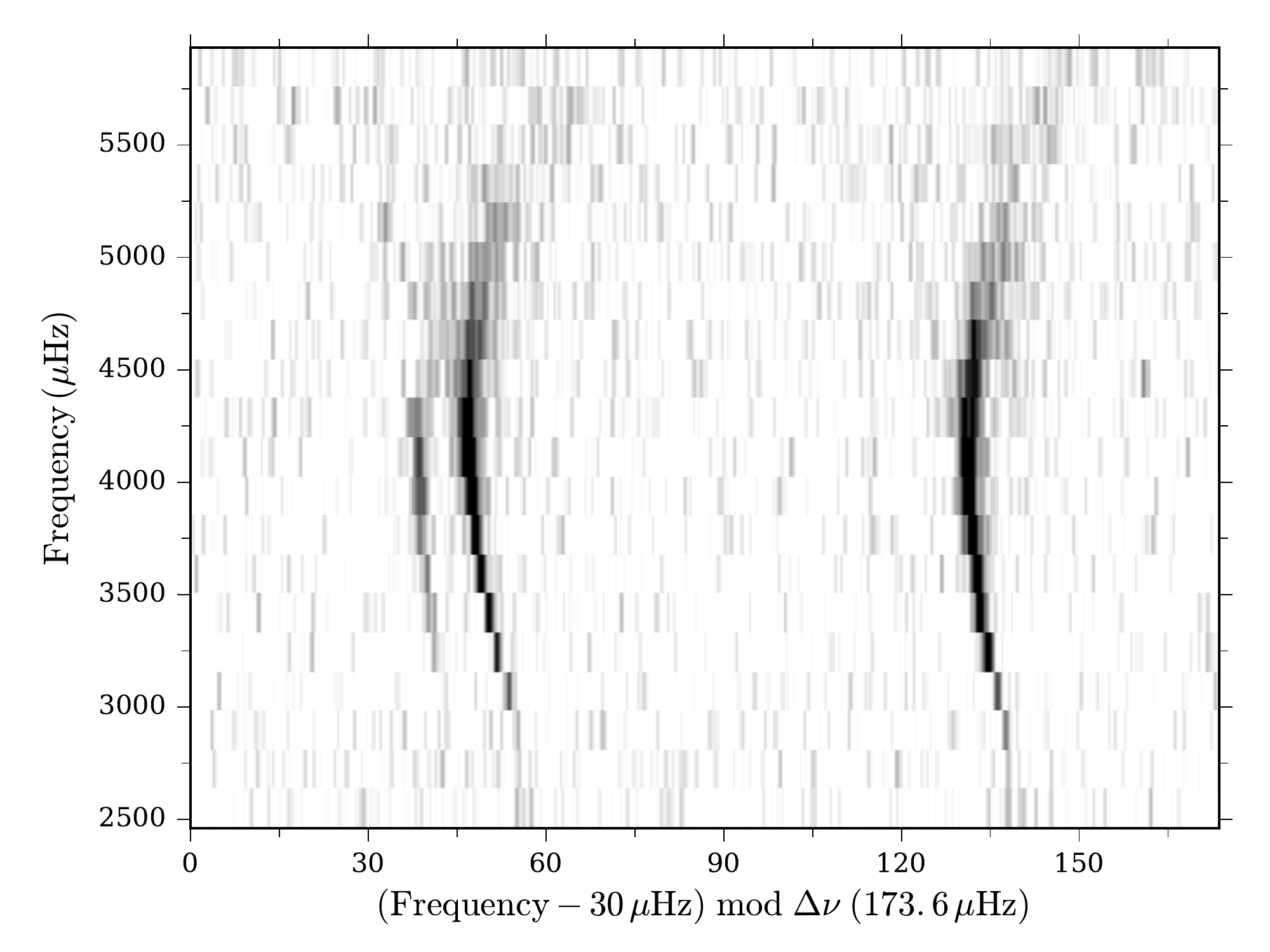}
    \end{subfigure}
    \caption{\'{E}chelle diagrams for three sample stars with different values of \numax; KIC1435467 (left), KIC 6106415 (middle), and KIC 7970740 (right). The color-scale goes from white (low power) to black (high power), and the power spectra have been background corrected and smoothed by a $2\, \mu\rm Hz$ Epanechnikov filter \citep[][]{16463184,hastie2009elements}. The \'{e}chelle spectra have been shifted along the abscissa for a better rendering (see individual labels), with the $l=1$ ridges being the right-most in each case and the $l=0,2$ ridges the left-most.  }
\label{fig:echelle}
\end{figure*}

Before the peak-bagging can commence, initial guesses must be defined for the mode-frequencies to include in \eqref{eq:spec}, and the modes must further be identified in terms of their angular degree $l$. For acoustic modes of high radial order $n$ and low angular degree $l$ the frequencies may be approximated by the asymptotic relation \citep[][]{1980ApJS...43..469T,1983SoPh...82...75S}:
\begin{equation}\label{eq:asymp}
\nu_{nl} \approx (n + \frac{l}{2} + \epsilon)\dnu - l(l+1)D_0\, .
\end{equation}
Here \dnu is the large separation, given by the average frequency spacing between consecutive overtones $n$ for modes of a given $l$; $\epsilon$ is a dimensionless offset sensitive to the surface layers \citep[see, \eg,][]{1986hmps.conf..117G,1998MNRAS.295..344P,2016A&A...585A..63R}; $D_0$ is sensitive to the sound-speed gradient near the stellar core \citep[][]{1983SoPh...82...75S,1993ASPC...42..347C}.
Mode identification was then, by and large, achieved via visual inspection of \'{e}chelle diagrams \citep[][]{1983SoPh...82...55G,2011arXiv1107.1723B}. Here, modes of a given $l$ will form vertical ridges for the correct average large separation. The identification of $l$ and radial order $n$ was checked against the relation for $\epsilon$ as a function of \teff (\fref{fig:dnu_epsilon}), where $\epsilon$ can be found from the \'{e}chelle diagram (\fref{fig:echelle}) by the vertical position of the radial degree ($l=0$) ridge \citep[see][]{2011ApJ...742L...3W,2011ApJ...743..161W}.

%%%%%%%%%%%%%%%%%%%%%%%%%%%%%%%%%%%%%%%%%%%%%%%%%%%%%%%%%%%%%%%%%%
%%%%%%%%%%%%%%%%%%%%%%%%%%%%%%%%%%%%%%%%%%%%%%%%%%%%%%%%%%%%%%%%%%
For this study, consisting of high S/N oscillation signals, the identification was relatively simple. Initial guesses for mode frequencies were primarily defined by hand from smoothed versions of the power density spectra. These were checked against frequencies returned from applying the pseudo-global fitting method of \citet[][]{2009ApJ...694..144F}.

The power spectrum was fitted in the range $\rm f_{\rm min}-5\dnu$ to $\rm f_{\rm max}+5\dnu$, where $\rm f_{\rm min}$ and $\rm f_{\rm max}$ denote the minimum and maximum mode frequency included in the peak-bagging. 
Before the peak-bagging fit, a background-only fit was performed in the range from $\rm5\, \mu Hz$ to the SC Nyquist frequency $\nu_{\rm nq}$ ($\rm {\sim} 8496\, \mu Hz$).
In this fit, the power from solar-like oscillations was accounted for by a Gaussian envelope centered at \numax. Using the posterior distributions from the background-only fit as priors in the peak-bagging allowed us to constrain the background in the relatively narrow frequency range included.

%%%%%%%%%%%%%%%%%%%%%%%%%%%%%%%%%%%%%%%%%%%%%%%%%%%%%%%%%%%%%%
%%%%%%%%%%%%%%%%%%%%%%%%%%%%%%%%%%%%%%%%%%%%%%%%%%%%%%%%%%%%%%

%%%%%%%%%%%%%%%%%%%%%%%%%%%%%%%%%%%%%%%%%%%%%%%%%%%%%%%%%%%%%%%%%%%%%%%%%%%%%%%%%%%%%%%%%%%%%%%%%%%%%%%%%%%%%%%%%%%%%%%%%%
\subsubsection{Prior functions}
\label{sec:prior}

For mode frequencies, we adopted $14\, \mu\rm Hz$ wide top-hat priors centered on the initial guesses of the frequencies. Top-hat priors were also adopted for the rotational frequency splitting $\nu_s$ and inclination $i_{\star}$, with the inclination sampled from the range $-90^{\circ}$ to $180^{\circ}$. The reason for the extended range in inclination is that if the solution is close to either $i_{\star}=0^{\circ}$ or $i_{\star}=90^{\circ}$ any sharp truncation from a prior at these values will make it difficult to properly sample these extreme values. The final posterior on the inclination was then obtained from folding the samples onto the range from $0^{\circ}$ to $90^{\circ}$.

For the amplitudes and line widths we adopted a modified Jeffrey's prior, given as \citep[see, \eg,][]{2011A&A...527A..56H}
\begin{equation}\label{eq:jeffreys}
\mathcal{F}(\theta)  =\left\{ 
\begin{array}{l l}
    \frac{1}{(\theta+\theta_{\rm uni})+ \ln \left[ (\theta_{\rm uni}+\theta_{\rm max})/\theta_{\rm uni} \right] } & ,\rm \quad 0 \leq \theta \leq \theta_{\rm max}\\ \vspace{0.1em}
    0 & ,\rm  \quad \text{otherwise}\\
  \end{array} \right.
\end{equation}
which behaves as a uniform prior when $\theta \ll \theta_{\rm uni}$ and a standard scale invariant Jeffrey's prior when $\theta \gg \theta_{\rm uni}$. The maximum of the prior occurs at $\theta_{\rm max}$.

For mode visibilities we adopted truncated Gaussian functions $\mathcal{N}(\theta_0, \sigma, \theta_{\rm min}, \theta_{\rm max})$ as priors, defined as:
\begin{equation}\label{eq:tgauss}
\mathcal{F}(\theta)  =\left\{ 
\begin{array}{l l}
    \frac{1}{D\, \sqrt{2\pi} \sigma}\exp\left( \frac{-(\theta - \theta_0)^2}{2\sigma^2} \right)   & ,\rm \quad \theta_{\rm min} \leq \theta \leq \theta_{\rm max}\\ \vspace{0.1em}
    0 & ,\rm  \quad \text{otherwise}\\
  \end{array} \right.
\end{equation}
with $D$ given as:
\begin{equation}
D = \frac{\erf\left( \frac{\theta_{\rm max} - \theta_0}{\sqrt{2}\sigma}\right) - \erf\left( \frac{\theta_{\rm min} - \theta_0}{\sqrt{2}\sigma}\right)}{2}\, . 
\end{equation}
Here, $\erf$ denotes the \emph{error function}, $\theta_0$ and $\sigma$ give the chosen mode value and width of the Gaussian, and $\theta_{\rm min}$ and $\theta_{\rm max}$ give the lower and upper truncations of the Gaussian. We specifically adopted $\mathcal{N}(1.5, 1.5, 0, 3)$ for $l=1$, $\mathcal{N}(0.5, 0.5, 0, 1)$ for $l=2$, and $\mathcal{N}(0.05, 0.05, 0, 0.5)$ for $l=3$.
Similarly, we adopted truncated Gaussian priors for the parameters of the background. Here we used as the Gaussian mode value ($\theta_0$) the median of the posteriors from the background-only fit (see \sref{sec:modeid}), and we adopted $\sigma=0.1\, \theta_0$, $\theta_{\rm min} = 0.1\, \theta_0$, and $\theta_{\rm max} = 10\, \theta_0$. 

To ensure that the mode identification did not swap for neighboring $l=0$ and $2$ modes, which is a risk especially at high frequency where the small frequency separation $\delta\nu_{02}=\nu_{n,0} - \nu_{n-1, 2}$ is small compared to the mode line width, we added the prior constraint that on $\delta\nu_{02}$ that it must be positive. In principle $\delta\nu_{02}$ could be negative in the event of bumped $l=2$ modes, however, because the stars were screened for bumped $l=1$ modes and the strength of an avoided $l=2$ crossing is expected to be lower than that of a $l=1$ mode due to the larger evanescent region \citep[see, \eg,][]{2010aste.book.....A,2011A&A...535A..91D}, we do not expect values of $\delta\nu_{02}<0$ for these stars.

%%%%%%%%%%%%%%%%%%%%%%%%%%%%%%%%%%%%%%%%%%%%%%%%%%%%%%%%%%%%%%%%%%%%%%%%%%%%%%%%%%%%%%%%%%%%%%%%%%%%%%%%%%%%%%%%%%%%%%%%%%
\subsubsection{Quality assurance}
\label{sec:quality}

For each fitted mode, we computed a metric for the quality of the fit in the same manner as detailed in \citet[][]{2016MNRAS.456.2183D}, see also \citet[][]{2004A&A...428.1039A} and \citet{2012A&A...543A..54A}.
Briefly, we first ran a fast null hypothesis ($H_0$) test to identify which modes had an unambiguous detection, and for which the probability of detection $p({\rm Det}_{n,l}|D)$ conditioned on the data $D$ needed to be explicitly determined and evaluated.
This was done because the explicit determination of $p({\rm Det}_{n,l}|D)$ is computationally expensive.
In the fast $H_0$ test it was assessed whether the S/N in the background corrected power spectrum ($D$) for a given proposed mode, with the power binned across a number of frequencies to account for the spread in power from the mode line width, was consistent with a pure noise spectrum or whether the $H_0$ hypothesis could be rejected at the $p(D|H_0)=0.001$ level \citep[][]{2003A&A...412..903A,2004A&A...428.1039A,2012MNRAS.427.1784L}.  
When the high S/N modes had been identified in this manner the probability of detection $p({\rm Det}_{n,l}|D)$ was computed for the remainder low-S/N modes.

In the computation of $p({\rm Det}_{n,l}|D)$ both the probability of $D$ assuming $H_0$, $p(D|H_0)$, and the probability of the alternative hypothesis $H_1$ of a detected mode, $p(D|H_1)$, need to be estimated.
The latter was assessed by integrating the probability of measuring the data given a model over a range of mode parameters $\boldsymbol \theta$ --- this integration was achieved by marginalizing over $p(D|H_1, \boldsymbol \theta)$, with the parameter space sampled over the posteriors from the peak-bagging using \texttt{emcee}. Specifically, a mixture-model was used in which both $p(D|H_0)$ and $p(D|H_1)$ were optimized simultaneously to give $p(D|\boldsymbol \theta, p_a)$, the probability of observing the data given the model of a given set of modes with parameters $\boldsymbol \theta$:
\begin{equation}
p(D|\boldsymbol \theta, p_a) = (1-p_a)p(D|H_0) + p_a p(D|H_1)\, .
\end{equation}
Here, the parameter $p_a$, ranging between 0 and 1, then gives the probability of the detection $p(D|{\rm Det}_l)$ of the given set of modes. This probability was kept free in the \texttt{emcee} run, and in the end was assessed from the posterior distribution of $p_a$ \citep[][]{2010arXiv1008.4686H,2015PhRvD..91b3005F}.  
We finally report the Bayes factor $K$, given as the median of the posterior probability distributions of the natural logarithm of the ratio of $p(D|{\rm Det}_l)$ over $p(D|H_0)$, as:
\begin{equation}
\ln K = \ln p(D|{\rm Det}_l ) - \ln p(D|H_0)\, .
\end{equation}
The value of $\ln K$ can then be assessed qualitatively on the \citet[][]{MR3363402} scale as follows:
\[
    \ln K = 
\begin{cases}
    \, <0						& \text{favours}\, H_0\\
    \, 0\, \text{to}\, 1		& \text{not worth more than a bare mention}\\
    \, 1\, \text{to}\, 3		& \text{positive}\\
    \, 3\, \text{to}\, 5		& \text{strong}\\
    \, >5						& \text{very strong}\\
\end{cases}
\]\label{eq:qual}
For a detailed account of the quality control we refer to \citet[][]{2016MNRAS.456.2183D}.

%%%%%%%%%%%%%%%%%%%%%%%%%%%%%%%%%%%%%%%%%%%%%%%%%%%%%%%%%%%%%%%%%%%%%%%%%%%%%%%%%%%%%%%%%%%%%%%%%%%%%%%%%%%%%%%%%%%%%%%%%%
%%%%%%%%%%%%%%%%%%%%%%%%%%%%%%%%%%%%%%%%%%%%%%%%%%%%%%%%%%%%%%%%%%%%%%%%%%%%%%%%%%%%%%%%%%%%%%%%%%%%%%%%%%%%%%%%%%%%%%%%%%
\subsection{Derived quantities and correlations}
\label{sec:deriv}

%\input{Figure_set_Fig5}
%%%%%%%%%%%%%%%%%%%%%%%%%%%%%%%%%%%%%%%%%%%%%%%%%%%%%%%%%%%%%%%%%%
\begin{figure}
	\centering
	\includegraphics[width=\columnwidth]{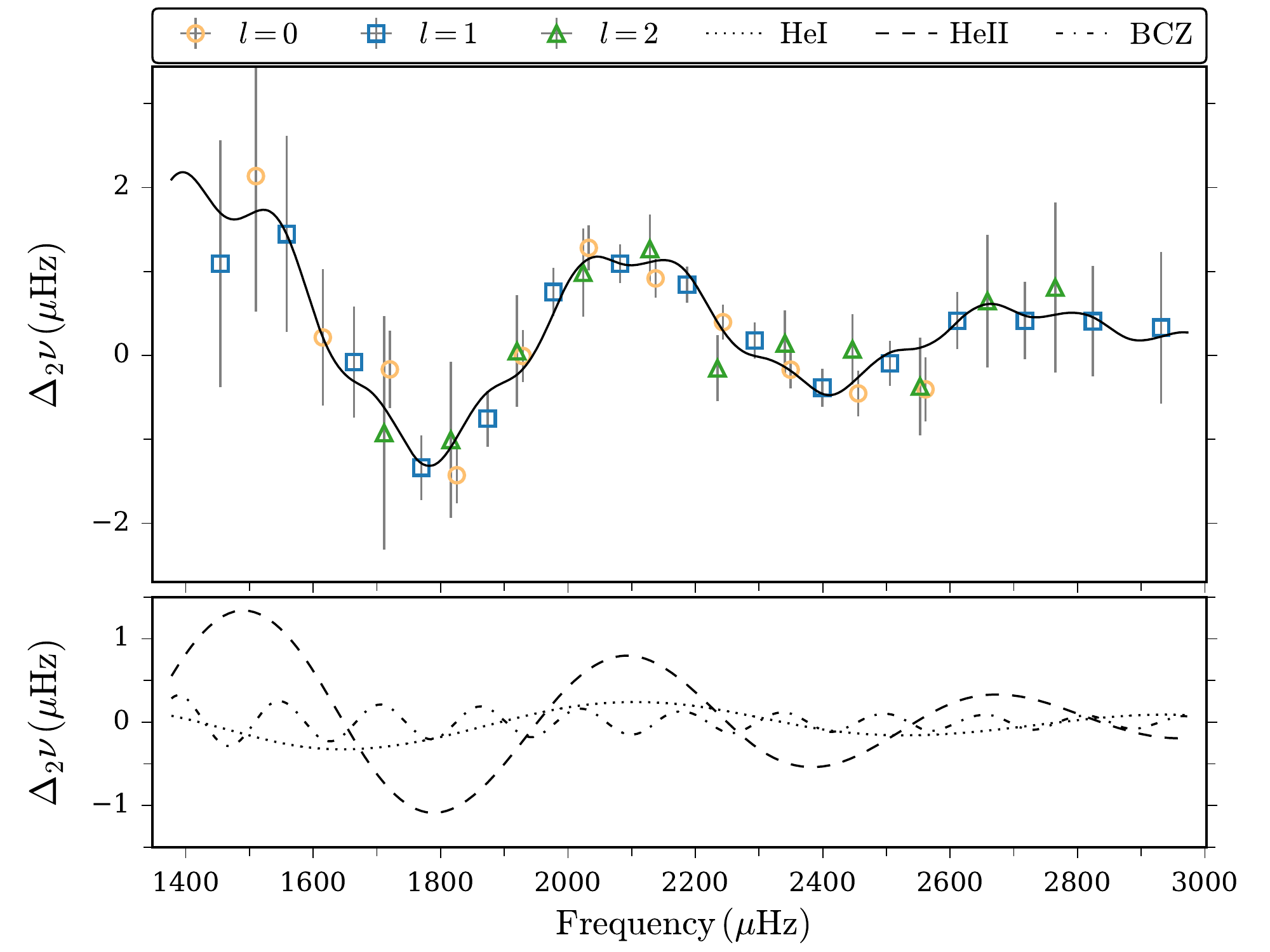}
    \caption{Top: Example of calculated second differences $\Delta_2 \nu(n,l)$ from \eqref{eq:twodiff}, for the star KIC 6225718 (Saxo2). The complete figure set (67 figures) is available in the online journal. The frequencies used to calculate $\Delta_2 \nu(n,l)$ are given in \tref{tab:mode_6225718}, and frequency uncertainties were taken as the average of the asymmetric uncertainties. A clear oscillation is seen in $\Delta_2 \nu$, indicating acoustic glitches. The full line gives the acoustic glitch fit to $\Delta_2 \nu(n,l)$ by Houdek et al. (in prep.). Only the values of $\Delta_2 \nu$ used in the fitting by Houdek et al. (in prep.) are shown. Bottom: Individual components from the acoustic glitches to $\Delta_2 \nu(n,l)$ (full line in top panel), showing the contributions from the first ($\rm HeI$) and second ($\rm HeII$) stages of Helium ionization, and the base of the convective zone ($\rm BCZ$).}
    \label{fig:secdiff}
\end{figure}
%%%%%%%%%%%%%%%%%%%%%%%%%%%%%%%%%%%%%%%%%%%%%%%%%%%%%%%%%%%%%%%%%%

Besides the parameters included in the model of the power spectrum, we computed parameters for derived quantities, such as frequency difference ratios.
Firstly, we derived the frequency ratios defined as \citep[][]{2003A&A...411..215R}
\begin{align}\label{eq:rat}
r_{01}(n) &= \frac{d_{01}(n)}{\Delta\nu_1(n)}, \quad r_{10}(n)=\frac{d_{10}(n)}{\Delta\nu_0(n+1)}\\ 
r_{02}(n) &=\frac{\nu_{n,0}-\nu_{n-1,2}}{\Delta\nu_1(n)} \nonumber \, .
\end{align}
Here, $d_{01}$ and $d_{10}$ are the smooth five-point small frequency separations defined as
\begin{align}
d_{01}(n) &= \tfrac{1}{8} \left( \nu_{n-1,0} - 4\nu_{n-1,1} + 6\nu_{n,0} -4\nu_{n,1} +\nu_{n+1,0} \right)\\
d_{10}(n) &= -\tfrac{1}{8} \left( \nu_{n-1,1} - 4\nu_{n,0} + 6\nu_{n,1} -4\nu_{n+1,0} +\nu_{n+1,1} \right)\, ,
\end{align} 
and the large separation is
\begin{equation}
\Delta\nu_l(n) = \Delta\nu_{n,l} - \Delta\nu_{n-1,l}\, .
\end{equation}
These ratios are useful for model fitting, where they can be used instead of individual frequencies \citep[][]{2014A&A...569A..21L,2015MNRAS.452.2127S} because they are largely insensitive to the stellar surface layers \citep[][]{2008ApJ...683L.175K,2014A&A...568A.123B,2015A&A...581A..58R,2016A&A...592A.159B}. For further details on the use of these ratios we refer to \citet[][]{2003A&A...411..215R,2013A&A...560A...2R}, \citet[][]{2005A&A...434..665R}, \citet{2005MNRAS.356..671O}, and \citet{2011A&A...529A..63S,2013ApJ...769..141S}. 

Secondly, we calculated the second differences:
\begin{equation}\label{eq:twodiff}
\Delta_2 \nu(n,l) = \nu_{n-1, l} - 2\nu_{n,l} + \nu_{n+1,l}\, ,
\end{equation}
which are useful for studying acoustic glitches from the base of the convection zone and the position of the second helium ionization zone \citep[see, \eg,][]{1994MNRAS.267..209B,2004MNRAS.350..277B,2007MNRAS.375..861H,2014ApJ...782...18M}. \fref{fig:secdiff} gives an example of the second differences for KIC 6225718 (Saxo2), together with the best-fitting glitch model from Houdek et al. (in prep.). The second differences shown in \fref{fig:secdiff} were computed using \eqref{eq:twodiff} on the frequencies given in \tref{tab:mode_6225718}, and frequency uncertainties were taken as the average of the asymmetric uncertainties.

In computing these derived quantities we used the full posterior probability distributions (PPDs) of the individual frequencies entering in the descriptions, rather than using simply the median value for the PPD of a given frequency. This ensured that any asymmetries, and deviations from a Gaussian shape in general, that might be for the PPDs of the individual frequencies were properly propagated to the description of the derived quantity. The final value and credible interval were then computed from the distribution of the quantity in the same manner as for the parameters describing the model power spectrum. Using the full distribution also allowed us to easily compute the correlations between the above frequency differences and ratios, such that these might be included as a covariance matrix in any fit to the quantities.

The parameter correlations were calculated in a robust way using the median absolute deviation (MAD) correlation coefficient $r_{\rm MAD}$ \citep[see][]{MR892849,robustcorr}. The MAD estimator is given by the median of the absolute deviation around the median. We opted for $r_{\rm MAD}$ instead of the standard Pearson product-moment correlation coefficient, because the latter would be very susceptible to even a single walker in the MCMC optimization straying away from the stationary solution. The $r_{\rm MAD}$ between two parameters $x$ and $y$ was calculated as follows:
\begin{equation}
r_{\rm MAD} = \frac{{\rm MAD}^2(u)-{\rm MAD}^2(v)}{{\rm MAD}^2(u)+{\rm MAD}^2(v)}\, ,
\end{equation}
where $u$ and $v$ are the robust principle variables for $x$ and $y$:
\begin{align}
u &= \frac{x-{\rm med}(x)}{\sqrt{2}\, {\rm MAD}(x)} + \frac{y-{\rm med}(y)}{\sqrt{2}\, {\rm MAD}(y)}\, ,\\
v &= \frac{x-{\rm med}(x)}{\sqrt{2}\, {\rm MAD}(x)} - \frac{y-{\rm med}(y)}{\sqrt{2}\, {\rm MAD}(y)}\, .
\end{align}

%%%%%%%%%%%%%%%%%%%%%%%%%%%%%%%%%%%%%%%%%%%%%%%%%%%%%%%%%%%%%%%%%%
\begin{figure}
	%trim={<left> <lower> <right> <upper>}, frame
    \centering
	\includegraphics[trim={0 3cm 0 1cm},clip,width=0.9\columnwidth]{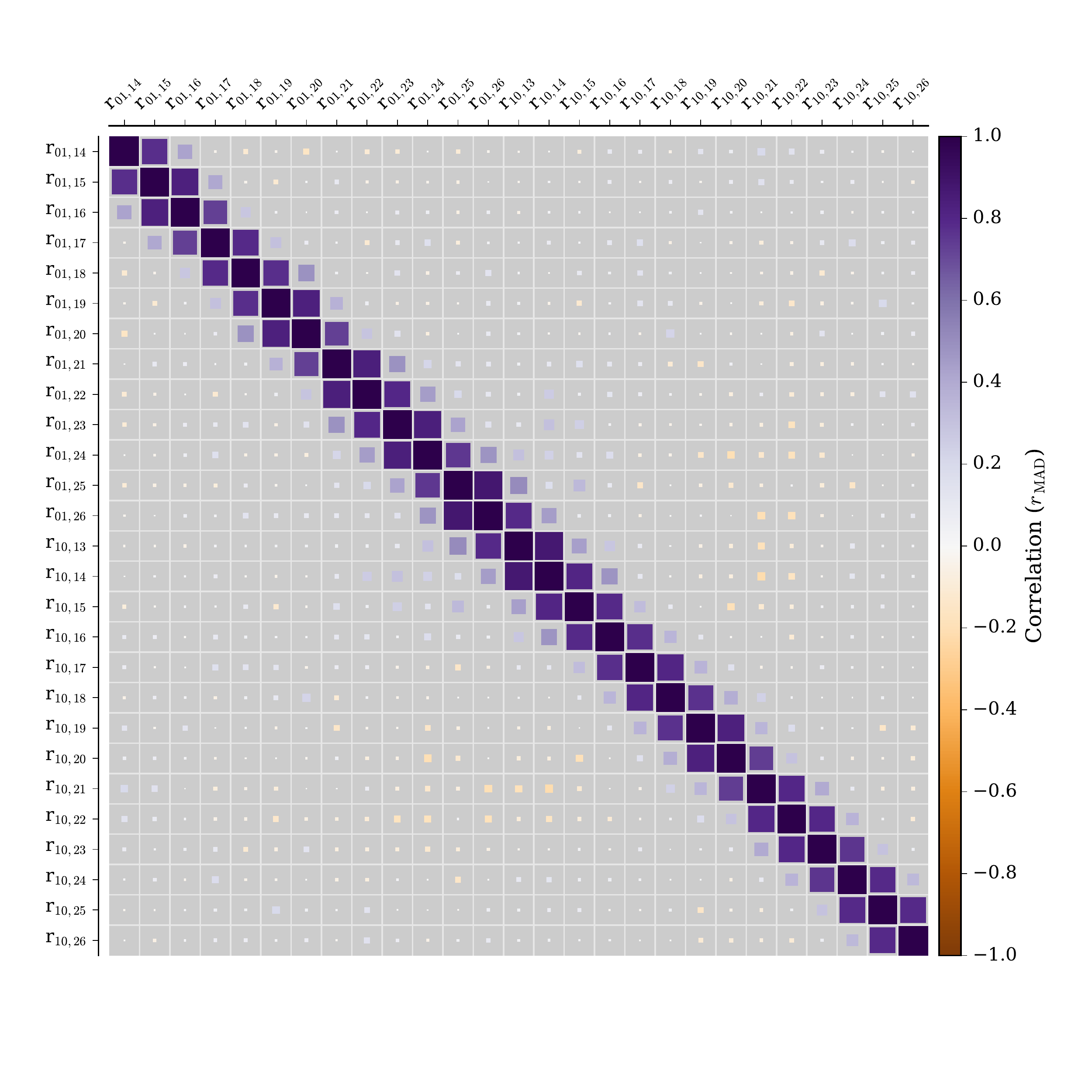}
    \caption{Hinton diagram showing the correlations between frequency difference ratios $r_{01,10}(n)$ of different radial order $n$ for the star KIC 6106415 (Perky). The size and color are proportional to the correlation $r_{\rm MAD}$.}
    \label{fig:correlation}
\end{figure}
%%%%%%%%%%%%%%%%%%%%%%%%%%%%%%%%%%%%%%%%%%%%%%%%%%%%%%%%%%%%%%%%%%
\fref{fig:correlation} gives an example of the correlation matrix between the $r_{01,10}$ frequency ratios for a given star in the form of a Hinton diagram \citep[][]{HintonDia}.

%%%%%%%%%%%%%%%%%%%%%%%%%%%%%%%%%%%%%%%%%%%%%%%%%%%%%%%%%%%%%%%%%%%%%%%%%%%%%%%%%%%%%%%%%%%%%%%%%%%%%%%%%%%%%%%%%%%%%%%%%%

%%%%%%%%%%%%%%%%%%%%%%%%%%%%%%%%%%%%%%%%%%%%%%%%%%%%%%%%%%%%%%%%%%%%%%%%%%%%%%%%%%%%%%%%%%%%%%%%%%%%%%%%%%%%%%%%%%%%%%%%%%
%%%%%%%%%%%%%%%%%%%%%%%%%%%%%%%%%%%%%%%%%%%%%%%%%%%%%%%%%%%%%%%%%%%%%%%%%%%%%%%%%%%%%%%%%%%%%%%%%%%%%%%%%%%%%%%%%%%%%%%%%%
%%%%%%%%%%%%%%%%%%%%%%%%%%%%%%%%%%%%%%%%%%%%%%%%%%%%%%%%%%%%%%%%%%%%%%%%%%%%%%%%%%%%%%%%%%%%%%%%%%%%%%%%%%%%%%%%%%%%%%%%%%
%%%%%%%%%%%%%%%%%%%%%%%%%%%%%%%%%%%%%%%%%%%%%%%%%%%%%%%%%%%%%%%%%%%%%%%%%%%%%%%%%%%%%%%%%%%%%%%%%%%%%%%%%%%%%%%%%%%%%%%%%%
\section{Results}
\label{sec:res}

Below we present some of the conclusions that can be drawn on the different parameters extracted from the peak-bagging.
Results on the rotational splittings and inclination angles will be presented in a separate paper.
All results will be made available on the KASOC database\footnoteref{note1}.

%%%%%%%%%%%%%%%%%%%%%%%%%%%%%%%%%%%%%%%%%%%%%%%%%%%%%%%%%%%%%%%%%%%%%%%%%%%%%%%%%%%%%%%%%%%%%%%%%%%%%%%%%%%%%%%%%%%%%%%%%%
%%%%%%%%%%%%%%%%%%%%%%%%%%%%%%%%%%%%%%%%%%%%%%%%%%%%%%%%%%%%%%%%%%%%%%%%%%%%%%%%%%%%%%%%%%%%%%%%%%%%%%%%%%%%%%%%%%%%%%%%%%
\subsection{Mode frequencies}
\label{sec:fre}

%%%%%%%%%%%%%%%%%%%%%%%%%%%%%%%%%%%%%%%%%%%%%%%%%%%%%%%%%%%%%%%%%%%%%%%%%%%%%%%%%%%%%%%%%%%%%%%%%%%%%%%%%%%%%%%%%%%%%%%%%%
\subsubsection{Frequency uncertainties}
\label{sec:errors}
%%%%%%%%%%%%%%%%%%%%%%%%%%%%%%%%%%%%%%%%%%%%%%%%%%%%%%%%%%%%%%
\begin{figure*}
    \centering
    \begin{subfigure}
        \centering
        \includegraphics[scale=0.347]{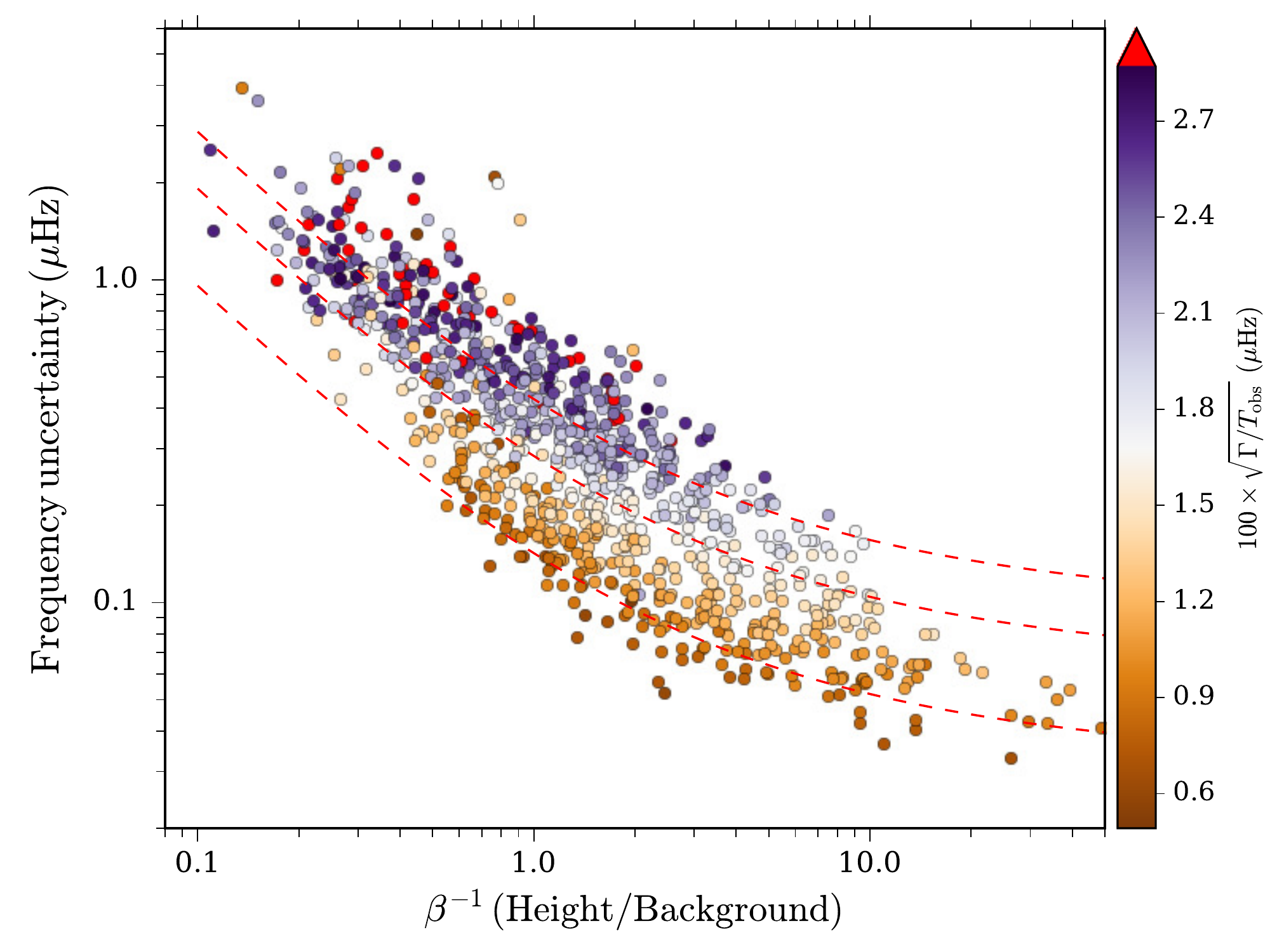}
    \end{subfigure}\hfill
    \begin{subfigure}
        \centering
        \includegraphics[scale=0.347]{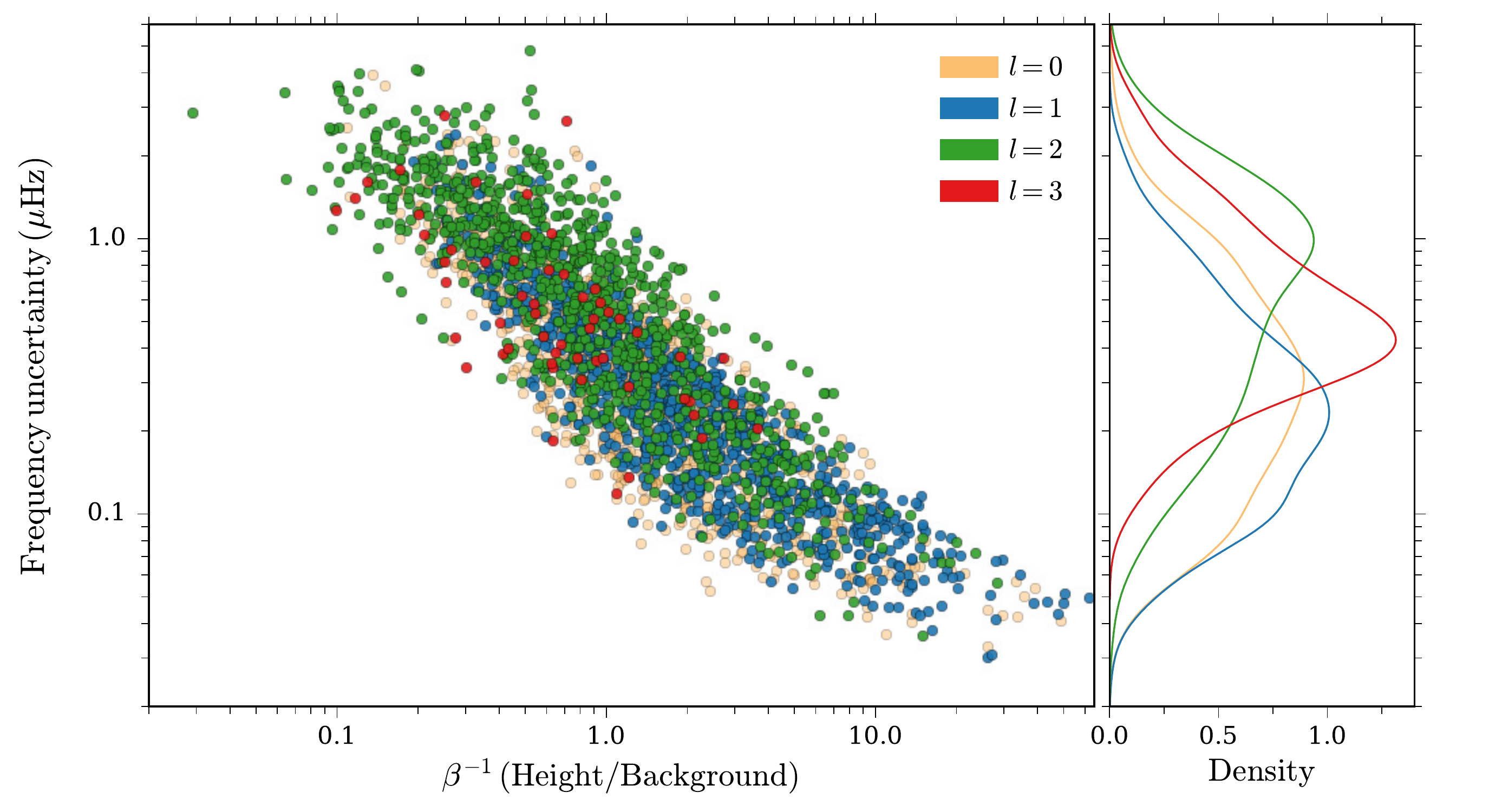}
    \end{subfigure}
    \caption{Left: Frequency uncertainties as a function of the height-to-background ratio ($\beta^{-1}$) for the radial modes ($l=0$). The color scale indicate the value of $\sqrt{\Gamma/T_{\rm obs}}$ which, for a given value of $\beta$, is expected to define the spread in the uncertainties (\eqref{eq:error}). Dashed red lines show the behavior of the $\beta$-dependent factor $\sqrt{f_0(\beta)}$ (\eqref{eq:error_f}) for the values $0.009$, $0.018$, and $0.027$ of $\sqrt{\Gamma/T_{\rm obs}}$. The $\sqrt{f_0(\beta)}$ lines nicely follow the points of a given $\sqrt{\Gamma/T_{\rm obs}}$, showing that the uncertainties behave as expected from an analytical approach (\sref{sec:errors}). Right: Uncertainties for all mode frequencies extracted from the peak-bagging. The color indicates the angular degree $l$ of the modes. The small right panel gives the Gaussian kernel density estimates of the uncertainties for each $l$. }
\label{fig:error}
\end{figure*}
%%%%%%%%%%%%%%%%%%%%%%%%%%%%%%%%%%%%%%%%%%%%%%%%%%%%%%%%%%%%%%
A proper understanding of the uncertainties on mode frequencies is important, because they will ultimately limit the precision with which stellar parameters can be estimated from modeling the individual frequencies.
We may compare the frequency uncertainties from the peak-bagging with expectations from an analytical maximum likelihood (ML) approach \citep[see, \eg,][]{1992ApJ...387..712L,1994A&A...289..649T,2008A&A...486..867B}.
It should be noted that the ML estimator (if unbiased) reaches the \emph{minimum variance bound}, in accordance with the Cram\'{e}r-Rao theorem \citep[][]{MR0016588,MR0015748}. Thus one should expect uncertainties at least as large as those from the ML estimator (MLE). 
The standard way of obtaining uncertainties on a ML estimator is from inverting the negative Hessian matrix, and therefore the standard parameters come from the diagonal elements of the resulting variance-covariance matrix.
The Hessian itself is obtained from the matrix of second derivatives of the log-likelihood function with respect to the parameters.
Assuming an isolated mode as described by \eqref{eq:lor} and a likelihood function as given in \eqref{eq:prob}, one may follow \citet[][]{1992ApJ...387..712L} and \citet{1994A&A...289..649T} in defining a theoretical Hessian corresponding to the average of a large number of realizations. From this, the predicted frequency uncertainty for an isolated mode of a given $l$ is given as \citep[][]{2008A&A...486..867B}:
\begin{equation}\label{eq:error}
\sigma_{\nu_{nl}} = \sqrt{\frac{1}{4\pi} \frac{\Gamma_{nl}}{T_{\rm obs}}\, f_l(\beta, x_s, i_{\star}) }\, ,
\end{equation}
where $\beta = B/H$ (the noise-to-signal ratio) is the level of the background divided by the mode height; $i_{\star}$ gives the stellar inclination; $x_s$ is the reduced splitting, given by $x_s=2\nu_s/\Gamma$; and $T_{\rm obs}$ is the observing duration. For radial modes ($l=0$) the factor $f_l$ only depends on $\beta$ and is given by \citep[][]{1992ApJ...387..712L}: 
\begin{equation}\label{eq:error_f}
f_0(\beta) = \sqrt{\beta + 1} \left( \sqrt{\beta + 1} + \sqrt{\beta} \right)^3\, .
\end{equation}
For the general version of $f_l(\beta, x_s, i_{\star})$, see \citet[][]{2008A&A...486..867B}.

In \fref{fig:error} we show the individual frequency uncertainties obtained from the peak-bagging as a function of $\beta^{-1}$, \ie, the signal-to-noise ratio of the mode.
The left panel shows the behavior for the radial ($l=0$) modes. We see that the uncertainties for a given value of $\sqrt{\Gamma/T_{\rm obs}}$ depend on $\beta$ as expected from \eqref{eq:error_f}, and for a given $\beta$ a clear dependence is seen as a function of $\sqrt{\Gamma/T_{\rm obs}}$. Given the relatively small spread in $T_{\rm obs}$, the vertical gradient largely depicts the gradient in $\Gamma$ and, as expected, the largest values of $\sqrt{\Gamma/T_{\rm obs}}$ are seen towards low $\beta^{-1}$ values because of the correlation between $\Gamma$ and mode heights. Comparing with \eqref{eq:error}, we find that the median uncertainties from our MCMC peak-bagging are ${\sim}1.2$ times larger than those predicted from the ML estimator.

The right panel of \fref{fig:error} shows the uncertainties for all extracted modes. The uncertainties for each angular degree again follow the expected trend against $\beta$. Some extra scatter is expected from $i_{\star}$ and $x_s$, but overall the $l>0$ modes are seen to follow the trend of the $l=0$ ones. The kernel density estimates of the uncertainties for a given $l$, obtained by representing each sample with a Gaussian kernel, show that the uncertainties of dipole ($l=1$) modes are overall the lowest, followed by $l=0$ and $l=2$. This is expected because $\tilde{V}^2_1>\tilde{V}^2_0>\tilde{V}^2_2$ (\eqref{eq:visi}), hence $l=1$ modes will typically have the highest S/N. The $l=3$ are seen not to follow this trend, probably because only the very highest amplitude $l=3$ were selected for fitting.   

It is reassuring to see that the measured uncertainties follow expectations to this level, also considering that \eqref{eq:error} is constructed for an isolated mode without factoring in potential contributions from close-by neighboring modes.
We note that the demonstration of the agreement is useful for predicting uncertainty yields for future missions like TESS \citep[][]{2014SPIE.9143E..20R} and PLATO \citep[][]{2013arXiv1310.0696R}, and could potentially be used to adjust uncertainties derived from a much faster MLE fitting \citep[see, \eg,][]{2012A&A...543A..54A} in cases where the sheer number of stars investigated would render the relatively slow MCMC approach impractical.

%%%%%%%%%%%%%%%%%%%%%%%%%%%%%%%%%%%%%%%%%%%%%%%%%%%%%%%%%%%%%%%%%%%%%%%%%%%%%%%%%%%%%%%%%%%%%%%%%%%%%%%%%%%%%%%%%%%%%%%%%%
\subsubsection{Average seismic parameters}
\label{sec:average}

%%%%%%%%%%%%%%%%%%%%%%%%%%%%%%%%%%%%%%%%%%%%%%%%%%%%%%%%%%%%%%%%%%%%%%%%%%%%%%%%%%%%%%%%%%%%%%%%%%%%%%%%%%%%%%%%%%%%%%%%%%
%%%%%%%%%%%%%%%%%%%%%%%%%%%%%%%%%%%%%%%%%%%%%%%%%%%%%%%%%%%%%%%%%%%%%%%%%%%%%%%%%%%%%%%%%%%%%%%%%%%%%%%%%%%%%%%%%%%%%%%%%%
%%%%%%%%%%%%%%%%%%%%%%%%%%%%%%%%%%%%%%%%%%%%%%%%%%%%%%%%%%%%%%%%%%%%%%%%%%%%%%%%%%%%%%%%%%%%%%%%%%%%%%%%%%%%%%%%%%%%%%%%%%

% Table made Wed Sep 28 16:55:26 2016
\begin{table*} 
\centering 
\resizebox*{!}{\textheight}{% 
\begin{threeparttable} 
\caption{Values from the fit of \eqref{eq:asym3} to the mode frequencies (see \fref{fig:asymp_curve_fit}). Note that \numax is obtained from the fit to the extracted modes.
    Plots of the fitted parameters are shown in Figures~\ref{fig:dnu_numax} to \ref{fig:d2d1}; the values for $\epsilon$ are shown in \fref{fig:dnu_epsilon}.} 
\label{tab:asym_fit_table} 
\begin{tabular}{@{}r@{\hskip 10ex}r@{\hskip 10ex}r@{\hskip 10ex}r@{\hskip 10ex}r@{\hskip 10ex}r@{\hskip 10ex}r@{\hskip 10ex}r@{\hskip 10ex}r@{\hskip 5ex}l} 
\toprule 
\multicolumn{1}{@{\hskip -8ex}c}{KIC} & \multicolumn{1}{@{\hskip -6ex}c}{$\numax$} & \multicolumn{1}{@{\hskip -5ex}c}{$\dnu$} & \multicolumn{1}{@{\hskip -4ex}c}{$\mathrm{d}\dnu/\mathrm{d}n$} & \multicolumn{1}{@{\hskip -5ex}c}{$\epsilon$} & \multicolumn{1}{@{\hskip -3ex}c}{$\delta\nu_{01}$} & \multicolumn{1}{@{\hskip -4ex}c}{$\mathrm{d}\delta\nu_{01}/\mathrm{d}n$} & \multicolumn{1}{@{\hskip -4ex}c}{$\delta\nu_{02}$} & \multicolumn{1}{@{\hskip 1ex}c}{$\mathrm{d}\delta\nu_{02}/\mathrm{d}n$} & \\   & \multicolumn{1}{@{\hskip -6ex}c}{$(\rm \mu Hz)$} & \multicolumn{1}{@{\hskip -5ex}c}{$(\rm \mu Hz)$} &   &   & \multicolumn{1}{@{\hskip -3ex}c}{$(\rm \mu Hz)$} &   & \multicolumn{1}{@{\hskip -4ex}c}{$(\rm \mu Hz)$} & & \\ 
\midrule 
$1435467$ &$1406.7$\rlap{$_{-8.4}^{+6.3}$} & $70.369$\rlap{$_{-0.033}^{+0.034}$} & $0.223$\rlap{$_{-0.009}^{+0.010}$} & $1.114$\rlap{$_{-0.009}^{+0.009}$} & $2.907$\rlap{$_{-0.125}^{+0.125}$} & $-0.126$\rlap{$_{-0.021}^{+0.020}$} & $5.682$\rlap{$_{-0.233}^{+0.252}$} & $-0.116$\rlap{$_{-0.040}^{+0.040}$} & \\ [0.5ex] 
$2837475$ &$1557.6$\rlap{$_{-9.2}^{+8.2}$} & $75.729$\rlap{$_{-0.042}^{+0.041}$} & $0.179$\rlap{$_{-0.012}^{+0.012}$} & $0.911$\rlap{$_{-0.011}^{+0.011}$} & $2.235$\rlap{$_{-0.188}^{+0.177}$} & $-0.017$\rlap{$_{-0.028}^{+0.028}$} & $6.417$\rlap{$_{-0.392}^{+0.408}$} & $-0.126$\rlap{$_{-0.048}^{+0.049}$} & \\ [0.5ex] 
$3427720$ &$2737.0$\rlap{$_{-17.7}^{+10.7}$} & $120.068$\rlap{$_{-0.032}^{+0.031}$} & $0.287$\rlap{$_{-0.008}^{+0.010}$} & $1.356$\rlap{$_{-0.006}^{+0.006}$} & $3.487$\rlap{$_{-0.080}^{+0.079}$} & $-0.053$\rlap{$_{-0.014}^{+0.014}$} & $10.186$\rlap{$_{-0.112}^{+0.100}$} & $-0.124$\rlap{$_{-0.022}^{+0.022}$} & \\ [0.5ex] 
$3456181$ &$970.0$\rlap{$_{-5.9}^{+8.3}$} & $52.264$\rlap{$_{-0.039}^{+0.041}$} & $0.216$\rlap{$_{-0.011}^{+0.011}$} & $0.988$\rlap{$_{-0.014}^{+0.013}$} & $3.551$\rlap{$_{-0.167}^{+0.158}$} & $-0.075$\rlap{$_{-0.023}^{+0.024}$} & $4.370$\rlap{$_{-0.248}^{+0.251}$} & $-0.030$\rlap{$_{-0.043}^{+0.043}$} & \\ [0.5ex] 
$3632418$ &$1166.8$\rlap{$_{-3.8}^{+3.0}$} & $60.704$\rlap{$_{-0.018}^{+0.019}$} & $0.232$\rlap{$_{-0.004}^{+0.004}$} & $1.114$\rlap{$_{-0.006}^{+0.005}$} & $3.697$\rlap{$_{-0.062}^{+0.064}$} & $-0.041$\rlap{$_{-0.010}^{+0.009}$} & $4.189$\rlap{$_{-0.115}^{+0.102}$} & $-0.053$\rlap{$_{-0.019}^{+0.017}$} & \\ [0.5ex] 
$3656476$ &$1925.0$\rlap{$_{-6.3}^{+7.0}$} & $93.194$\rlap{$_{-0.020}^{+0.018}$} & $0.207$\rlap{$_{-0.008}^{+0.009}$} & $1.445$\rlap{$_{-0.004}^{+0.004}$} & $4.608$\rlap{$_{-0.042}^{+0.041}$} & $-0.095$\rlap{$_{-0.011}^{+0.010}$} & $4.554$\rlap{$_{-0.051}^{+0.055}$} & $-0.157$\rlap{$_{-0.014}^{+0.016}$} & \\ [0.5ex] 
$3735871$ &$2862.6$\rlap{$_{-26.5}^{+16.6}$} & $123.049$\rlap{$_{-0.046}^{+0.047}$} & $0.260$\rlap{$_{-0.019}^{+0.019}$} & $1.325$\rlap{$_{-0.009}^{+0.008}$} & $3.823$\rlap{$_{-0.124}^{+0.125}$} & $-0.004$\rlap{$_{-0.029}^{+0.027}$} & $11.028$\rlap{$_{-0.208}^{+0.202}$} & $-0.077$\rlap{$_{-0.053}^{+0.047}$} & \\ [0.5ex] 
$4914923$ &$1817.0$\rlap{$_{-5.2}^{+6.3}$} & $88.531$\rlap{$_{-0.019}^{+0.019}$} & $0.233$\rlap{$_{-0.008}^{+0.007}$} & $1.377$\rlap{$_{-0.004}^{+0.004}$} & $4.534$\rlap{$_{-0.042}^{+0.048}$} & $-0.119$\rlap{$_{-0.010}^{+0.009}$} & $5.378$\rlap{$_{-0.074}^{+0.068}$} & $-0.118$\rlap{$_{-0.021}^{+0.022}$} & \\ [0.5ex] 
$5184732$ &$2089.3$\rlap{$_{-4.1}^{+4.4}$} & $95.545$\rlap{$_{-0.023}^{+0.024}$} & $0.216$\rlap{$_{-0.006}^{+0.005}$} & $1.374$\rlap{$_{-0.005}^{+0.005}$} & $2.772$\rlap{$_{-0.047}^{+0.044}$} & $-0.104$\rlap{$_{-0.009}^{+0.009}$} & $5.863$\rlap{$_{-0.065}^{+0.063}$} & $-0.097$\rlap{$_{-0.015}^{+0.015}$} & \\ [0.5ex] 
$5773345$ &$1101.2$\rlap{$_{-6.6}^{+5.7}$} & $57.303$\rlap{$_{-0.027}^{+0.030}$} & $0.257$\rlap{$_{-0.007}^{+0.007}$} & $1.077$\rlap{$_{-0.010}^{+0.009}$} & $1.754$\rlap{$_{-0.101}^{+0.122}$} & $-0.093$\rlap{$_{-0.016}^{+0.016}$} & $4.131$\rlap{$_{-0.261}^{+0.268}$} & $-0.047$\rlap{$_{-0.046}^{+0.047}$} & \\ [0.5ex] 
$5950854$ &$1926.7$\rlap{$_{-20.4}^{+21.9}$} & $96.629$\rlap{$_{-0.107}^{+0.102}$} & $0.208$\rlap{$_{-0.029}^{+0.032}$} & $1.431$\rlap{$_{-0.020}^{+0.023}$} & $4.988$\rlap{$_{-0.178}^{+0.215}$} & $-0.212$\rlap{$_{-0.038}^{+0.038}$} & $4.713$\rlap{$_{-0.235}^{+0.273}$} & $-0.232$\rlap{$_{-0.074}^{+0.078}$} & \\ [0.5ex] 
$6106415$ &$2248.6$\rlap{$_{-3.9}^{+4.6}$} & $104.074$\rlap{$_{-0.026}^{+0.023}$} & $0.254$\rlap{$_{-0.005}^{+0.005}$} & $1.343$\rlap{$_{-0.005}^{+0.005}$} & $3.422$\rlap{$_{-0.047}^{+0.049}$} & $-0.114$\rlap{$_{-0.008}^{+0.008}$} & $6.881$\rlap{$_{-0.070}^{+0.066}$} & $-0.118$\rlap{$_{-0.013}^{+0.014}$} & \\ [0.5ex] 
$6116048$ &$2126.9$\rlap{$_{-5.0}^{+5.5}$} & $100.754$\rlap{$_{-0.017}^{+0.017}$} & $0.258$\rlap{$_{-0.005}^{+0.005}$} & $1.336$\rlap{$_{-0.003}^{+0.003}$} & $3.687$\rlap{$_{-0.052}^{+0.048}$} & $-0.143$\rlap{$_{-0.009}^{+0.009}$} & $6.034$\rlap{$_{-0.070}^{+0.068}$} & $-0.155$\rlap{$_{-0.016}^{+0.014}$} & \\ [0.5ex] 
$6225718$ &$2364.2$\rlap{$_{-4.6}^{+4.9}$} & $105.695$\rlap{$_{-0.017}^{+0.018}$} & $0.274$\rlap{$_{-0.005}^{+0.005}$} & $1.225$\rlap{$_{-0.004}^{+0.004}$} & $3.207$\rlap{$_{-0.060}^{+0.061}$} & $-0.053$\rlap{$_{-0.010}^{+0.011}$} & $8.741$\rlap{$_{-0.088}^{+0.085}$} & $-0.062$\rlap{$_{-0.017}^{+0.016}$} & \\ [0.5ex] 
$6508366$ &$958.3$\rlap{$_{-3.6}^{+4.6}$} & $51.553$\rlap{$_{-0.047}^{+0.046}$} & $0.223$\rlap{$_{-0.009}^{+0.009}$} & $1.006$\rlap{$_{-0.017}^{+0.017}$} & $2.880$\rlap{$_{-0.138}^{+0.132}$} & $-0.088$\rlap{$_{-0.019}^{+0.019}$} & $2.535$\rlap{$_{-0.199}^{+0.205}$} & $-0.108$\rlap{$_{-0.028}^{+0.030}$} & \\ [0.5ex] 
$6603624$ &$2384.0$\rlap{$_{-5.6}^{+5.4}$} & $110.128$\rlap{$_{-0.012}^{+0.012}$} & $0.201$\rlap{$_{-0.004}^{+0.004}$} & $1.492$\rlap{$_{-0.002}^{+0.002}$} & $2.801$\rlap{$_{-0.027}^{+0.029}$} & $-0.159$\rlap{$_{-0.006}^{+0.006}$} & $4.944$\rlap{$_{-0.034}^{+0.031}$} & $-0.201$\rlap{$_{-0.008}^{+0.008}$} & \\ [0.5ex] 
$6679371$ &$941.8$\rlap{$_{-5.0}^{+5.1}$} & $50.601$\rlap{$_{-0.029}^{+0.029}$} & $0.181$\rlap{$_{-0.007}^{+0.008}$} & $0.880$\rlap{$_{-0.010}^{+0.011}$} & $2.861$\rlap{$_{-0.116}^{+0.123}$} & $-0.016$\rlap{$_{-0.016}^{+0.017}$} & $3.143$\rlap{$_{-0.266}^{+0.265}$} & $0.044$\rlap{$_{-0.034}^{+0.034}$} & \\ [0.5ex] 
$6933899$ &$1389.9$\rlap{$_{-3.6}^{+3.9}$} & $72.135$\rlap{$_{-0.018}^{+0.018}$} & $0.255$\rlap{$_{-0.005}^{+0.005}$} & $1.319$\rlap{$_{-0.005}^{+0.004}$} & $5.314$\rlap{$_{-0.041}^{+0.042}$} & $0.013$\rlap{$_{-0.009}^{+0.008}$} & $4.910$\rlap{$_{-0.054}^{+0.054}$} & $-0.063$\rlap{$_{-0.015}^{+0.014}$} & \\ [0.5ex] 
$7103006$ &$1167.9$\rlap{$_{-6.9}^{+7.2}$} & $59.658$\rlap{$_{-0.030}^{+0.029}$} & $0.211$\rlap{$_{-0.008}^{+0.007}$} & $0.978$\rlap{$_{-0.009}^{+0.010}$} & $2.504$\rlap{$_{-0.107}^{+0.140}$} & $-0.045$\rlap{$_{-0.018}^{+0.019}$} & $4.471$\rlap{$_{-0.295}^{+0.354}$} & $0.031$\rlap{$_{-0.048}^{+0.045}$} & \\ [0.5ex] 
$7106245$ &$2397.9$\rlap{$_{-28.7}^{+24.0}$} & $111.376$\rlap{$_{-0.061}^{+0.063}$} & $0.246$\rlap{$_{-0.032}^{+0.025}$} & $1.392$\rlap{$_{-0.012}^{+0.011}$} & $3.489$\rlap{$_{-0.118}^{+0.110}$} & $-0.113$\rlap{$_{-0.031}^{+0.032}$} & $6.529$\rlap{$_{-0.167}^{+0.189}$} & $-0.265$\rlap{$_{-0.068}^{+0.069}$} & \\ [0.5ex] 
$7206837$ &$1652.5$\rlap{$_{-11.7}^{+10.6}$} & $79.131$\rlap{$_{-0.039}^{+0.037}$} & $0.253$\rlap{$_{-0.011}^{+0.010}$} & $1.054$\rlap{$_{-0.010}^{+0.011}$} & $2.106$\rlap{$_{-0.147}^{+0.140}$} & $-0.054$\rlap{$_{-0.023}^{+0.023}$} & $6.619$\rlap{$_{-0.417}^{+0.419}$} & $-0.094$\rlap{$_{-0.073}^{+0.079}$} & \\ [0.5ex] 
$7296438$ &$1847.8$\rlap{$_{-12.6}^{+8.5}$} & $88.698$\rlap{$_{-0.036}^{+0.040}$} & $0.242$\rlap{$_{-0.015}^{+0.015}$} & $1.358$\rlap{$_{-0.008}^{+0.009}$} & $4.505$\rlap{$_{-0.081}^{+0.073}$} & $-0.055$\rlap{$_{-0.022}^{+0.020}$} & $5.079$\rlap{$_{-0.098}^{+0.088}$} & $-0.135$\rlap{$_{-0.028}^{+0.029}$} & \\ [0.5ex] 
$7510397$ &$1189.1$\rlap{$_{-4.4}^{+3.4}$} & $62.249$\rlap{$_{-0.020}^{+0.020}$} & $0.258$\rlap{$_{-0.004}^{+0.004}$} & $1.112$\rlap{$_{-0.006}^{+0.006}$} & $3.971$\rlap{$_{-0.059}^{+0.072}$} & $0.003$\rlap{$_{-0.009}^{+0.009}$} & $4.370$\rlap{$_{-0.086}^{+0.086}$} & $-0.016$\rlap{$_{-0.017}^{+0.015}$} & \\ [0.5ex] 
$7680114$ &$1709.1$\rlap{$_{-6.5}^{+7.1}$} & $85.145$\rlap{$_{-0.043}^{+0.039}$} & $0.238$\rlap{$_{-0.007}^{+0.007}$} & $1.368$\rlap{$_{-0.009}^{+0.010}$} & $5.039$\rlap{$_{-0.054}^{+0.050}$} & $-0.043$\rlap{$_{-0.011}^{+0.011}$} & $4.980$\rlap{$_{-0.072}^{+0.074}$} & $-0.108$\rlap{$_{-0.016}^{+0.018}$} & \\ [0.5ex] 
$7771282$ &$1465.1$\rlap{$_{-18.7}^{+27.0}$} & $72.463$\rlap{$_{-0.079}^{+0.069}$} & $0.368$\rlap{$_{-0.039}^{+0.042}$} & $1.117$\rlap{$_{-0.019}^{+0.021}$} & $3.527$\rlap{$_{-0.238}^{+0.241}$} & $-0.118$\rlap{$_{-0.050}^{+0.049}$} & $5.058$\rlap{$_{-0.445}^{+0.484}$} & $-0.020$\rlap{$_{-0.104}^{+0.089}$} & \\ [0.5ex] 
$7871531$ &$3455.9$\rlap{$_{-26.5}^{+19.3}$} & $151.329$\rlap{$_{-0.023}^{+0.025}$} & $0.285$\rlap{$_{-0.008}^{+0.009}$} & $1.504$\rlap{$_{-0.004}^{+0.003}$} & $2.142$\rlap{$_{-0.068}^{+0.071}$} & $-0.149$\rlap{$_{-0.014}^{+0.013}$} & $7.350$\rlap{$_{-0.169}^{+0.155}$} & $-0.143$\rlap{$_{-0.049}^{+0.044}$} & \\ [0.5ex] 
$7940546$ &$1116.6$\rlap{$_{-3.6}^{+3.3}$} & $58.762$\rlap{$_{-0.029}^{+0.029}$} & $0.217$\rlap{$_{-0.005}^{+0.004}$} & $1.075$\rlap{$_{-0.009}^{+0.009}$} & $3.985$\rlap{$_{-0.071}^{+0.072}$} & $-0.002$\rlap{$_{-0.011}^{+0.011}$} & $4.346$\rlap{$_{-0.123}^{+0.133}$} & $0.012$\rlap{$_{-0.019}^{+0.019}$} & \\ [0.5ex] 
$7970740$ &$4197.4$\rlap{$_{-18.4}^{+21.2}$} & $173.541$\rlap{$_{-0.068}^{+0.060}$} & $0.272$\rlap{$_{-0.005}^{+0.005}$} & $1.455$\rlap{$_{-0.008}^{+0.010}$} & $2.356$\rlap{$_{-0.084}^{+0.083}$} & $-0.097$\rlap{$_{-0.010}^{+0.011}$} & $7.901$\rlap{$_{-0.165}^{+0.169}$} & $-0.268$\rlap{$_{-0.026}^{+0.025}$} & \\ [0.5ex] 
$8006161$ &$3574.7$\rlap{$_{-10.5}^{+11.4}$} & $149.427$\rlap{$_{-0.014}^{+0.015}$} & $0.195$\rlap{$_{-0.005}^{+0.005}$} & $1.547$\rlap{$_{-0.002}^{+0.002}$} & $3.061$\rlap{$_{-0.046}^{+0.041}$} & $-0.084$\rlap{$_{-0.007}^{+0.008}$} & $9.680$\rlap{$_{-0.063}^{+0.070}$} & $-0.150$\rlap{$_{-0.012}^{+0.012}$} & \\ [0.5ex] 
$8150065$ &$1876.9$\rlap{$_{-32.4}^{+38.1}$} & $89.264$\rlap{$_{-0.121}^{+0.134}$} & $0.403$\rlap{$_{-0.047}^{+0.048}$} & $1.163$\rlap{$_{-0.030}^{+0.029}$} & $3.027$\rlap{$_{-0.203}^{+0.198}$} & $0.036$\rlap{$_{-0.063}^{+0.079}$} & $6.357$\rlap{$_{-0.348}^{+0.342}$} & $0.012$\rlap{$_{-0.128}^{+0.135}$} & \\ [0.5ex] 
$8179536$ &$2074.9$\rlap{$_{-12.0}^{+13.8}$} & $95.090$\rlap{$_{-0.054}^{+0.058}$} & $0.277$\rlap{$_{-0.019}^{+0.019}$} & $1.153$\rlap{$_{-0.013}^{+0.012}$} & $3.137$\rlap{$_{-0.164}^{+0.171}$} & $-0.082$\rlap{$_{-0.032}^{+0.034}$} & $8.245$\rlap{$_{-0.315}^{+0.352}$} & $-0.041$\rlap{$_{-0.070}^{+0.073}$} & \\ [0.5ex] 
$8228742$ &$1190.5$\rlap{$_{-3.7}^{+3.4}$} & $62.071$\rlap{$_{-0.021}^{+0.022}$} & $0.244$\rlap{$_{-0.005}^{+0.005}$} & $1.158$\rlap{$_{-0.007}^{+0.006}$} & $4.371$\rlap{$_{-0.061}^{+0.057}$} & $0.007$\rlap{$_{-0.010}^{+0.010}$} & $4.517$\rlap{$_{-0.087}^{+0.082}$} & $-0.048$\rlap{$_{-0.020}^{+0.019}$} & \\ [0.5ex] 
$8379927$ &$2795.3$\rlap{$_{-5.7}^{+6.0}$} & $120.288$\rlap{$_{-0.018}^{+0.017}$} & $0.232$\rlap{$_{-0.005}^{+0.005}$} & $1.311$\rlap{$_{-0.003}^{+0.003}$} & $3.676$\rlap{$_{-0.062}^{+0.062}$} & $-0.058$\rlap{$_{-0.011}^{+0.011}$} & $10.932$\rlap{$_{-0.083}^{+0.096}$} & $-0.062$\rlap{$_{-0.020}^{+0.022}$} & \\ [0.5ex] 
$8394589$ &$2396.7$\rlap{$_{-9.4}^{+10.5}$} & $109.488$\rlap{$_{-0.035}^{+0.034}$} & $0.234$\rlap{$_{-0.011}^{+0.011}$} & $1.267$\rlap{$_{-0.007}^{+0.006}$} & $3.382$\rlap{$_{-0.091}^{+0.089}$} & $-0.061$\rlap{$_{-0.021}^{+0.022}$} & $7.979$\rlap{$_{-0.161}^{+0.164}$} & $-0.106$\rlap{$_{-0.045}^{+0.045}$} & \\ [0.5ex] 
$8424992$ &$2533.7$\rlap{$_{-28.1}^{+27.0}$} & $120.584$\rlap{$_{-0.064}^{+0.062}$} & $0.120$\rlap{$_{-0.038}^{+0.035}$} & $1.517$\rlap{$_{-0.010}^{+0.012}$} & $2.678$\rlap{$_{-0.103}^{+0.100}$} & $-0.179$\rlap{$_{-0.032}^{+0.033}$} & $5.190$\rlap{$_{-0.143}^{+0.122}$} & $-0.282$\rlap{$_{-0.055}^{+0.046}$} & \\ [0.5ex] 
$8694723$ &$1470.5$\rlap{$_{-4.1}^{+3.7}$} & $75.112$\rlap{$_{-0.021}^{+0.019}$} & $0.296$\rlap{$_{-0.005}^{+0.005}$} & $1.113$\rlap{$_{-0.005}^{+0.005}$} & $5.339$\rlap{$_{-0.067}^{+0.068}$} & $0.005$\rlap{$_{-0.009}^{+0.010}$} & $5.879$\rlap{$_{-0.108}^{+0.111}$} & $0.012$\rlap{$_{-0.021}^{+0.020}$} & \\ [0.5ex] 
$8760414$ &$2455.3$\rlap{$_{-8.3}^{+9.1}$} & $117.230$\rlap{$_{-0.018}^{+0.022}$} & $0.295$\rlap{$_{-0.007}^{+0.007}$} & $1.400$\rlap{$_{-0.004}^{+0.003}$} & $4.403$\rlap{$_{-0.053}^{+0.045}$} & $-0.280$\rlap{$_{-0.012}^{+0.009}$} & $5.132$\rlap{$_{-0.059}^{+0.063}$} & $-0.291$\rlap{$_{-0.015}^{+0.014}$} & \\ [0.5ex] 
$8938364$ &$1675.1$\rlap{$_{-5.8}^{+5.2}$} & $85.684$\rlap{$_{-0.020}^{+0.018}$} & $0.235$\rlap{$_{-0.006}^{+0.007}$} & $1.444$\rlap{$_{-0.004}^{+0.004}$} & $6.491$\rlap{$_{-0.035}^{+0.044}$} & $-0.046$\rlap{$_{-0.009}^{+0.010}$} & $5.184$\rlap{$_{-0.052}^{+0.048}$} & $-0.119$\rlap{$_{-0.014}^{+0.013}$} & \\ [0.5ex] 
$9025370$ &$2988.6$\rlap{$_{-16.9}^{+20.0}$} & $132.628$\rlap{$_{-0.024}^{+0.030}$} & $0.205$\rlap{$_{-0.015}^{+0.016}$} & $1.475$\rlap{$_{-0.005}^{+0.004}$} & $3.099$\rlap{$_{-0.062}^{+0.065}$} & $-0.066$\rlap{$_{-0.017}^{+0.018}$} & $9.141$\rlap{$_{-0.119}^{+0.113}$} & $-0.126$\rlap{$_{-0.030}^{+0.037}$} & \\ [0.5ex] 
$9098294$ &$2314.7$\rlap{$_{-10.4}^{+9.2}$} & $108.894$\rlap{$_{-0.022}^{+0.023}$} & $0.251$\rlap{$_{-0.008}^{+0.009}$} & $1.439$\rlap{$_{-0.005}^{+0.004}$} & $3.331$\rlap{$_{-0.057}^{+0.053}$} & $-0.185$\rlap{$_{-0.011}^{+0.012}$} & $5.265$\rlap{$_{-0.088}^{+0.086}$} & $-0.228$\rlap{$_{-0.025}^{+0.024}$} & \\ [0.5ex] 
$9139151$ &$2690.4$\rlap{$_{-9.0}^{+14.5}$} & $117.294$\rlap{$_{-0.032}^{+0.031}$} & $0.240$\rlap{$_{-0.010}^{+0.011}$} & $1.337$\rlap{$_{-0.006}^{+0.006}$} & $3.557$\rlap{$_{-0.077}^{+0.083}$} & $-0.019$\rlap{$_{-0.020}^{+0.017}$} & $10.050$\rlap{$_{-0.158}^{+0.162}$} & $-0.093$\rlap{$_{-0.057}^{+0.055}$} & \\ [0.5ex] 
$9139163$ &$1729.8$\rlap{$_{-5.9}^{+6.2}$} & $81.170$\rlap{$_{-0.036}^{+0.042}$} & $0.241$\rlap{$_{-0.005}^{+0.005}$} & $1.007$\rlap{$_{-0.010}^{+0.010}$} & $2.079$\rlap{$_{-0.098}^{+0.109}$} & $-0.024$\rlap{$_{-0.010}^{+0.012}$} & $6.213$\rlap{$_{-0.215}^{+0.218}$} & $0.040$\rlap{$_{-0.028}^{+0.026}$} & \\ [0.5ex] 
$9206432$ &$1866.4$\rlap{$_{-14.9}^{+10.3}$} & $84.926$\rlap{$_{-0.051}^{+0.046}$} & $0.135$\rlap{$_{-0.013}^{+0.013}$} & $0.958$\rlap{$_{-0.012}^{+0.012}$} & $3.235$\rlap{$_{-0.210}^{+0.207}$} & $0.001$\rlap{$_{-0.031}^{+0.027}$} & $7.115$\rlap{$_{-0.411}^{+0.388}$} & $0.004$\rlap{$_{-0.055}^{+0.056}$} & \\ [0.5ex] 
$9353712$ &$934.3$\rlap{$_{-8.3}^{+11.1}$} & $51.467$\rlap{$_{-0.104}^{+0.091}$} & $0.254$\rlap{$_{-0.011}^{+0.012}$} & $1.095$\rlap{$_{-0.031}^{+0.038}$} & $3.536$\rlap{$_{-0.159}^{+0.146}$} & $-0.038$\rlap{$_{-0.018}^{+0.018}$} & $3.907$\rlap{$_{-0.250}^{+0.236}$} & $-0.090$\rlap{$_{-0.031}^{+0.031}$} & \\ [0.5ex] 
$9410862$ &$2278.8$\rlap{$_{-16.6}^{+31.2}$} & $107.390$\rlap{$_{-0.053}^{+0.050}$} & $0.223$\rlap{$_{-0.021}^{+0.020}$} & $1.343$\rlap{$_{-0.010}^{+0.009}$} & $3.625$\rlap{$_{-0.121}^{+0.116}$} & $-0.181$\rlap{$_{-0.032}^{+0.027}$} & $6.098$\rlap{$_{-0.204}^{+0.201}$} & $-0.239$\rlap{$_{-0.086}^{+0.081}$} & \\ [0.5ex] 
$9414417$ &$1155.3$\rlap{$_{-4.6}^{+6.1}$} & $60.115$\rlap{$_{-0.024}^{+0.024}$} & $0.237$\rlap{$_{-0.007}^{+0.006}$} & $1.045$\rlap{$_{-0.008}^{+0.008}$} & $3.572$\rlap{$_{-0.092}^{+0.106}$} & $-0.029$\rlap{$_{-0.014}^{+0.015}$} & $4.648$\rlap{$_{-0.200}^{+0.211}$} & $-0.006$\rlap{$_{-0.030}^{+0.031}$} & \\ [0.5ex] 
$9812850$ &$1255.2$\rlap{$_{-7.0}^{+9.1}$} & $64.746$\rlap{$_{-0.068}^{+0.067}$} & $0.240$\rlap{$_{-0.011}^{+0.012}$} & $1.067$\rlap{$_{-0.020}^{+0.021}$} & $2.654$\rlap{$_{-0.166}^{+0.156}$} & $-0.134$\rlap{$_{-0.025}^{+0.025}$} & $4.418$\rlap{$_{-0.307}^{+0.297}$} & $-0.032$\rlap{$_{-0.045}^{+0.048}$} & \\ [0.5ex] 
$9955598$ &$3616.8$\rlap{$_{-29.6}^{+21.2}$} & $153.283$\rlap{$_{-0.032}^{+0.029}$} & $0.195$\rlap{$_{-0.011}^{+0.011}$} & $1.529$\rlap{$_{-0.004}^{+0.005}$} & $2.796$\rlap{$_{-0.076}^{+0.073}$} & $-0.095$\rlap{$_{-0.017}^{+0.016}$} & $8.941$\rlap{$_{-0.126}^{+0.143}$} & $-0.172$\rlap{$_{-0.030}^{+0.029}$} & \\ [0.5ex] 
$9965715$ &$2079.3$\rlap{$_{-10.4}^{+9.2}$} & $97.236$\rlap{$_{-0.042}^{+0.041}$} & $0.373$\rlap{$_{-0.015}^{+0.016}$} & $1.139$\rlap{$_{-0.009}^{+0.009}$} & $3.685$\rlap{$_{-0.129}^{+0.132}$} & $-0.102$\rlap{$_{-0.027}^{+0.024}$} & $7.958$\rlap{$_{-0.269}^{+0.254}$} & $-0.105$\rlap{$_{-0.058}^{+0.063}$} & \\ [0.5ex] 
$10068307$ &$995.1$\rlap{$_{-2.7}^{+2.8}$} & $53.945$\rlap{$_{-0.020}^{+0.019}$} & $0.247$\rlap{$_{-0.003}^{+0.004}$} & $1.131$\rlap{$_{-0.006}^{+0.007}$} & $4.220$\rlap{$_{-0.058}^{+0.055}$} & $0.014$\rlap{$_{-0.008}^{+0.008}$} & $3.799$\rlap{$_{-0.088}^{+0.083}$} & $-0.041$\rlap{$_{-0.014}^{+0.014}$} & \\ [0.5ex] 
$10079226$ &$2653.0$\rlap{$_{-44.3}^{+47.7}$} & $116.345$\rlap{$_{-0.052}^{+0.059}$} & $0.264$\rlap{$_{-0.028}^{+0.027}$} & $1.350$\rlap{$_{-0.011}^{+0.010}$} & $3.236$\rlap{$_{-0.199}^{+0.205}$} & $-0.014$\rlap{$_{-0.046}^{+0.043}$} & $9.387$\rlap{$_{-0.371}^{+0.401}$} & $0.098$\rlap{$_{-0.098}^{+0.098}$} & \\ [0.5ex] 
$10162436$ &$1052.0$\rlap{$_{-4.2}^{+4.0}$} & $55.725$\rlap{$_{-0.039}^{+0.035}$} & $0.242$\rlap{$_{-0.004}^{+0.004}$} & $1.106$\rlap{$_{-0.012}^{+0.013}$} & $3.746$\rlap{$_{-0.080}^{+0.078}$} & $-0.033$\rlap{$_{-0.009}^{+0.009}$} & $3.706$\rlap{$_{-0.117}^{+0.115}$} & $-0.031$\rlap{$_{-0.018}^{+0.017}$} & \\ [0.5ex] 
$10454113$ &$2357.2$\rlap{$_{-9.1}^{+8.2}$} & $105.063$\rlap{$_{-0.033}^{+0.031}$} & $0.283$\rlap{$_{-0.009}^{+0.009}$} & $1.206$\rlap{$_{-0.007}^{+0.007}$} & $3.059$\rlap{$_{-0.110}^{+0.095}$} & $-0.063$\rlap{$_{-0.020}^{+0.016}$} & $9.426$\rlap{$_{-0.165}^{+0.178}$} & $-0.079$\rlap{$_{-0.033}^{+0.033}$} & \\ [0.5ex] 
$10516096$ &$1689.8$\rlap{$_{-5.8}^{+4.6}$} & $84.424$\rlap{$_{-0.025}^{+0.022}$} & $0.257$\rlap{$_{-0.008}^{+0.009}$} & $1.318$\rlap{$_{-0.006}^{+0.005}$} & $5.001$\rlap{$_{-0.061}^{+0.054}$} & $-0.058$\rlap{$_{-0.012}^{+0.013}$} & $5.248$\rlap{$_{-0.083}^{+0.084}$} & $-0.089$\rlap{$_{-0.022}^{+0.023}$} & \\ [0.5ex] 
$10644253$ &$2899.7$\rlap{$_{-22.8}^{+21.3}$} & $123.080$\rlap{$_{-0.055}^{+0.056}$} & $0.250$\rlap{$_{-0.016}^{+0.016}$} & $1.313$\rlap{$_{-0.011}^{+0.010}$} & $3.863$\rlap{$_{-0.139}^{+0.140}$} & $0.004$\rlap{$_{-0.030}^{+0.029}$} & $11.378$\rlap{$_{-0.155}^{+0.192}$} & $-0.138$\rlap{$_{-0.049}^{+0.049}$} & \\ [0.5ex] 
$10730618$ &$1282.1$\rlap{$_{-12.7}^{+14.6}$} & $66.333$\rlap{$_{-0.064}^{+0.061}$} & $0.239$\rlap{$_{-0.017}^{+0.016}$} & $1.032$\rlap{$_{-0.018}^{+0.018}$} & $2.651$\rlap{$_{-0.218}^{+0.199}$} & $0.019$\rlap{$_{-0.037}^{+0.040}$} & $4.556$\rlap{$_{-0.454}^{+0.428}$} & $0.113$\rlap{$_{-0.080}^{+0.073}$} & \\ [0.5ex] 
$10963065$ &$2203.7$\rlap{$_{-6.3}^{+6.7}$} & $103.179$\rlap{$_{-0.027}^{+0.027}$} & $0.297$\rlap{$_{-0.008}^{+0.008}$} & $1.275$\rlap{$_{-0.005}^{+0.005}$} & $3.567$\rlap{$_{-0.071}^{+0.072}$} & $-0.081$\rlap{$_{-0.014}^{+0.014}$} & $7.083$\rlap{$_{-0.096}^{+0.103}$} & $-0.058$\rlap{$_{-0.020}^{+0.022}$} & \\ [0.5ex] 
$11081729$ &$1968.3$\rlap{$_{-12.6}^{+11.0}$} & $90.116$\rlap{$_{-0.047}^{+0.048}$} & $0.242$\rlap{$_{-0.017}^{+0.016}$} & $1.020$\rlap{$_{-0.011}^{+0.011}$} & $3.056$\rlap{$_{-0.264}^{+0.251}$} & $-0.103$\rlap{$_{-0.026}^{+0.031}$} & $6.602$\rlap{$_{-0.664}^{+0.605}$} & $0.010$\rlap{$_{-0.084}^{+0.087}$} & \\ [0.5ex] 
$11253226$ &$1590.6$\rlap{$_{-6.8}^{+10.6}$} & $76.858$\rlap{$_{-0.030}^{+0.026}$} & $0.183$\rlap{$_{-0.008}^{+0.008}$} & $0.920$\rlap{$_{-0.008}^{+0.008}$} & $1.748$\rlap{$_{-0.136}^{+0.155}$} & $-0.132$\rlap{$_{-0.018}^{+0.020}$} & $6.973$\rlap{$_{-0.396}^{+0.435}$} & $0.039$\rlap{$_{-0.049}^{+0.048}$} & \\ [0.5ex] 
$11772920$ &$3674.7$\rlap{$_{-36.1}^{+55.1}$} & $157.746$\rlap{$_{-0.033}^{+0.032}$} & $0.238$\rlap{$_{-0.015}^{+0.015}$} & $1.516$\rlap{$_{-0.005}^{+0.004}$} & $2.366$\rlap{$_{-0.084}^{+0.078}$} & $-0.082$\rlap{$_{-0.019}^{+0.017}$} & $7.849$\rlap{$_{-0.209}^{+0.200}$} & $-0.131$\rlap{$_{-0.063}^{+0.059}$} & \\ [0.5ex] 
$12009504$ &$1865.6$\rlap{$_{-6.2}^{+7.7}$} & $88.217$\rlap{$_{-0.025}^{+0.026}$} & $0.289$\rlap{$_{-0.007}^{+0.008}$} & $1.200$\rlap{$_{-0.006}^{+0.006}$} & $3.533$\rlap{$_{-0.079}^{+0.078}$} & $-0.072$\rlap{$_{-0.014}^{+0.014}$} & $6.117$\rlap{$_{-0.134}^{+0.133}$} & $-0.066$\rlap{$_{-0.034}^{+0.032}$} & \\ [0.5ex] 
$12069127$ &$884.7$\rlap{$_{-8.0}^{+10.1}$} & $48.400$\rlap{$_{-0.048}^{+0.048}$} & $0.204$\rlap{$_{-0.014}^{+0.012}$} & $1.061$\rlap{$_{-0.018}^{+0.018}$} & $3.399$\rlap{$_{-0.173}^{+0.153}$} & $-0.037$\rlap{$_{-0.025}^{+0.023}$} & $3.650$\rlap{$_{-0.291}^{+0.244}$} & $-0.102$\rlap{$_{-0.039}^{+0.039}$} & \\ [0.5ex] 
$12069424$ &$2188.5$\rlap{$_{-3.0}^{+4.6}$} & $103.277$\rlap{$_{-0.020}^{+0.021}$} & $0.246$\rlap{$_{-0.004}^{+0.004}$} & $1.437$\rlap{$_{-0.004}^{+0.004}$} & $3.392$\rlap{$_{-0.039}^{+0.039}$} & $-0.152$\rlap{$_{-0.006}^{+0.007}$} & $5.274$\rlap{$_{-0.047}^{+0.051}$} & $-0.181$\rlap{$_{-0.010}^{+0.010}$} & \\ [0.5ex] 
$12069449$ &$2561.3$\rlap{$_{-5.6}^{+5.0}$} & $116.929$\rlap{$_{-0.013}^{+0.012}$} & $0.201$\rlap{$_{-0.004}^{+0.004}$} & $1.461$\rlap{$_{-0.002}^{+0.002}$} & $2.707$\rlap{$_{-0.031}^{+0.032}$} & $-0.117$\rlap{$_{-0.006}^{+0.006}$} & $6.045$\rlap{$_{-0.044}^{+0.037}$} & $-0.133$\rlap{$_{-0.010}^{+0.009}$} & \\ [0.5ex] 
$12258514$ &$1512.7$\rlap{$_{-2.9}^{+3.3}$} & $74.799$\rlap{$_{-0.015}^{+0.016}$} & $0.209$\rlap{$_{-0.004}^{+0.004}$} & $1.281$\rlap{$_{-0.004}^{+0.004}$} & $4.227$\rlap{$_{-0.043}^{+0.042}$} & $-0.061$\rlap{$_{-0.008}^{+0.007}$} & $4.827$\rlap{$_{-0.052}^{+0.061}$} & $-0.056$\rlap{$_{-0.011}^{+0.011}$} & \\ [0.5ex] 
$12317678$ &$1212.4$\rlap{$_{-4.9}^{+5.5}$} & $63.464$\rlap{$_{-0.024}^{+0.025}$} & $0.231$\rlap{$_{-0.005}^{+0.005}$} & $0.928$\rlap{$_{-0.008}^{+0.006}$} & $3.883$\rlap{$_{-0.115}^{+0.112}$} & $-0.032$\rlap{$_{-0.013}^{+0.013}$} & $5.273$\rlap{$_{-0.194}^{+0.188}$} & $-0.061$\rlap{$_{-0.025}^{+0.023}$} & \\ [0.5ex] 
\bottomrule
\end{tabular} 
 \end{threeparttable}% 
 } 
\end{table*}

%%%%%%%%%%%%%%%%%%%%%%%%%%%%%%%%%%%%%%%%%%%%%%%%%%%%%%%%%%%%%%%%%%%%%%%%%%%%%%%%%%%%%%%%%%%%%%%%%%%%%%%%%%%%%%%%%%%%%%%%%%
%%%%%%%%%%%%%%%%%%%%%%%%%%%%%%%%%%%%%%%%%%%%%%%%%%%%%%%%%%%%%%%%%%%%%%%%%%%%%%%%%%%%%%%%%%%%%%%%%%%%%%%%%%%%%%%%%%%%%%%%%%
%%%%%%%%%%%%%%%%%%%%%%%%%%%%%%%%%%%%%%%%%%%%%%%%%%%%%%%%%%%%%%%%%%%%%%%%%%%%%%%%%%%%%%%%%%%%%%%%%%%%%%%%%%%%%%%%%%%%%%%%%%

For each star we computed the average seismic parameters, including \numax and \dnu, from the mode frequencies (see \tref{tab:asym_fit_table}).
The value of \numax was obtained by fitting a Gaussian function to the extracted amplitudes as a function of frequency. The value found for the solar \numax of $3078 \pm 13\,\rm \mu Hz$ is in agreement with the value of $\nu_{\rm max,\odot}=3090 \pm 30\,\rm \mu Hz$ by \citet[][]{2011ApJ...743..143H}. We obtained \dnu and $\epsilon$ from a fit to the extracted mode frequencies with an extended version of the asymptotic relation, as in \citet[][]{2014ApJ...782....2L} \citep[see also][]{2011A&A...525L...9M, 2013A&A...550A.126M}. This fit was made in a Bayesian manner using \texttt{emcee}, with a likelihood function assuming Gaussian frequency errors, and with final parameter values and uncertainties given by the posterior medians and HPDs.
We initially adopted the following formula:
\begin{align}\label{eq:asym2}
\nu_{nl} \simeq & \left(n + \frac{l}{2} + \epsilon \right) \Delta\nu_0 - l(l+1) D_0\\
 &-l(l+1) \frac{\mathrm{d} D_0}{\mathrm{d} n}(n-n_{\numax, l})\notag \\
&+  \frac{\mathrm{d} \Delta\nu / \mathrm{d} n }{2} (n-n_{\numax, l})^2\, . \notag
\end{align}
Here $\Delta\nu_0$ gives the value of the large separation at \numax. The latter corresponds to the radial order $n_{\numax}$, which may take on a non-integer value. Specifically, $n_{\numax, l}$ is obtained for a given $l$ from interpolating the measured mode frequencies ($\nu_l$) against radial order ($n_l$) to the frequency of \numax.
This description gives a direct estimate of, for instance, the small frequency separations $\delta\nu_{01}$, given as the amount by which $l=1$ modes are offset from the the neighboring $l=0$ modes ($\delta\nu_{01} = (\nu_{n,0} - \nu_{n+1,0})/2 - \nu_{n,1} = 2D_0$), and $\delta\nu_{02} =\nu_{n,l=0} - \nu_{n-1,l=2}= 6D_0$ \citep[see, \eg,][]{2011arXiv1107.1723B}. The value of $\delta\nu_{02}$ is a good probe of the evolutionary state of the star because it is sensitive to the sound-speed gradient of the core, which in turn varies with the composition. 
The frequency dependence of these frequency separations is captured by the change in $D_0$, assumed to change linearly with $n$ (or frequency), as found, for instance, for the Sun by \citet[][]{1990Natur.347..536E} \citep[see also][]{1992A&A...257..287T,1992A&A...255..363A}. Lastly, the change in the large separation, assumed quadratic in $n$, is included and mimics the overall curvature of the ridges in the \'{e}chelle diagram. 

\begin{figure}
	\centering
	\includegraphics[width=\columnwidth]{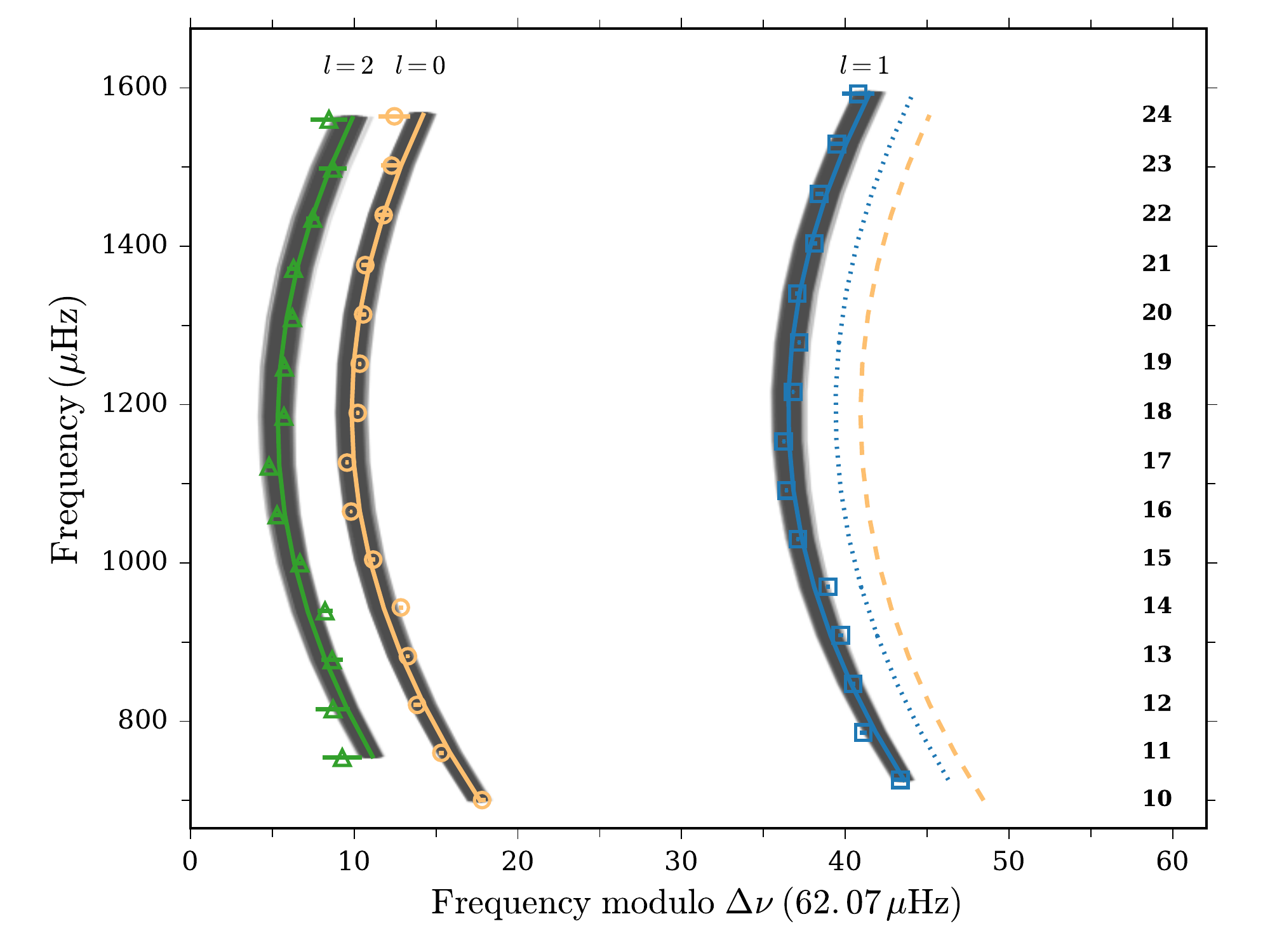}
	\caption{Example of a fit of \eqref{eq:asym3} to individual frequencies of KIC 8228742 (Horace), displayed in \'{e}chelle diagram format. The angular degrees of the ridges are indicated in the top of the plot; the numbers on the right side of the plot give the radial orders $n$ of the radial $l=0$ modes. The full lines give the solution from the median of the posteriors distributions of the fit; the dark-grey lines give 500 solutions with parameters drawn from the posterior distributions. The small separation $\delta\nu_{02}$ is given by the difference between the $l=0$ and $2$ ridges at \numax. The dashed line gives the $l=0$ ridge offset by $\dnu/2$, hence $\delta\nu_{01}$ is given by the difference between this line and the $l=1$ ridge at \numax. The dotted line gives the expected position of the $l=1$ ridge from the asymptotic relation (\eqref{eq:asymp}), where $\delta\nu_{01} = \delta\nu_{02} / 3$. A clear oscillatory behavior from acoustic glitches is seen for the frequencies around the median solution.}
\label{fig:asymp_curve_fit}
\end{figure}

We found this description to perform poorly across the range of stars in the sample, which is to be expected for more evolved stars \citep[see, \eg,][]{1989A&A...226..278G,Christensen-Dalsgaard:226720}. To that end we modified the formula as follows:
\begin{align}\label{eq:asym3}
\nu_{nl} \simeq & \left(n + \frac{l}{2} + \epsilon \right) \Delta\nu_0 - \delta\nu_{0l} \\
&-\frac{\mathrm{d} \delta\nu_{0l}}{\mathrm{d} n}(n-n_{\numax,l})\notag \\
&+ \frac{\mathrm{d} \Delta\nu / \mathrm{d} n }{2} (n-n_{\numax,l})^2 \, ,\notag 
\end{align}
where the term $l(l+1)D_0$ has been replaced by $\delta\nu_{0l}$, which takes on independent values for $l>0$. Thereby we optimize independently for the separations $\delta\nu_{02}$ and $\delta\nu_{01}$. In the fit we also included \numax with a Gaussian prior from the fit to the mode amplitudes; the values of \numax, \dnu, and $\epsilon$ correlate strongly, but we include \numax to properly marginalize over the uncertainty of the pivoting $n_{\numax,l}$.
\fref{fig:asymp_curve_fit} gives an example of the fit of \eqref{eq:asym3} to KIC 8228742 (Horace).
All parameters from the fits are listed in \tref{tab:asym_fit_table}.
%%%%%%%%%%%%%%%%%%%%%%%%%%%%%%%%%%%%%%%%%%%%%%%%%%%%%%%%%%%%%%%
%%%%%%%%%%%%%%%%%%%%%%%%%%%%%%%%%%%%%%%%%%%%%%%%%%%%%%%%%%%%%%
%%%%%%%%%%%%%%%%%%%%%%%%%%%%%%%%%%%%%%%%%%%%%%%%%%%%%%%%%%%%%%
\begin{figure*}
    \centering
    \begin{subfigure}
        \centering
        \includegraphics[width=\columnwidth]{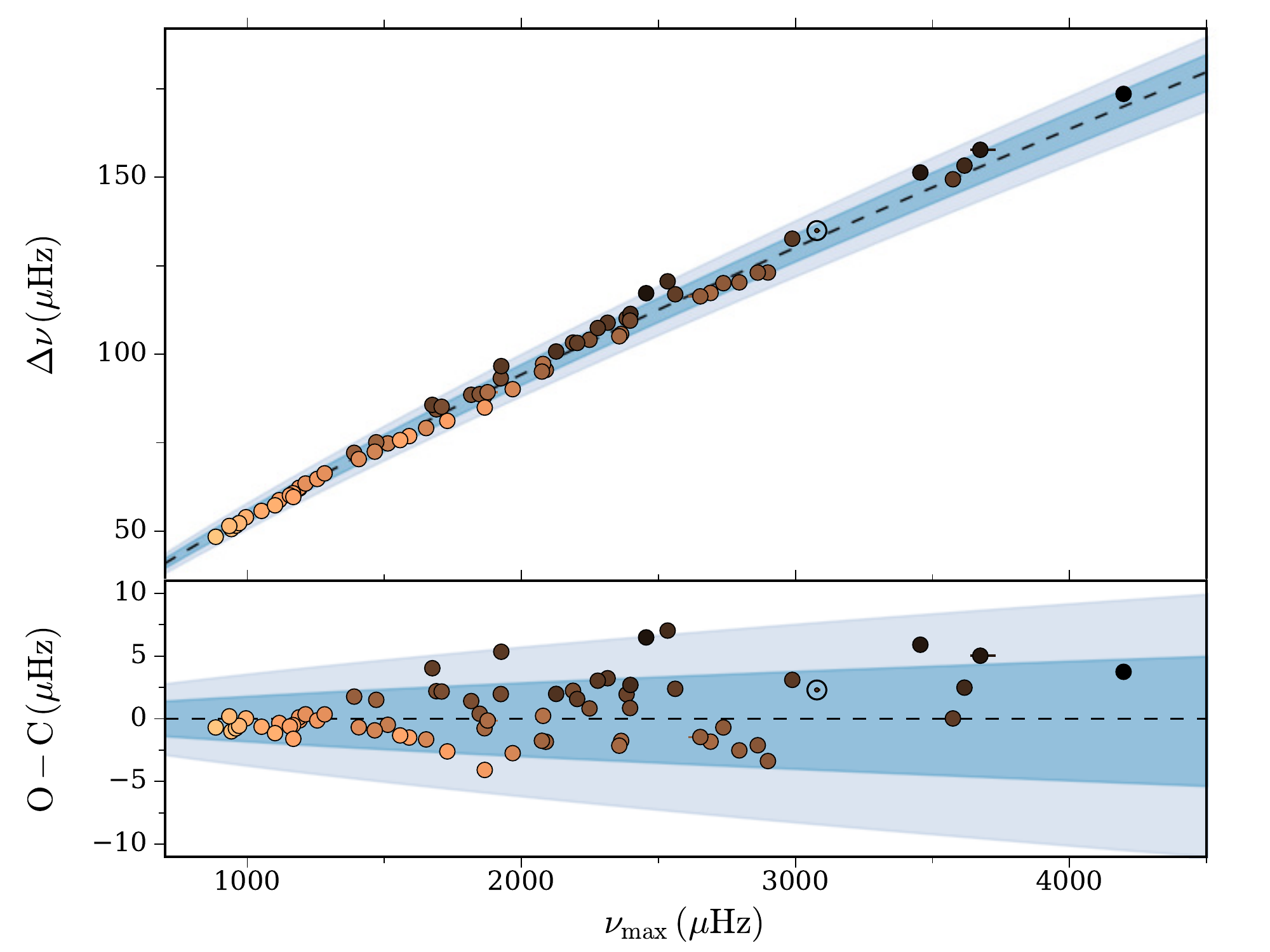}
    \end{subfigure}\hfill
    \begin{subfigure}
        \centering
        \includegraphics[width=\columnwidth]{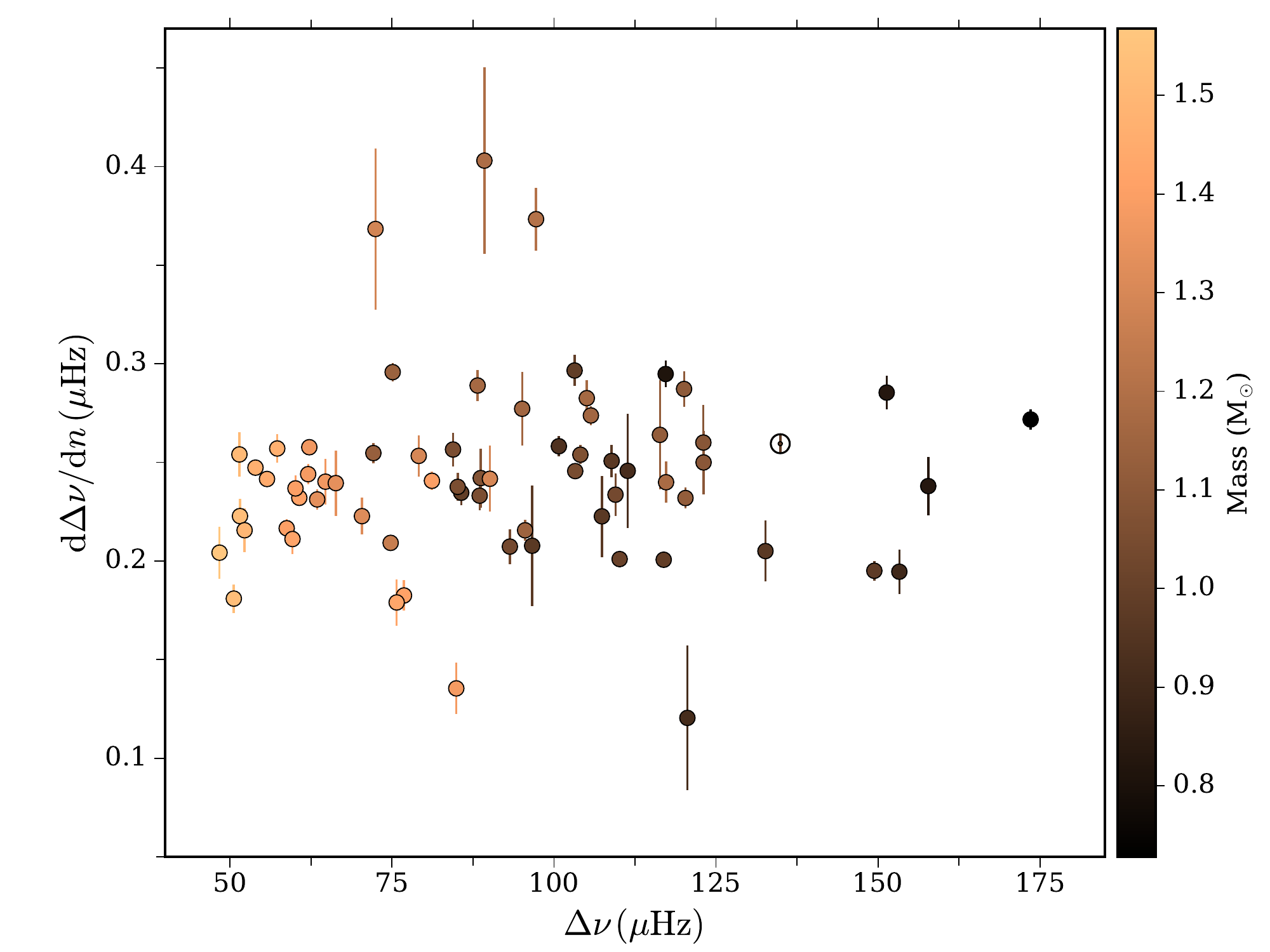}
    \end{subfigure}
    \caption{Left: Measured values of \dnu against \numax from fitting of \eqref{eq:asym3}, with the color indicating the modeled mass. The dashed line shows the empirical relation by \citet{2011ApJ...743..143H}, with the $1$ and $2\sigma$ uncertainties on the relation given by the dark and light blue regions. From our values we calculate a correlation of $\rho_{\alpha, \beta}=-0.99913$ between then $\alpha$ and $\beta$ parameters of the \citet{2011ApJ...743..143H} relation --- this was included to estimate the uncertainty regions. The bottom panel shows the residuals of the measured values to the relation. Measured values are seen to overall agree well with expectations, with a modest mass gradient across the residuals. Right: Measured gradient of \dnu with radial order, $\mathrm{d} \Delta\nu / \mathrm{d} n$, defining the overall curvature of the ridges in the \'{e}chelle diagram against \dnu.}
\label{fig:dnu_numax}
\end{figure*}
%%%%%%%%%%%%%%%%%%%%%%%%%%%%%%%%%%%%%%%%%%%%%%%%%%%%%%%%%%%%%%

\fref{fig:dnu_numax} shows the estimated values for \dnu vs. \numax, together with the empirical relation by \citet{2011ApJ...743..143H}, with which we find an excellent agreement. We can even see the expected mass gradient in the residuals, with higher mass stars having slightly lower \dnu for a given \numax. Also shown is the change in the large separation, which is found to be near-constant at $\mathrm{d} \Delta\nu / \mathrm{d} n\approx 0.25\, \rm \mu Hz$ and thus always with the same concavity sign, corresponding to a positive gradient of the large separation with frequency.
Here one should remember that a constant value of $\mathrm{d} \Delta\nu / \mathrm{d} n$ against $\Delta\nu_0$ would correspond to a linear change in $\Delta\nu$ with 
frequency. 

For the Sun we obtain a value of $\dnu_{\odot}$ of $134.91\pm 0.02\,\rm \mu Hz$; comparing this to the value of $\dnu_{\odot}=135.1\pm 0.1\,\rm \mu Hz$ by \citet[][]{2011ApJ...743..143H} we would expect a slightly smaller value, because the curvature is incorporated in our estimation. However, the value of $\dnu_{\odot}=134.92\pm 0.02\,\rm \mu Hz$ by \citet[][]{1992A&A...257..287T}, who also used a second-order version of the asymptotic relation, compares very well to our estimate.

The estimated values of $\epsilon$ have already been shown in \fref{fig:dnu_epsilon}. For the estimation of $\epsilon$ it should be noted that this value is strongly anti-correlated with \dnu, and a small change in \dnu can thus induce a significant change in $\epsilon$ \citep[][]{2011ApJ...742L...3W}. This will especially take effect if the estimate of \numax is off, because \numax defines the pivoting point for the change in $\delta\nu_{01}$, $\delta\nu_{02}$, and the curvatures. Like \citet[][]{2011ApJ...742L...3W}, we see an offset between the estimated $\epsilon$ and those obtained from model tracks, which is ascribed to the effects of the incorrect modeling of the stellar surface layers. In \fref{fig:dnu_epsilon} we show the $\epsilon$-tracks adopted from \citet[][]{2011ApJ...743..161W}, which were computed from evolutionary tracks from the Aarhus STellar Evolution Code \citep[ASTEC;][]{2008Ap&SS.316...13C}, neglecting diffusion and core overshoot and with $Z_0=0.017$. It should also be noted that the $\epsilon$ from models were derived in a slightly different manner than used here, as described by \citet[][]{2011ApJ...743..161W}.
We therefore also estimated the values for \dnu and $\epsilon$ using the \citet[][]{2011ApJ...743..161W} method, namely from a weighted fit of the asymptotic function in \eqref{eq:asymp} to the radial mode frequencies as a function of radial order $n$, and with weights given by a Gaussian with a FWHM of $0.25\, \numax$. Comparing the values for \dnu and $\epsilon$ from the fit of \eqref{eq:asymp} versus \eqref{eq:asym3} we only find minor differences in estimated values. For \dnu the maximum absolute difference was ${\sim}0.36\, \rm \mu Hz$, with no systematic differences; for $\epsilon$ a maximum difference of ${\sim}0.14$ was found, and again with no systematic differences.

%%%%%%%%%%%%%%%%%%%%%%%%%%%%%%%%%%%%%%%%%%%%%%%%%%%%%%%%%%%%%%
\begin{figure*}
    \centering
    \begin{subfigure}
        \centering
        \includegraphics[width=\columnwidth]{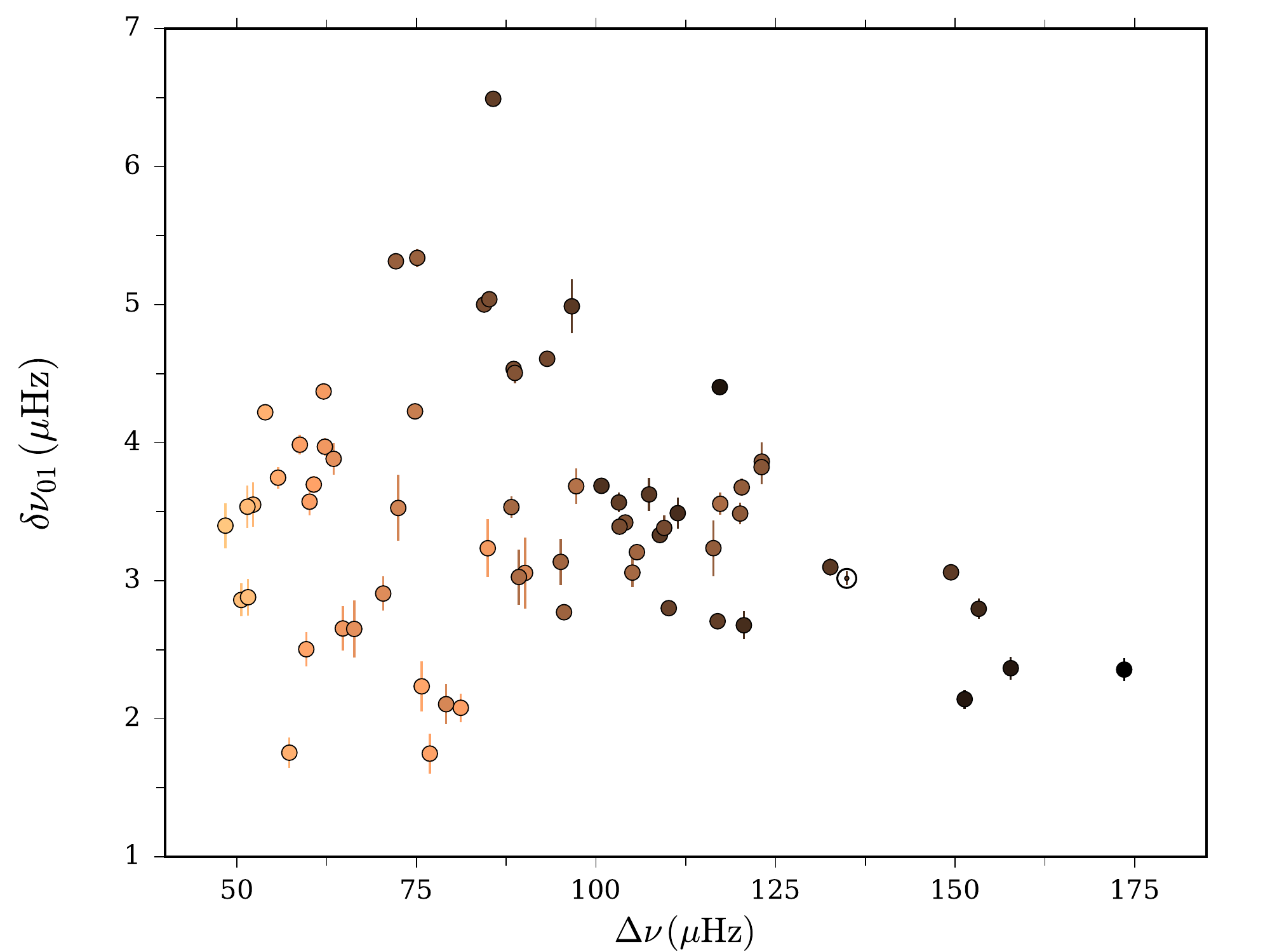}
    \end{subfigure}\hfill
    \begin{subfigure}
        \centering
        \includegraphics[width=\columnwidth]{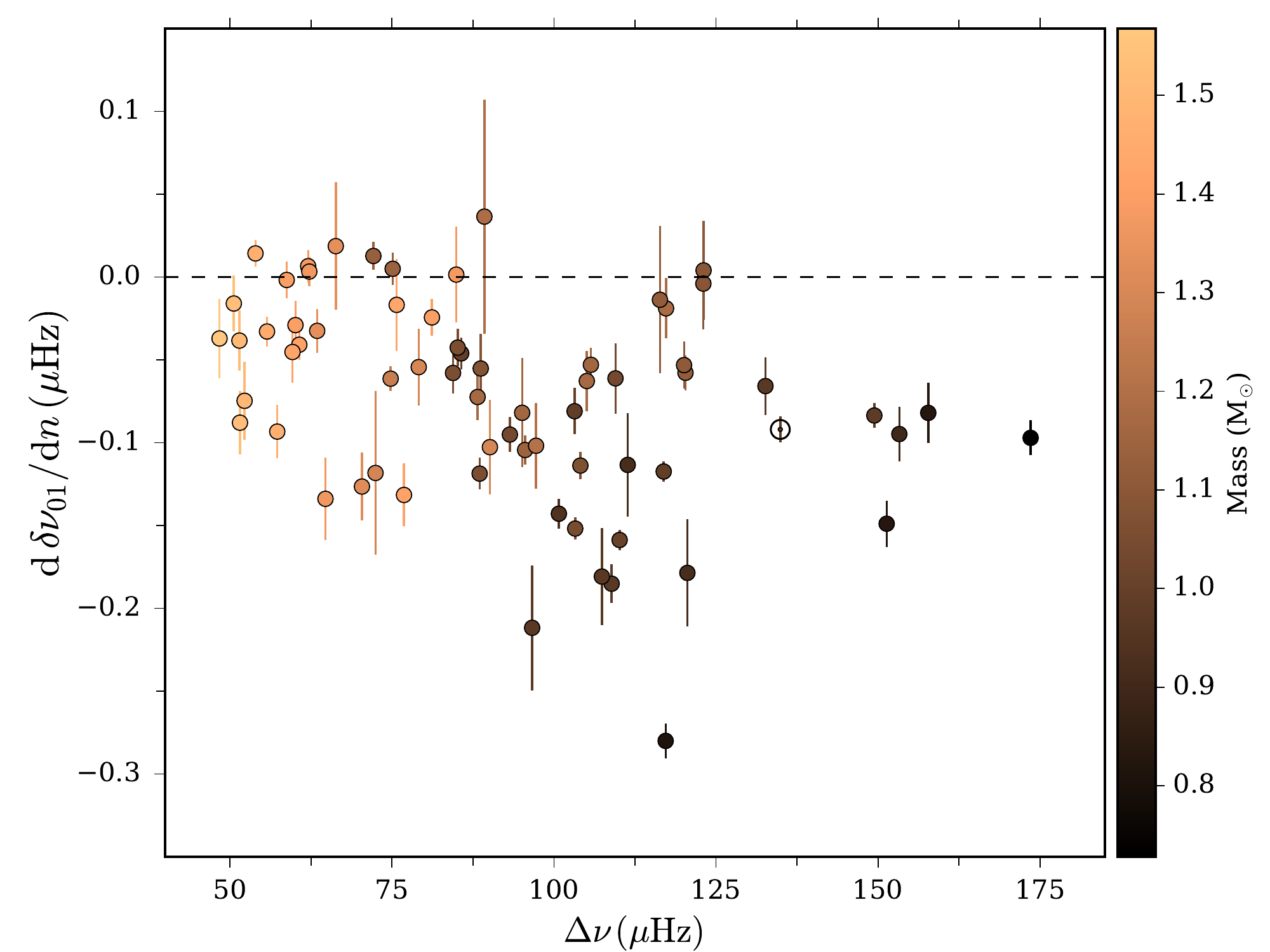}
    \end{subfigure}
    \caption{Left: Measured values of $\delta\nu_{01}$ against \dnu from fitting of \eqref{eq:asym3}, with the color indicating the modeled mass. Right: Measured change in $\delta\nu_{01}$ with radial order, $\mathrm{d} \delta\nu_{01}/\mathrm{d} n$ against \dnu. All slopes in $\delta\nu_{01}$ are seen to be either negative or consistent with zero.}
\label{fig:d1}
\end{figure*}
%%%%%%%%%%%%%%%%%%%%%%%%%%%%%%%%%%%%%%%%%%%%%%%%%%%%%%%%%%%%%%
%%%%%%%%%%%%%%%%%%%%%%%%%%%%%%%%%%%%%%%%%%%%%%%%%%%%%%%%%%%%%%
\begin{figure*}
    \centering
    \begin{subfigure}
        \centering
        \includegraphics[width=\columnwidth]{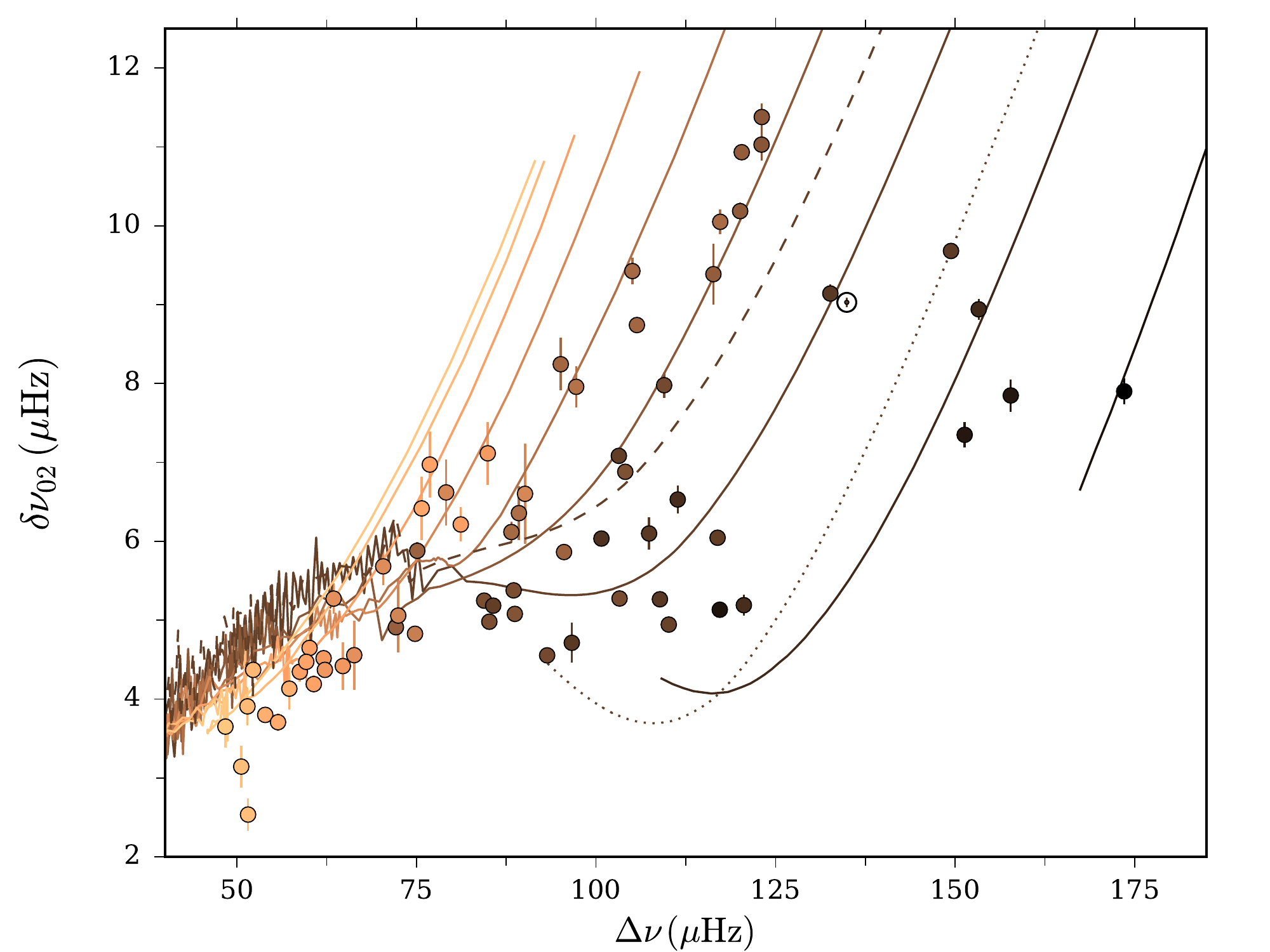}
    \end{subfigure}\hfill
    \begin{subfigure}
        \centering
        \includegraphics[width=\columnwidth]{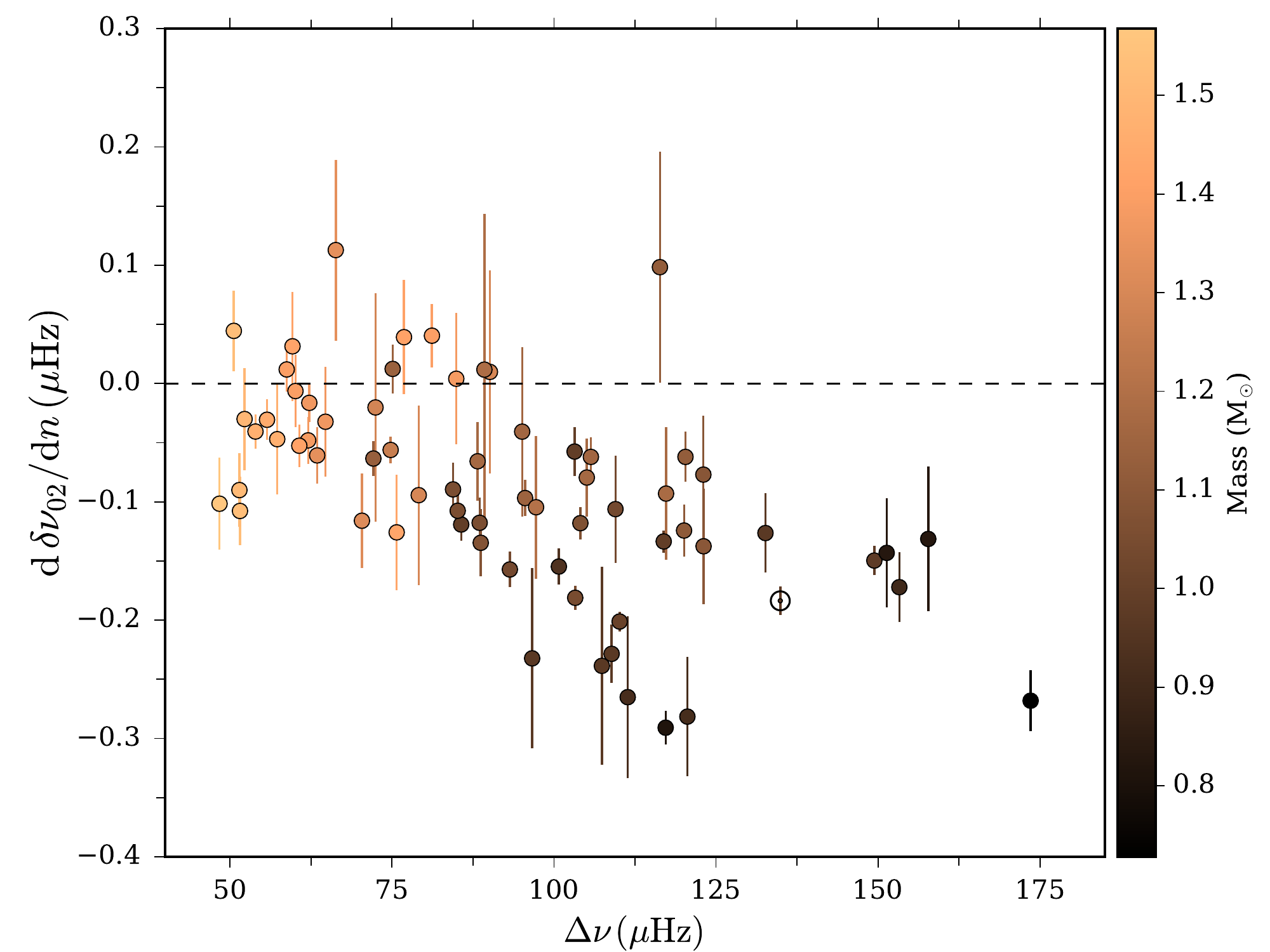}
    \end{subfigure}
    \caption{Left: Measured values of $\delta\nu_{02}$ against \dnu from fitting of \eqref{eq:asym3}, with the color indicating the modeled mass. Shown are also the $\delta\nu_{02}$-evolutionary tracks from \citet[][]{2011ApJ...743..161W}, calculated from ASTEC evolutionary tracks with $Z_0=0.017$ \citep[][]{2008Ap&SS.316...13C}, for masses going from $0.8\, \rm M_{\odot}$ to $1.6\, \rm M_{\odot}$ in steps of $0.1\, \rm M_{\odot}$. For the $M=1\, \rm M_{\odot}$ track we have indicated the effect of changing the metallicity to $Z_0 = 0.011$ (dashed) and $Z_0 = 0.028$ (dotted). Right: Measured change in $\delta\nu_{02}$ with radial order, $\mathrm{d} \delta\nu_{02}/\mathrm{d} n$ against \dnu. All slopes in $\delta\nu_{02}$ are seen to be either negative or consistent with zero.}
\label{fig:d2}
\end{figure*}
\begin{figure*}
    \centering
    \begin{subfigure}
        \centering
        \includegraphics[width=\columnwidth]{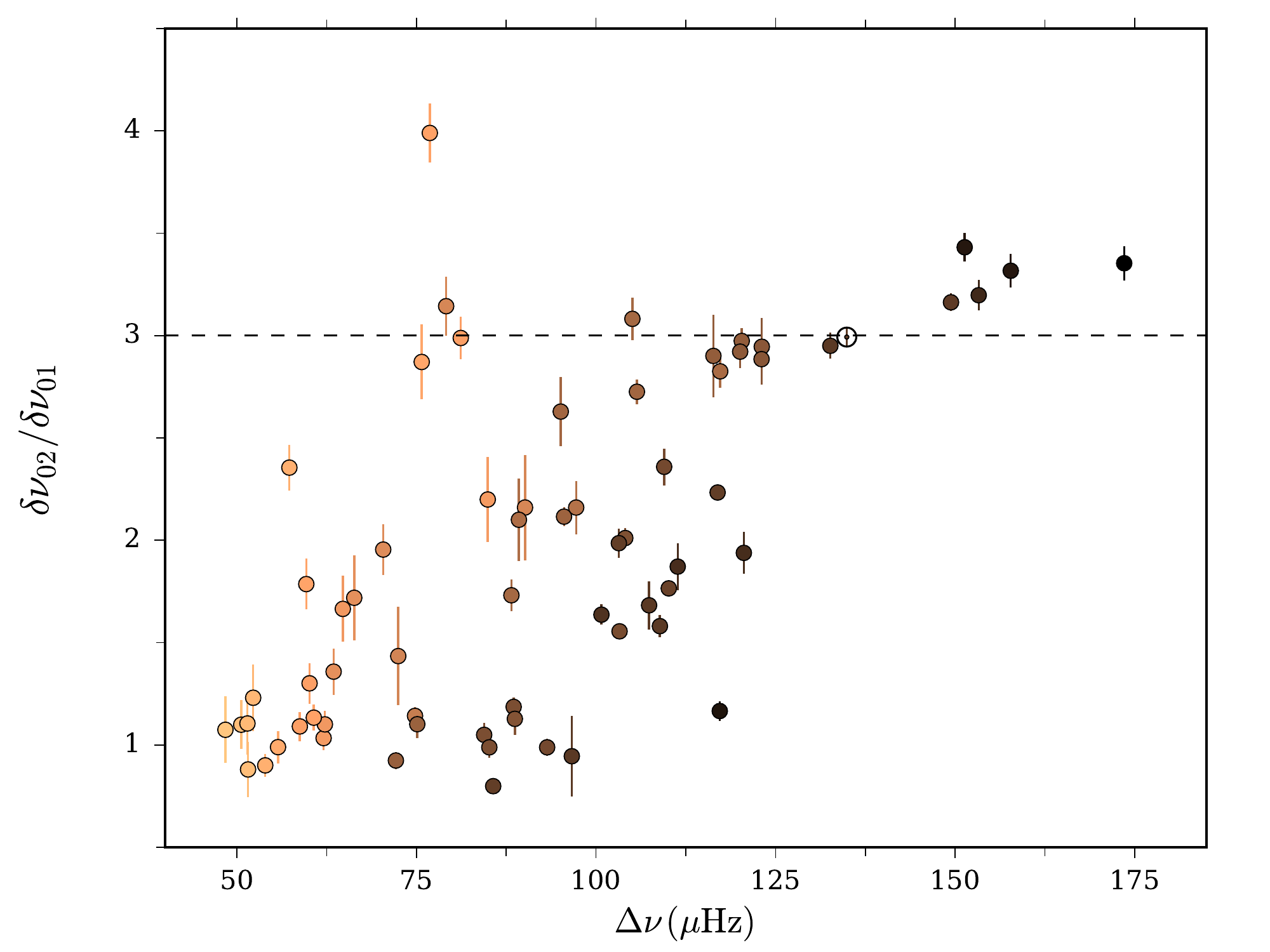}
    \end{subfigure}\hfill
    \begin{subfigure}
        \centering
        \includegraphics[width=\columnwidth]{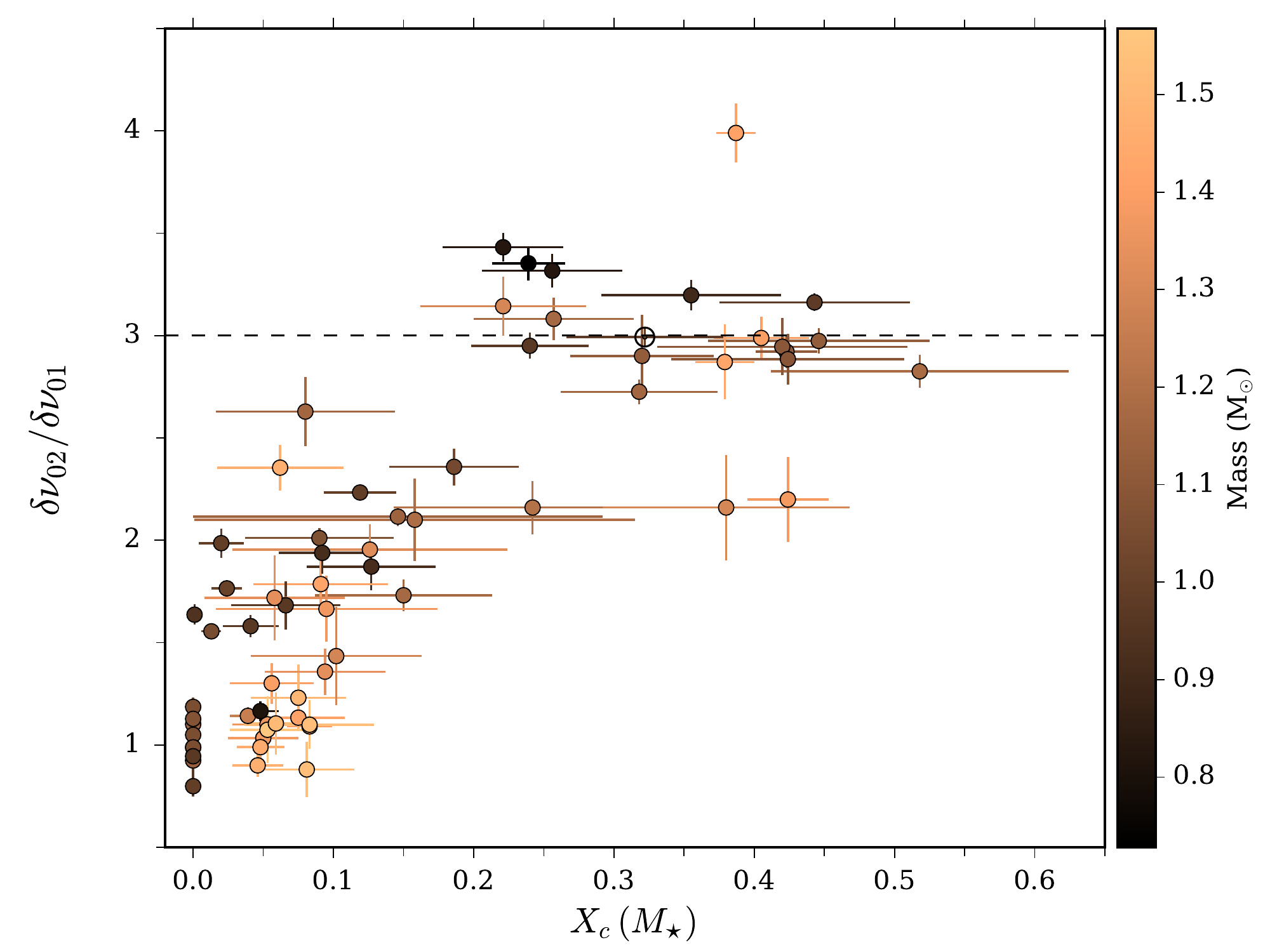}
    \end{subfigure}
    \caption{Left: Ratio between measured values of $\delta\nu_{02}$ and $\delta\nu_{01}$ against \dnu, with the color indicating the modeled mass. The dashed line indicates the expected value of $3$ from the asymptotic relation \eqref{eq:asymp} where $\delta\nu_{0l} = l(l+1)D_0$. Right: Ratio plotted against the modeled value for central hydrogen content $X_c$, which is a good indicator for the evolutionary state. More evolved stars with low $X_c$ are seen to deviate more from the asymptotic expectation than lesser evolved stars. }
\label{fig:d2d1}
\end{figure*}

%%%%%%%%%%%%%%%%%%%%%%%%%%%%%%%%%%%%%%%%%%%%%%%%%%%%%%%%%%%%%%
In Figures~\ref{fig:d1} and \ref{fig:d2} we show the estimated values for the separations $\delta\nu_{01}$ and $\delta\nu_{02}$, together with their gradients in $n$ (or frequency); the results for $\delta\nu_{02}$ in \fref{fig:d2} are given in the form of a modified C-D diagram \citep[see][]{1993ASPC...42..347C,2011ApJ...742L...3W,2011ApJ...743..161W,2012ApJ...751L..36W}. Because all values for the changes in $\delta\nu_{01}$ and $\delta\nu_{02}$ are either zero or negative, we see that the small separations are virtually all decreasing functions of frequency. In \fref{fig:d2} we show again the ASTEC tracks from \citet[][]{2011ApJ...743..161W}.
It is interesting to see the degree to which the asymptotic relations to first (\eqref{eq:asymp}) and second (\eqref{eq:asym2}) order are satisfied.
From these one would expect a ratio of $\delta\nu_{02} / \delta\nu_{01} = 3$ from assuming $\delta\nu_{0l} = l(l+1)D_0$ -- in \fref{fig:d2d1} we show this ratio for the sample.
%%%%%%%%%%%%%%%%%%%%%%%%%%%%%%%%%%%%%%%%%%%%%%%%%%%%%%%%%%%%%%
%%%%%%%%%%%%%%%%%%%%%%%%%%%%%%%%%%%%%%%%%%%%%%%%%%%%%%%%%%%%%%
As seen, only a few stars, including the Sun, actually adhere to the expectation from the asymptotic relation. This was similarly found by \citet[][]{2010Ap&SS.328...51C}\footnote{expressed in the units $D_{n0}^{(1)}=\delta\nu_{01}$ and $D_{n0}=\delta\nu_{02}/3$.} for $\alpha$ Cen A \citep[][]{2004ApJ...614..380B}, while $\alpha$ Cen B \citep[][]{2005ApJ...635.1281K} was found to fulfill the asymptotic relation. As explained by \citet[][]{2010Ap&SS.328...51C} this reflects a rapid variation of the sound speed in the core of $\alpha$ Cen A from its more evolved state due to its higher mass compared to the B component. 
Similarly, \citet[][]{2014ApJ...794..114V} found an increasing departure from the asymptotic description of the oscillation frequencies while fitting the signatures of the acoustic glitches. The departure becomes noticeable when a peak in the Brunt-V{\"a}is{\"a}l{\"a} frequency $N$ develops just outside the stellar core towards the end of the MS and becomes comparable or higher than the lowest frequency fitted. To check this evolutionary explanation, we plot in the right panel of \fref{fig:d2d1} the $\delta\nu_{02} / \delta\nu_{01}$ against the central hydrogen content $X_c$, because this serves as a good probe for the evolutionary state. As seen, stars with high $X_c$, \ie, stars that are less evolved, adhere better to the asymptotic regime than the evolved stars with low $X_c$. The values of $X_c$ were obtained as part of the modeling with the BASTA pipeline; we refer to Paper~II for further details.
While a difference from a ratio of $\delta\nu_{02} / \delta\nu_{01} = 3$ can be explained from an evolutionary viewpoint, it is interesting to note that the majority of the stars analyzed indeed do not follow the asymptotic relation. This is a cautionary note to the use of \eqref{eq:asymp} and/or \eqref{eq:asym2} for extracting average seismic parameters as here, or if used for predicting the location of oscillation modes frequencies in the power spectrum from $\Delta\nu_0$ measured in an independent manner. 
For details on the physics behind the curvatures we refer to \citet[][]{1980ApJS...43..469T, 2007MNRAS.375..861H, 2007ApJ...666..413C}, and \citet[][]{2013A&A...550A.126M} and references therein.

%%%%%%%%%%%%%%%%%%%%%%%%%%%%%%%%%%%%%%%%%%%%%%%%%%%%%%%%%%%%%%%%%%%%%%%%%%%%%%%%%%%%%%%%%%%%%%%%%%%%%%%%%%%%%%%%%%%%%%%%%%
%%%%%%%%%%%%%%%%%%%%%%%%%%%%%%%%%%%%%%%%%%%%%%%%%%%%%%%%%%%%%%%%%%%%%%%%%%%%%%%%%%%%%%%%%%%%%%%%%%%%%%%%%%%%%%%%%%%%%%%%%%
\subsection{Mode amplitudes}
\label{sec:ampl}

%%%%%%%%%%%%%%%%%%%%%%%%%%%%%%%%%%%%%%%%%%%%%%%%%%%%%%%%%%%%%%%%%%%%%%%%%%%%%%%%%%%%%%%%%%%%%%%%%%%%%%%%%%%%%%%%%%%%%%%%%%
%%%%%%%%%%%%%%%%%%%%%%%%%%%%%%%%%%%%%%%%%%%%%%%%%%%%%%%%%%%%%%%%%%%%%%%%%%%%%%%%%%%%%%%%%%%%%%%%%%%%%%%%%%%%%%%%%%%%%%%%%%
%%%%%%%%%%%%%%%%%%%%%%%%%%%%%%%%%%%%%%%%%%%%%%%%%%%%%%%%%%%%%%%%%%%%%%%%%%%%%%%%%%%%%%%%%%%%%%%%%%%%%%%%%%%%%%%%%%%%%%%%%%

% Table made Sat Oct  1 22:01:50 2016
\begin{table*} 
\centering 
\resizebox*{!}{\textheight}{% 
\begin{threeparttable} 
\caption{Values for the line width at \numax (\fref{fig:gam_vs_teff}) from the fit of \eqref{eq:gam_v_fre}; $A_{\rm max}$ gives the mode amplitude
    at \numax from a fit to the individual mode amplitudes against frequency; $A_{\rm max, \, smo}\,(c=3.04)$ and $A_{\rm max, \, smo}\,(c=\tilde{V}_{\rm tot}^2)$
    give the amplitudes from the smoothing method by \citet[][]{2008ApJ...682.1370K} (\fref{fig:amax_vs_asmo}), with different values for the effective number of modes per radial order $c$;
    $\alpha$, $\Gamma_{\alpha}$, $\Delta\Gamma_{\rm dip}$, $W_{\rm dip}$, and $\nu_{\rm dip}$ give the obtained parameters from the fit of \eqref{eq:gam_v_fre}. The values of 
    $\Delta\Gamma_{\rm dip}$ and $W_{\rm dip}$ are only given if the inclusion of the Lorentzian in \eqref{eq:gam_v_fre} gave a better fit than the power law only fit.
    $\rm FWHM_{dip}$ gives the full-width-half-maximum of the Lorentzian line width dip.
    } 
\label{tab:width_amp_max_table} 
\begin{tabular}{@{}r@{\hskip 5ex}r@{\hskip 10ex}r@{\hskip 13ex}r@{\hskip 17ex}r@{\hskip 9ex}r@{\hskip 10ex}r@{\hskip 9ex}r@{\hskip 9ex}r@{\hskip 9ex}r@{\hskip 5ex}r} 
\toprule 
\multicolumn{1}{@{\hskip -2ex}c}{KIC} & \multicolumn{1}{@{\hskip -5ex}c}{$\Gamma\, @\, \numax$} & \multicolumn{1}{@{\hskip -8ex}c}{$A_{\rm max}$} & \multicolumn{1}{@{\hskip -12ex}c}{$A_{\rm max, \, smo}\,(c=3.04)$} & \multicolumn{1}{@{\hskip -5ex}c}{$A_{\rm max, \, smo}\,(c=\tilde{V}_{\rm tot}^2)$} & \multicolumn{1}{@{\hskip -4ex}c}{$\alpha$} &  \multicolumn{1}{@{\hskip -4ex}c}{$\Gamma_{\alpha}$} & \multicolumn{1}{@{\hskip -5ex}c}{$\Delta\Gamma_{\rm dip}$} & \multicolumn{1}{@{\hskip -5ex}c}{$W_{\rm dip}$} & \multicolumn{1}{@{\hskip -4ex}c}{$\nu_{\rm dip}$} & \multicolumn{1}{c}{$\rm FWHM_{dip}$}\\   & \multicolumn{1}{@{\hskip -5ex}c}{$(\rm \mu Hz)$} & \multicolumn{1}{@{\hskip -8ex}c}{$(\rm ppm)$} & \multicolumn{1}{@{\hskip -15ex}c}{$(\rm ppm)$} & \multicolumn{1}{@{\hskip -6ex}c}{$(\rm ppm)$} &  &  \multicolumn{1}{@{\hskip -4ex}c}{$(\rm \mu Hz)$} & \multicolumn{1}{@{\hskip -5ex}c}{$(\rm \mu Hz)$} & \multicolumn{1}{@{\hskip -5ex}c}{$(\rm \mu Hz)$} & \multicolumn{1}{@{\hskip -4ex}c}{$(\rm \mu Hz)$} & \multicolumn{1}{c}{$(\rm \mu Hz)$}\\ 
\midrule 
$1435467$ &$5.18 \pm 0.17$ & $4.37$\rlap{$_{-0.07}^{+0.10}$} & $4.30$ & $4.25$ & $0.54$\rlap{$_{-0.22}^{+0.21}$} & $5.19 \pm 0.18$ &   &   &   &    \\ [0.5ex] 
$2837475$ &$6.39 \pm 0.2$ & $3.87$\rlap{$_{-0.05}^{+0.06}$} & $3.67$ & $3.71$ & $0.91$\rlap{$_{-0.19}^{+0.19}$} & $6.42 \pm 0.21$ &   &   &   &    \\ [0.5ex] 
$3427720$ &$1.88 \pm 0.15$ & $2.62$\rlap{$_{-0.07}^{+0.04}$} & $2.55$ & $2.38$ & $2.49$\rlap{$_{-0.96}^{+0.80}$} & $5.0 \pm 2.96$ & $0.36$\rlap{$_{-0.26}^{+0.13}$} & $4428$\rlap{$_{-1660}^{+605}$} & $2695$\rlap{$_{-100}^{+126}$} & $1309 \pm 711$  \\ [0.5ex] 
$3456181$ &$4.07 \pm 0.19$ & $5.60$\rlap{$_{-0.14}^{+0.14}$} & $5.92$ & $5.48$ & $1.11$\rlap{$_{-0.26}^{+0.27}$} & $4.06 \pm 0.2$ &   &   &   &    \\ [0.5ex] 
$3632418$ &$2.72 \pm 0.16$ & $5.04$\rlap{$_{-0.07}^{+0.06}$} & $5.04$ & $4.76$ & $1.76$\rlap{$_{-0.22}^{+0.18}$} & $4.67 \pm 0.71$ & $0.56$\rlap{$_{-0.07}^{+0.09}$} & $1698$\rlap{$_{-304}^{+210}$} & $1188$\rlap{$_{-34}^{+26}$} & $448 \pm 183$  \\ [0.5ex] 
$3656476$ &$0.69 \pm 0.01$ & $4.70$\rlap{$_{-0.13}^{+0.13}$} & $4.13$ & $4.12$ & $2.50$\rlap{$_{-0.10}^{+0.13}$} & $0.69 \pm 0.01$ &   &   &   &    \\ [0.5ex] 
$3735871$ &$2.25 \pm 0.25$ & $2.52$\rlap{$_{-0.07}^{+0.08}$} & $2.38$ & $2.30$ & $2.79$\rlap{$_{-1.24}^{+1.18}$} & $7.34 \pm 4.29$ & $0.28$\rlap{$_{-0.18}^{+0.10}$} & $4717$\rlap{$_{-1110}^{+798}$} & $2825$\rlap{$_{-167}^{+148}$} & $1426 \pm 595$  \\ [0.5ex] 
$4914923$ &$1.16 \pm 0.08$ & $4.62$\rlap{$_{-0.08}^{+0.12}$} & $4.19$ & $4.15$ & $3.52$\rlap{$_{-0.57}^{+0.48}$} & $6.2 \pm 2.4$ & $0.18$\rlap{$_{-0.07}^{+0.07}$} & $2600$\rlap{$_{-270}^{+201}$} & $1847$\rlap{$_{-25}^{+24}$} & $666 \pm 170$  \\ [0.5ex] 
$5184732$ &$1.36 \pm 0.06$ & $4.08$\rlap{$_{-0.06}^{+0.06}$} & $3.84$ & $3.88$ & $3.52$\rlap{$_{-0.41}^{+0.38}$} & $7.65 \pm 2.61$ & $0.17$\rlap{$_{-0.06}^{+0.06}$} & $3247$\rlap{$_{-305}^{+243}$} & $2117$\rlap{$_{-28}^{+26}$} & $941 \pm 184$  \\ [0.5ex] 
$5773345$ &$3.37 \pm 0.13$ & $5.69$\rlap{$_{-0.11}^{+0.13}$} & $5.17$ & $5.52$ & $0.58$\rlap{$_{-0.23}^{+0.26}$} & $3.37 \pm 0.14$ &   &   &   &    \\ [0.5ex] 
$5950854$ &$0.83 \pm 0.01$ & $4.12$\rlap{$_{-0.27}^{+0.37}$} & $3.77$ & $3.87$ & $0.56$\rlap{$_{-0.03}^{+0.03}$} & $0.83 \pm 0.01$ &   &   &   &    \\ [0.5ex] 
$6106415$ &$1.64 \pm 0.07$ & $3.86$\rlap{$_{-0.04}^{+0.06}$} & $3.60$ & $3.63$ & $3.49$\rlap{$_{-0.25}^{+0.25}$} & $5.53 \pm 1.08$ & $0.29$\rlap{$_{-0.05}^{+0.06}$} & $3242$\rlap{$_{-265}^{+201}$} & $2274$\rlap{$_{-26}^{+23}$} & $837 \pm 166$  \\ [0.5ex] 
$6116048$ &$1.62 \pm 0.09$ & $4.02$\rlap{$_{-0.06}^{+0.06}$} & $3.75$ & $3.67$ & $3.26$\rlap{$_{-0.28}^{+0.32}$} & $5.74 \pm 1.35$ & $0.28$\rlap{$_{-0.06}^{+0.06}$} & $3102$\rlap{$_{-319}^{+207}$} & $2098$\rlap{$_{-28}^{+27}$} & $796 \pm 182$  \\ [0.5ex] 
$6225718$ &$2.58 \pm 0.12$ & $3.53$\rlap{$_{-0.03}^{+0.04}$} & $3.39$ & $3.29$ & $2.32$\rlap{$_{-0.22}^{+0.24}$} & $5.77 \pm 0.67$ & $0.43$\rlap{$_{-0.05}^{+0.05}$} & $3076$\rlap{$_{-180}^{+139}$} & $2316$\rlap{$_{-23}^{+26}$} & $611 \pm 121$  \\ [0.5ex] 
$6508366$ &$5.22 \pm 0.17$ & $5.02$\rlap{$_{-0.07}^{+0.09}$} & $5.21$ & $4.75$ & $1.95$\rlap{$_{-0.12}^{+0.13}$} & $5.21 \pm 0.15$ &   &   &   &    \\ [0.5ex] 
$6603624$ &$0.56 \pm 0.02$ & $4.15$\rlap{$_{-0.07}^{+0.10}$} & $3.55$ & $3.58$ & $4.42$\rlap{$_{-0.24}^{+0.19}$} & $0.57 \pm 0.02$ &   &   &   &    \\ [0.5ex] 
$6679371$ &$4.53 \pm 0.18$ & $5.56$\rlap{$_{-0.08}^{+0.07}$} & $5.33$ & $5.28$ & $0.45$\rlap{$_{-0.45}^{+0.19}$} & $5.51 \pm 0.74$ & $0.69$\rlap{$_{-0.16}^{+0.14}$} & $1339$\rlap{$_{-389}^{+209}$} & $833$\rlap{$_{-48}^{+44}$} & $295 \pm 190$  \\ [0.5ex] 
$6933899$ &$1.3 \pm 0.07$ & $5.55$\rlap{$_{-0.10}^{+0.09}$} & $5.07$ & $4.94$ & $3.15$\rlap{$_{-0.69}^{+0.56}$} & $6.0 \pm 2.75$ & $0.21$\rlap{$_{-0.11}^{+0.08}$} & $2218$\rlap{$_{-304}^{+233}$} & $1422$\rlap{$_{-33}^{+30}$} & $671 \pm 178$  \\ [0.5ex] 
$7103006$ &$5.09 \pm 0.16$ & $4.74$\rlap{$_{-0.09}^{+0.11}$} & $4.29$ & $4.58$ & $1.17$\rlap{$_{-0.18}^{+0.18}$} & $5.09 \pm 0.17$ &   &   &   &    \\ [0.5ex] 
$7106245$ &$1.65 \pm 0.14$ & $3.39$\rlap{$_{-0.15}^{+0.11}$} & $2.96$ & $2.91$ & $1.71$\rlap{$_{-0.97}^{+0.93}$} & $1.64 \pm 0.14$ &   &   &   &    \\ [0.5ex] 
$7206837$ &$4.34 \pm 0.32$ & $4.32$\rlap{$_{-0.06}^{+0.10}$} & $3.85$ & $4.18$ & $0.75$\rlap{$_{-0.62}^{+0.36}$} & $8.25 \pm 2.83$ & $0.50$\rlap{$_{-0.21}^{+0.16}$} & $2854$\rlap{$_{-830}^{+545}$} & $1720$\rlap{$_{-101}^{+92}$} & $952 \pm 433$  \\ [0.5ex] 
$7296438$ &$1.22 \pm 0.12$ & $4.59$\rlap{$_{-0.22}^{+0.20}$} & $4.24$ & $4.11$ & $3.87$\rlap{$_{-1.11}^{+1.10}$} & $8.39 \pm 4.56$ & $0.14$\rlap{$_{-0.08}^{+0.05}$} & $2794$\rlap{$_{-336}^{+272}$} & $1835$\rlap{$_{-48}^{+49}$} & $764 \pm 205$  \\ [0.5ex] 
$7510397$ &$2.42 \pm 0.15$ & $3.62$\rlap{$_{-0.04}^{+0.05}$} & $3.52$ & $3.40$ & $1.90$\rlap{$_{-0.31}^{+0.25}$} & $4.18 \pm 0.71$ & $0.55$\rlap{$_{-0.08}^{+0.11}$} & $1663$\rlap{$_{-270}^{+199}$} & $1226$\rlap{$_{-36}^{+30}$} & $413 \pm 176$  \\ [0.5ex] 
$7680114$ &$1.16 \pm 0.09$ & $4.71$\rlap{$_{-0.10}^{+0.10}$} & $4.40$ & $4.40$ & $3.70$\rlap{$_{-0.77}^{+0.66}$} & $4.68 \pm 1.95$ & $0.24$\rlap{$_{-0.11}^{+0.09}$} & $2438$\rlap{$_{-311}^{+215}$} & $1736$\rlap{$_{-34}^{+32}$} & $620 \pm 191$  \\ [0.5ex] 
$7771282$ &$3.29 \pm 0.28$ & $4.20$\rlap{$_{-0.21}^{+0.12}$} & $4.14$ & $3.96$ & $0.71$\rlap{$_{-0.71}^{+0.66}$} & $3.29 \pm 0.29$ &   &   &   &    \\ [0.5ex] 
$7871531$ &$1.21 \pm 0.12$ & $1.91$\rlap{$_{-0.07}^{+0.09}$} & $1.71$ & $1.77$ & $1.94$\rlap{$_{-1.53}^{+0.91}$} & $4.83 \pm 3.18$ & $0.20$\rlap{$_{-0.15}^{+0.09}$} & $6026$\rlap{$_{-1405}^{+967}$} & $3161$\rlap{$_{-178}^{+145}$} & $1780 \pm 653$  \\ [0.5ex] 
$7940546$ &$2.92 \pm 0.14$ & $5.29$\rlap{$_{-0.07}^{+0.07}$} & $5.24$ & $5.51$ & $1.56$\rlap{$_{-0.19}^{+0.19}$} & $5.15 \pm 0.69$ & $0.53$\rlap{$_{-0.06}^{+0.08}$} & $1578$\rlap{$_{-194}^{+137}$} & $1177$\rlap{$_{-23}^{+23}$} & $409 \pm 126$  \\ [0.5ex] 
$7970740$ &$1.99 \pm 0.15$ & $1.65$\rlap{$_{-0.03}^{+0.03}$} & $1.35$ & $1.51$ & $4.25$\rlap{$_{-0.72}^{+0.92}$} & $6.85 \pm 3.6$ & $0.26$\rlap{$_{-0.17}^{+0.10}$} & $6875$\rlap{$_{-1384}^{+1110}$} & $3910$\rlap{$_{-151}^{+177}$} & $1949 \pm 735$  \\ [0.5ex] 
$8006161$ &$1.17 \pm 0.06$ & $1.94$\rlap{$_{-0.04}^{+0.03}$} & $1.69$ & $1.73$ & $4.75$\rlap{$_{-0.62}^{+0.68}$} & $6.0 \pm 3.27$ & $0.19$\rlap{$_{-0.13}^{+0.06}$} & $5581$\rlap{$_{-861}^{+621}$} & $3507$\rlap{$_{-71}^{+73}$} & $1575 \pm 478$  \\ [0.5ex] 
$8150065$ &$2.37 \pm 0.3$ & $3.55$\rlap{$_{-0.17}^{+0.25}$} & $3.38$ & $3.23$ & $2.01$\rlap{$_{-1.53}^{+1.47}$} & $2.36 \pm 0.28$ &   &   &   &    \\ [0.5ex] 
$8179536$ &$3.51 \pm 0.26$ & $3.42$\rlap{$_{-0.07}^{+0.09}$} & $3.35$ & $3.25$ & $1.88$\rlap{$_{-0.89}^{+0.78}$} & $7.05 \pm 3.12$ & $0.47$\rlap{$_{-0.30}^{+0.14}$} & $3940$\rlap{$_{-535}^{+1129}$} & $2112$\rlap{$_{-194}^{+195}$} & $1378 \pm 486$  \\ [0.5ex] 
$8228742$ &$2.05 \pm 0.13$ & $5.30$\rlap{$_{-0.06}^{+0.08}$} & $5.20$ & $4.94$ & $2.39$\rlap{$_{-0.30}^{+0.26}$} & $4.19 \pm 0.68$ & $0.48$\rlap{$_{-0.07}^{+0.09}$} & $1640$\rlap{$_{-191}^{+127}$} & $1189$\rlap{$_{-26}^{+21}$} & $383 \pm 117$  \\ [0.5ex] 
$8379927$ &$2.43 \pm 0.11$ & $2.21$\rlap{$_{-0.03}^{+0.03}$} & $2.09$ & $2.05$ & $2.45$\rlap{$_{-0.27}^{+0.31}$} & $6.12 \pm 1.22$ & $0.39$\rlap{$_{-0.07}^{+0.08}$} & $4091$\rlap{$_{-435}^{+324}$} & $2731$\rlap{$_{-43}^{+49}$} & $1046 \pm 258$  \\ [0.5ex] 
$8394589$ &$2.15 \pm 0.16$ & $3.49$\rlap{$_{-0.05}^{+0.08}$} & $3.33$ & $3.24$ & $1.71$\rlap{$_{-0.60}^{+0.59}$} & $5.79 \pm 1.69$ & $0.28$\rlap{$_{-0.08}^{+0.09}$} & $3185$\rlap{$_{-356}^{+231}$} & $2233$\rlap{$_{-34}^{+41}$} & $637 \pm 208$  \\ [0.5ex] 
$8424992$ &$1.14 \pm 0.1$ & $2.97$\rlap{$_{-0.19}^{+0.18}$} & $2.77$ & $2.67$ & $2.99$\rlap{$_{-1.42}^{+1.46}$} & $1.14 \pm 0.1$ &   &   &   &    \\ [0.5ex] 
$8694723$ &$3.08 \pm 0.16$ & $5.38$\rlap{$_{-0.05}^{+0.09}$} & $5.08$ & $5.07$ & $1.98$\rlap{$_{-0.20}^{+0.21}$} & $5.42 \pm 0.79$ & $0.56$\rlap{$_{-0.07}^{+0.09}$} & $2129$\rlap{$_{-323}^{+220}$} & $1472$\rlap{$_{-34}^{+34}$} & $548 \pm 192$  \\ [0.5ex] 
$8760414$ &$1.25 \pm 0.08$ & $3.96$\rlap{$_{-0.06}^{+0.11}$} & $3.49$ & $3.52$ & $2.97$\rlap{$_{-0.61}^{+0.77}$} & $5.06 \pm 2.23$ & $0.23$\rlap{$_{-0.12}^{+0.09}$} & $3659$\rlap{$_{-486}^{+359}$} & $2349$\rlap{$_{-45}^{+52}$} & $943 \pm 277$  \\ [0.5ex] 
$8938364$ &$0.8 \pm 0.01$ & $5.22$\rlap{$_{-0.14}^{+0.12}$} & $4.61$ & $4.43$ & $2.49$\rlap{$_{-0.13}^{+0.10}$} & $0.8 \pm 0.01$ &   &   &   &    \\ [0.5ex] 
$9025370$ &$1.39 \pm 0.08$ & $1.64$\rlap{$_{-0.05}^{+0.04}$} & $1.45$ & $1.45$ & $1.60$\rlap{$_{-0.64}^{+0.62}$} & $1.38 \pm 0.09$ &   &   &   &    \\ [0.5ex] 
$9098294$ &$1.3 \pm 0.08$ & $3.55$\rlap{$_{-0.09}^{+0.09}$} & $3.21$ & $3.16$ & $2.81$\rlap{$_{-1.17}^{+1.09}$} & $7.81 \pm 4.74$ & $0.16$\rlap{$_{-0.10}^{+0.06}$} & $3986$\rlap{$_{-650}^{+517}$} & $2317$\rlap{$_{-95}^{+86}$} & $1275 \pm 355$  \\ [0.5ex] 
$9139151$ &$1.99 \pm 0.14$ & $2.88$\rlap{$_{-0.06}^{+0.05}$} & $2.72$ & $2.66$ & $3.03$\rlap{$_{-0.82}^{+0.71}$} & $7.78 \pm 3.84$ & $0.25$\rlap{$_{-0.15}^{+0.07}$} & $4136$\rlap{$_{-697}^{+494}$} & $2697$\rlap{$_{-70}^{+72}$} & $1169 \pm 399$  \\ [0.5ex] 
$9139163$ &$5.28 \pm 0.13$ & $3.76$\rlap{$_{-0.04}^{+0.05}$} & $3.64$ & $3.81$ & $1.79$\rlap{$_{-0.14}^{+0.13}$} & $5.28 \pm 0.13$ &   &   &   &    \\ [0.5ex] 
$9206432$ &$5.87 \pm 0.27$ & $3.54$\rlap{$_{-0.09}^{+0.12}$} & $3.47$ & $3.48$ & $0.88$\rlap{$_{-0.29}^{+0.31}$} & $5.88 \pm 0.28$ &   &   &   &    \\ [0.5ex] 
$9353712$ &$3.22 \pm 0.22$ & $5.54$\rlap{$_{-0.16}^{+0.18}$} & $5.68$ & $5.46$ & $1.20$\rlap{$_{-0.36}^{+0.38}$} & $3.2 \pm 0.21$ &   &   &   &    \\ [0.5ex] 
$9410862$ &$1.89 \pm 0.25$ & $3.75$\rlap{$_{-0.10}^{+0.13}$} & $3.60$ & $3.67$ & $2.40$\rlap{$_{-0.81}^{+0.82}$} & $7.2 \pm 3.8$ & $0.24$\rlap{$_{-0.14}^{+0.07}$} & $3337$\rlap{$_{-769}^{+460}$} & $2204$\rlap{$_{-63}^{+65}$} & $846 \pm 414$  \\ [0.5ex] 
$9414417$ &$3.47 \pm 0.23$ & $5.55$\rlap{$_{-0.10}^{+0.08}$} & $5.18$ & $5.29$ & $1.04$\rlap{$_{-0.35}^{+0.30}$} & $5.81 \pm 1.31$ & $0.57$\rlap{$_{-0.13}^{+0.14}$} & $1910$\rlap{$_{-552}^{+374}$} & $1191$\rlap{$_{-65}^{+56}$} & $605 \pm 299$  \\ [0.5ex] 
$9812850$ &$5.68 \pm 0.19$ & $4.61$\rlap{$_{-0.08}^{+0.11}$} & $4.51$ & $4.85$ & $0.69$\rlap{$_{-0.18}^{+0.19}$} & $5.67 \pm 0.2$ &   &   &   &    \\ [0.5ex] 
$9955598$ &$0.77 \pm 0.03$ & $2.07$\rlap{$_{-0.05}^{+0.06}$} & $1.79$ & $1.94$ & $4.98$\rlap{$_{-0.21}^{+0.26}$} & $0.77 \pm 0.01$ &   &   &   &    \\ [0.5ex] 
$9965715$ &$3.14 \pm 0.34$ & $3.89$\rlap{$_{-0.07}^{+0.07}$} & $3.54$ & $3.59$ & $1.82$\rlap{$_{-0.57}^{+0.47}$} & $5.92 \pm 1.9$ & $0.47$\rlap{$_{-0.16}^{+0.15}$} & $3024$\rlap{$_{-802}^{+687}$} & $2103$\rlap{$_{-84}^{+78}$} & $792 \pm 528$  \\ [0.5ex] 
$10068307$ &$2.15 \pm 0.13$ & $5.98$\rlap{$_{-0.09}^{+0.09}$} & $5.86$ & $5.54$ & $2.15$\rlap{$_{-0.18}^{+0.16}$} & $3.91 \pm 0.45$ & $0.53$\rlap{$_{-0.05}^{+0.07}$} & $1388$\rlap{$_{-165}^{+110}$} & $1015$\rlap{$_{-22}^{+19}$} & $340 \pm 102$  \\ [0.5ex] 
$10079226$ &$2.05 \pm 0.35$ & $2.81$\rlap{$_{-0.15}^{+0.14}$} & $2.63$ & $2.63$ & $3.36$\rlap{$_{-1.44}^{+1.49}$} & $6.73 \pm 4.24$ & $0.27$\rlap{$_{-0.19}^{+0.12}$} & $4248$\rlap{$_{-1099}^{+743}$} & $2601$\rlap{$_{-165}^{+156}$} & $1236 \pm 585$  \\ [0.5ex] 
$10162436$ &$3.08 \pm 0.08$ & $5.38$\rlap{$_{-0.09}^{+0.11}$} & $5.40$ & $5.42$ & $1.82$\rlap{$_{-0.15}^{+0.16}$} & $3.08 \pm 0.09$ &   &   &   &    \\ [0.5ex] 
$10454113$ &$4.12 \pm 0.15$ & $3.19$\rlap{$_{-0.04}^{+0.05}$} & $2.87$ & $3.01$ & $1.60$\rlap{$_{-0.26}^{+0.26}$} & $4.13 \pm 0.15$ &   &   &   &    \\ [0.5ex] 
$10516096$ &$1.56 \pm 0.1$ & $4.74$\rlap{$_{-0.11}^{+0.08}$} & $4.40$ & $4.26$ & $3.44$\rlap{$_{-0.54}^{+0.48}$} & $5.33 \pm 2.47$ & $0.28$\rlap{$_{-0.13}^{+0.12}$} & $2629$\rlap{$_{-539}^{+345}$} & $1723$\rlap{$_{-39}^{+38}$} & $768 \pm 298$  \\ [0.5ex] 
$10644253$ &$2.21 \pm 0.22$ & $2.22$\rlap{$_{-0.05}^{+0.06}$} & $2.15$ & $2.06$ & $3.16$\rlap{$_{-1.06}^{+0.99}$} & $11.06 \pm 4.51$ & $0.19$\rlap{$_{-0.09}^{+0.05}$} & $4253$\rlap{$_{-456}^{+364}$} & $2852$\rlap{$_{-69}^{+74}$} & $1099 \pm 282$  \\ [0.5ex] 
$10730618$ &$5.07 \pm 0.25$ & $4.64$\rlap{$_{-0.13}^{+0.14}$} & $4.24$ & $4.49$ & $0.41$\rlap{$_{-0.40}^{+0.39}$} & $5.09 \pm 0.27$ &   &   &   &    \\ [0.5ex] 
$10963065$ &$2.16 \pm 0.11$ & $3.72$\rlap{$_{-0.05}^{+0.09}$} & $3.62$ & $3.45$ & $3.36$\rlap{$_{-0.59}^{+0.54}$} & $8.44 \pm 3.9$ & $0.25$\rlap{$_{-0.14}^{+0.07}$} & $3921$\rlap{$_{-761}^{+568}$} & $2158$\rlap{$_{-71}^{+64}$} & $1261 \pm 383$  \\ [0.5ex] 
$11081729$ &$6.19 \pm 0.3$ & $3.11$\rlap{$_{-0.06}^{+0.07}$} & $2.99$ & $3.01$ & $1.79$\rlap{$_{-0.31}^{+0.30}$} & $6.19 \pm 0.3$ &   &   &   &    \\ [0.5ex] 
$11253226$ &$5.8 \pm 0.17$ & $4.20$\rlap{$_{-0.05}^{+0.07}$} & $3.62$ & $3.98$ & $0.40$\rlap{$_{-0.17}^{+0.17}$} & $5.8 \pm 0.17$ &   &   &   &    \\ [0.5ex] 
$11772920$ &$0.77 \pm 0.03$ & $1.83$\rlap{$_{-0.07}^{+0.07}$} & $1.49$ & $1.55$ & $2.69$\rlap{$_{-0.11}^{+0.12}$} & $0.77 \pm 0.01$ &   &   &   &    \\ [0.5ex] 
$12009504$ &$2.38 \pm 0.14$ & $3.94$\rlap{$_{-0.06}^{+0.08}$} & $3.84$ & $3.65$ & $2.31$\rlap{$_{-0.49}^{+0.46}$} & $6.06 \pm 2.12$ & $0.38$\rlap{$_{-0.13}^{+0.14}$} & $2915$\rlap{$_{-621}^{+420}$} & $1821$\rlap{$_{-57}^{+49}$} & $819 \pm 334$  \\ [0.5ex] 
$12069127$ &$3.6 \pm 0.23$ & $6.00$\rlap{$_{-0.17}^{+0.18}$} & $6.00$ & $5.79$ & $0.93$\rlap{$_{-0.36}^{+0.38}$} & $3.61 \pm 0.23$ &   &   &   &    \\ [0.5ex] 
$12069424$ &$0.98 \pm 0.04$ & $3.96$\rlap{$_{-0.06}^{+0.06}$} & $3.55$ & $3.56$ & $3.33$\rlap{$_{-0.29}^{+0.28}$} & $5.47 \pm 1.11$ & $0.18$\rlap{$_{-0.03}^{+0.04}$} & $3187$\rlap{$_{-171}^{+139}$} & $2181$\rlap{$_{-18}^{+20}$} & $825 \pm 108$  \\ [0.5ex] 
$12069449$ &$0.91 \pm 0.05$ & $3.49$\rlap{$_{-0.06}^{+0.07}$} & $3.05$ & $3.08$ & $4.10$\rlap{$_{-0.34}^{+0.33}$} & $6.06 \pm 1.56$ & $0.15$\rlap{$_{-0.04}^{+0.04}$} & $3716$\rlap{$_{-233}^{+185}$} & $2579$\rlap{$_{-22}^{+23}$} & $966 \pm 148$  \\ [0.5ex] 
$12258514$ &$1.69 \pm 0.09$ & $4.74$\rlap{$_{-0.06}^{+0.07}$} & $4.52$ & $4.35$ & $2.92$\rlap{$_{-0.18}^{+0.17}$} & $4.01 \pm 0.41$ & $0.42$\rlap{$_{-0.04}^{+0.04}$} & $1939$\rlap{$_{-116}^{+87}$} & $1510$\rlap{$_{-13}^{+12}$} & $376 \pm 79$  \\ [0.5ex] 
$12317678$ &$5.83 \pm 0.16$ & $4.75$\rlap{$_{-0.06}^{+0.08}$} & $4.98$ & $4.57$ & $1.10$\rlap{$_{-0.12}^{+0.14}$} & $5.82 \pm 0.15$ &   &   &   &    \\ [0.5ex] 
\bottomrule
\end{tabular} 
 \end{threeparttable}% 
 } 
\end{table*}

%%%%%%%%%%%%%%%%%%%%%%%%%%%%%%%%%%%%%%%%%%%%%%%%%%%%%%%%%%%%%%%%%%%%%%%%%%%%%%%%%%%%%%%%%%%%%%%%%%%%%%%%%%%%%%%%%%%%%%%%%%
%%%%%%%%%%%%%%%%%%%%%%%%%%%%%%%%%%%%%%%%%%%%%%%%%%%%%%%%%%%%%%%%%%%%%%%%%%%%%%%%%%%%%%%%%%%%%%%%%%%%%%%%%%%%%%%%%%%%%%%%%%
%%%%%%%%%%%%%%%%%%%%%%%%%%%%%%%%%%%%%%%%%%%%%%%%%%%%%%%%%%%%%%%%%%%%%%%%%%%%%%%%%%%%%%%%%%%%%%%%%%%%%%%%%%%%%%%%%%%%%%%%%%

\begin{figure*}
    \centering
    \begin{subfigure}
        \centering
        \includegraphics[width=\columnwidth]{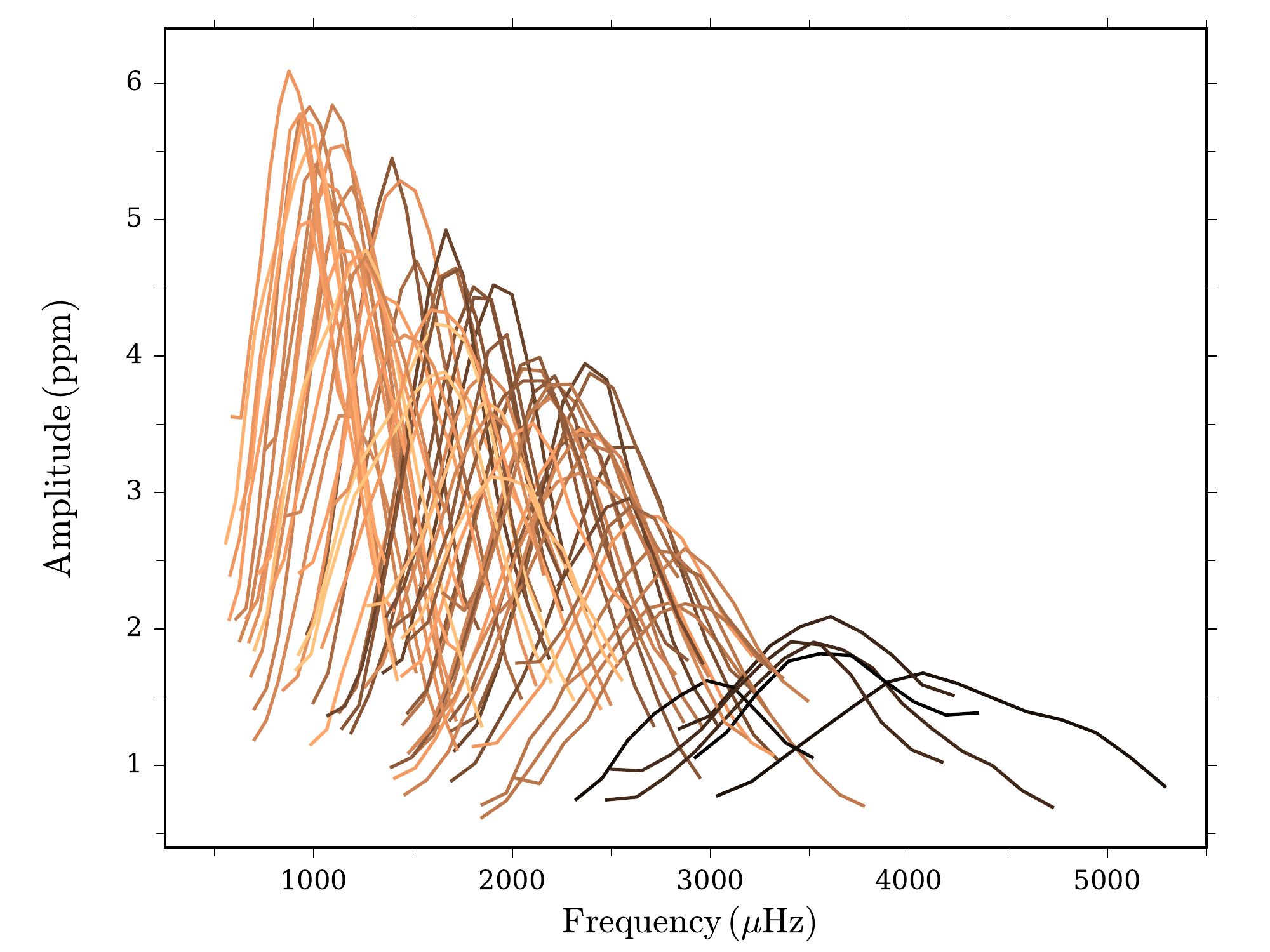}
    \end{subfigure}\hfill
    \begin{subfigure}
        \centering
        \includegraphics[width=\columnwidth]{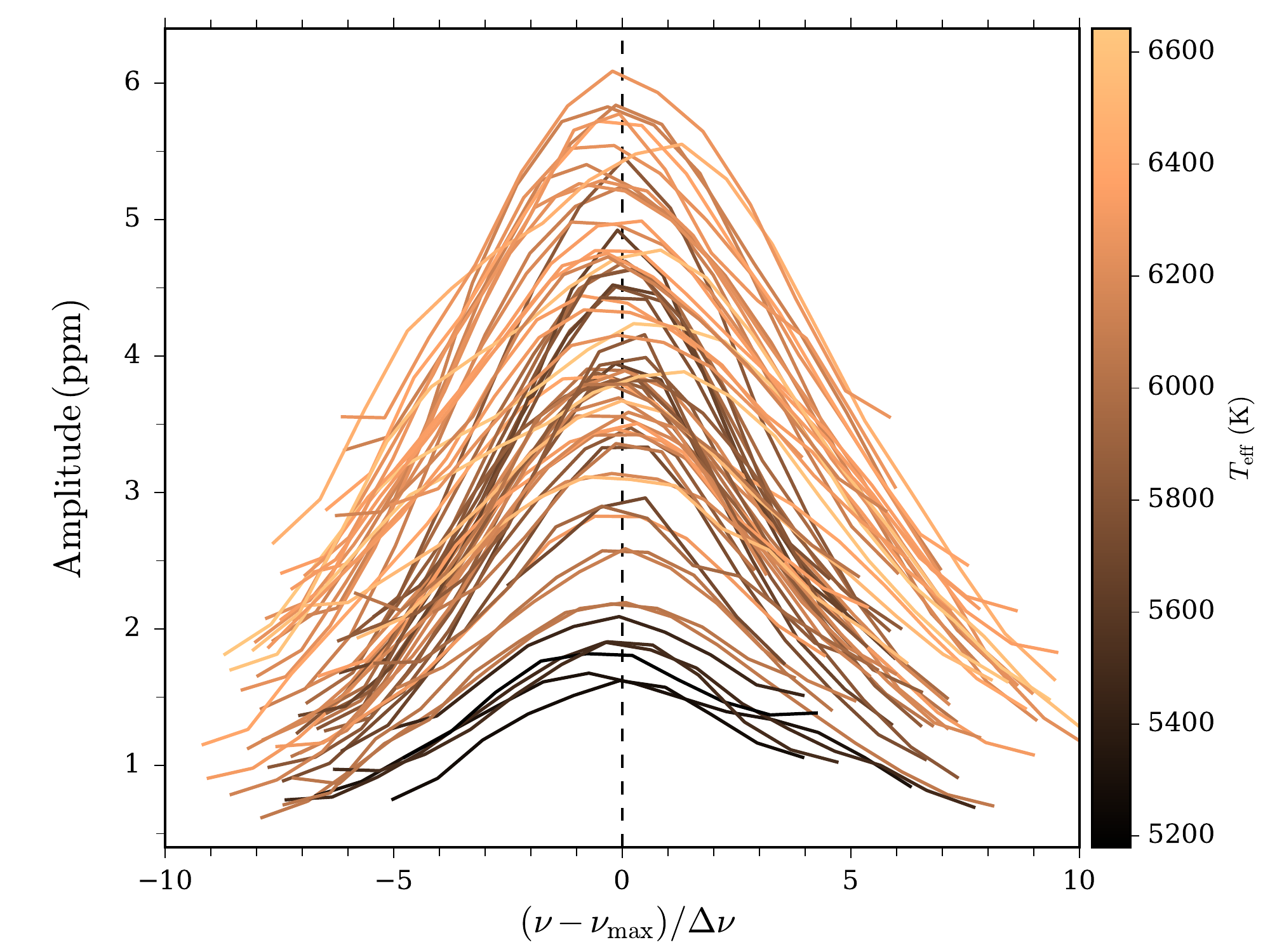}
    \end{subfigure}
    \caption{Left: Radial mode amplitude envelopes against frequency, with the colour indicating the \teff. For a better visualization the amplitudes have been smoothed with 5-point Epanechnikov filter. The dashed envelope gives the results obtained for the Sun. Right: Amplitudes centered on \numax and plotted against a proxy for the radial order.}
\label{fig:amplitude2}
\end{figure*}

Mode amplitudes were measured in the peak-bagging as described in \sref{sec:osc_model}. 
\fref{fig:amplitude2} displays the radial degree 5-point Epanechnikov \citep[][]{16463184,hastie2009elements} smoothed amplitudes for all the stars in the sample against frequency (left) and normalized against distance from \numax (right). One clearly sees the overall decrease in amplitude with increasing \numax \citep[][]{2005ApJ...635.1281K, 2008ApJ...687.1180A}, and thus decreasing \teff \citep[][]{1995A&A...293...87K,2011A&A...529L...8K,2014A&A...566A..20A}.
%%%%%%%%%%%%%%%%%%%%%%%%%%%%%%%%%%%%%%%%%%%%%%%%%%%%%%%%%%%%%%
%%%%%%%%%%%%%%%%%%%%%%%%%%%%%%%%%%%%%%%%%%%%%%%%%%%%%%%%%%%%%%
The value of \numax was estimated from the amplitudes as the frequency where the modes of oscillation show their maximum amplitude, $A_{\rm max}$. 

\begin{figure*}
    \centering
    \begin{subfigure}
        \centering
        \includegraphics[width=\columnwidth]{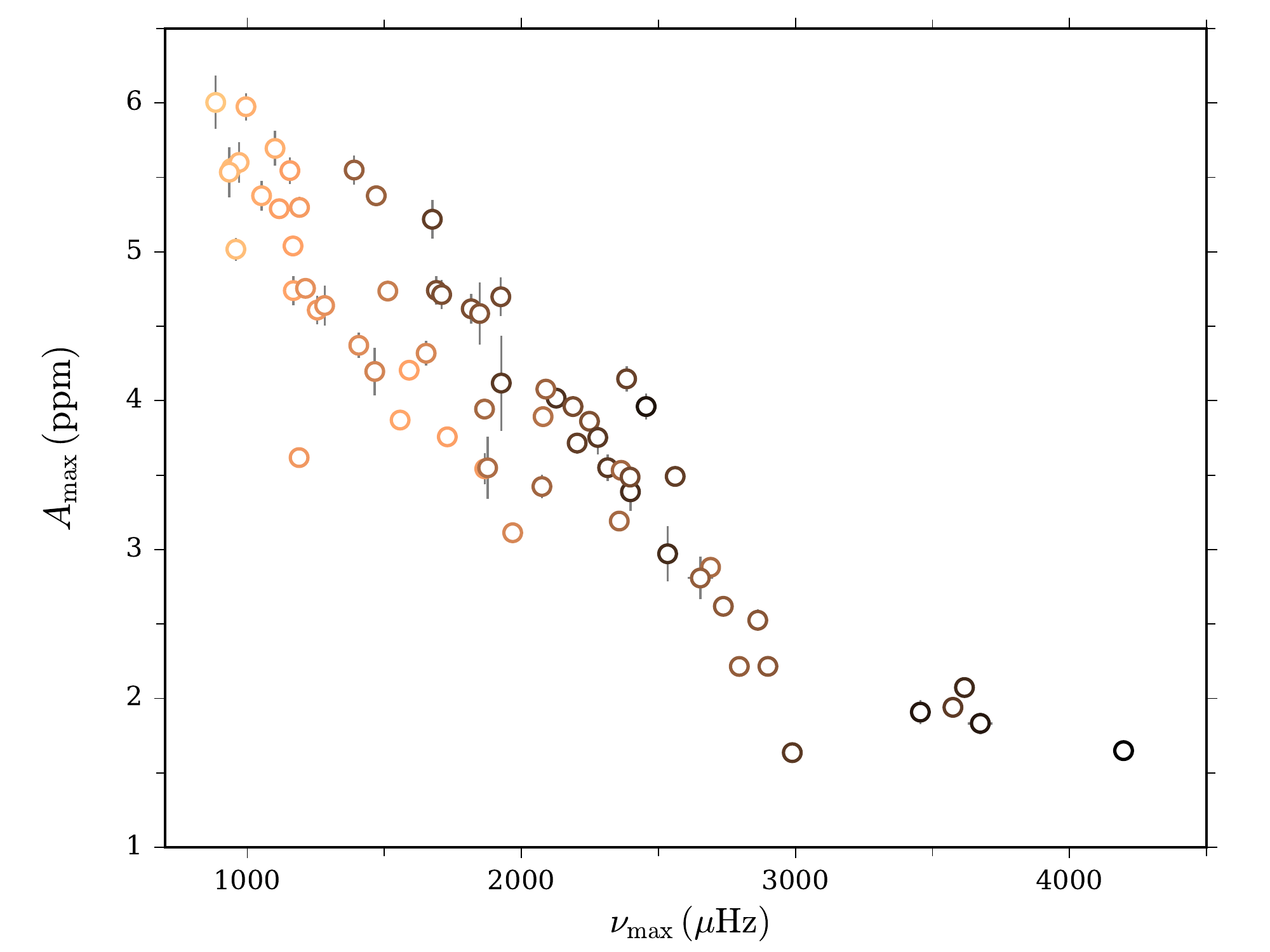}
    \end{subfigure}\hfill
    \begin{subfigure}
        \centering
        \includegraphics[width=\columnwidth]{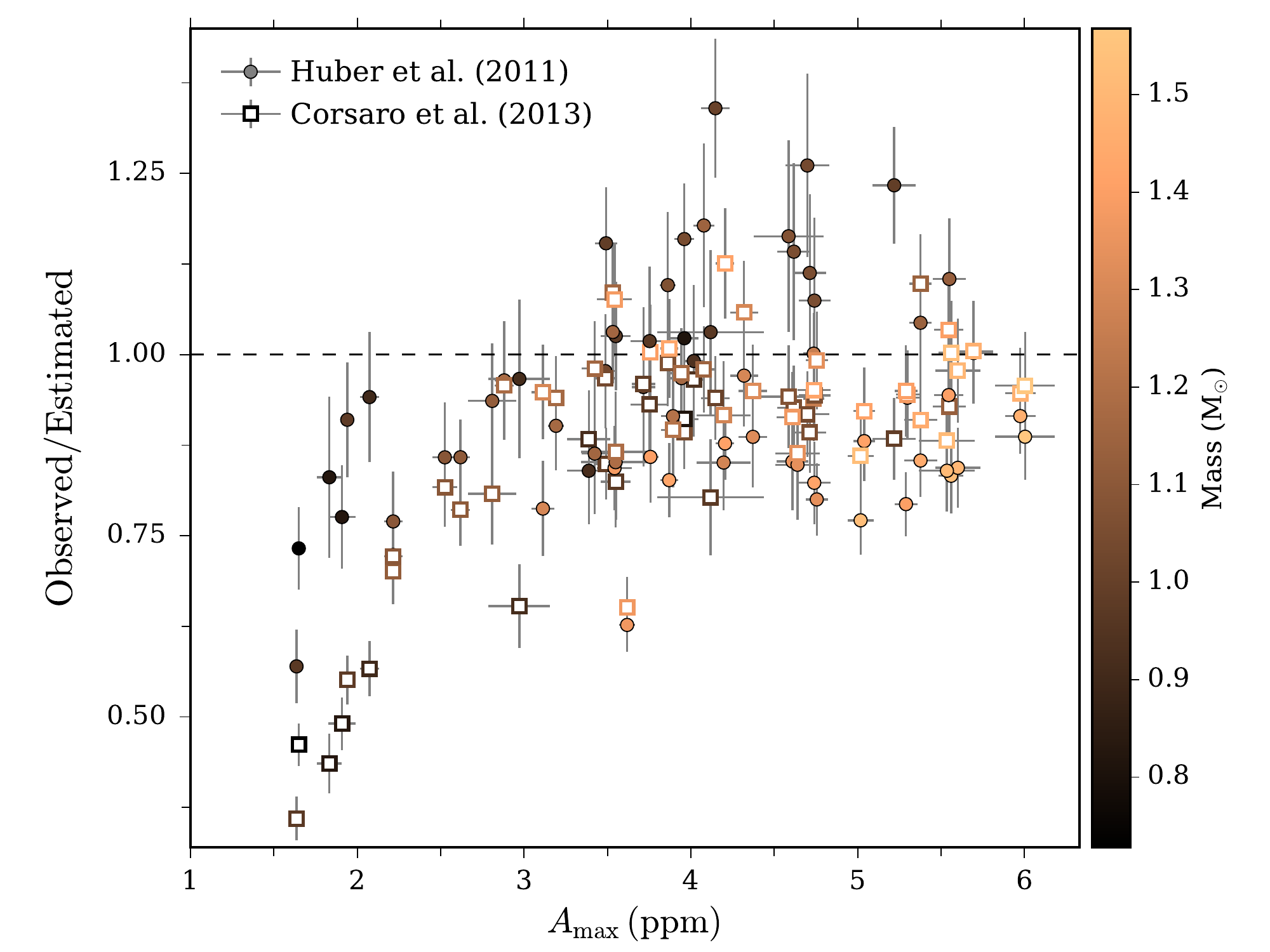}
    \end{subfigure}
    \caption{Left: Estimates of radial mode amplitudes at \numax as a function of \numax; the color indicates the modeled mass. The $A_{\rm max}$ values are obtained from a fit of the individual mode amplitudes against frequency. Right: Fractional differences between observed values of $A_{\rm max}$ and those estimated by the relations of \citet[][]{2011ApJ...743..143H} and \citet[][]{2013MNRAS.430.2313C}.}
\label{fig:amp_vs_numax}
\end{figure*}

\fref{fig:amp_vs_numax} gives the measured values for $A_{\rm max}$ against \numax (\tref{tab:width_amp_max_table}). We see both the expected change with \numax, in addition to a mass gradient across the overall decrease in $A_{\rm max}$ with \numax \citep[][]{2011ApJ...743..143H}.
%%%%%%%%%%%%%%%%%%%%%%%%%%%%%%%%%%%%%%%%%%%%%%%%%%%%%%%%%%%%%%
%%%%%%%%%%%%%%%%%%%%%%%%%%%%%%%%%%%%%%%%%%%%%%%%%%%%%%%%%%%%%%
In \fref{fig:amp_vs_numax} we also give a comparison to the amplitudes estimated from the scaling relations of \citet[][]{2011ApJ...743..143H}:
\begin{equation}
A_{\rm Kp} = \left(\frac{L}{\rm L_{\odot}}\right)^s \left(\frac{M}{\rm M_{\odot}}\right)^{-t}  \left(\frac{\teff}{\rm T_{\rm eff,\odot}}\right)^{1-r} \frac{A_{\rm bol, \odot}}{c_{\rm K}(T_{\rm eff})}\, ,
\end{equation}
with $t=1.32\pm 0.02$, $s=0.838\pm 0.002$, and $r=2$ and \citet[][]{2013MNRAS.430.2313C}:
\begin{equation}
\resizebox{.85\hsize}{!}{$A_{\rm Kp} =\beta \left(\frac{\numax}{\nu_{\rm max,\odot}} \right)^{2s-3t}  \left(\frac{\dnu}{\dnu_{\odot}} \right)^{4t-4s} \left(\frac{\teff}{T_{\rm eff,\odot}} \right)^{5s-1.5t-r+0.2}\frac{A_{\rm bol, \odot}}{c_{\rm K}(T_{\rm eff})}\, , $}
\end{equation}
with $s=0.748\pm0.015$, $t=1.27\pm0.04$, $r=3.47\pm0.09$, and $\ln\beta = 0.321\pm0.020$.
In converting the bolometric amplitudes from these relations to the \kp bandpass we used the root-mean-square value of $A_{\rm bol, \odot}=2.53\pm 0.11\, \rm ppm$ from \citet[][]{2009A&A...495..979M} and the temperature dependent bolometric correction $c_{\rm K}(T_{\rm eff})$ by \citet[][]{2011A&A...531A.124B}, which is specific to the \kp spectral response function. 
Overall we see a reasonable agreement with a scatter within approximately $25\%$ as also found by \citet[][]{2011ApJ...743..143H}. At low amplitudes the \citet[][]{2011ApJ...743..143H} relation is seen to provide the best agreement, while the \citet[][]{2013MNRAS.430.2313C} relation has an overall lower scatter across the amplitude range. We note that the relations tested here were calibrated against $A_{\rm max}$ values obtained using the method of \citet[][]{2005ApJ...635.1281K,2008ApJ...682.1370K} and the high \numax range occupied by the stars in the current analysis is sparsely covered in the calibrations of the scaling relations, so a better relationship than observed cannot readily be expected (see more below). In addition, amplitudes will have natural scatter due to the impact of activity \citep[][]{2011ApJ...732L...5C}.

%%%%%%%%%%%%%%%%%%%%%%%%%%%%%%%%%%%%%%%%%%%%%%%%%%%%%%%%%%%%%%%%%%%%%%%%%%%%%%%%%%%%%%%%%%%%%%%%%%%%%%%%%%%%%%%%%%%%%%%%%%
\subsubsection{Amplitudes from smoothed amplitude spectra}
\label{sec:smo_amp}

With amplitudes measured from individual modes it is interesting to see how these compare to those obtained from the often adopted method by \citet[][]{2005ApJ...635.1281K,2008ApJ...682.1370K}. This is especially worthwhile because it is often amplitudes from this method that are extracted by automated analysis pipelines \citep[see, \eg,][]{2009CoAst.160...74H,2010MNRAS.402.2049H,2010A&A...511A..46M}, and thus used in calibrating scaling relations \citep[see, \eg,][]{2011ApJ...737L..10S,2011ApJ...743..143H,2013MNRAS.430.2313C}.

Amplitudes were estimated following the prescription of \citet[][]{2008ApJ...682.1370K}. Here the power density spectrum is first convolved with a Gaussian filter with a FWHM of $4\dnu$, to produce a spectrum with a single power hump from the oscillations. A noise background is then fitted to the smoothed spectrum and subtracted, after which the spectrum is multiplied by $\dnu/c$ and the square-root is taken to convert to amplitude. The multiplication of $\dnu/c$ converts to power per radial mode, with $c$ giving the effective number of radial modes per order, \ie, $c\approx \tilde{V}^2_{\rm tot}$ (see \eqref{eq:visi}).   
Concerning the order of the different steps in the method we subtract instead first the fitted background function from the peak-bagging, and then apply the smoothing --- this removes the potential bias on both the amplitude and central frequency of the smoothed power hump from mixing background and oscillation power, and then try to fit the now smoothed underlying background with a non-smoothed background function. 

The value of $c$ is often kept fixed rather than estimated for a given star following, for instance, the results by \citet[][]{2011A&A...531A.124B}. \citet[][]{2010ApJ...713..935B} estimated a value of $c=3.04$ for the Sun as seen by \kp using the method of \citet[][]{1996MNRAS.280.1155B} and \citet{2008ApJ...682.1370K} and adopting a mean observing wavelength of $\lambda=650\rm\, nm$ --- this value has since been used by, \eg, \citet[][]{2011ApJ...737L..10S}, \citet{2011ApJ...743..143H}, and \citet{2013MNRAS.430.2313C}. In our estimation of amplitudes we tried both the value of $c=3.04$ and the one obtained from the visibilities measured in the peak-bagging $c=\tilde{V}^2_{\rm tot}$.

%%%%%%%%%%%%%%%%%%%%%%%%%%%%%%%%%%%%%%%%%%%%%%%%%%%%%%%%%%%%%%%%
\begin{figure}
\centering
\includegraphics[width=\columnwidth]{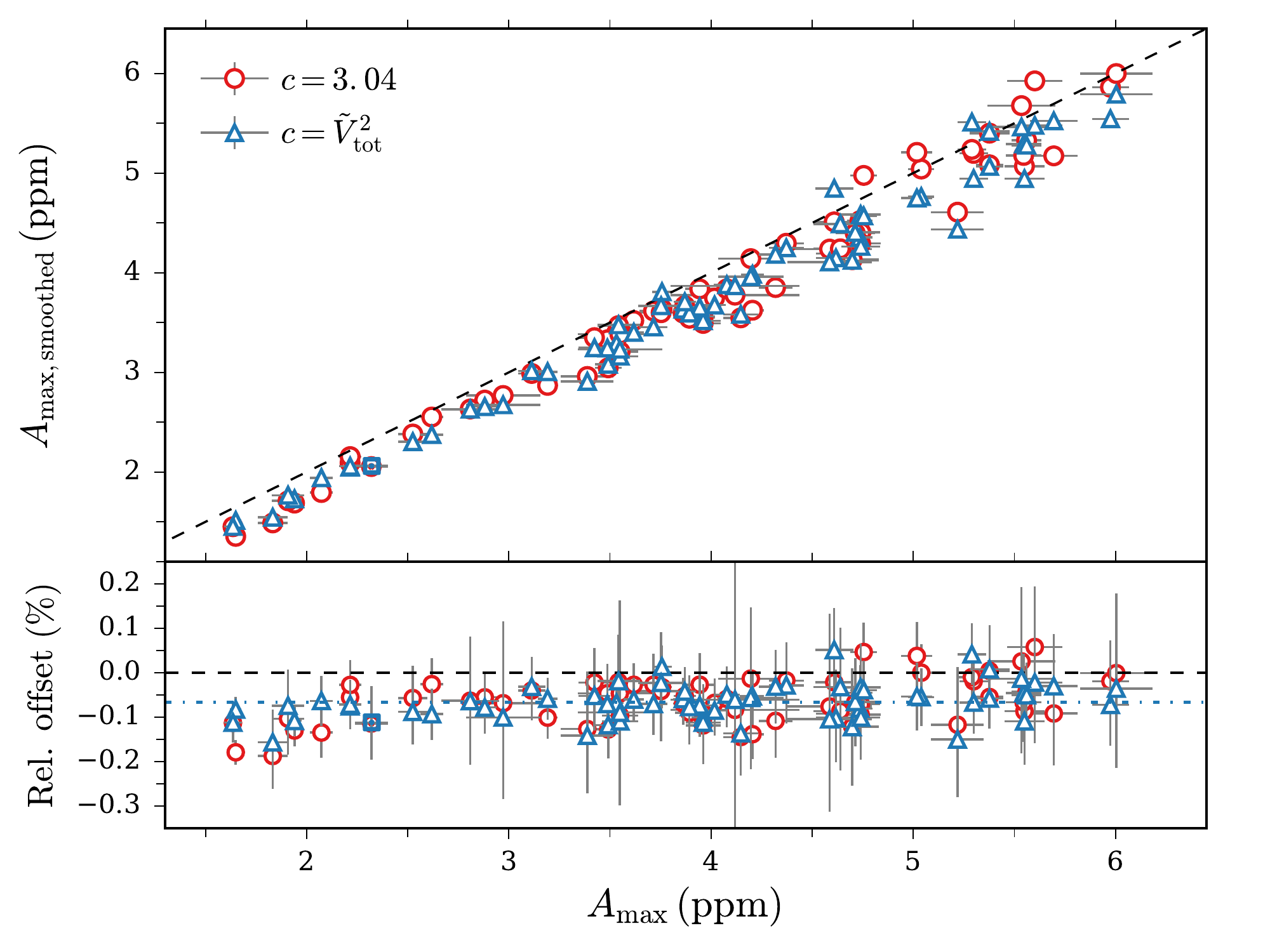}
\caption{Top: Comparison between $A_{\rm max}$ obtained from the smoothed amplitude spectrum following the procedure by \citet[][]{2008ApJ...682.1370K} with those obtained from the individual mode amplitudes. The different markers indicate the value of $c$ used to convert the maximum smoothed amplitude to amplitude per radial degree mode. The dashed line gives the $1:1$ relation. Bottom: Relative offset between the two amplitude measures, with the zero offset given by the black dashed line. The blue dash-dot line indicates the median $-6\%$ offset between the amplitudes, which is found to be the same for both values of $c$.}
\label{fig:amax_vs_asmo}
\end{figure}
%%%%%%%%%%%%%%%%%%%%%%%%%%%%%%%%%%%%%%%%%%%%%%%%%%%%%%%%%%%%%%%
\fref{fig:amax_vs_asmo} shows the result from this exercise; we find that the $A_{\rm max}$ values from the above method are systematically offset (fixed bias) from those estimated from individual frequencies, with a median relative offset of $-6\%$; no proportional bias is seen. This offset could in part originate from the somewhat arbitrary choice of $4\dnu$ for the smoothing window --- however, if this was the only contributor one might expect a proportional offset rather than a constant one. The identified offset fully corroborates the results by \citet[][]{2011MNRAS.415.3539V} who found a systematic offset between $-15$ and $-2\%$ from pipeline analysis of simulated data. We observe the same median offset for the different values of $c$, but the scatter is slightly lower for the values using $c$ from the measured visibilities. Considering that $\tilde{V}^2_{\rm tot}$ and thus $c$ depends on various stellar parameters, such as \teff, \feh, and \logg, it should be expected that adopting a constant $c$ adds to the scatter. The values of $A_{\rm max}$ from the smoothing method, with both values of $c$ are given in \tref{tab:width_amp_max_table}.

%%%%%%%%%%%%%%%%%%%%%%%%%%%%%%%%%%%%%%%%%%%%%%%%%%%%%%%%%%%%%%%%%%%%%%%%%%%%%%%%%%%%%%%%%%%%%%%%%%%%%%%%%%%%%%%%%%%%%%%%%%
%%%%%%%%%%%%%%%%%%%%%%%%%%%%%%%%%%%%%%%%%%%%%%%%%%%%%%%%%%%%%%%%%%%%%%%%%%%%%%%%%%%%%%%%%%%%%%%%%%%%%%%%%%%%%%%%%%%%%%%%%%
\subsection{Mode line widths}
\label{sec:lws}

%%%%%%%%%%%%%%%%%%%%%%%%%%%%%%%%%%%%%%%%%%%%%%%%%%%%%%%%%%%%%%%
\begin{figure}
\centering
\includegraphics[width=\columnwidth]{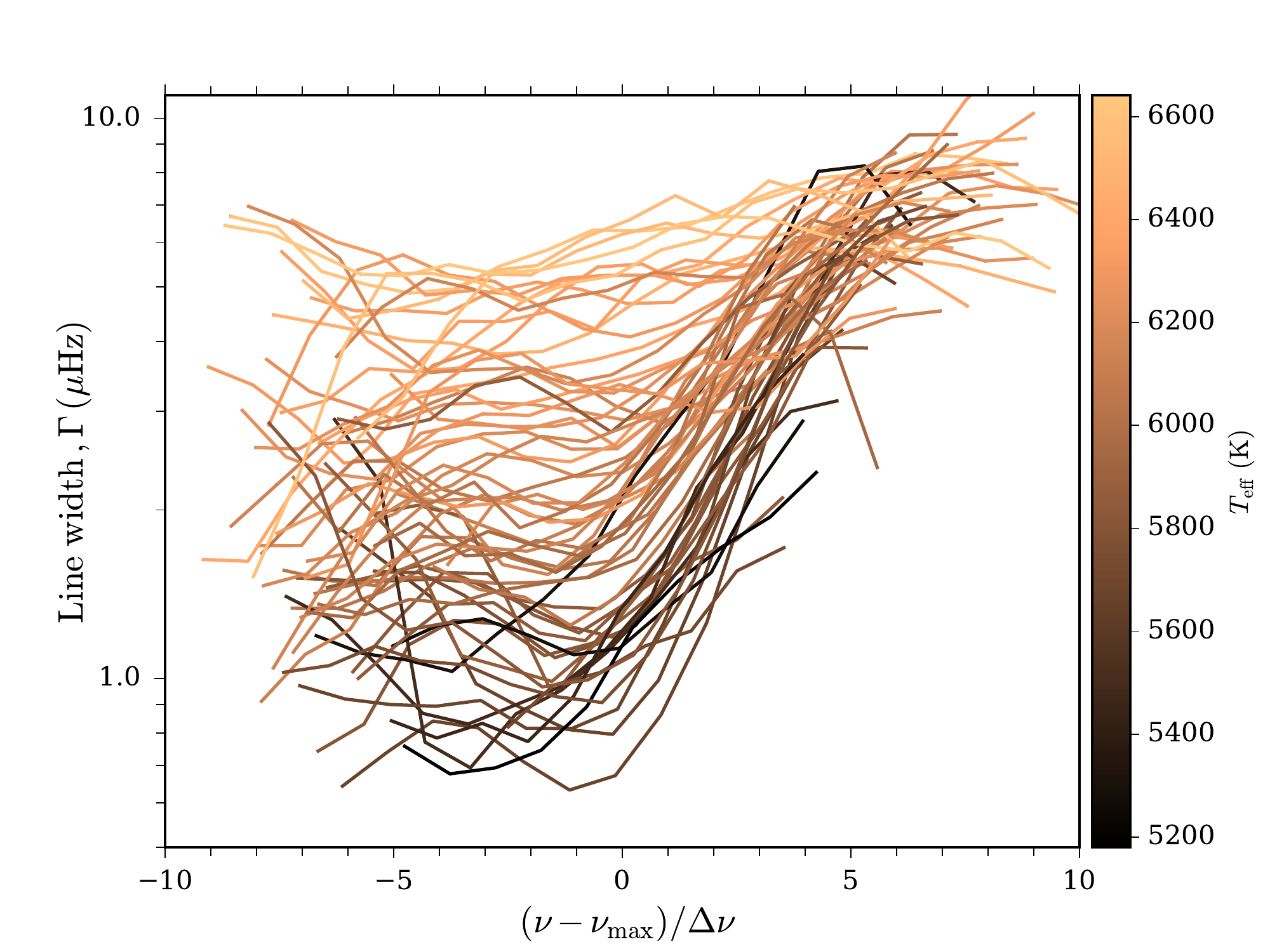}
\caption{Radial mode line widths against a proxy for the radial order, with the color indicating the \teff. For a better visualization the line widths have been smoothed with 5-point Epanechnikov filter.}
\label{fig:relfre_vs_gam}
\end{figure}
%%%%%%%%%%%%%%%%%%%%%%%%%%%%%%%%%%%%%%%%%%%%%%%%%%%%%%%%%%%%%%
Following \citet[][]{2014A&A...566A..20A} we adopt the following relation for the line widths $\Gamma$ against mode frequency:
\begin{equation}\label{eq:gam_v_fre}
\ln(\Gamma) = (\alpha \ln(\nu/\numax) + \ln \Gamma_{\alpha} ) + \left(\frac{\ln\Delta\Gamma_{\rm dip}}{1 + \left(\frac{2\ln(\nu/\nu_{\rm dip})}{\ln(W_{\rm dip}/\numax)} \right)^2} \right)\, ,
\end{equation}
where the first part describes a power law trend with an exponent $\alpha$ and a value of $\Gamma_{\alpha}$ at \numax. The second part describes a Lorentzian dip in the line widths in $\ln\nu$, centered on $\nu_{\rm dip}$ with a width $W_{\rm dip}$ and an depth $\Delta\Gamma_{\rm dip}$. For the fit the values of $\ln(\Gamma)$ were calculated from the $\Gamma$ posterior distributions; the $\ln(\Gamma)$-distribution is typically better approximated by a normal distribution \citep[][]{1994A&A...289..649T}. We note that while \eqref{eq:gam_v_fre} matches that reported by \citet[][]{2014A&A...566A..20A}, these authors did in fact use a formula with the Lorentzian subtracted rather than added as in \eqref{eq:gam_v_fre} \citep[][]{2016A&A...595C...2A}. Both formulations can be used, but the reported parameters will naturally differ; $\Delta\Gamma_{\rm dip}$ in \eqref{eq:gam_v_fre} is for instance constrained to a value between 0 and 1 if a dip in $\Gamma$ is to be produced, whereas it can take on any value above 1 if the Lorentzian is instead subtracted. We chose the formulation where the Lorentzian is added in log-space, because we found that this provided better constrained fits from a lower correlation between the $\Gamma_{\alpha}$, $\Delta\Gamma_{\rm dip}$, and $W_{\rm dip}$ parameters.
We note that the parameter $W_{\rm dip}$ has two solutions, one higher than \numax, and one lower --- in our fits we chose the $W_{\rm dip}>\numax$ solution with a prior on $W_{\rm dip}$.  
All stars were fitted using both the full version of \eqref{eq:gam_v_fre} and using only the first power law component --- \fref{fig:relfre_vs_gam} shows for all stars in the sample the measured line widths against a proxy for the radial order. The fit was made using \texttt{emcee} \citep[][]{2013PASP..125..306F}, with the parameter values and uncertainties given by the posterior distribution medians and $68\%$ HPD intervals. We evaluate which of the two types of fit is the best based on a visual inspection of the fits.

%%%%%%%%%%%%%%%%%%%%%%%%%%%%%%%%%%%%%%%%%%%%%%%%%%%%%%%%%%%%%%%%
\begin{figure}
\centering 
\includegraphics[width=\columnwidth]{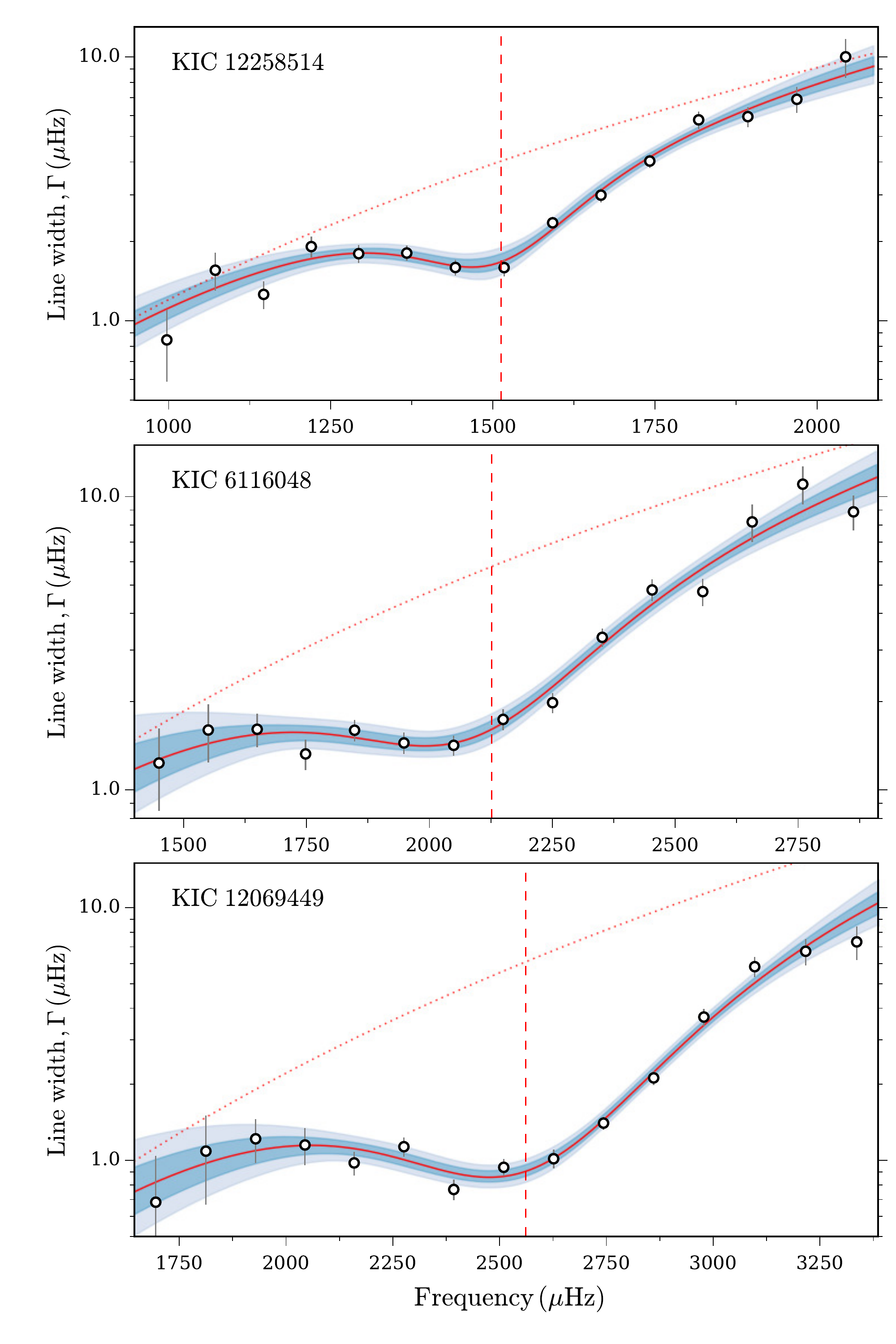}
\caption{Radial mode line widths as a function of frequency for KIC 12258514 (top), KIC 6116048 (middle), and KIC 12069449 (bottom). For all three stars the line widths are best fit with the Lorentzian included in \eqref{eq:gam_v_fre}, this fit is indicated by the full red lines; the dotted line gives the power law component of the fit. The shaded dark and light blue regions indicate the 1 and 2$\sigma$ credible regions of the fits. The dashed vertical lines give the \numax values.}
\label{fig:gam_fit_example}
\end{figure}
%%%%%%%%%%%%%%%%%%%%%%%%%%%%%%%%%%%%%%%%%%%%%%%%%%%%%%%%%%%%%%%
\fref{fig:gam_fit_example} shows three examples where the full version of \eqref{eq:gam_v_fre} was deemed the better --- as seen the line widths show a clear depression around \numax.
We omitted fitting the solar line widths from the degraded VIRGO data, because in computing relations for the fit parameters we are interested in as accurate and precise data as possible.

\begin{figure*}
\centering 
\includegraphics[width=\textwidth]{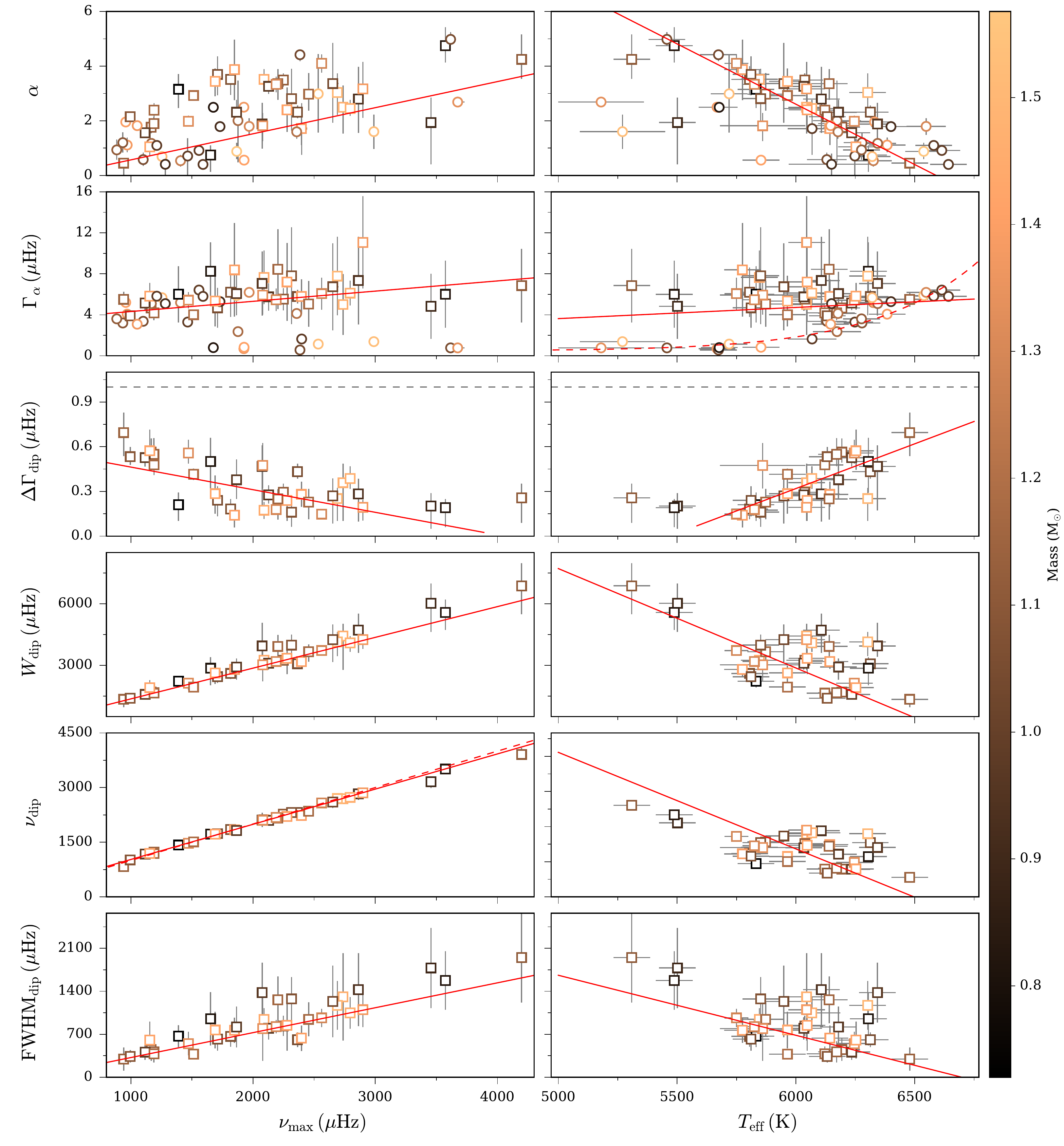}
\caption{Parameters from the fit of \eqref{eq:gam_v_fre} to the radial mode line widths of the sample stars against \numax (left panels) and \teff (right panels). The color indicates the modeled mass. Stars fitted using only the first part of \eqref{eq:gam_v_fre}, \ie, with parameters $\alpha$ and $\Gamma_{\alpha}$, are given by circles ($\circ$); stars where also the Lorentzian component was included, \ie, using also $\Delta\Gamma_{\rm dip}$, $W_{\rm dip}$, and $\nu_{\rm dip}$, are plotted with squares ($\square$). The coefficient for the fitted linear relations (solid lines) are given in \tref{tab:lin_coeff}. The dashed line in the $\Gamma_{\alpha}$ against \teff panel gives the relation for $\Gamma$ at \numax by \citet[][]{2012A&A...537A.134A}.}
\label{fig:fig_gam_fre}
\end{figure*}

The parameters from the fits are given in \tref{tab:width_amp_max_table} and shown in \fref{fig:fig_gam_fre} as a function of \numax and \teff. For the dependencies of the FWHM of the Lorentzian line width dip in frequency units we used the transformation $\rm FWHM_{\rm dip} = \nu_{\rm dip} (|\sqrt{W_{\rm dip}/\numax}-\sqrt{\numax/W_{\rm dip}}|)$ \citep[][]{2016A&A...595C...2A}. The amplitude of the Lorentzian dip may be calculated using the transformation $A_{\rm dip} = \exp(|\ln \Delta\Gamma_{\rm dip}|)$\footnote{Note that this $A_{\rm dip}$, where the Lorentzian in \eqref{eq:gam_v_fre} is added rather than subtracted, is different from that given by \citet[][]{2014A&A...566A..20A} who instead has $A_{\rm dip} = \Delta\Gamma_{\rm dip}$.}. 
Like \citet[][]{2014A&A...566A..20A} for their sample of 22 \kp stars, we find a clear correlation between the Lorentzian width ($W_{\rm dip}$ or $\rm FWHM_{\rm dip}$) and \numax, while the power law exponent ($\alpha$) and the dip amplitude ($\Delta\Gamma_{\rm dip}$) are found to correlate most strongly with \teff. Only for the fits without the Lorentzian component does the value of $\Gamma_{\alpha}$ correspond to the line width at \numax, which is seen to correlate with \teff. We return to the overall behavior of the line width at \numax against \teff below. Our estimates of $\rm FWHM_{\rm dip}$ agree well with updated values from \citet[][]{2016A&A...595C...2A}. We make linear fits to the different parameters ($P$) against \numax as $P=a\, (\numax/3090)+b$ and \teff as $P=a\, (\teff/5777)+b$. Specifically, we perform a Deming regression \citep[][]{MR0009819} where uncertainties in both the dependent ($\sigma_{x}$) and independent variables ($\sigma_{y}$) are considered via the following merit function \citep[][]{Press:1993:NRF:563041}:
\begin{equation}
M(a,b) = \sum_{i=1}^N \frac{(y_i - a - b x_i)^2}{\sigma_{y,i}^2 + b^2\sigma_{x,i}^2}\, ,
\end{equation}
which serves as our log-likelihood function in the \texttt{emcee} optimization. The coefficients from the different fits are given in \tref{tab:lin_coeff}.
For the fit of $\rm \Gamma_{\alpha}$ against \teff we include only the values from the fit including the Lorentzian in \eqref{eq:gam_v_fre}. Some of the scatter in the relations will likely be due to activity; the dip in line widths for the Sun has for instance been found to depend on the solar cycle \citep[see, \eg,][]{2000ApJ...531.1094K,2000ApJ...543..472K}.
We note that the fit coefficients in \tref{tab:lin_coeff} may well be used to define prior functions and initial guesses in future peak-bagging exercises.

\begin{figure}
\centering 
\includegraphics[width=\columnwidth]{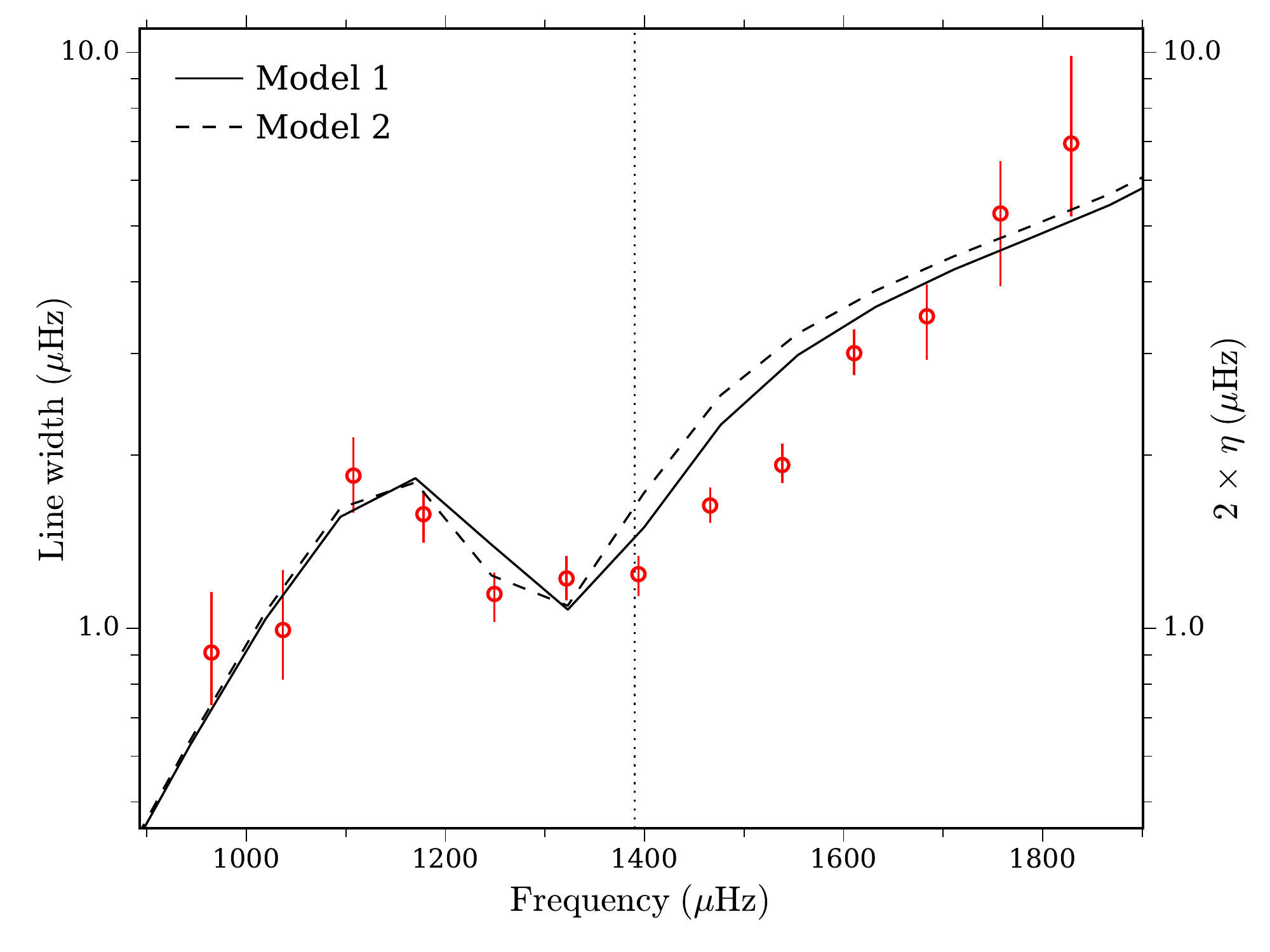}
\caption{Line widths (left axis) for KIC 6933899 (Fred) and model calculated damping-rates ($\eta$) by Houdek et al. (submitted) (right axis), multiplied by 2 to correspond to the mode FWHM.
The full and dashed lines gives the results for two model calculations of mode damping-rates. Model 1 assumes a constant value for the velocity anisotropy $\Phi$; Model 2 has a depth-dependent velocity anisotropy, guided by 3D simulation results from \citet[][]{2014MNRAS.442..805T}. Stellar parameters from the best fitting ASTFIT model (Paper~II) were used in the damping-rate calculations. The vertical dotted line indicates the value of \numax.}
\label{fig:gam_fit_GH}
\end{figure}

For a star with a temperature of the Sun we find that it should have an exponent of $\alpha\approx 3.6$ --- this comes close to that measured by \citet[][]{2000ApJ...543..472K} of $\alpha \approx 3.31$ for frequencies below $2450\, \rm \mu Hz$, which is about the lower limit of what one would observe for a similar star with \kp. In general the exponent for the Sun has been measured below the dip where higher values have been found, \eg, $\alpha \approx 5$ by \citet[][]{1991Sci...253..152L}, $\alpha \approx 7$ by \citet[][]{1997MNRAS.288..623C}, and $\alpha \approx 8$ by \citet[][]{2014MNRAS.439.2025D} --- it is therefore not surprising that a lower value of $\alpha$ is found when the measurement is done across the dip.
Like \citet[][]{2014A&A...566A..20A} we find that the amplitude of the line width dip decreases with \teff. This behavior is contrary to theoretical damping-rate estimates \citep[][]{1992MNRAS.255..603B,1996PhDT........80H} which assume solar-calibrated convection parameters, such as the mixing-length and anisotropy of the turbulent velocity field, for damping-rate calculations in other stars. Updated calculations of mode damping rates (Houdek et al., in prep.) are, however, able to capture the overall behavior of the line widths for stars in the current study spanning a large range in \teff -- these new calculations adopt, in addition to the standard ingredients described in \citet[][]{1999A&A...351..582H} and \citet{2006astro.ph.12024H}, turbulent pressure profiles and $T-\tau$ relations calibrated to 3D hydrodynamical simulation results by \citet[][]{2014MNRAS.442..805T}. An example of such an improved damping-rate computation provides \fref{fig:gam_fit_GH}, which compares estimated mode line widths with observations for KIC 6933899 (Fred), using stellar parameters from the best fitting ASTFIT model (Paper~II).
%%%%%%%%%%%%%%%%%%%%%%%%%%%%%%%%%%%%%%%%%%%%%%%%%%%%%%%%%%%%%%%%
%%%%%%%%%%%%%%%%%%%%%%%%%%%%%%%%%%%%%%%%%%%%%%%%%%%%%%%%%%%%%%%

%%%%%%%%%%%%%%%%%%%%%%%%%%%%%%%%%
%%%%%%%%%%%%%%%%%%%%%%%%%%%%%%%%%
%%%%%%%%%%%%%%%%%%%%%%%%%%%%%%%%%
\begin{table} 
\mysize
\centering 
\begin{threeparttable}
\caption{Parameters from linear fits to the fitted values of \eqref{eq:gam_v_fre} against \numax and \teff. The linear fits are indicated as red lines in \fref{fig:fig_gam_fre}. For the fits of $\Gamma_{\alpha}$ only values from the full fit of \eqref{eq:gam_v_fre} were included.} 
\label{tab:lin_coeff}
\begin{tabular}{cccccc}  
\cmidrule[1.0pt](lr){1-6}\\[-1.8ex]
& \multicolumn{2}{c}{$P=a\, (\numax/3090)+ b$} &   & \multicolumn{2}{c}{$P=a\, (\teff/5777) + b$}\\ [0.5ex]
\cmidrule(lr){2-3}\cmidrule(lr){5-6}\\ [-1.8ex]
$P\, \rm(\mu Hz)$ & $b$ & $a$ &   & $b$ & $a$\\ 
\cmidrule(lr){1-6}\\ [-1.8ex]
$\alpha$ 						& $2.95\pm 0.16$  & $0.39\pm 0.08$   &  & $-25.5 \pm 1.4$ &  $ 29.1\pm 1.5$ \\
$\Gamma_{\alpha}$ 				& $3.08\pm 0.98$  & $3.32\pm 0.50$   &  & $6.3 \pm 6.6$   &  $-1.8 \pm 6.9$ \\
$\Delta\Gamma_{\rm dip}$ 		& $-0.47\pm 0.06$ & $0.62\pm 0.04$   &  & $3.5 \pm 0.4$   &  $-3.3 \pm 0.4$  \\
$W_{\rm dip}$ 					& $4637 \pm 237$  & $-141 \pm 138$  &  & $-28021 \pm 2964$   &  $31971 \pm 3105$  \\
$\nu_{\rm dip}$ 					& $2984 \pm 31$   & $60\pm 18$      &  & $-23818 \pm 1909$    &  $26785 \pm 1983$  \\
$\rm FWHM_{\rm dip}$ 			& $1253 \pm 162$  & $-85 \pm 96$ &  & $-5649 \pm 1093$ &  $6550 \pm 1151$  \\ [1ex]
\cmidrule[1.0pt](lr){1-6}
\end{tabular} 
\end{threeparttable}  
\end{table} 
%%%%%%%%%%%%%%%%%%%%%%%%%%%%%%%%%%%%%%%%%%%%
%%%%%%%%%%%%%%%%%%%%%%%%%%%%%%%%%%%%%%%%%%%%
%%%%%%%%%%%%%%%%%%%%%%%%%%%%%%%%%%%%%%%%%%%%

%%%%%%%%%%%%%%%%%%%%%%%%%%%%%%%%%%%%%%%%%%%%%%%%%%%%%%%%%%%%%%%%
%%%%%%%%%%%%%%%%%%%%%%%%%%%%%%%%%%%%%%%%%%%%%%%%%%%%%%%%%%%%%%%

\begin{figure}
\centering
\includegraphics[width=\columnwidth]{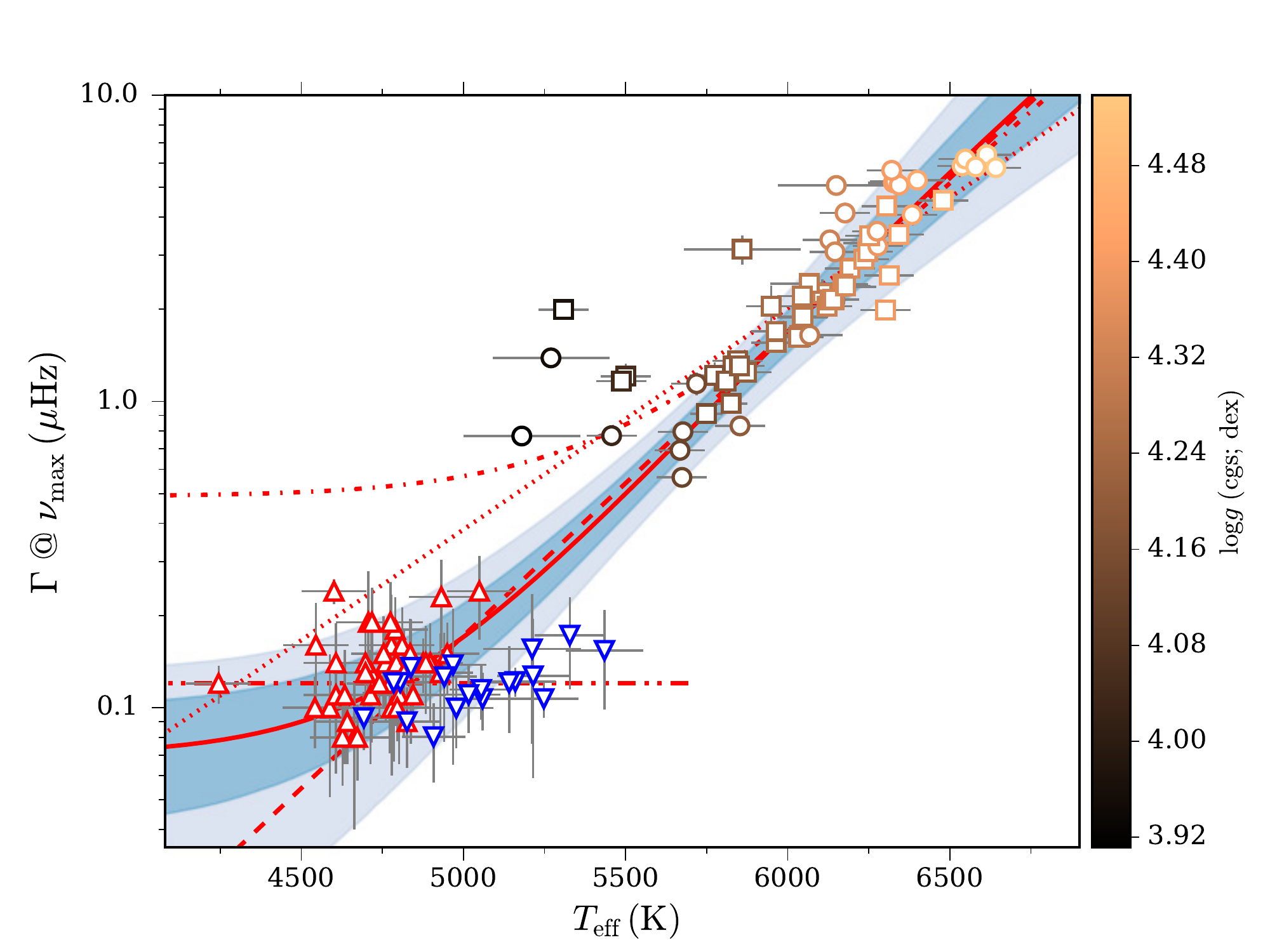}
\caption{Line width at \numax against \teff with the colour indicating the value of \logg. See \tref{tab:width_amp_max_table} for the plotted values, where the line widths are estimated from the fit of \eqref{eq:gam_v_fre}. The red upward triangles indicate the line widths from the peak-bagging of 42 giants in NGC 6819 by Handberg et al. (2016, submitted); the blue downward triangles give the line widths from \citet[][]{2015A&A...579A..83C} for 19 red giants. The full red line gives the fit of \eqref{eq:gam_teff} with the shaded dark and light blue regions indicating the $1$ and $2\sigma$ intervals of the fit; the dash-dot line gives the corresponding fit of \eqref{eq:gam_teff} by \citet[][]{2012A&A...537A.134A} (their Table~3); the dashed line gives the fit of \eqref{eq:gam_teff2}; the dotted line gives the corresponding fit of \eqref{eq:gam_teff2} by \citet[][]{2012ApJ...757..190C}; the dash-dot-dot line gives the constant fit to only the red giants. The markers indicate if the full fit of \eqref{eq:gam_v_fre} is preferred ($\square$), or only the power law component ($\circ$).}
\label{fig:gam_vs_teff}
\end{figure}

The estimate of $\Gamma$ at \numax was obtained by a Monte Carlo sampling from the posteriors of the fit parameters in \eqref{eq:gam_v_fre} and the estimate of \numax from the mode amplitudes. The values obtained are plotted against \teff in \fref{fig:gam_vs_teff} and given in \tref{tab:width_amp_max_table}. 
We have fitted two relations for the line width at \numax, namely, the power-law relation by \citet[][]{2012A&A...537A.134A}:
\begin{equation}\label{eq:gam_teff}
\Gamma = \Gamma_0 + \alpha \left( \frac{T_{\rm eff}}{5777} \right)^{\beta}\, {\rm \mu Hz}\, ,
\end{equation}
and the exponential relation used, for instance, by \citet[][]{2012ApJ...757..190C}:
\begin{equation}\label{eq:gam_teff2}
\Gamma = \Gamma_0\exp\left( \frac{T_{\rm eff}-5777}{T_0} \right)\, {\rm \mu Hz}\, .
\end{equation}
In the fits of Equations~\ref{eq:gam_teff} and \ref{eq:gam_teff2} we complement our set of line widths with those from peak-bagging of 42 giants in NGC 6819 by Handberg et al. (2016, submitted) where we adopt an uncertainty of $100\, \rm K$ on \teff for all stars, and 19 red giants by \citet[][]{2015A&A...579A..83C} where we adopt \teff values from \citet[][]{2012ApJS..199...30P} (from the infra-red flux method). The line widths from Handberg et al. (2016, submitted) are given by the average of the radial modes; those from \citet[][]{2015A&A...579A..83C} are given as the average over three radial modes centered in the mode with the highest amplitude.
%%%%%%%%%%%%%%%%%%%%%%%%%%%%%%%%%%%%%%%%%%%%%%%%%%%%%%%%%%%%%%%
%%%%%%%%%%%%%%%%%%%%%%%%%%%%%%%%%%%%%%%%%%%%%%%%%%%%%%%%%%%%%%
We fit both relations to the line widths using an orthogonal distance regression \citep[ODR;][]{MR1087109} in order to take into account both the uncertainties on $\Gamma$ and \teff. This procedure is employed in a Monte Carlo (MC) run where we in each iteration draw at random 50 stars to include in the fit. For \eqref{eq:gam_teff} we obtain from the MC distributions the following values for the fitting parameters: $\Gamma_0\approx0.07\pm 0.03\, {\rm \mu Hz}$, $\alpha\approx0.91\pm 0.13\, {\rm \mu Hz}$, and $\beta\approx15.3\pm 1.9$; for \eqref{eq:gam_teff2} we obtain: $\Gamma_0\approx 1.02\pm 0.07\, {\rm \mu Hz}$ and $T_0\approx 436\pm 24\, {\rm K}$. If we fit only the red giants by a constant value for the line width we find a value of $\Gamma_0\approx 0.12\pm 0.01\, {\rm \mu Hz}$. For the parameter uncertainties we have added in quadrature the median from the internal uncertainties from the ODR and the standardized MAD of the best fit ODR values from the MC run. We note here that neglecting the uncertainties in \teff from using, for instance, an ordinary least squares (OLS) approach will affect the fit and the parameters in \eqref{eq:gam_teff}. Both relations perform reasonably well, but the fit of \eqref{eq:gam_teff} gives a slightly lower $\chi^2$ value. The resulting relations are displayed in \fref{fig:gam_vs_teff}. Also shown in \fref{fig:gam_vs_teff} are the fits obtained by \citet[][]{2012A&A...537A.134A} (parameters from their Table 3) and \citet[][]{2012ApJ...757..190C}. 

We see that the cooler MS targets appear to be outliers to the overall relation between line width and temperature. From \fref{fig:fig_gam_fre} we can see that the value of $\nu_{\rm dip}$ has a proportional bias with respect to \numax, hence the higher the \numax the further below \numax does the dip in line widths appear. For \fref{fig:gam_vs_teff} the line widths were estimated at \numax and thus away from the line width dip for the highest \numax stars, it is therefore not surprising that these deviate from the overall relation. An offset could potentially also be caused by frequency shifts from stellar activity cycles \citep[][]{2007MNRAS.377...17C,2008MNRAS.384.1668C}, the evidence of which is currently being studied by Santos et al. (in prep.). 
The line width for KIC 7970740 was omitted in the fits as it appears to be a particularly strong outlier at $\teff\approx 5300$ K. 

We find that our estimates agree well with those from \citet[][]{2012A&A...537A.134A} with the main difference seen at low temperatures where the fit is the least constrained if no line widths from giants are included. The estimates from \citet[][]{2012ApJ...757..190C} appear to be higher at low temperatures than the line widths from Handberg et al. (2016, submitted) --- this could be due to the method of estimating $\Gamma$ in \citet[][]{2012ApJ...757..190C} from collapsed \'{e}chelle diagrams for a group of stars. Some scatter is to be expected, because the line widths will depend on parameters besides \teff --- \citet[][]{2012A&A...540L...7B,2013EPJWC..4303009B} specifically suggest a scaling with \teff and \logg. As noted by \citet[][]{2012A&A...537A.134A}, the uncertainty on \teff currently sets the limitations for obtaining a more well constrained prediction for $\Gamma$ at \numax, which may then be confronted with theoretical calculations.
As for the dip amplitude, activity and stellar cycles will play a role in the scatter around the mean relation for $\Gamma$ at \numax \citep[][]{2000ApJ...543..472K,2008MNRAS.384.1668C}. For other analysis of the variation of $\Gamma$ at \numax we refer to 
\citet[][]{1991ApJ...374..366G,1999A&A...351..582H,2009A&A...500L..21C,2011A&A...529A..84B}, and \citet{2012A&A...540L...7B,2013EPJWC..4303009B}.

%%%%%%%%%%%%%%%%%%%%%%%%%%%%%%%%%%%%%%%%%%%%%%%%%%%%%%%%%%%%%%%%%%%%%%%%%%%%%%%%%%%%%%%%%%%%%%%%%%%%%%%%%%%%%%%%%%%%%%%%%%
%%%%%%%%%%%%%%%%%%%%%%%%%%%%%%%%%%%%%%%%%%%%%%%%%%%%%%%%%%%%%%%%%%%%%%%%%%%%%%%%%%%%%%%%%%%%%%%%%%%%%%%%%%%%%%%%%%%%%%%%%%
\subsection{Mode visibilities}
\label{sec:visibilities}

%%%%%%%%%%%%%%%%%%%%%%%%%%%%%%%%%%%%%%%%%%%%%%%%%%%%%%%%%%%%%%%%%%%%%%%%%%%%%%%%%%%%%%%%%%%%%%%%%%%%%%%%%%%%%%%%%%%%%%%%%%
%%%%%%%%%%%%%%%%%%%%%%%%%%%%%%%%%%%%%%%%%%%%%%%%%%%%%%%%%%%%%%%%%%%%%%%%%%%%%%%%%%%%%%%%%%%%%%%%%%%%%%%%%%%%%%%%%%%%%%%%%%
%%%%%%%%%%%%%%%%%%%%%%%%%%%%%%%%%%%%%%%%%%%%%%%%%%%%%%%%%%%%%%%%%%%%%%%%%%%%%%%%%%%%%%%%%%%%%%%%%%%%%%%%%%%%%%%%%%%%%%%%%%

% Table made Wed Sep 28 16:49:50 2016
\begin{table} 
\centering 
\resizebox*{!}{\textheight}{% 
\begin{threeparttable} 
\caption{Extracted mode visibilities for angular degrees $l=1-3$. Total visibilities have been constructed from combining the MCMC chains of the individual visibilities. Uncertainties are obtained from the 68\% HPD intervals of the posterior PDFs.} 
\label{tab:vis} 
\begin{tabular}{@{}r@{\hskip 8ex}c@{\hskip 8ex}c@{\hskip 8ex}c@{\hskip 8ex}c@{}} 
\toprule 
\multicolumn{1}{@{\hskip -5ex}c}{KIC} & $\tilde{V}_1^2$ & $\tilde{V}_2^2$ & $\tilde{V}_3^2$ & $\tilde{V}_{\rm tot}^2$ \\ 
\midrule 
$1435467$ & $1.52_{-0.09}^{+0.07}$ & $0.58_{-0.05}^{+0.09}$ &   & $3.08_{-0.12}^{+0.17}$ \\[0.5ex] 
$2837475$ & $1.46_{-0.08}^{+0.06}$ & $0.53_{-0.07}^{+0.05}$ &   & $3.00_{-0.14}^{+0.10}$ \\[0.5ex] 
$3427720$ & $1.75_{-0.08}^{+0.11}$ & $0.74_{-0.05}^{+0.05}$ &   & $3.50_{-0.13}^{+0.13}$ \\[0.5ex] 
$3456181$ & $1.74_{-0.12}^{+0.12}$ & $0.80_{-0.08}^{+0.13}$ &   & $3.56_{-0.21}^{+0.23}$ \\[0.5ex] 
$3632418$ & $1.65_{-0.06}^{+0.06}$ & $0.73_{-0.04}^{+0.07}$ &   & $3.41_{-0.12}^{+0.10}$ \\[0.5ex] 
$3656476$ & $1.35_{-0.06}^{+0.09}$ & $0.62_{-0.05}^{+0.04}$ & $0.07_{-0.01}^{+0.02}$ & $3.08_{-0.15}^{+0.08}$ \\ [0.5ex] 
$3735871$ & $1.55_{-0.09}^{+0.11}$ & $0.69_{-0.06}^{+0.06}$ &   & $3.21_{-0.11}^{+0.17}$ \\[0.5ex] 
$4914923$ & $1.43_{-0.06}^{+0.10}$ & $0.58_{-0.05}^{+0.02}$ & $0.08_{-0.02}^{+0.01}$ & $3.11_{-0.14}^{+0.10}$ \\ [0.5ex] 
$5184732$ & $1.57_{-0.04}^{+0.05}$ & $0.68_{-0.03}^{+0.03}$ & $0.08_{-0.01}^{+0.01}$ & $3.32_{-0.06}^{+0.09}$ \\ [0.5ex] 
$5773345$ & $1.27_{-0.06}^{+0.09}$ & $0.38_{-0.06}^{+0.07}$ &   & $2.65_{-0.13}^{+0.13}$ \\[0.5ex] 
$5950854$ & $1.32_{-0.17}^{+0.17}$ & $0.62_{-0.10}^{+0.11}$ &   & $2.92_{-0.20}^{+0.29}$ \\[0.5ex] 
$6106415$ & $1.50_{-0.05}^{+0.03}$ & $0.63_{-0.02}^{+0.03}$ & $0.10_{-0.01}^{+0.01}$ & $3.21_{-0.04}^{+0.08}$ \\ [0.5ex] 
$6116048$ & $1.45_{-0.06}^{+0.04}$ & $0.62_{-0.02}^{+0.03}$ & $0.09_{-0.01}^{+0.01}$ & $3.14_{-0.06}^{+0.09}$ \\ [0.5ex] 
$6225718$ & $1.54_{-0.04}^{+0.03}$ & $0.62_{-0.02}^{+0.02}$ & $0.07_{-0.01}^{+0.01}$ & $3.24_{-0.06}^{+0.05}$ \\ [0.5ex] 
$6508366$ & $1.80_{-0.07}^{+0.04}$ & $0.99_{-0.05}^{+0.01}$ &   & $3.77_{-0.10}^{+0.07}$ \\[0.5ex] 
$6603624$ & $1.33_{-0.05}^{+0.06}$ & $0.59_{-0.03}^{+0.04}$ & $0.06_{-0.01}^{+0.01}$ & $2.98_{-0.08}^{+0.08}$ \\ [0.5ex] 
$6679371$ & $1.44_{-0.07}^{+0.10}$ & $0.64_{-0.07}^{+0.08}$ &   & $3.07_{-0.12}^{+0.19}$ \\[0.5ex] 
$6933899$ & $1.55_{-0.06}^{+0.07}$ & $0.64_{-0.04}^{+0.03}$ &   & $3.20_{-0.09}^{+0.10}$ \\[0.5ex] 
$7103006$ & $1.27_{-0.07}^{+0.09}$ & $0.37_{-0.06}^{+0.08}$ &   & $2.66_{-0.15}^{+0.14}$ \\[0.5ex] 
$7106245$ & $1.48_{-0.13}^{+0.12}$ & $0.66_{-0.08}^{+0.09}$ &   & $3.14_{-0.18}^{+0.19}$ \\[0.5ex] 
$7206837$ & $1.25_{-0.04}^{+0.09}$ & $0.29_{-0.05}^{+0.05}$ &   & $2.58_{-0.12}^{+0.08}$ \\[0.5ex] 
$7296438$ & $1.56_{-0.12}^{+0.10}$ & $0.67_{-0.06}^{+0.06}$ &   & $3.24_{-0.16}^{+0.15}$ \\[0.5ex] 
$7510397$ & $1.58_{-0.05}^{+0.11}$ & $0.64_{-0.04}^{+0.06}$ &   & $3.23_{-0.10}^{+0.15}$ \\[0.5ex] 
$7680114$ & $1.56_{-0.08}^{+0.07}$ & $0.65_{-0.04}^{+0.04}$ & $0.09_{-0.02}^{+0.03}$ & $3.32_{-0.13}^{+0.09}$ \\ [0.5ex] 
$7771282$ & $1.48_{-0.13}^{+0.15}$ & $0.98_{-0.21}^{+0.02}$ &   & $3.35_{-0.23}^{+0.24}$ \\[0.5ex] 
$7871531$ & $1.45_{-0.11}^{+0.10}$ & $0.40_{-0.05}^{+0.05}$ &   & $2.86_{-0.14}^{+0.11}$ \\[0.5ex] 
$7940546$ & $1.63_{-0.08}^{+0.05}$ & $0.74_{-0.06}^{+0.06}$ &   & $3.33_{-0.11}^{+0.15}$ \\[0.5ex] 
$7970740$ & $1.11_{-0.04}^{+0.05}$ & $0.32_{-0.03}^{+0.03}$ &   & $2.42_{-0.05}^{+0.09}$ \\[0.5ex] 
$8006161$ & $1.37_{-0.04}^{+0.05}$ & $0.45_{-0.02}^{+0.02}$ & $0.07_{-0.01}^{+0.01}$ & $2.88_{-0.05}^{+0.07}$ \\ [0.5ex] 
$8150065$ & $1.65_{-0.24}^{+0.17}$ & $0.72_{-0.14}^{+0.11}$ &   & $3.38_{-0.36}^{+0.24}$ \\[0.5ex] 
$8179536$ & $1.54_{-0.08}^{+0.10}$ & $0.68_{-0.07}^{+0.06}$ &   & $3.23_{-0.14}^{+0.14}$ \\[0.5ex] 
$8228742$ & $1.63_{-0.05}^{+0.06}$ & $0.72_{-0.04}^{+0.04}$ &   & $3.35_{-0.09}^{+0.10}$ \\[0.5ex] 
$8379927$ & $1.52_{-0.03}^{+0.04}$ & $0.65_{-0.03}^{+0.02}$ &   & $3.17_{-0.05}^{+0.05}$ \\[0.5ex] 
$8394589$ & $1.57_{-0.06}^{+0.08}$ & $0.62_{-0.04}^{+0.04}$ &   & $3.18_{-0.08}^{+0.12}$ \\[0.5ex] 
$8424992$ & $1.57_{-0.22}^{+0.17}$ & $0.70_{-0.12}^{+0.11}$ &   & $3.22_{-0.25}^{+0.29}$ \\[0.5ex] 
$8694723$ & $1.48_{-0.04}^{+0.06}$ & $0.57_{-0.02}^{+0.05}$ &   & $3.07_{-0.09}^{+0.08}$ \\[0.5ex] 
$8760414$ & $1.38_{-0.06}^{+0.05}$ & $0.55_{-0.03}^{+0.03}$ & $0.08_{-0.02}^{+0.01}$ & $3.01_{-0.10}^{+0.08}$ \\ [0.5ex] 
$8938364$ & $1.57_{-0.07}^{+0.09}$ & $0.64_{-0.04}^{+0.04}$ & $0.05_{-0.02}^{+0.02}$ & $3.29_{-0.12}^{+0.09}$ \\ [0.5ex] 
$9025370$ & $1.41_{-0.07}^{+0.11}$ & $0.61_{-0.06}^{+0.05}$ &   & $3.03_{-0.12}^{+0.12}$ \\[0.5ex] 
$9098294$ & $1.51_{-0.08}^{+0.07}$ & $0.62_{-0.04}^{+0.04}$ &   & $3.12_{-0.10}^{+0.11}$ \\[0.5ex] 
$9139151$ & $1.60_{-0.08}^{+0.05}$ & $0.59_{-0.03}^{+0.04}$ &   & $3.18_{-0.08}^{+0.10}$ \\[0.5ex] 
$9139163$ & $1.46_{-0.05}^{+0.06}$ & $0.52_{-0.05}^{+0.06}$ &   & $2.99_{-0.12}^{+0.10}$ \\[0.5ex] 
$9206432$ & $1.44_{-0.08}^{+0.12}$ & $0.56_{-0.09}^{+0.10}$ &   & $2.99_{-0.15}^{+0.24}$ \\[0.5ex] 
$9353712$ & $1.62_{-0.14}^{+0.16}$ & $0.74_{-0.14}^{+0.15}$ &   & $3.37_{-0.27}^{+0.29}$ \\[0.5ex] 
$9410862$ & $1.34_{-0.10}^{+0.09}$ & $0.57_{-0.06}^{+0.07}$ &   & $2.95_{-0.16}^{+0.11}$ \\[0.5ex] 
$9414417$ & $1.36_{-0.08}^{+0.09}$ & $0.52_{-0.06}^{+0.08}$ &   & $2.85_{-0.11}^{+0.19}$ \\[0.5ex] 
$9812850$ & $1.42_{-0.10}^{+0.09}$ & $0.58_{-0.10}^{+0.10}$ &   & $2.99_{-0.18}^{+0.20}$ \\[0.5ex] 
$9955598$ & $1.20_{-0.09}^{+0.06}$ & $0.41_{-0.04}^{+0.04}$ &   & $2.58_{-0.08}^{+0.11}$ \\[0.5ex] 
$9965715$ & $1.38_{-0.10}^{+0.06}$ & $0.59_{-0.06}^{+0.05}$ &   & $2.97_{-0.16}^{+0.11}$ \\[0.5ex] 
$10068307$ & $1.68_{-0.08}^{+0.08}$ & $0.72_{-0.07}^{+0.05}$ &   & $3.40_{-0.14}^{+0.13}$ \\[0.5ex] 
$10079226$ & $1.52_{-0.20}^{+0.15}$ & $0.52_{-0.08}^{+0.10}$ &   & $3.01_{-0.21}^{+0.25}$ \\[0.5ex] 
$10162436$ & $1.69_{-0.09}^{+0.09}$ & $0.73_{-0.08}^{+0.08}$ &   & $3.39_{-0.15}^{+0.18}$ \\[0.5ex] 
$10454113$ & $1.21_{-0.04}^{+0.05}$ & $0.49_{-0.02}^{+0.03}$ & $0.06_{-0.01}^{+0.02}$ & $2.77_{-0.06}^{+0.07}$ \\ [0.5ex] 
$10516096$ & $1.58_{-0.06}^{+0.06}$ & $0.66_{-0.04}^{+0.04}$ &   & $3.23_{-0.09}^{+0.09}$ \\[0.5ex] 
$10644253$ & $1.59_{-0.09}^{+0.11}$ & $0.73_{-0.05}^{+0.06}$ &   & $3.32_{-0.12}^{+0.15}$ \\[0.5ex] 
$10730618$ & $1.29_{-0.09}^{+0.13}$ & $0.38_{-0.09}^{+0.11}$ &   & $2.68_{-0.19}^{+0.21}$ \\[0.5ex] 
$10963065$ & $1.62_{-0.07}^{+0.06}$ & $0.72_{-0.03}^{+0.04}$ &   & $3.31_{-0.07}^{+0.11}$ \\[0.5ex] 
$11081729$ & $1.43_{-0.11}^{+0.04}$ & $0.60_{-0.10}^{+0.05}$ &   & $2.98_{-0.15}^{+0.13}$ \\[0.5ex] 
$11253226$ & $1.18_{-0.04}^{+0.06}$ & $0.31_{-0.03}^{+0.06}$ &   & $2.48_{-0.05}^{+0.12}$ \\[0.5ex] 
$11772920$ & $1.42_{-0.12}^{+0.14}$ & $0.39_{-0.07}^{+0.06}$ &   & $2.80_{-0.15}^{+0.18}$ \\[0.5ex] 
$12009504$ & $1.63_{-0.05}^{+0.07}$ & $0.72_{-0.05}^{+0.04}$ &   & $3.34_{-0.09}^{+0.11}$ \\[0.5ex] 
$12069127$ & $1.53_{-0.12}^{+0.14}$ & $0.69_{-0.10}^{+0.16}$ &   & $3.27_{-0.27}^{+0.24}$ \\[0.5ex] 
$12069424$ & $1.38_{-0.03}^{+0.05}$ & $0.59_{-0.02}^{+0.02}$ & $0.08_{-0.01}^{+0.00}$ & $3.07_{-0.07}^{+0.05}$ \\ [0.5ex] 
$12069449$ & $1.37_{-0.05}^{+0.05}$ & $0.53_{-0.02}^{+0.03}$ & $0.07_{-0.00}^{+0.01}$ & $2.98_{-0.08}^{+0.07}$ \\ [0.5ex] 
$12258514$ & $1.60_{-0.05}^{+0.04}$ & $0.68_{-0.03}^{+0.03}$ &   & $3.29_{-0.08}^{+0.06}$ \\[0.5ex] 
$12317678$ & $1.77_{-0.07}^{+0.12}$ & $0.81_{-0.08}^{+0.10}$ &   & $3.53_{-0.17}^{+0.20}$ \\[0.5ex] 
\bottomrule
\end{tabular} 
 \end{threeparttable}% 
 } 
\end{table}

%%%%%%%%%%%%%%%%%%%%%%%%%%%%%%%%%%%%%%%%%%%%%%%%%%%%%%%%%%%%%%%%%%%%%%%%%%%%%%%%%%%%%%%%%%%%%%%%%%%%%%%%%%%%%%%%%%%%%%%%%%
%%%%%%%%%%%%%%%%%%%%%%%%%%%%%%%%%%%%%%%%%%%%%%%%%%%%%%%%%%%%%%%%%%%%%%%%%%%%%%%%%%%%%%%%%%%%%%%%%%%%%%%%%%%%%%%%%%%%%%%%%%
%%%%%%%%%%%%%%%%%%%%%%%%%%%%%%%%%%%%%%%%%%%%%%%%%%%%%%%%%%%%%%%%%%%%%%%%%%%%%%%%%%%%%%%%%%%%%%%%%%%%%%%%%%%%%%%%%%%%%%%%%%

\begin{figure*}
\centering
\includegraphics[width=\textwidth]{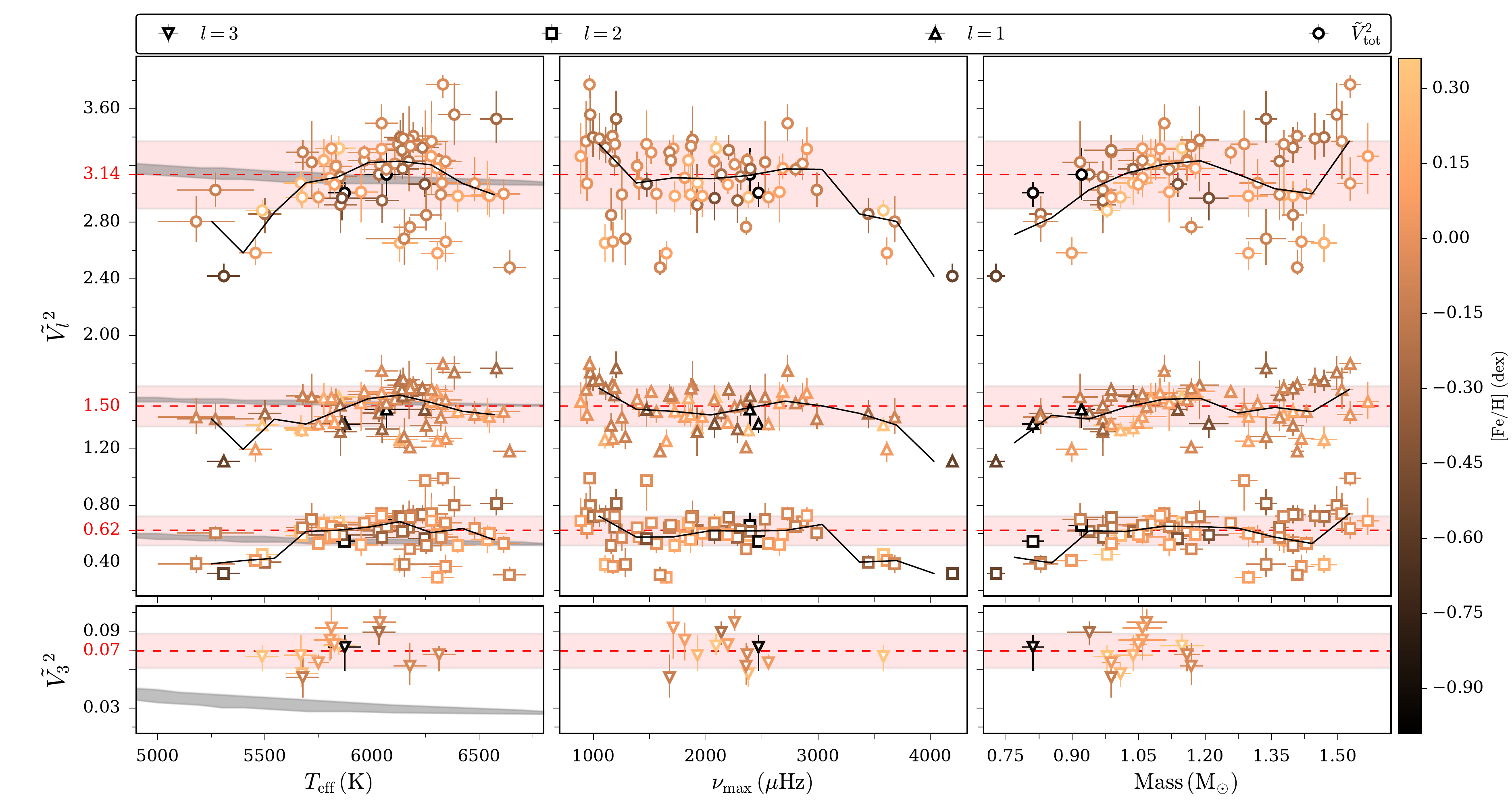}
\caption{Visibilities as a function of \teff (left), \numax (middle), and mass (right). The color indicates the metallicity; the markers indicate the angular degree (see legend), and shown are also the total visibilities $\tilde{V}_{\rm tot}^{\, 2}$ --- the bottom panels show the $l=3$ visibilities on an expanded ordinate scale. The red dashed horizontal lines give the medians for the respective visibilities, with the values indicated in red on the ordinate; the shaded red regions give the standardized median-absolute-deviation (MAD) around the median values, given as 1.4826 times the MAD. The shaded grey regions in the left panel indicate the expected theoretical values from \citet[][]{2011A&A...531A.124B} for $\logg=4.0$ and \feh in the range $\pm 1$. The continuous black lines give the median binned values where the span of the parameters in the different panels have been divided into 10 bins. }
\label{fig:vis1}
\end{figure*}

Mode visibilities are estimated as part of the peak-bagging (see \eqref{eq:visi}); these are given in \tref{tab:vis}.
\fref{fig:vis1} shows the obtained visibilities as a function of \teff, \numax, and mass. The total visibilities are computed from the combination of the MCMC chains for the individual visibilities. We see no strong correlation with \teff, \feh (not shown), or \logg (not shown), and some scatter is observed for the visibilities about their median values. From median binned values there are slight indications of structure in the visibilities against the plotted parameters; the visibilities do, for instance, seem to peak around a temperature of ${\sim}6100\, \rm K$ and have a depression around a mass of ${\sim}1.4\, \rm M_{\odot}$. None of the dependencies are, however, very strong, but they do appear similar in shape for different angular degrees.
%%%%%%%%%%%%%%%%%%%%%%%%%%%%%%%%%%%%%%%%%%%%%%%%%%%%%%%%%%%%%%%
%%%%%%%%%%%%%%%%%%%%%%%%%%%%%%%%%%%%%%%%%%%%%%%%%%%%%%%%%%%%%%

\begin{figure*}
\includegraphics[width=\textwidth]{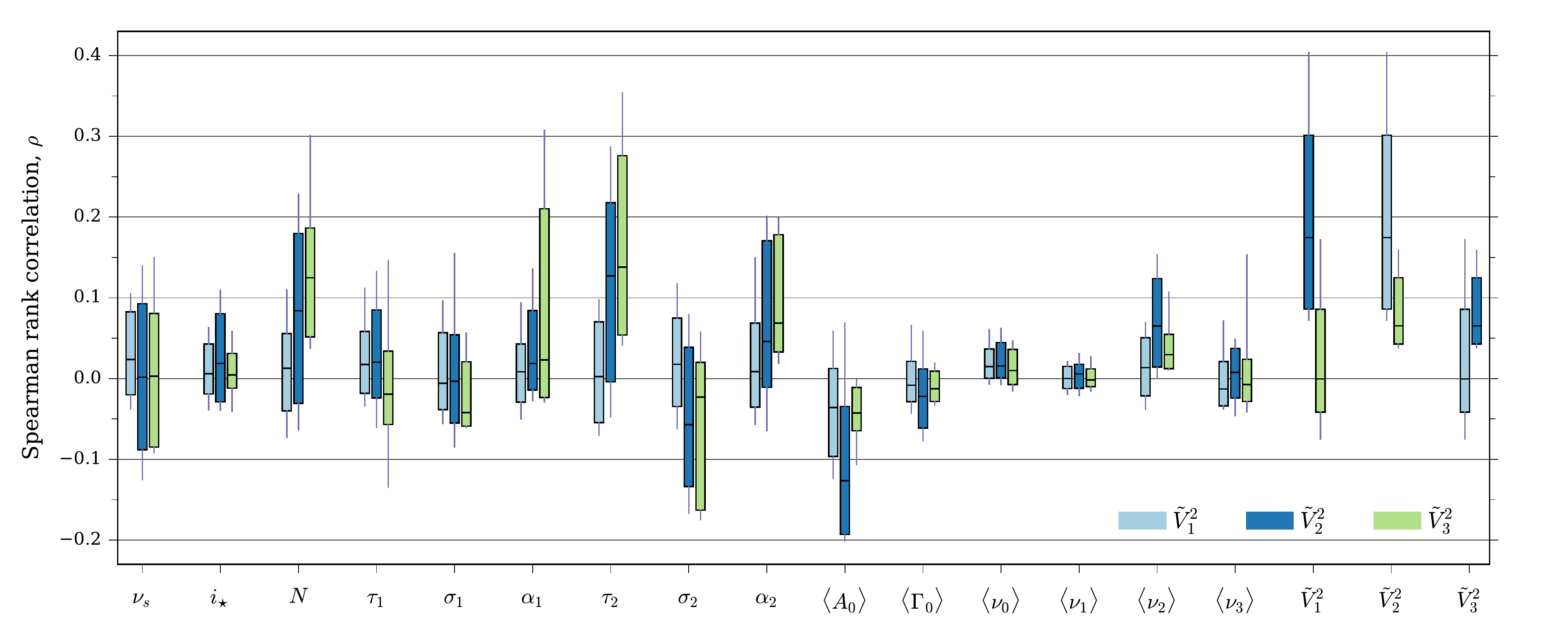}
\caption{Box-plot of Spearman's rank correlation $\rho$ between the parameters optimized in the peak-bagging and the mode visibilities. The box whiskers give the 15th and 85th percentiles. }
\label{fig:visi_correlations}
\end{figure*}

In comparing to the theoretically estimated values by \citet[][]{2011A&A...531A.124B}, which are calculated considering the \kp bandpass, we see that in median $l=1$ visibilities are lower than expected, those for $l=2$ and $l=3$ are larger than expected, and the total visibilities (up to $l=3$) are only slightly lower than expected. The difference is most pronounced for the $l=3$ modes. It is important to note that because $l=3$ visibilities have only been measured for a subset of stars, it might be that the remainder of the stars (with a S/N too low for a visual detection of $l=3$ modes) have visibilities in agreement with theory.
The comparisons do, however, qualitatively match those obtained by \citet[][]{2012A&A...537A..30M} for giants observed by \kp; the $l=3$ modes disagree most with theory in their results too. 
Similar discrepancies with theory have also been observed for individual CoRoT and \kp targets analyzed by \citet[][]{2010A&A...515A..87D}, \citet[][]{2013A&A...549A..12M}, and \citet[][]{2014ApJ...782....2L}, and the Sun by \citet[][]{2011A&A...528A..25S}. Given these discrepancies we discourage the adoption of fixed mode visibilities in peak-bagging exercises.
To assess which other parameters in the peak-bagging contribute the most to the uncertainties on the visibilities, and thus which parameters will be most affected by adopting fixed visibilities, we show in \fref{fig:visi_correlations} with a box-plot the correlations between the visibilities and the remainder of the fitted parameters. Correlations have been estimated using Spearman's rank correlation \citep[][]{spearman04} from the MCMC chains, because this better catches dependencies which are monotonic but not necessarily linear \citep[as done in Pearson's correlation;][]{pearson1895note}.
%%%%%%%%%%%%%%%%%%%%%%%%%%%%%%%%%%%%%%%%%%%%%%%%%%%%%%%%%%%%%%%
%%%%%%%%%%%%%%%%%%%%%%%%%%%%%%%%%%%%%%%%%%%%%%%%%%%%%%%%%%%%%%
As seen, the correlations are overall quite small with median values within $\rho = \pm 0.2$ in all cases. The visibilities are seen to primarily correlate with the parameters of the noise background, as might be expected with an increasing correlation with increasing angular degree $l$; a small change in, say, $N$ has a relatively larger impact on the visibility of $l=3$ modes compared to that of $l=1$ modes and one should therefore expect a larger correlation. This suggests that fixing the visibilities might bias the fit of the background and vice versa for fixed backgrounds. One should however also be cautious with the extracted visibilities, because an inappropriate model for the background might bias the measured values.

Considering the total visibility we note that \citet[][]{2011A&A...531A.124B} include modes up to $l=4$, but this should not significantly affect any observed discrepancy.
The median value obtained for $\tilde{V}_{\rm tot}^{\, 2}$ is $3.07$, which is close to the value of $c=3.04$ \citep[][]{2010ApJ...713..935B} often adopted in estimating radial mode amplitudes via the method of \citet[][]{2005ApJ...635.1281K,2008ApJ...682.1370K}. In \sref{sec:smo_amp} we did indeed also find that the amplitudes from the smoothing method were equal in median from using $c=3.04$ and $c=\tilde{V}_{\rm tot}^{\, 2}$, with only a slightly reduced scatter from using the measured total visibility. The value of $c=3.04$ (or alternatively $c=3.07$) should therefore serve as a reasonably good choice for analysis of amplitudes for a large sample of stars, but the small systematic offset found in \sref{sec:smo_amp} should be remembered when comparing with theory. We note also that \citet[][]{2012A&A...537A..30M} find a mean value of $c=3.06$ from \kp giants with \teff between approximately 4000 and 5100 K, but with a larger scatter than for our values. This does, however, indicate that the visibilities do not increase with decreasing temperature as suggested by theory. If a trend exists with \teff and/or \feh, which should be the main parameters determining the visibilities, they cannot be clearly discerned from the observations.

Some of the discrepancies, and scatter in observed values, can likely be explained by some of the simplifications adopted in the calculations by \citet[][]{2011A&A...531A.124B} and in general. 
These include, for instance, the neglect of non-adiabatic effects and a height dependence on mode amplitudes in the stellar atmosphere \citep[see, \eg,][]{2012ApJ...760L...1B,2014IAUS..301..481S,2015A&A...580L..11S}. Furthermore, phenomena such as spots and other local surface features will influence the measured visibility.  
An effect will also come from the way in which the stellar limb-darkening (LD) is described; the calculation of LD coefficients (LDCs) for parametrized LD laws will, \eg, depend on the description of convection in the adopted atmosphere models, the method used for integrating the specific flux, and the resolution used when fitting a parametrized LD law to these flux values. \citet[][]{2011A&A...531A.124B} compared their visibilities with those obtained using LDC from \citet[][]{2010A&A...510A..21S} (specific to the \kp bandpass) and found that their values were generally higher. Similarly, \citet[][]{2011A&A...529A..75C} compared their \kp LDCs to those of \citet[][]{2010A&A...510A..21S} and found differences. This indicates that some systematic uncertainty should be added to the theoretically derived visibilities. In any case \citet[][]{2010A&A...510A..21S} showed from fits to planetary transit curves that model computations of LDC generally disagree with those derived empirically. The treatment of the LD is, however, likely a secondary effect --- as shown in \citet[][]{2014ApJ...782....2L} the specific LD law adopted (neglecting the linear one) only has minor effects on the visibilities, and the shape has to be changed by a large amount away from the limb in order to take effect. This, and the fact that discrepancies between measured and modeled visibilities are found for the Sun \citep[][]{2011A&A...528A..25S}, where the LD is well known as a function of wavelength \citep[][]{1994SoPh..153...91N}, indicates that the simplified assumptions concerning the mode physics likely are the main contributor to the discrepancies. We note, however, that modes of $l=3$ would be relatively more affected by details of the LD, because of the considerably stronger cancellation for $l=3$ modes (total in the absence of LD), compared to modes of $l=1, 2$ \citep[][]{2014ApJ...782....2L}.
It could also be questioned if the assumption of equipartition of energy between modes of different angular degree holds true.

%%%%%%%%%%%%%%%%%%%%%%%%%%%%%%%%%%%%%%%%%%%%%%%%%%%%%%%%%%%%%%%%%%%%%%%%%%%%%%%%%%%%%%%%%%%%%%%%%%%%%%%%%%%%%%%%%%%%%%%%%%
\subsubsection{Detection of $l\geqslant3$ modes}

For solar-like oscillators observed by \kp the highest angular degree of modes is typically $l=2$. Only for the highest S/N cases can higher degree modes be identified --- in the current sample octupole $l=3$ modes were identified and included in the peak-bagging in 14 such targets. It is, however, possible to obtain information on the combined signal for $l=3$ modes in a given star. Such a signal could, for instance, be used in estimating the small separation $\delta\nu_{1,3}$ which can contribute a constraint in modeling efforts \citep[see, \eg,][]{2016arXiv160702137B}.

\begin{figure*}
\includegraphics[width=1\textwidth]{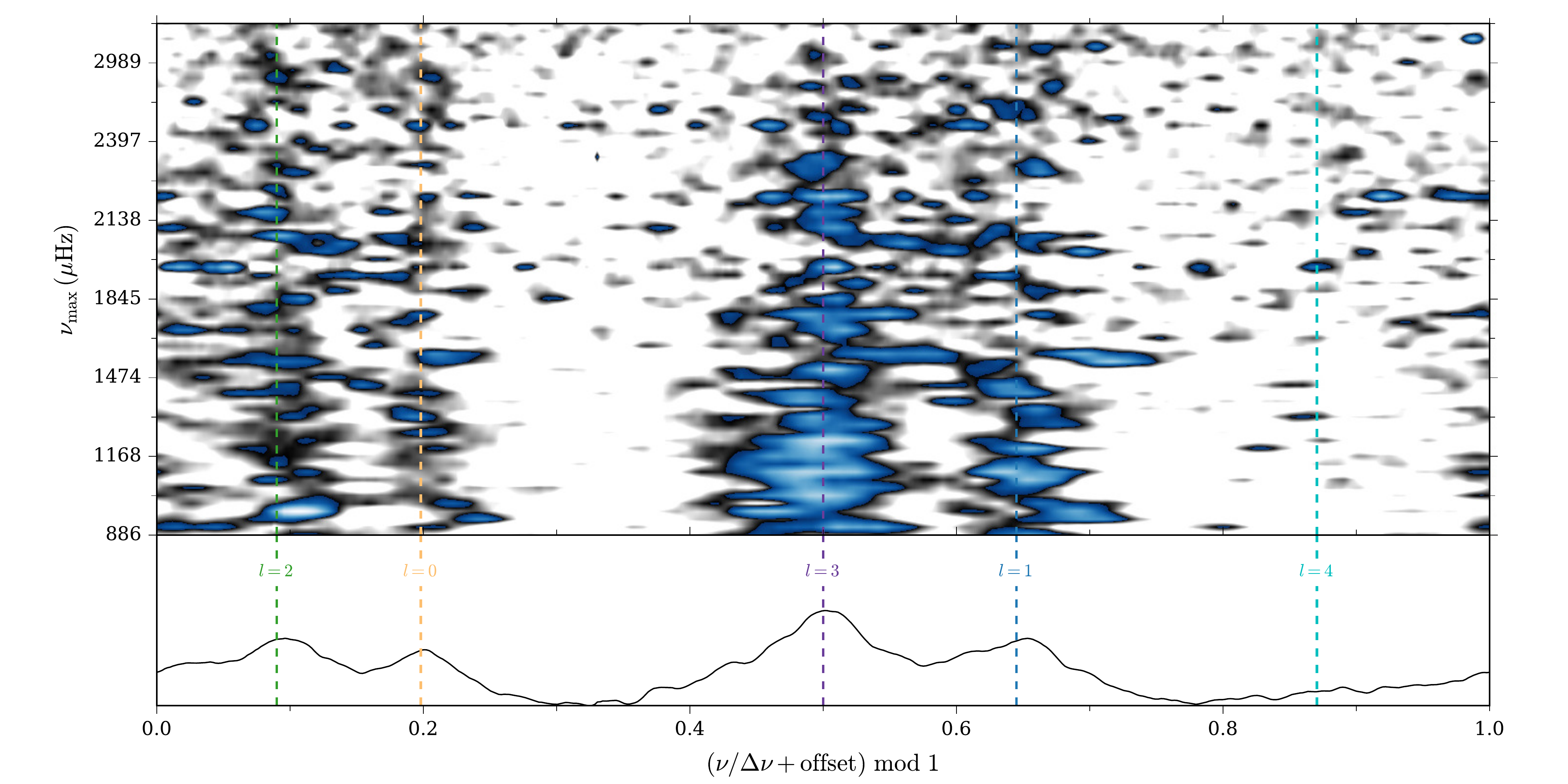}
\caption{Collapsed \'{e}chelle diagrams optimised for $l=3$ using the method of \citet[][]{2014ApJ...782....2L}. The residual power spectra were used where the peak-bagging model has been divided out. Each horizontal line in the top part of the plot corresponds to a given star, ordered in terms of \numax. The scale goes white (low power) to black (high power) below the 90th percentile, and from blue (low power) to white (high power) above it. The lower panel shows the average collapsed \'{e}chelle diagram from all stars. Vertical dashed lines have been added to guide the eye to the contribution from different angular degrees. For an optimization of the $l=3$ signal, modes from $l\neq 3$ will not form a straight ridge in the \'{e}chelle diagram and will consequently give a wide signal in the collapsed spectrum. An offset has been added to the frequencies for each of the collapsed spectra to align all $l=3$ signals at $0.5$. }
\label{fig:highl3}
\end{figure*}

To optimize the combined signal from modes of a given angular degree we applied the method outlined in \citet[][]{2014ApJ...782....2L}. Briefly, the \'{e}chelle diagram is collapsed along the vertical direction after having first stretched the frequency scale such that modes of a given $l$ form a straight ridge --- this ensures that all mode power from the particular $l$-values is co-added with as little spread as possible. Individual frequencies from best fitting ASTFIT models were used to stretch the frequency scale (see Paper~II). Before the power is co-added the power spectrum is smoothed to account also for the mode line widths and a potential rotational splitting. In \fref{fig:highl3} we show in a contour plot the collapsed \'{e}chelle diagrams for all stars, optimized to increase the signal of $l=3$ modes. To further increase the visibility of these modes we first divided out the model of the power spectrum from the peak-bagging, including also fitted $l=3$ modes. We see that for most stars a clearly detectable signal from $l=3$ is present. It is beyond the scope of this paper to estimate $\delta\nu_{1,3}$ for the stars and test the statistical significance of the signals, but \fref{fig:highl3} should indicate that obtaining this information would be possible. Some of the signal seen around $l<3$ modes in the residual spectra is likely due contamination from higher degree modes ($l\geq5$) or deviations in the mode shape from the assumed simple Lorentzian --- asymmetry of the modes would, for instance, leave some unaccounted for power in the residual spectrum.
%%%%%%%%%%%%%%%%%%%%%%%%%%%%%%%%%%%%%%%%%%%%%%%%%%%%%%%%%%%%%%%
%%%%%%%%%%%%%%%%%%%%%%%%%%%%%%%%%%%%%%%%%%%%%%%%%%%%%%%%%%%%%%

For the highest S/N targets one may further look for indications of hexadecapole $l=4$ modes.
%%%%%%%%%%%%%%%%%%%%%%%%%%%%%%%%%%%%%%%%%%%%%%%%%%%%%%%%%%%%%%
\begin{figure*}
	\centering
    \begin{subfigure}
        \centering
        \includegraphics[width=0.32\textwidth]{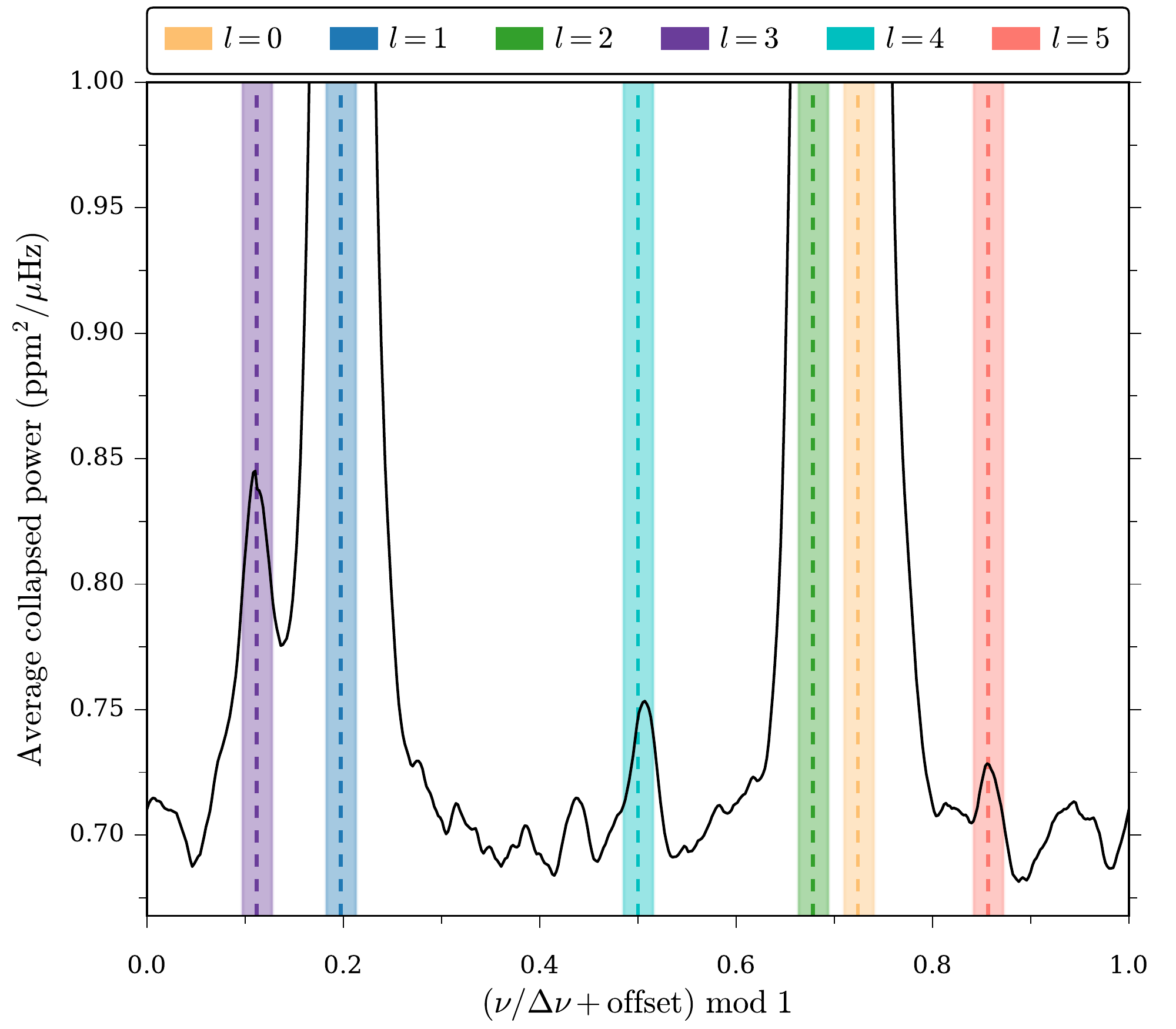}
    \end{subfigure}\hfill
    \begin{subfigure}
        \centering
        \includegraphics[width=0.32\textwidth]{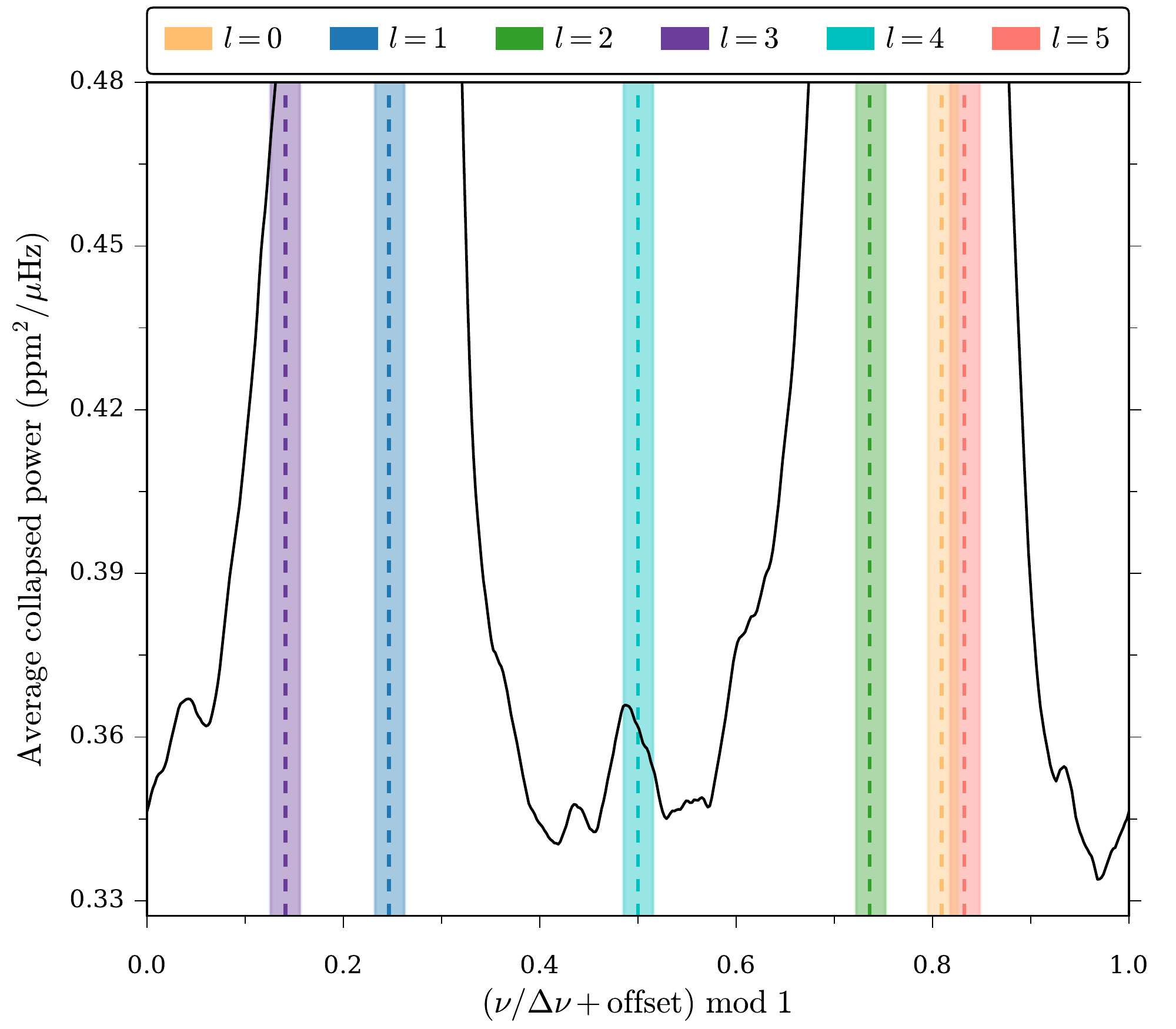}
    \end{subfigure}\hfill
    \begin{subfigure}
        \centering
        \includegraphics[width=0.32\textwidth]{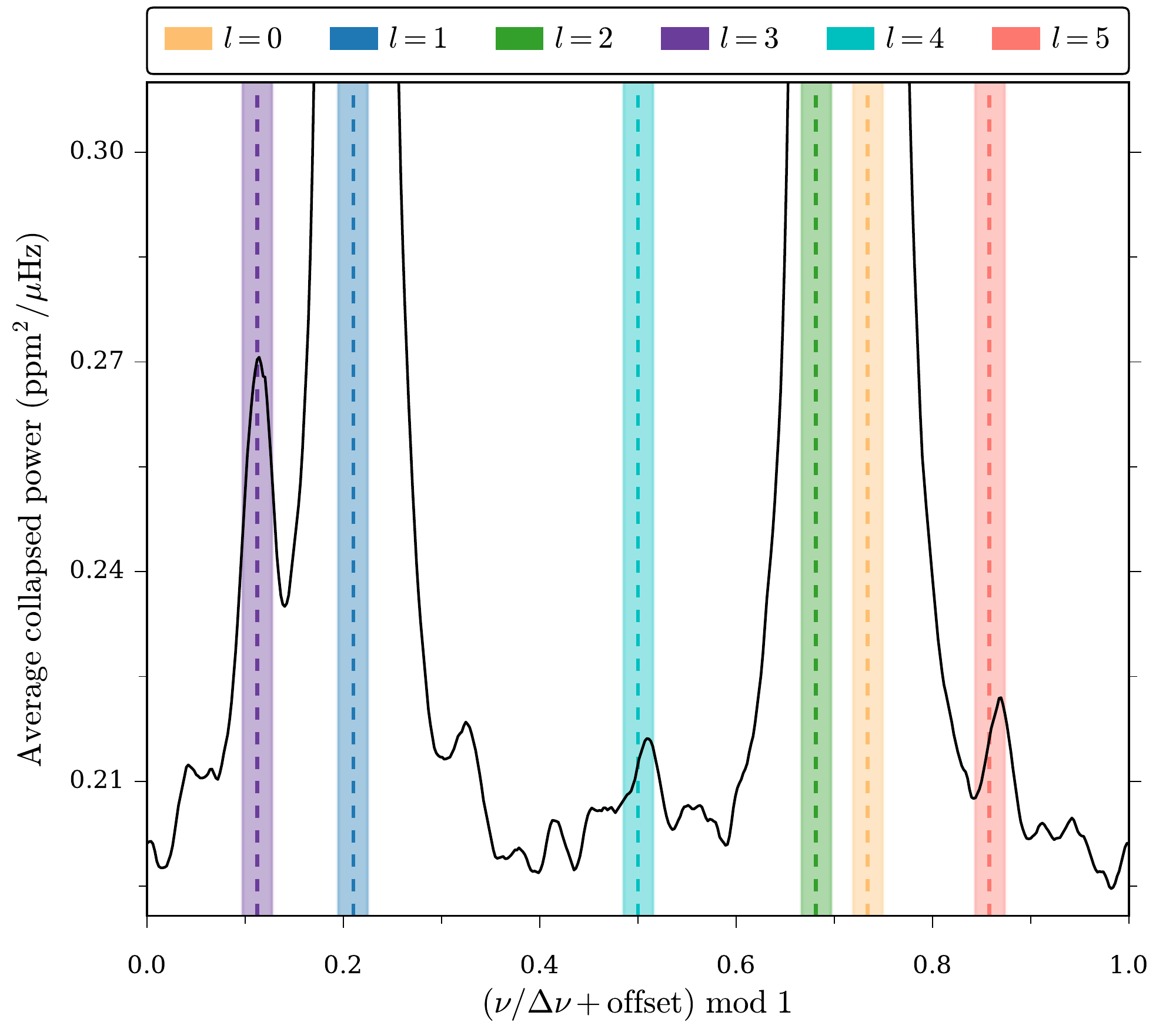}
    \end{subfigure}
    \caption{Examples of collapsed \'{e}chelle diagrams optimised for the detection of $l=4$. Shown are the spectra for KIC 6603624 (left; Saxo), KIC 7510397 (middle), and KIC 12069449 (right; 16 Cyg B), these represent some of the cases with the strongest apparent $l=4$ signals. An offset has been added to place the position of the expected $l=4$ signal at $0.5$.}
\label{fig:highl4}
\end{figure*}
%%%%%%%%%%%%%%%%%%%%%%%%%%%%%%%%%%%%%%%%%%%%%%%%%%%%%%%%%%%%%%
In \fref{fig:highl4} we show the collapsed \'{e}chelle diagrams optimized for the detection of $l=4$ for three targets with the strongest apparent $l=4$ signals, namely, KIC 6603624 (Saxo), KIC 7510397 (middle), and KIC 12069449 (16 Cyg B).
As seen an excess in collapsed power in each of these cases falls close to the expected position at \numax from the ASTFIT model. The signal seen in 16 Cyg A (not shown) and B corresponds well to the signal found in \citet[][]{2014ApJ...782....2L} from a shorter \kp data set. Curiously, the peak around ${\sim}0.85$ seen in Saxo and 16 Cyg B appears to coincide with the expected position of dotriacontapole $l=5$ modes --- whether this signal truly is from $l=5$ modes requires further investigation. In any case, it is clear that for a star like KIC 7510397 the $l=0,2$ modes will be polluted by $l=5$ modes; similarly will $l=6,9$ modes pollute the $l=1$ signal, etc. \citep[][]{1998ESASP.418...99A,2014ApJ...782....2L}.

%%%%%%%%%%%%%%%%%%%%%%%%%%%%%%%%%%%%%%%%%%%%%%%%%%%%%%%%%%%%%%%%%%%%%%%%%%%%%%%%%%%%%%%%%%%%%%%%%%%%%%%%%%%%%%%%%%%%%%%%%%
%%%%%%%%%%%%%%%%%%%%%%%%%%%%%%%%%%%%%%%%%%%%%%%%%%%%%%%%%%%%%%%%%%%%%%%%%%%%%%%%%%%%%%%%%%%%%%%%%%%%%%%%%%%%%%%%%%%%%%%%%%
%%%%%%%%%%%%%%%%%%%%%%%%%%%%%%%%%%%%%%%%%%%%%%%%%%%%%%%%%%%%%%%%%%%%%%%%%%%%%%%%%%%%%%%%%%%%%%%%%%%%%%%%%%%%%%%%%%%%%%%%%%
%%%%%%%%%%%%%%%%%%%%%%%%%%%%%%%%%%%%%%%%%%%%%%%%%%%%%%%%%%%%%%%%%%%%%%%%%%%%%%%%%%%%%%%%%%%%%%%%%%%%%%%%%%%%%%%%%%%%%%%%%%
\section{Example output}
\label{sec:exout}

For each of the 66 stars in the sample we have provided a set of outputs from the peak-bagging.
Tables and plots for all stars will be available in the online appendices. The extracted parameters may in addition be obtained from the results section of the KASOC data base\footnote{\url{http://kasoc.phys.au.dk/results/}.}.

%%%%%%%%%%%%%%%%%%%%%%%%%%%%%%%%%%%%%%%%%%%%%%%%%%%%%%%%%%%%%%%%%%%%%%%%%%%%%%%%%%%%%%%%%%%%%%%%%%%%%%%%%%%%%%%%%%%%%%%%%%
%%%%%%%%%%%%%%%%%%%%%%%%%%%%%%%%%%%%%%%%%%%%%%%%%%%%%%%%%%%%%%%%%%%%%%%%%%%%%%%%%%%%%%%%%%%%%%%%%%%%%%%%%%%%%%%%%%%%%%%%%%
%%%%%%%%%%%%%%%%%%%%%%%%%%%%%%%%%%%%%%%%%%%%%%%%%%%%%%%%%%%%%%%%%%%%%%%%%%%%%%%%%%%%%%%%%%%%%%%%%%%%%%%%%%%%%%%%%%%%%%%%%%
%\input{Table_set_Table6}
% Table made Fri Sep 30 15:53:35 2016
\begin{table} 
\centering 
\resizebox*{!}{\textheight}{% 
\begin{threeparttable} 
\caption{Extracted mode parameters and quality control (\eqref{eq:qual}) for KIC 6225718 (Saxo2). The complete table set (66 tables) is available in the online journal. } 
\label{tab:mode_6225718} 
\begin{tabular}{cc@{\hskip 6ex}c@{\hskip 7ex}c@{\hskip 7ex}c@{\hskip 7ex}c} 
\toprule 
$n$ & $l$ & \multicolumn{1}{@{\hskip 0ex}c}{$\rm Frequency$} & \multicolumn{1}{@{\hskip 0ex}c}{$\rm Amplitude$} & \multicolumn{1}{@{\hskip 0ex}c}{$\rm Line\,\, width$} & $\ln K$  \\   &  & \multicolumn{1}{@{\hskip 0ex}c}{$(\rm \mu Hz)$} & \multicolumn{1}{@{\hskip 0ex}c}{$(\rm ppm)$} &   \multicolumn{1}{@{\hskip 0ex}c}{$(\rm \mu Hz)$} &  \\ 
\midrule 
$11$ & $1 $ & $1351.15$\rlap{$_{-0.70}^{+0.59}$} &   &   & 2.2 \\ [0.5ex] 
$12$ & $0 $ & $1407.23$\rlap{$_{-1.18}^{+0.95}$} & $0.79$\rlap{$_{-0.13}^{+0.09}$} & $2.50$\rlap{$_{-1.34}^{+2.31}$} & >6 \\ [0.5ex] 
$12$ & $1 $ & $1454.25$\rlap{$_{-0.70}^{+0.53}$} &   &   & >6 \\ [0.5ex] 
$13$ & $0 $ & $1510.10$\rlap{$_{-0.48}^{+0.70}$} & $0.99$\rlap{$_{-0.15}^{+0.11}$} & $2.50$\rlap{$_{-0.95}^{+2.66}$} & >6 \\ [0.5ex] 
$13$ & $1 $ & $1558.45$\rlap{$_{-0.42}^{+0.54}$} &   &   & >6 \\ [0.5ex] 
$13$ & $2 $ & $1605.68$\rlap{$_{-0.81}^{+0.74}$} &   &   & 1.45 \\ [0.5ex] 
$14$ & $0 $ & $1615.12$\rlap{$_{-0.29}^{+0.24}$} & $1.16$\rlap{$_{-0.07}^{+0.07}$} & $2.60$\rlap{$_{-0.51}^{+0.67}$} & >6 \\ [0.5ex] 
$14$ & $1 $ & $1664.09$\rlap{$_{-0.23}^{+0.21}$} &   &   & >6 \\ [0.5ex] 
$14$ & $2 $ & $1711.40$\rlap{$_{-0.60}^{+0.50}$} &   &   & 3.14 \\ [0.5ex] 
$15$ & $0 $ & $1720.35$\rlap{$_{-0.17}^{+0.18}$} & $1.46$\rlap{$_{-0.07}^{+0.06}$} & $2.30$\rlap{$_{-0.37}^{+0.35}$} & >6 \\ [0.5ex] 
$15$ & $1 $ & $1769.65$\rlap{$_{-0.14}^{+0.15}$} &   &   & >6 \\ [0.5ex] 
$15$ & $2 $ & $1816.19$\rlap{$_{-0.36}^{+0.34}$} &   &   & >6 \\ [0.5ex] 
$16$ & $0 $ & $1825.41$\rlap{$_{-0.13}^{+0.12}$} & $1.89$\rlap{$_{-0.07}^{+0.06}$} & $2.19$\rlap{$_{-0.22}^{+0.34}$} & >6 \\ [0.5ex] 
$16$ & $1 $ & $1873.88$\rlap{$_{-0.14}^{+0.13}$} &   &   & >6 \\ [0.5ex] 
$16$ & $2 $ & $1919.97$\rlap{$_{-0.26}^{+0.26}$} &   &   & >6 \\ [0.5ex] 
$17$ & $0 $ & $1929.05$\rlap{$_{-0.14}^{+0.12}$} & $2.29$\rlap{$_{-0.06}^{+0.06}$} & $2.84$\rlap{$_{-0.20}^{+0.30}$} & >6 \\ [0.5ex] 
$17$ & $1 $ & $1977.35$\rlap{$_{-0.12}^{+0.11}$} &   &   & >6 \\ [0.5ex] 
$17$ & $2 $ & $2023.80$\rlap{$_{-0.21}^{+0.22}$} &   &   & >6 \\ [0.5ex] 
$18$ & $0 $ & $2032.68$\rlap{$_{-0.11}^{+0.11}$} & $2.77$\rlap{$_{-0.05}^{+0.07}$} & $2.67$\rlap{$_{-0.18}^{+0.23}$} & >6 \\ [0.5ex] 
$18$ & $1 $ & $2081.57$\rlap{$_{-0.09}^{+0.09}$} &   &   & >6 \\ [0.5ex] 
$18$ & $2 $ & $2128.62$\rlap{$_{-0.16}^{+0.15}$} &   &   & >6 \\ [0.5ex] 
$19$ & $0 $ & $2137.59$\rlap{$_{-0.09}^{+0.10}$} & $3.14$\rlap{$_{-0.06}^{+0.06}$} & $2.50$\rlap{$_{-0.15}^{+0.21}$} & >6 \\ [0.5ex] 
$19$ & $1 $ & $2186.89$\rlap{$_{-0.09}^{+0.08}$} &   &   & >6 \\ [0.5ex] 
$19$ & $2 $ & $2234.70$\rlap{$_{-0.16}^{+0.16}$} &   &   & >6 \\ [0.5ex] 
$20$ & $0 $ & $2243.42$\rlap{$_{-0.08}^{+0.08}$} & $3.47$\rlap{$_{-0.06}^{+0.07}$} & $2.22$\rlap{$_{-0.12}^{+0.16}$} & >6 \\ [0.5ex] 
$19$ & $3 $ & $2281.61$\rlap{$_{-3.36}^{+1.97}$} &   &   & 3.01 \\ [0.5ex] 
$20$ & $1 $ & $2293.05$\rlap{$_{-0.09}^{+0.09}$} &   &   & >6 \\ [0.5ex] 
$20$ & $2 $ & $2340.63$\rlap{$_{-0.16}^{+0.17}$} &   &   & >6 \\ [0.5ex] 
$21$ & $0 $ & $2349.64$\rlap{$_{-0.09}^{+0.08}$} & $3.46$\rlap{$_{-0.06}^{+0.07}$} & $2.61$\rlap{$_{-0.17}^{+0.16}$} & >6 \\ [0.5ex] 
$20$ & $3 $ & $2385.57$\rlap{$_{-1.16}^{+0.93}$} &   &   & 3.94 \\ [0.5ex] 
$21$ & $1 $ & $2399.39$\rlap{$_{-0.10}^{+0.08}$} &   &   & >6 \\ [0.5ex] 
$21$ & $2 $ & $2446.71$\rlap{$_{-0.15}^{+0.16}$} &   &   & >6 \\ [0.5ex] 
$22$ & $0 $ & $2455.69$\rlap{$_{-0.10}^{+0.11}$} & $3.46$\rlap{$_{-0.06}^{+0.07}$} & $3.03$\rlap{$_{-0.23}^{+0.19}$} & >6 \\ [0.5ex] 
$21$ & $3 $ & $2493.08$\rlap{$_{-1.64}^{+1.24}$} &   &   & 3.66 \\ [0.5ex] 
$22$ & $1 $ & $2505.34$\rlap{$_{-0.11}^{+0.10}$} &   &   & >6 \\ [0.5ex] 
$22$ & $2 $ & $2552.85$\rlap{$_{-0.24}^{+0.22}$} &   &   & >6 \\ [0.5ex] 
$23$ & $0 $ & $2561.29$\rlap{$_{-0.14}^{+0.15}$} & $3.18$\rlap{$_{-0.06}^{+0.06}$} & $4.01$\rlap{$_{-0.18}^{+0.25}$} & >6 \\ [0.5ex] 
$22$ & $3 $ & $2598.55$\rlap{$_{-1.66}^{+1.55}$} &   &   & 3.12 \\ [0.5ex] 
$23$ & $1 $ & $2611.20$\rlap{$_{-0.14}^{+0.13}$} &   &   & >6 \\ [0.5ex] 
$23$ & $2 $ & $2658.63$\rlap{$_{-0.32}^{+0.33}$} &   &   & >6 \\ [0.5ex] 
$24$ & $0 $ & $2666.49$\rlap{$_{-0.22}^{+0.24}$} & $2.72$\rlap{$_{-0.04}^{+0.06}$} & $5.23$\rlap{$_{-0.33}^{+0.26}$} & >6 \\ [0.5ex] 
$24$ & $1 $ & $2717.47$\rlap{$_{-0.17}^{+0.18}$} &   &   & >6 \\ [0.5ex] 
$24$ & $2 $ & $2765.05$\rlap{$_{-0.40}^{+0.40}$} &   &   & >6 \\ [0.5ex] 
$25$ & $0 $ & $2773.06$\rlap{$_{-0.31}^{+0.30}$} & $2.41$\rlap{$_{-0.06}^{+0.05}$} & $6.75$\rlap{$_{-0.51}^{+0.38}$} & >6 \\ [0.5ex] 
$25$ & $1 $ & $2824.15$\rlap{$_{-0.26}^{+0.27}$} &   &   & >6 \\ [0.5ex] 
$25$ & $2 $ & $2872.28$\rlap{$_{-0.54}^{+0.55}$} &   &   & >6 \\ [0.5ex] 
$26$ & $0 $ & $2879.34$\rlap{$_{-0.50}^{+0.56}$} & $1.96$\rlap{$_{-0.06}^{+0.06}$} & $7.60$\rlap{$_{-0.81}^{+0.63}$} & >6 \\ [0.5ex] 
$26$ & $1 $ & $2931.24$\rlap{$_{-0.34}^{+0.35}$} &   &   & >6 \\ [0.5ex] 
$26$ & $2 $ & $2978.49$\rlap{$_{-0.84}^{+0.80}$} &   &   & 3.78 \\ [0.5ex] 
$27$ & $0 $ & $2987.15$\rlap{$_{-0.49}^{+0.49}$} & $1.70$\rlap{$_{-0.06}^{+0.04}$} & $7.53$\rlap{$_{-0.74}^{+0.87}$} & >6 \\ [0.5ex] 
$27$ & $1 $ & $3038.67$\rlap{$_{-0.53}^{+0.51}$} &   &   & >6 \\ [0.5ex] 
$27$ & $2 $ & $3084.55$\rlap{$_{-1.59}^{+1.37}$} &   &   & 1.45 \\ [0.5ex] 
$28$ & $0 $ & $3092.80$\rlap{$_{-0.92}^{+0.88}$} & $1.46$\rlap{$_{-0.06}^{+0.07}$} & $8.85$\rlap{$_{-1.11}^{+1.44}$} & >6 \\ [0.5ex] 
$28$ & $1 $ & $3145.65$\rlap{$_{-0.62}^{+0.61}$} &   &   & 4.92 \\ [0.5ex] 
$28$ & $2 $ & $3194.64$\rlap{$_{-1.27}^{+1.68}$} &   &   & 2.18 \\ [0.5ex] 
$29$ & $0 $ & $3204.41$\rlap{$_{-0.93}^{+0.88}$} & $0.98$\rlap{$_{-0.10}^{+0.08}$} & $5.87$\rlap{$_{-1.41}^{+2.75}$} & 3.81 \\ [0.5ex] 
$29$ & $1 $ & $3251.96$\rlap{$_{-1.47}^{+1.71}$} &   &   & 4.2 \\ [0.5ex] 
$29$ & $2 $ & $3302.59$\rlap{$_{-2.57}^{+2.16}$} &   &   & 1.22 \\ [0.5ex] 
$30$ & $0 $ & $3314.17$\rlap{$_{-2.02}^{+2.07}$} & $1.12$\rlap{$_{-0.08}^{+0.07}$} & $11.64$\rlap{$_{-1.65}^{+1.55}$} & 3.81 \\ [0.5ex] 
\bottomrule
\end{tabular} 
 \end{threeparttable}% 
 } 
\end{table} 

%%%%%%%%%%%%%%%%%%%%%%%%%%%%%%%%%%%%%%%%%%%%%%%%%%%%%%%%%%%%%%%%%%%%%%%%%%%%%%%%%%%%%%%%%%%%%%%%%%%%%%%%%%%%%%%%%%%%%%%%%%
%%%%%%%%%%%%%%%%%%%%%%%%%%%%%%%%%%%%%%%%%%%%%%%%%%%%%%%%%%%%%%%%%%%%%%%%%%%%%%%%%%%%%%%%%%%%%%%%%%%%%%%%%%%%%%%%%%%%%%%%%%
%%%%%%%%%%%%%%%%%%%%%%%%%%%%%%%%%%%%%%%%%%%%%%%%%%%%%%%%%%%%%%%%%%%%%%%%%%%%%%%%%%%%%%%%%%%%%%%%%%%%%%%%%%%%%%%%%%%%%%%%%%

The outputs include first of all a table with the mode information from the MCMC peak-bagging, like the one given in \tref{tab:mode_6225718} for KIC 6225718 (Saxo2). The table gives for a given mode the angular degree, radial order, frequency, amplitude, line width, and the natural logarithm of the Bayes factor $K$ from the quality control in \sref{sec:quality}. The uncertainties on the mode parameters are obtained from the $68\%$ HPD interval of the posterior probability distributions.
As amplitudes and line widths are only fitted to radial $l=0$ modes we only give these values for these modes. The visibilities for a given star can be found in \tref{tab:vis}.
We note that the radial orders given are obtained from matching $\epsilon$ to the expected value as a function of \teff (see \sref{sec:modeid}) --- we suggest that this be used with some caution in modeling efforts and checked independently.
A table is also included with the derived frequency difference ratios $r_{01,10,02}$ and second differences $\Delta_2 \nu(n,l)$ (see \sref{sec:deriv}), examples of these are given in Tables~\ref{tab:ratio_6225718} and \ref{tab:secd_6225718}.
%For each of these we also provide a variance-covariance matrix as the different ratios and differences for a given star are inherently correlated.
%%%%%%%%%%%%%%%%%%%%%%%%%%%%%%%%%%%%%%%%%%%%%%%%%%%%%%%%%%%%%%%%%%%%%%%%%%%%%%%%%%%%%%%%%%%%%%%%%%%%%%%%%%%%%%%%%%%%%%%%%%
%%%%%%%%%%%%%%%%%%%%%%%%%%%%%%%%%%%%%%%%%%%%%%%%%%%%%%%%%%%%%%%%%%%%%%%%%%%%%%%%%%%%%%%%%%%%%%%%%%%%%%%%%%%%%%%%%%%%%%%%%%
%%%%%%%%%%%%%%%%%%%%%%%%%%%%%%%%%%%%%%%%%%%%%%%%%%%%%%%%%%%%%%%%%%%%%%%%%%%%%%%%%%%%%%%%%%%%%%%%%%%%%%%%%%%%%%%%%%%%%%%%%%

%\input{Table_set_Table7}
% Table made Fri Sep 30 15:44:54 2016
\begin{table} 
\centering 
\begin{threeparttable} 
\caption{Example of calculated mode frequency difference ratios $r_{01,10,02}(n)$ (\eqref{eq:rat}) for KIC 6225718 (Saxo2). The complete table set (66 tables) is available in the online journal. } 
\label{tab:ratio_6225718} 
\begin{tabular}{c@{\hskip 8ex}c@{\hskip 10ex}c} 
\toprule 
$\rm Ratio\, type$ & $n$ & $\rm Ratio$ \\ 
\midrule 
$r_{02}$ & $14$ & $0.0888_{-0.0155}^{+0.0153}$ \\ [0.5ex] 
$r_{02}$ & $15$ & $0.0856_{-0.0115}^{+0.0092}$ \\ [0.5ex] 
$r_{02}$ & $16$ & $0.0883_{-0.0091}^{+0.0074}$ \\ [0.5ex] 
$r_{02}$ & $17$ & $0.0877_{-0.0066}^{+0.0081}$ \\ [0.5ex] 
$r_{02}$ & $18$ & $0.0853_{-0.0075}^{+0.0064}$ \\ [0.5ex] 
$r_{02}$ & $19$ & $0.0854_{-0.0049}^{+0.0041}$ \\ [0.5ex] 
$r_{02}$ & $20$ & $0.0821_{-0.0056}^{+0.0039}$ \\ [0.5ex] 
$r_{02}$ & $21$ & $0.0845_{-0.0039}^{+0.0062}$ \\ [0.5ex] 
$r_{02}$ & $22$ & $0.0849_{-0.0069}^{+0.0049}$ \\ [0.5ex] 
$r_{02}$ & $23$ & $0.0797_{-0.0062}^{+0.0063}$ \\ [0.5ex] 
$r_{02}$ & $24$ & $0.0742_{-0.0087}^{+0.0064}$ \\ [0.5ex] 
$r_{02}$ & $25$ & $0.0748_{-0.0101}^{+0.0081}$ \\ [0.5ex] 
$r_{02}$ & $26$ & $0.0656_{-0.0121}^{+0.0125}$ \\ [0.5ex] 
$r_{02}$ & $27$ & $0.0781_{-0.0135}^{+0.0161}$ \\ [0.5ex] 
$r_{02}$ & $28$ & $0.0754_{-0.0204}^{+0.0188}$ \\ [0.5ex] 
$r_{02}$ & $29$ & $0.0883_{-0.0257}^{+0.0260}$ \\ [0.5ex] 
$r_{10}$ & $12$ & $0.0413_{-0.0202}^{+0.0197}$ \\ [0.5ex] 
$r_{01}$ & $13$ & $0.0387_{-0.0208}^{+0.0117}$ \\ [0.5ex] 
$r_{10}$ & $13$ & $0.0374_{-0.0152}^{+0.0105}$ \\ [0.5ex] 
$r_{01}$ & $14$ & $0.0366_{-0.0083}^{+0.0070}$ \\ [0.5ex] 
$r_{10}$ & $14$ & $0.0349_{-0.0060}^{+0.0067}$ \\ [0.5ex] 
$r_{01}$ & $15$ & $0.0328_{-0.0042}^{+0.0040}$ \\ [0.5ex] 
$r_{10}$ & $15$ & $0.0323_{-0.0040}^{+0.0046}$ \\ [0.5ex] 
$r_{01}$ & $16$ & $0.0332_{-0.0043}^{+0.0034}$ \\ [0.5ex] 
$r_{10}$ & $16$ & $0.0331_{-0.0034}^{+0.0041}$ \\ [0.5ex] 
$r_{01}$ & $17$ & $0.0330_{-0.0053}^{+0.0044}$ \\ [0.5ex] 
$r_{10}$ & $17$ & $0.0329_{-0.0053}^{+0.0045}$ \\ [0.5ex] 
$r_{01}$ & $18$ & $0.0323_{-0.0043}^{+0.0049}$ \\ [0.5ex] 
$r_{10}$ & $18$ & $0.0327_{-0.0027}^{+0.0029}$ \\ [0.5ex] 
$r_{01}$ & $19$ & $0.0330_{-0.0031}^{+0.0034}$ \\ [0.5ex] 
$r_{10}$ & $19$ & $0.0331_{-0.0053}^{+0.0033}$ \\ [0.5ex] 
$r_{01}$ & $20$ & $0.0329_{-0.0041}^{+0.0022}$ \\ [0.5ex] 
$r_{10}$ & $20$ & $0.0325_{-0.0051}^{+0.0030}$ \\ [0.5ex] 
$r_{01}$ & $21$ & $0.0319_{-0.0042}^{+0.0022}$ \\ [0.5ex] 
$r_{10}$ & $21$ & $0.0313_{-0.0046}^{+0.0023}$ \\ [0.5ex] 
$r_{01}$ & $22$ & $0.0309_{-0.0025}^{+0.0029}$ \\ [0.5ex] 
$r_{10}$ & $22$ & $0.0300_{-0.0038}^{+0.0034}$ \\ [0.5ex] 
$r_{01}$ & $23$ & $0.0281_{-0.0033}^{+0.0037}$ \\ [0.5ex] 
$r_{10}$ & $23$ & $0.0252_{-0.0039}^{+0.0036}$ \\ [0.5ex] 
$r_{01}$ & $24$ & $0.0221_{-0.0052}^{+0.0050}$ \\ [0.5ex] 
$r_{10}$ & $24$ & $0.0211_{-0.0053}^{+0.0054}$ \\ [0.5ex] 
$r_{01}$ & $25$ & $0.0203_{-0.0077}^{+0.0058}$ \\ [0.5ex] 
$r_{10}$ & $25$ & $0.0182_{-0.0072}^{+0.0080}$ \\ [0.5ex] 
$r_{01}$ & $26$ & $0.0166_{-0.0076}^{+0.0085}$ \\ [0.5ex] 
$r_{10}$ & $26$ & $0.0177_{-0.0087}^{+0.0085}$ \\ [0.5ex] 
$r_{01}$ & $27$ & $0.0177_{-0.0118}^{+0.0088}$ \\ [0.5ex] 
$r_{10}$ & $27$ & $0.0136_{-0.0117}^{+0.0104}$ \\ [0.5ex] 
$r_{01}$ & $28$ & $0.0148_{-0.0137}^{+0.0104}$ \\ [0.5ex] 
$r_{10}$ & $28$ & $0.0285_{-0.0138}^{+0.0107}$ \\ [0.5ex] 
$r_{01}$ & $29$ & $0.0516_{-0.0198}^{+0.0185}$ \\ [0.5ex] 
\bottomrule
\end{tabular} 
 \end{threeparttable} 
\end{table} 

%\input{Table_set_Table8}
% Table made Fri Sep 30 15:44:55 2016
\begin{table} 
\centering 
\begin{threeparttable} 
\caption{Example of calculated second differences $\Delta_2 \nu(n,l)$ (\eqref{eq:twodiff}) for KIC 6225718 (Saxo2). The complete table set (66 tables) is available in the online journal. } 
\label{tab:secd_6225718} 
\begin{tabular}{c@{\hskip 10ex}c@{\hskip 10ex}r@{\hskip 5ex}l} 
\toprule 
$n$ & $l$ & \multicolumn{1}{@{\hskip 1ex}c}{$\Delta_2 \nu$} &\\   &  & \multicolumn{1}{@{\hskip 1ex}c}{$(\rm \mu Hz)$}&\\ 
\midrule 
$12$ & $1 $ & $1.49$\rlap{$_{-1.87}^{+1.43}$} &\\ [0.5ex] 
$13$ & $0 $ & $1.64$\rlap{$_{-1.65}^{+2.97}$} &\\ [0.5ex] 
$13$ & $1 $ & $1.20$\rlap{$_{-1.23}^{+1.49}$} &\\ [0.5ex] 
$14$ & $0 $ & $0.37$\rlap{$_{-1.09}^{+0.83}$} &\\ [0.5ex] 
$14$ & $1 $ & $0.13$\rlap{$_{-0.91}^{+0.54}$} &\\ [0.5ex] 
$14$ & $2 $ & $-1.08$\rlap{$_{-1.28}^{+1.55}$} &\\ [0.5ex] 
$15$ & $0 $ & $-0.23$\rlap{$_{-0.41}^{+0.52}$} &\\ [0.5ex] 
$15$ & $1 $ & $-1.27$\rlap{$_{-0.45}^{+0.34}$} &\\ [0.5ex] 
$15$ & $2 $ & $-0.96$\rlap{$_{-0.97}^{+0.92}$} &\\ [0.5ex] 
$16$ & $1 $ & $-0.76$\rlap{$_{-0.32}^{+0.35}$} &\\ [0.5ex] 
$16$ & $2 $ & $0.14$\rlap{$_{-0.78}^{+0.61}$} &\\ [0.5ex] 
$16$ & $0 $ & $-1.35$\rlap{$_{-0.40}^{+0.25}$} &\\ [0.5ex] 
$17$ & $0 $ & $0.02$\rlap{$_{-0.32}^{+0.27}$} &\\ [0.5ex] 
$17$ & $2 $ & $0.97$\rlap{$_{-0.51}^{+0.57}$} &\\ [0.5ex] 
$17$ & $1 $ & $0.69$\rlap{$_{-0.20}^{+0.38}$} &\\ [0.5ex] 
$18$ & $2 $ & $1.25$\rlap{$_{-0.41}^{+0.40}$} &\\ [0.5ex] 
$18$ & $1 $ & $1.13$\rlap{$_{-0.28}^{+0.21}$} &\\ [0.5ex] 
$18$ & $0 $ & $1.30$\rlap{$_{-0.29}^{+0.25}$} &\\ [0.5ex] 
$19$ & $2 $ & $-0.15$\rlap{$_{-0.42}^{+0.39}$} &\\ [0.5ex] 
$19$ & $1 $ & $0.82$\rlap{$_{-0.19}^{+0.26}$} &\\ [0.5ex] 
$19$ & $0 $ & $0.89$\rlap{$_{-0.21}^{+0.25}$} &\\ [0.5ex] 
$20$ & $2 $ & $0.13$\rlap{$_{-0.37}^{+0.41}$} &\\ [0.5ex] 
$20$ & $3 $ & $-1.08$\rlap{$_{-1.28}^{+1.55}$} &\\ [0.5ex] 
$20$ & $1 $ & $0.15$\rlap{$_{-0.18}^{+0.24}$} &\\ [0.5ex] 
$20$ & $0 $ & $0.42$\rlap{$_{-0.23}^{+0.19}$} &\\ [0.5ex] 
$21$ & $3 $ & $-0.96$\rlap{$_{-0.97}^{+0.92}$} &\\ [0.5ex] 
$21$ & $1 $ & $-0.37$\rlap{$_{-0.24}^{+0.20}$} &\\ [0.5ex] 
$21$ & $2 $ & $0.16$\rlap{$_{-0.48}^{+0.33}$} &\\ [0.5ex] 
$21$ & $0 $ & $-0.16$\rlap{$_{-0.23}^{+0.21}$} &\\ [0.5ex] 
$22$ & $1 $ & $-0.09$\rlap{$_{-0.27}^{+0.27}$} &\\ [0.5ex] 
$22$ & $2 $ & $-0.42$\rlap{$_{-0.50}^{+0.63}$} &\\ [0.5ex] 
$22$ & $0 $ & $-0.45$\rlap{$_{-0.26}^{+0.27}$} &\\ [0.5ex] 
$23$ & $2 $ & $0.43$\rlap{$_{-0.53}^{+1.06}$} &\\ [0.5ex] 
$23$ & $0 $ & $-0.44$\rlap{$_{-0.35}^{+0.42}$} &\\ [0.5ex] 
$23$ & $1 $ & $0.38$\rlap{$_{-0.30}^{+0.37}$} &\\ [0.5ex] 
$24$ & $0 $ & $1.38$\rlap{$_{-0.61}^{+0.54}$} &\\ [0.5ex] 
$24$ & $2 $ & $0.72$\rlap{$_{-0.96}^{+1.08}$} &\\ [0.5ex] 
$24$ & $1 $ & $0.40$\rlap{$_{-0.41}^{+0.47}$} &\\ [0.5ex] 
$25$ & $0 $ & $-0.20$\rlap{$_{-0.96}^{+0.73}$} &\\ [0.5ex] 
$25$ & $2 $ & $-1.27$\rlap{$_{-1.17}^{+1.56}$} &\\ [0.5ex] 
$25$ & $1 $ & $0.47$\rlap{$_{-0.69}^{+0.59}$} &\\ [0.5ex] 
$26$ & $2 $ & $-0.15$\rlap{$_{-1.96}^{+2.13}$} &\\ [0.5ex] 
$26$ & $0 $ & $1.51$\rlap{$_{-1.26}^{+1.17}$} &\\ [0.5ex] 
$26$ & $1 $ & $0.36$\rlap{$_{-0.90}^{+0.85}$} &\\ [0.5ex] 
$27$ & $0 $ & $-1.98$\rlap{$_{-1.70}^{+1.11}$} &\\ [0.5ex] 
$27$ & $1 $ & $-0.35$\rlap{$_{-1.34}^{+1.18}$} &\\ [0.5ex] 
$27$ & $2 $ & $4.80$\rlap{$_{-4.22}^{+2.77}$} &\\ [0.5ex] 
$28$ & $2 $ & $-4.34$\rlap{$_{-3.40}^{+6.80}$} &\\ [0.5ex] 
$28$ & $1 $ & $-0.51$\rlap{$_{-2.18}^{+1.68}$} &\\ [0.5ex] 
$28$ & $0 $ & $6.76$\rlap{$_{-2.89}^{+1.55}$} &\\ [0.5ex] 
$29$ & $0 $ & $-1.46$\rlap{$_{-2.99}^{+2.12}$} &\\ [0.5ex] 
\bottomrule
\end{tabular} 
 \end{threeparttable} 
\end{table} 

%%%%%%%%%%%%%%%%%%%%%%%%%%%%%%%%%%%%%%%%%%%%%%%%%%%%%%%%%%%%%%%%%%%%%%%%%%%%%%%%%%%%%%%%%%%%%%%%%%%%%%%%%%%%%%%%%%%%%%%%%%
%%%%%%%%%%%%%%%%%%%%%%%%%%%%%%%%%%%%%%%%%%%%%%%%%%%%%%%%%%%%%%%%%%%%%%%%%%%%%%%%%%%%%%%%%%%%%%%%%%%%%%%%%%%%%%%%%%%%%%%%%%
%%%%%%%%%%%%%%%%%%%%%%%%%%%%%%%%%%%%%%%%%%%%%%%%%%%%%%%%%%%%%%%%%%%%%%%%%%%%%%%%%%%%%%%%%%%%%%%%%%%%%%%%%%%%%%%%%%%%%%%%%%

\begin{figure*}
\centering
\includegraphics[width=0.96\textwidth]{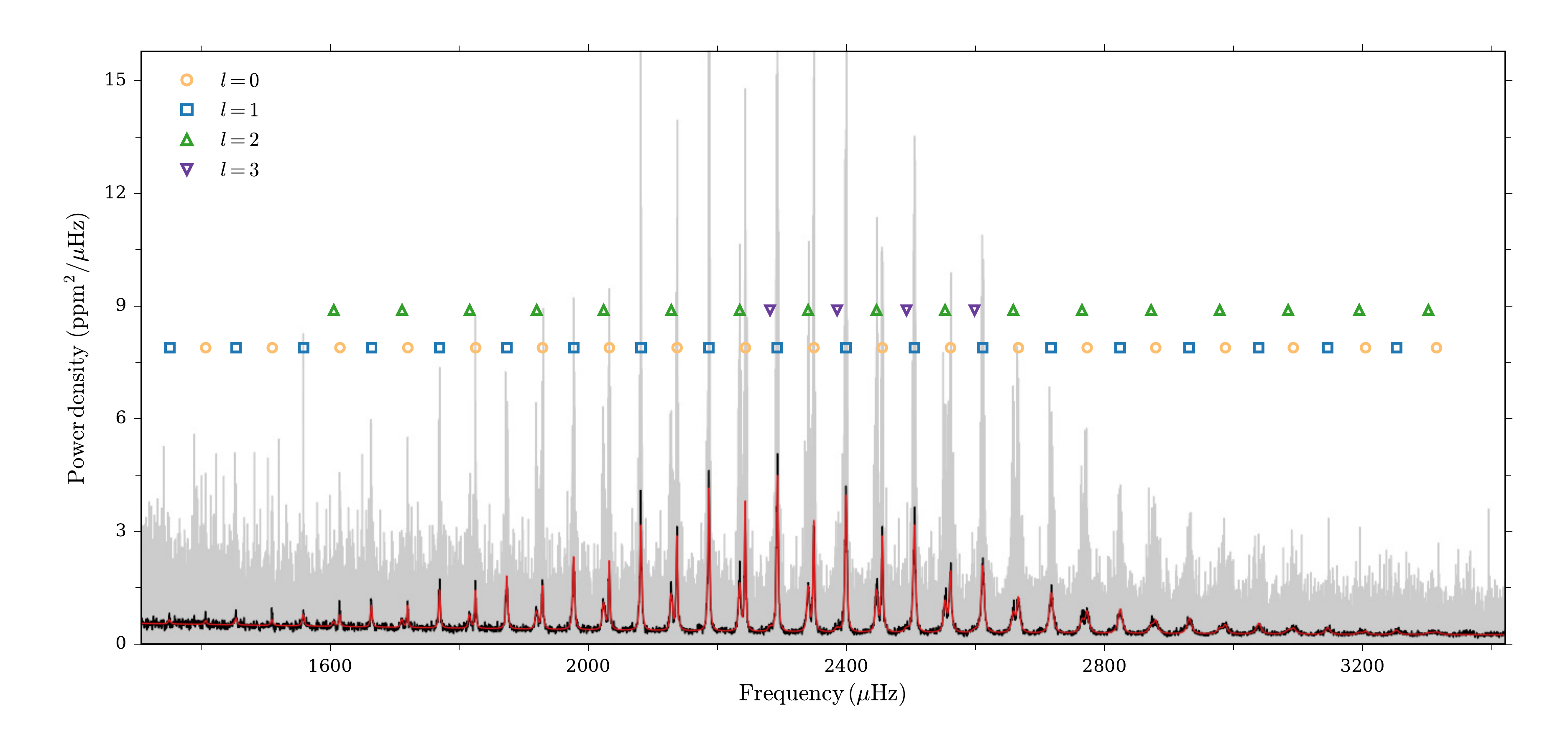}
\caption{Example of the peak-bagging fit for KIC 6225718 (Saxo2). The complete figure set (66 figures) is available in the online journal. The power density spectrum is shown in grey with a $\rm 1\, \mu Hz$ Epanechnikov smoothed version overlain in black, and with the fitted power spectrum model given by the red curve. The markers indicate the frequency and angular degree of the fitted modes.}
\label{fig:6225718_modelfit}
\end{figure*}

\begin{figure*}
	\centering
    \begin{subfigure}
        \centering
        \includegraphics[width=\columnwidth]{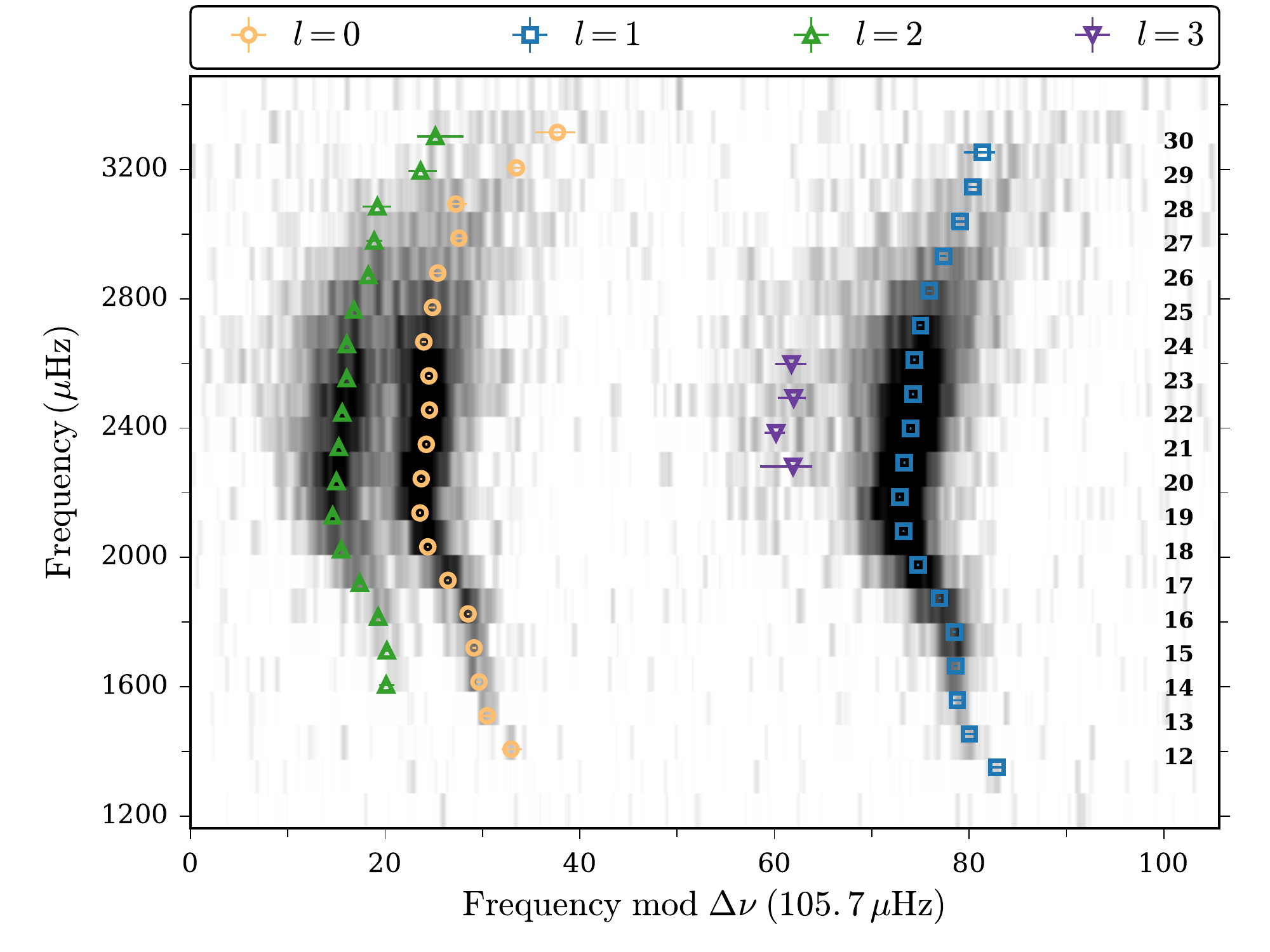}
    \end{subfigure}\hfill
    \begin{subfigure}
        \centering
        \includegraphics[width=\columnwidth]{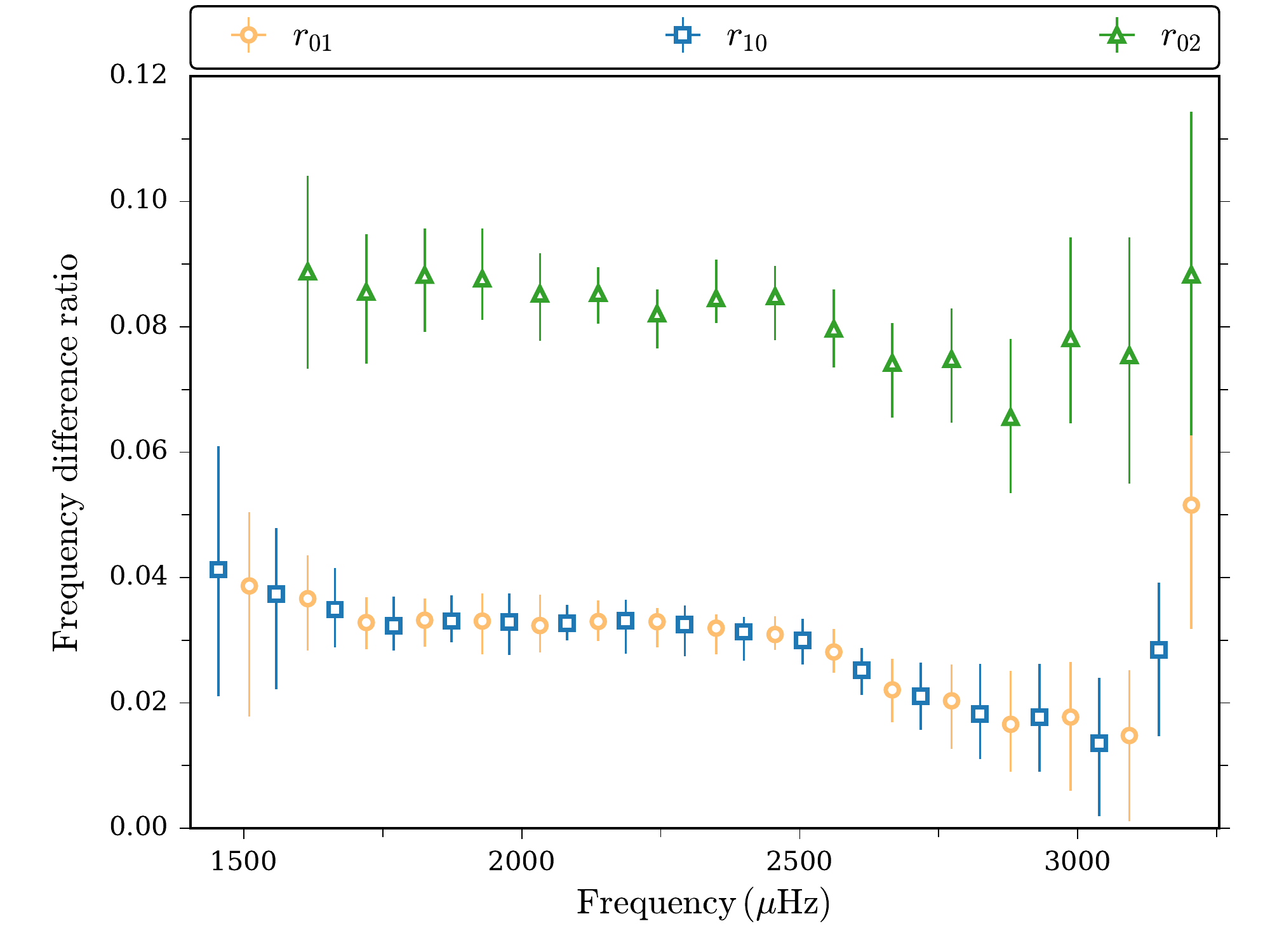}
    \end{subfigure}
    \caption{Left: example of \'{e}chelle diagram for KIC 6225718 (Saxo2). The complete figure set (66 figures) is available in the online journal. The power spectrum used has been background corrected and smoothed by a $\rm 1\, \mu Hz$ Epanechnikov filter. The color scale indicates the power to background level, going from white at low level to black at high levels. The number on the right hand of the plot gives the radial order $n$ of the $l=0$ modes. Right: frequency difference ratios for KIC 6225718 as a function of the central frequency of the respective ratios (see \eqref{eq:rat}).}
\label{fig:examp_out}
\end{figure*}

\begin{figure}
\centering
\includegraphics[width=\columnwidth]{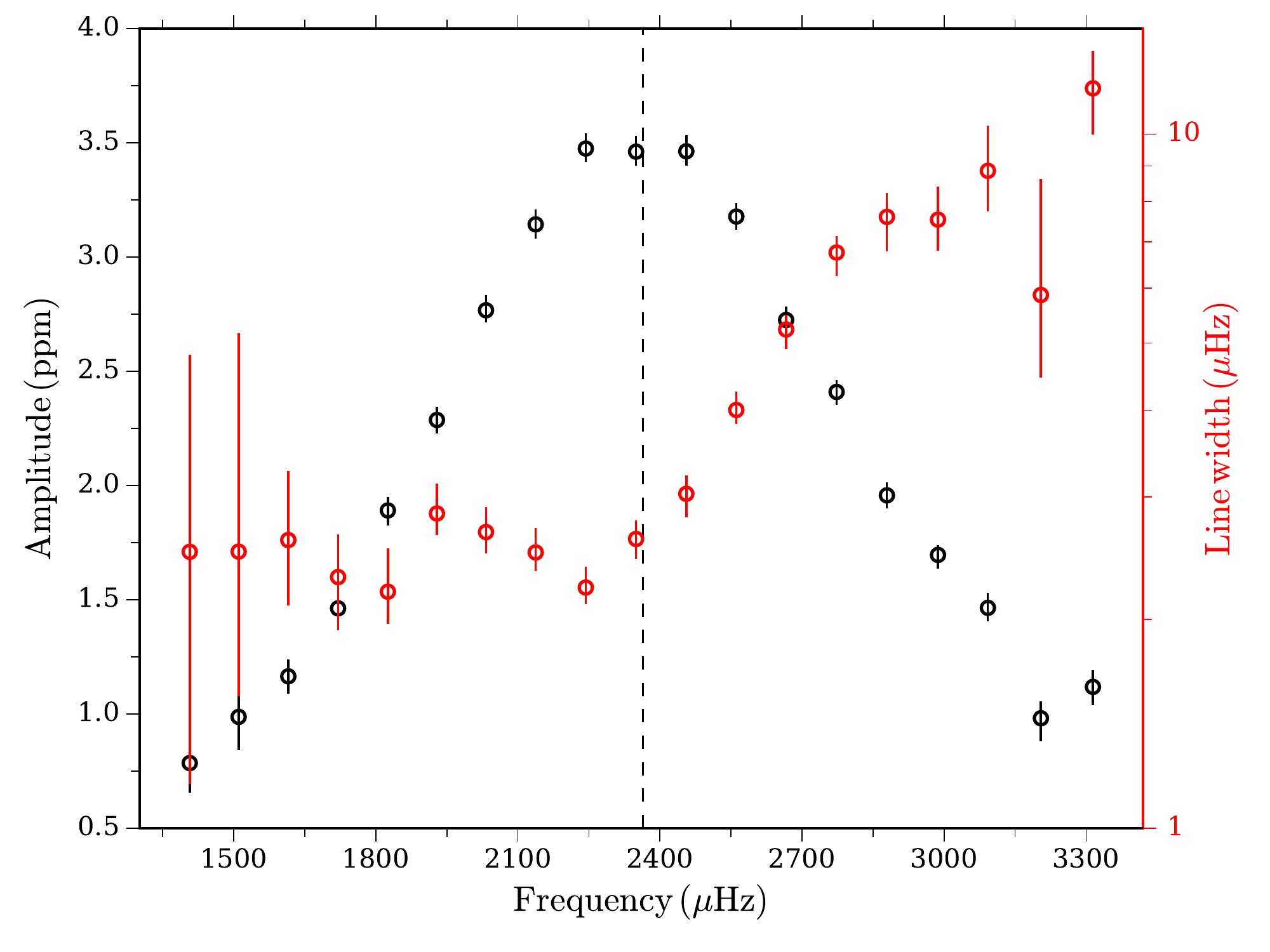}
\caption{Example of the extracted mode amplitudes and line widths for KIC 6225718 (Saxo2). The complete figure set (66 figures) is available in the online journal. Plotted in black are the amplitudes (left ordinate) and in red the line widths (right ordinate). The vertical dashed line gives the value for \numax.}
\label{fig:6225718_amp_lw}
\end{figure}

For each star a number of plots are also prepared. These include (1) a visualization of the obtained `best fit' from the peak-bagging as in \fref{fig:6225718_modelfit}, given by a plot of the power spectrum overlain with the best fit model and with an indication of the extracted modes; (2) an \'{e}chelle diagram overlain with the extracted frequencies as in the left panel of \fref{fig:examp_out}; (3) a plot of the derived frequency difference ratios $r_{01,10,02}$ as shown in the right panel of \fref{fig:examp_out}; (4) a plot of the derived second differences $\Delta_2 \nu(n,l)$ as shown in \fref{fig:secdiff} for KIC 6225718 (Saxo2), but without model fits; (5) a plot of the extracted mode amplitudes and line widths as shown in \fref{fig:6225718_amp_lw}.
%(6) a plot of the correlations between the different $r_{01,10,02}$ and $\Delta_2 \nu(n,l)$ values as shown in \fref{fig:correlation}; 

%\input{Figure_set_Fig25}
%%%%%%%%%%%%%%%%%%%%%%%%%%%%%%%%%%%%%%%%%%%%%%%%%%%%%%%%%%%%%%%
%%%%%%%%%%%%%%%%%%%%%%%%%%%%%%%%%%%%%%%%%%%%%%%%%%%%%%%%%%%%%%
%\input{Figure_set_Fig26}
%%%%%%%%%%%%%%%%%%%%%%%%%%%%%%%%%%%%%%%%%%%%%%%%%%%%%%%%%%%%%%
%%%%%%%%%%%%%%%%%%%%%%%%%%%%%%%%%%%%%%%%%%%%%%%%%%%%%%%%%%%%%%
%\input{Figure_set_Fig27}
%%%%%%%%%%%%%%%%%%%%%%%%%%%%%%%%%%%%%%%%%%%%%%%%%%%%%%%%%%%%%%%%
%%%%%%%%%%%%%%%%%%%%%%%%%%%%%%%%%%%%%%%%%%%%%%%%%%%%%%%%%%%%%%%

%%%%%%%%%%%%%%%%%%%%%%%%%%%%%%%%%%%%%%%%%%%%%%%%%%%%%%%%%%%%%%%%%%%%%%%%%%%%%%%%%%%%%%%%%%%%%%%%%%%%%%%%%%%%%%%%%%%%%%%%%%
%%%%%%%%%%%%%%%%%%%%%%%%%%%%%%%%%%%%%%%%%%%%%%%%%%%%%%%%%%%%%%%%%%%%%%%%%%%%%%%%%%%%%%%%%%%%%%%%%%%%%%%%%%%%%%%%%%%%%%%%%%
%%%%%%%%%%%%%%%%%%%%%%%%%%%%%%%%%%%%%%%%%%%%%%%%%%%%%%%%%%%%%%%%%%%%%%%%%%%%%%%%%%%%%%%%%%%%%%%%%%%%%%%%%%%%%%%%%%%%%%%%%%
%%%%%%%%%%%%%%%%%%%%%%%%%%%%%%%%%%%%%%%%%%%%%%%%%%%%%%%%%%%%%%%%%%%%%%%%%%%%%%%%%%%%%%%%%%%%%%%%%%%%%%%%%%%%%%%%%%%%%%%%%%
\section{Conclusions}
\label{sec:con}

In this paper we have presented the mode parameters for a sample of 66 MS solar-like oscillators, the frequencies of which are modeled in Paper~II.
In addition to the individual mode frequencies we have constructed frequency difference ratios and their correlations for the use in modeling efforts. We also report for each star the values for the mode line widths, amplitudes, and visibilities. For each of these quantities we have derived summary parameters and descriptions, such as the average seismic parameters \dnu and \numax and the behavior of mode line widths against frequency.

The reported parameters were derived through peak-bagging of the power spectra using an MCMC optimization scheme. This resulted in posterior probability distributions for each of the included parameters, from which credible intervals and correlations were directly obtained.
We found that the derived frequency uncertainties adhere to the expectations from theory in terms of S/N, observing time, and line widths, and are only higher by a factor of ${\sim}1.2$ compared to estimates from MLE. This observation will be useful for predicting seismic modeling yields for future missions such as TESS and PLATO. 
As a quality control on the detection of the reported modes, we performed a Bayesian hypothesis testing that for each mode gave the probability of detection \citep[][]{2016MNRAS.456.2183D,2012A&A...543A..54A}.
Our main conclusions are as follows:

\begin{itemize}
\item[$\circ$] The derived values for \dnu and \numax agree with empirically derived relations from the \kp mission. We derived parameters from an extended version of the standard asymptotic frequency relations, including mode ridge curvatures and variations of \dnu and small frequency separations. From the small frequency separations we further found that most stars deviate from the asymptotic description by an amount that correlates with the evolutionary state of the star, \ie, the central hydrogen content. 
\item[$\circ$] The measured amplitudes at \numax for our sample largely follow the expected trend from empirical relations from \kp. We also identified a systematic offset of approximately $-6\%$ between the maximum amplitudes obtained from the modes and those obtained from the smoothing method by \citet[][]{2008ApJ...682.1370K}. This corroborates the findings by \citet[][]{2011MNRAS.415.3539V}. This systematic offset should be corrected for whenever the two methods are compared. 
\item[$\circ$] For the line widths we adopted the frequency dependence of \citet[][]{2014A&A...566A..20A} given by an overall power law dependence and a Lorentzian dip near \numax. We fitted this relation for all stars and were able to derive simple relations between the parameters of the fit and \teff and \numax. These were found to confirm the results by \citet[][]{2014A&A...566A..20A}. Such relations will be useful for future simulations of solar-like oscillators and may be compared to theory. We also obtained a fit for the \teff dependence of the line width at \numax, complementing our values with line widths from 42 giants in NGC 6819 (Handberg et al., submitted). The obtained dependence largely agreed with that found by \citet[][]{2012A&A...537A.134A}, except for stars with low \teff. 
\item[$\circ$] Concerning the estimated mode visibilities, we found that those for $l=1$ were slightly lower than expected from the theoretical calculations by \citet[][]{2011A&A...531A.124B}, whereas those for $l=2$ and $3$ were larger than expected, especially for the $l=3$ modes. We found no overall dependence on \teff, which is also evident from the fact that \citet[][]{2012A&A...537A..30M} found a mean value of the total visibility at nearly the same level as here for \kp giants, which have \teff values lower by about $1000-2000$ K. Some structure was observed in the visibilities against \teff and mass, but it was not possible to say directly if this is simply due to scatter from the measurements or if they have some underlying physical explanation. Applying the method of \citet[][]{2014ApJ...782....2L} enabled us to identify power from $l=3$ modes in most stars, and for some high S/N targets even for $l=4$.

\end{itemize}

%%%%%%%%%%%%%%%%%%%%%%%%%%%%%%%%%%%%%%%%%%%%%%%%%%%%%%%%%%%%%%%%%%%%%%%%%%%%%%%%%%%%%%%%%%%%%%%%%%%%%%%%%%%%%%%%%%%%%%%%%%
%%%%%%%%%%%%%%%%%%%%%%%%%%%%%%%%%%%%%%%%%%%%%%%%%%%%%%%%%%%%%%%%%%%%%%%%%%%%%%%%%%%%%%%%%%%%%%%%%%%%%%%%%%%%%%%%%%%%%%%%%%
%%%%%%%%%%%%%%%%%%%%%%%%%%%%%%%%%%%%%%%%%%%%%%%%%%%%%%%%%%%%%%%%%%%%%%%%%%%%%%%%%%%%%%%%%%%%%%%%%%%%%%%%%%%%%%%%%%%%%%%%%%
%%%%%%%%%%%%%%%%%%%%%%%%%%%%%%%%%%%%%%%%%%%%%%%%%%%%%%%%%%%%%%%%%%%%%%%%%%%%%%%%%%%%%%%%%%%%%%%%%%%%%%%%%%%%%%%%%%%%%%%%%%
\section*{Acknowledgments}
\footnotesize

Funding for this Discovery mission is provided by NASA's Science Mission Directorate.
The authors acknowledge the dedicated team behind the \kp and K2 missions, without whom this work would not have been possible.
We thank Thierry Appourchaux for useful discussions on the properties of the line width fits, Ian Roxburgh for discussions on mode frequencies, and Enrico Corsaro for supplying line widths for a set of 19 red giants. We thank Jens Jessen-Hansen for inspiration to the paper title.
Funding for the Stellar Astrophysics Centre (SAC) is provided by The Danish National Research Foundation (Grant DNRF106). The research was supported by the ASTERISK project (ASTERoseismic Investigations with SONG and \kp) funded by the European Research Council (Grant agreement no.: 267864).
MNL acknowledges the support of The Danish Council for Independent Research | Natural Science (Grant DFF-4181-00415).
VSA and TRW acknowledges support from VILLUM FONDEN (research grant 10118).
GRD and WJC acknowledge the support of the UK Science and Technology Facilities Council (STFC).
SB is partially supported by NSF grant AST-1514676 and NASA grant NNX16A109G. 
DH acknowledges support by the Australian Research Council's Discovery Projects funding scheme (project number DE140101364) and support by the National Aeronautics and Space Administration under Grant NNX14AB92G issued through the \kp Participating Scientist Program.
WHB acknowledges research funding by Deutsche Forschungsgemeinschaft (DFG) under grant SFB 963/1 ``Astrophysical flow instabilities and turbulence'' (Project A18).

%%%%%%%%%%%%%%%%%%%%%%%%%%%%%%%%%%%%%%%%%%%%%%%%%%%%%%%%%%%%%%%%%%%%%%%%%%%%%%%%%%%%%%%%%%%%%%%%%%%%%%%%%%%%%%%%%%%%%%%%%%
%%%%%%%%%%%%%%%%%%%%%%%%%%%%%%%%%%%%%%%%%%%%%%%%%%%%%%%%%%%%%%%%%%%%%%%%%%%%%%%%%%%%%%%%%%%%%%%%%%%%%%%%%%%%%%%%%%%%%%%%%%
%%%%%%%%%%%%%%%%%%%%%%%%%%%%%%%%%%%%%%%%%%%%%%%%%%%%%%%%%%%%%%%%%%%%%%%%%%%%%%%%%%%%%%%%%%%%%%%%%%%%%%%%%%%%%%%%%%%%%%%%%%
%%%%%%%%%%%%%%%%%%%%%%%%%%%%%%%%%%%%%%%%%%%%%%%%%%%%%%%%%%%%%%%%%%%%%%%%%%%%%%%%%%%%%%%%%%%%%%%%%%%%%%%%%%%%%%%%%%%%%%%%%%
\vspace{1cm}
\small
\bibliography{MasterBIB}
\label{lastpage}

\end{document}